\documentclass[11pt]{report}
\usepackage{bbold}
\usepackage{geometry}
\usepackage{amsmath}
\DeclareMathOperator{\sech}{sech}

\DeclareMathOperator{\arcsinh}{arcsinh}

\usepackage{amsfonts}
\usepackage{amssymb}
\usepackage{graphicx}
\usepackage{caption}
\usepackage{subcaption}
\usepackage{hyperref}
\usepackage{float}
\usepackage{booktabs}
\usepackage{array}
\usepackage{braket}
\usepackage{multicol}
\usepackage{multirow}
\usepackage{mathrsfs}
\usepackage{titlepic}
\usepackage[nottoc,notlot,notlof]{tocbibind}
\usepackage{titlesec}
\usepackage{cite}
\usepackage{colortbl}
\usepackage[dvipsnames]{xcolor}
\usepackage{tcolorbox}
\usepackage{setspace}
\usepackage{fancyhdr}

\input{tcilatex}

\fancypagestyle{preliminary}{
	\fancyhf{}% Clear header/footer
	\fancyfoot[C]{\thepage}% Footer
}

\fancypagestyle{mainmatter}{
	\fancyhf{}% Clear header/footer
	\fancyfoot[C]{\thepage}% Footer
}

\begin{document}
	
\begin{titlepage}

    \begin{center}
	\vspace*{2mm}
	
	%Full title of the thesis, as approved by the Board of Studies, which should describe the contents accurately and concisely
	\begin{spacing}{1.2}
		\Huge
		\textbf{Time-dependence in non-Hermitian quantum systems}
	\end{spacing}
	
	%Full name of the author
	\vspace{1cm}
	\Large
	\textbf{Thomas David Frith}
	
	\vspace{1.5cm}
	
	%Qualification for which the thesis is submitted
	\begin{spacing}{1}
		Doctor of Philosophy\\
	\end{spacing}
	
	\vspace{2cm} 
	
	%City University logo
	\includegraphics[width=0.35\textwidth]{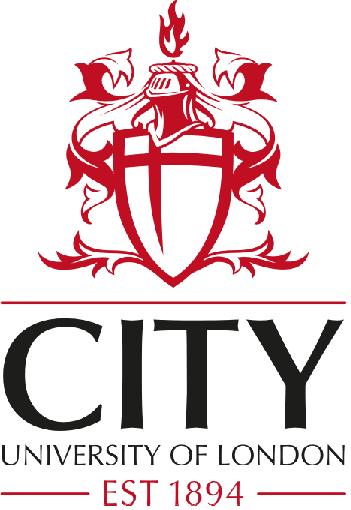}
	
	\vspace{1cm}
	
	%Name of the institution to which the thesis is submitted
	\textbf{City, University of London}\\
	%Department or organisation in which the research was conducted
	\textbf{Department of Mathematics}\\
	
	\vfill
	%Month and year of submission
	August 2019
	
	\end{center}	

\end{titlepage}
	
\pagestyle{preliminary}
\pagenumbering{roman}

\newpage

\begin{figure}
	\centering
\includegraphics[scale=0.5]{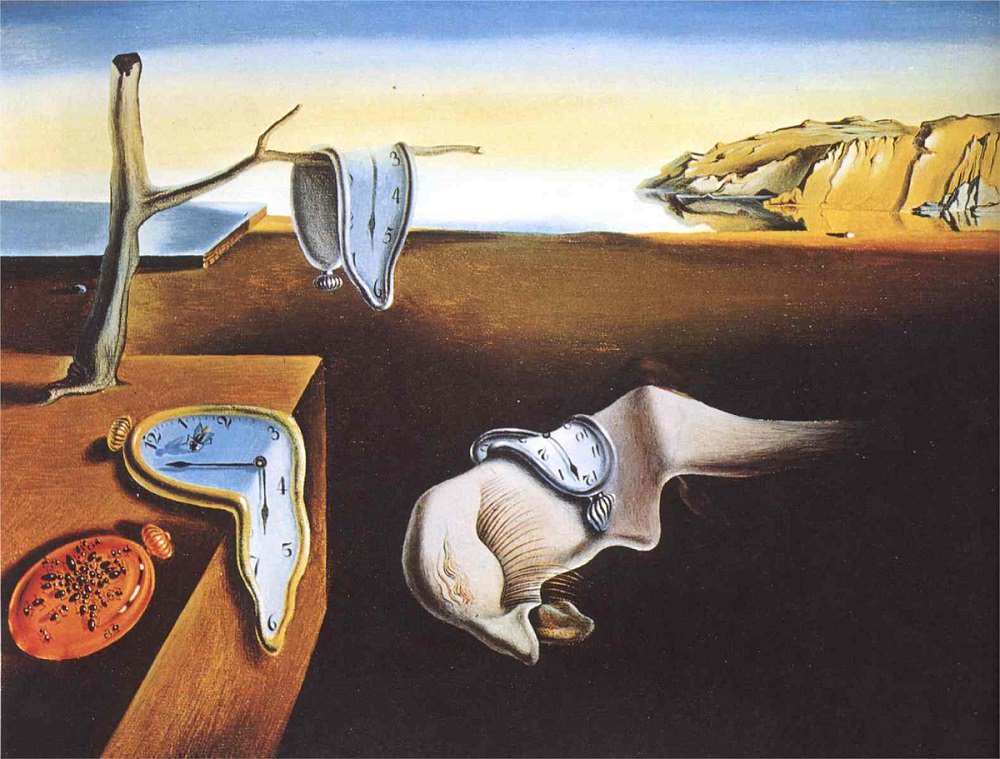}
\caption*{The Persistence of Memory by Salvador Dali.}
\end{figure}

\addcontentsline{toc}{chapter}{Contents}
\tableofcontents

\cleardoublepage
\phantomsection
\addcontentsline{toc}{chapter}{\listfigurename}
\listoffigures
 
\cleardoublepage
\phantomsection
\addcontentsline{toc}{chapter}{\listtablename}
\listoftables

\chapter*{Acknowledgements}
\addcontentsline{toc}{chapter}{Acknowledgements}

\hrulefill
\vspace{0.5cm}

Thank you to City, University of London for funding me throughout my doctoral degree with a Research Fellowship.

I want to also thank my supervisor Professor Andreas Fring for providing me with all the advice and support I could ask for. He has been the prime example of what it means to be an academic and I have thoroughly enjoyed working with him and learning from him.

Furthermore, I wish to sincerely thank my friends and family for their help over the course of my degree. Many have listened patiently as I have talked through ideas presented in this thesis.

\vspace{0.5cm}
\noindent\hrulefill

\chapter*{Declaration}
\addcontentsline{toc}{chapter}{Declaration}

\hrulefill
\vspace{0.5cm}

I declare that this thesis has been composed solely by myself and that it has not been submitted, in whole or in part, in any previous application for a degree. Except where stated otherwise by reference or acknowledgment, the work presented is entirely my own.

\vspace{0.5cm}
\noindent\hrulefill

\chapter*{Abstract}
\addcontentsline{toc}{chapter}{Abstract}

\hrulefill
\vspace{0.5cm}

In this thesis we present a coherent and consistent framework for explicit time-dependence in non-Hermitian quantum mechanics. The area of non-Hermitian quantum mechanics has been growing rapidly over the past twenty years \cite{jorge2015top10}. This has been driven by the fact that $\mathcal{PT}$-symmetric non-Hermitian systems exhibit real energy eigenvalues and unitary time evolution \cite{bender1998real,bender2007making,mostafazadeh2010pseudo}.

Historically, the introduction of time into the world of non-Hermitian quantum mechanics has been a conceptually difficult problem to address \cite{mostafazadeh2007time,znojil2008time}, as it requires the Hamiltonian to become unobservable. However, we solve this issue with the introduction of a new observable energy operator \cite{AndTom3}. We explain why its instigation is a necessary and natural progression in this setting.

For the first time, the introduction of time has allowed us to make sense of the parameter regime in which the $\mathcal{PT}$-symmetry is spontaneously broken. Ordinarily, in the time-independent setting, the energy eigenvalues become complex and the wave functions are asymptotically unbounded. However, we demonstrate that in the time-dependent setting this broken symmetry can be mended and analysis on the spontaneously broken $\mathcal{PT}$ regime is indeed possible. We provide many examples of this mending on a wide range of different systems, beginning with a $2\times2$ matrix model \cite{AndTom1} and extending to higher dimensional matrix models \cite{AndTom2} and coupled harmonic oscillator systems with infinite Hilbert space \cite{AndTom4,fring2018tdm}. Furthermore, we use the framework to perform analysis on time-dependent quasi-exactly solvable models \cite{AndTom5}. 

The ability to make sense of the spontaneously broken $\mathcal{PT}$ regime has revealed a vast array of new and exotic effects. We present the "eternal life" of entropy \cite{fring2019eternal} in this thesis. Ordinarily, for entangled quantum systems coupled to the environments, the entropy decays rapidly to zero. However, in the spontaneously broken regime, we find the entropy decays asymptotically to a non-zero value.

Finally, we create an elegant framework for Darboux and Darboux/Crum transformations for time-dependent non-Hermitian Hamiltonians \cite{cen2019time}. This combines the area of non-Hermitian quantum mechanics with non linear differential equations and solitons.

\vspace{0.5cm}
\noindent\hrulefill

\chapter{Introduction}
\pagestyle{mainmatter}
\pagenumbering{arabic}

Quantum mechanics is the science of matter at microscopic scales. It describes how atoms and subatomic particles behave and interact with extraordinary elegance and beauty. At such small scales, classical theories break down and fail to predict many of the wonderful phenomena observed. This became apparent in the 19th and early 20th century with the black-body radiation problem and the photoelectric effect. The simple yet revolutionary resolution was to hypothesise that energy is radiated and absorbed in discrete packets or "quanta". With this new idea, the framework of quantum mechanics was laid down in the early 20th century and the experimental observations were matched with theory.

At the heart of the quantum mechanical framework is the description of particles and quantum systems in terms of a wave function $\ket{\Psi}$. This mathematical object contains all the information about the evolution of a system and is calculated using the time-dependent Schr\"odinger equation with an initial starting state

\begin{equation}
i\hbar\partial_t{\ket{\Psi\left(t\right)}}=H\ket{\Psi\left(t\right)},
\end{equation}
where $\ket{\Psi\left(t\right)}$ is the wave function and $\hbar\equiv1.05457...\times10^{-34}Js$ is the reduced Planck constant. In this thesis we use natural units by setting $\hbar=c=1$, where $c$ is the speed of light in a vacuum. $H$ is the Hamiltonian of the system. If the Hamiltonian is absent of any explicit time-dependence then the time-dependent Schr\"odinger equation reduces to the time-independent Schr\"odinger equation

\begin{equation}
H\ket{\psi}=E\ket{\psi},
\end{equation}
which is an eigenvalue equation with $E$ as the energy of the system. In this case one can form a solution to the time-dependent Schr\"odinger equation using the time-independent eigenfunctions of the Hamiltonian $\ket{\Psi\left(t\right)}=e^{-iEt}\ket{\psi}$. Once a solution for the wave function has been obtained, one can proceed with calculating observables of the particular system with quantum mechanical operators 

\begin{equation}\label{Observables}
O\left(t\right)=\left\langle\Psi\left(t\right)|\mathcal{O}\left(t\right)\Psi\left(t\right)\right\rangle
\end{equation}
Energy, position, momentum and spin are all examples of such observables and so obtaining solutions for the wave functions is vital for calculating such quantities.

In all standard quantum mechanics textbooks, the authors will insist on the Hamiltonian and any observable being a self-adjoint operator (or more widely referred to as Hermitian). This ensures that the observables can act from both sides in (\ref{Observables}) equivalently. Furthermore, the Hermiticity of the Hamiltonian ensures that the energy observables are real and that the time evolution is unitary, both of which are needed in order to proceed with a viable quantum mechanical theory set on a well-defined Hilbert space.

\begin{figure}[H]
\centering
\includegraphics[scale=0.8]{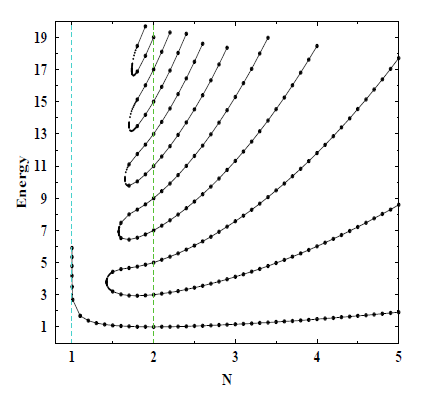}
\caption{Energy eigenvalues for the Hamiltonian $H\protect=p\protect^2-\left(ix\right)\protect^{N}$ for varying $N$, taken from \cite{bender1998real}.}\label{BBFigure}
\end{figure}

However, since 1998 \cite{bender1998real} it has been known that the condition of Hermiticity is not required for real energy eigenvalues and unitary time evolution. Mathematically, this had been realised before 1998 but it was \cite{bender1998real} that drew together, interpreted and presented these results. In fact, it is possible for the Hamiltonian to be non-Hermitian and still possess these important qualities if there exists an anti-linear symmetry which leaves the Hamiltonian invariant. The most common of these anti-linear symmetries is parity-time reversal ($\mathcal{PT}$) symmetry which can take many forms. For example, for a one dimensional Hamiltonian depending on momentum $p$ and position $x$, one particular symmetry takes the form
\begin{equation}\label{PTSymmetry}
PT: \quad p\rightarrow p \quad x\rightarrow-x \quad i\rightarrow-i.
\end{equation}
In the position representation the momentum operator is $p=-i\partial_x$. It is clear that one can form many Hamiltonians under this symmetry. The simplest example is the system

\begin{equation}\label{BBHamiltonian}
H=p^2-\left(ix\right)^{N}, \quad N>0
\end{equation}
with $N\in\mathbb{R}$. The Hamiltonian (\ref{BBHamiltonian}) is invariant under the $\mathcal{PT}$-symmetry (\ref{PTSymmetry}). Therefore we may expect the energy spectrum to be real. In parts, this is indeed the case as was shown in \cite{bender1998real} in figure \ref{BBFigure}.

As is clear, the energies are real for all values of $N>2$. In fact for $N=2$ the systems reduces to the harmonic oscillator for which the spectrum is $E=2n+1$. For $N<2$ the energy values begin to coalesce at what is commonly referred to as an exceptional point. The explanation for this is that the $\mathcal{PT}$-symmetry is spontaneously broken and the energies appear in complex conjugate pairs. The Hamiltonian remains $\mathcal{PT}$-symmetric, however the eigenstates cease to be an eigenstates of the $\mathcal{PT}$-symmetry operator \cite{EW} with a phase for an eigenvalue. This demonstrates the two part requirement for unbroken $\mathcal{PT}$-symmetry, one must have an antilinear operator that commutes with the Hamiltonian and shares given eigenstates with the Hamiltonian (with the eigenvalue being a phase). If both these requirements are met, then the energy eigenvalues will be real and the time-evolution will be unitary. 

\begin{equation}
\left[ \mathcal{PT},H\right] =0,\qquad \text{and\qquad }\mathcal{PT}\varphi
\left(t\right)=e^{i\phi }\varphi\left(t\right).
\end{equation}

\begin{figure}[H]
\centering

\includegraphics[scale=0.4]{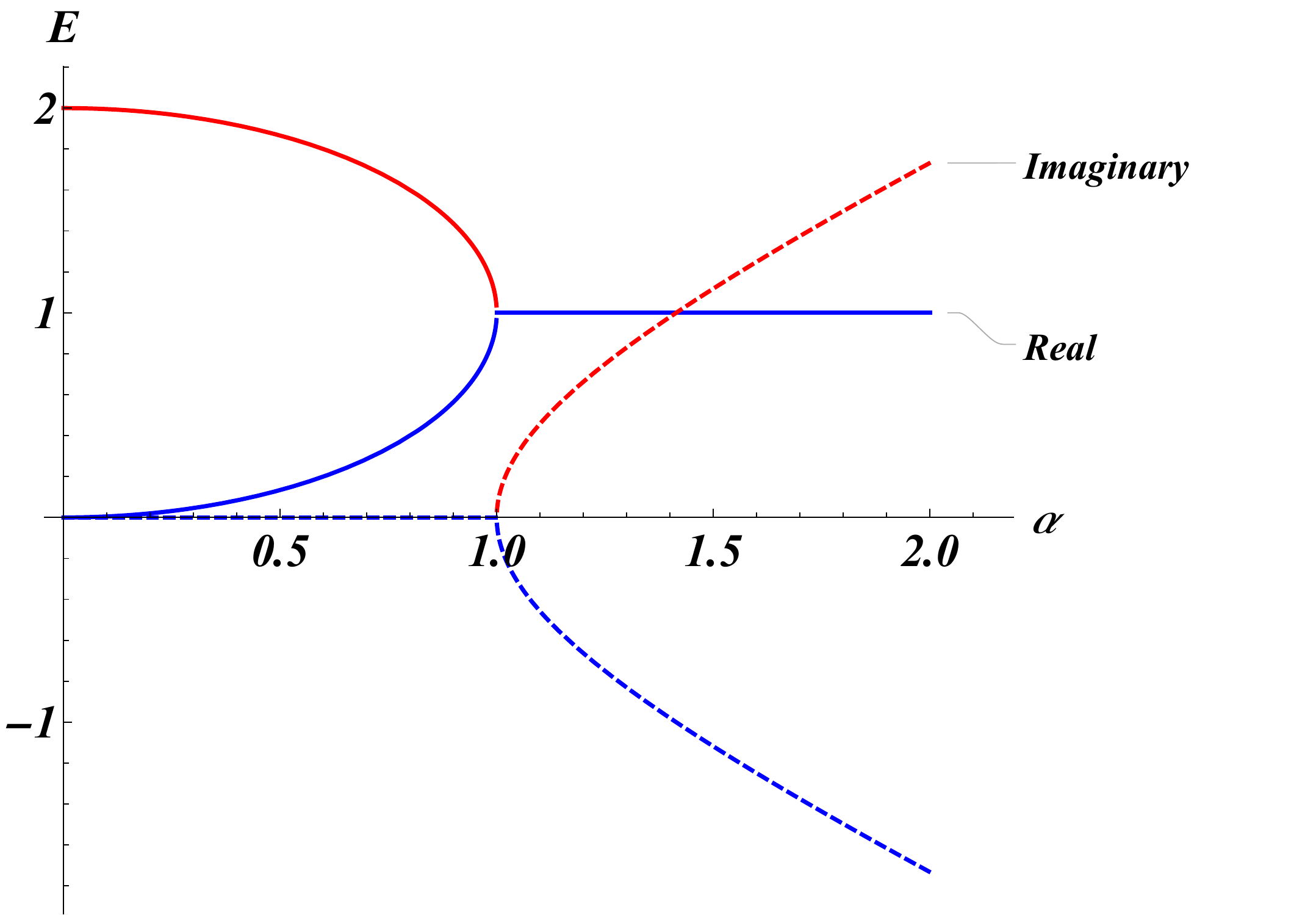}

\caption{The effect of spontaneous $\protect\mathcal{PT}$-symmetry breaking on the energies of a non-Hermitian quantum system $H=\protect\sigma\protect_z+i\protect\alpha\protect\sigma\protect_x$. The solid lines show the real part of the energy and the dashed lines show the imaginary part. The exceptional point at $\protect\alpha=1$ is the transition between real and complex energy.}\label{complexeigen}

\end{figure}

This feature of spontaneously broken $\mathcal{PT}$ symmetry breaking is one of the most interesting areas of non-Hermitian quantum mechanics. Figure \ref{complexeigen} is a ubiquitous example in this area of spontaneous symmetry breaking. It shows a pair of energy eigenvalues for the non-Hermitian Hamiltonian $H=\sigma_z+i\alpha\sigma_x$. The system exhibits an exceptional point at $\alpha=1$ beyond which the energy appears in complex conjugate pairs.

Non-Hermitian quantum mechanics became popular in 1998 with the paper by Bender and Boettcher \cite{bender1998real} in which they studied the Hamiltonian (\ref{BBHamiltonian}). The work was then backed up with more mathematical rigour by \cite{dorey2001spectral}. Before these works there had been acknowledgements of the utility of non-Hermitian Hamiltonians and even the basis for performing time-independent analysis had been formed \cite{scholtz1992quasi}. These initial works were quickly followed by a large body of work as researchers attempted to understand the framework and mathematical underpinning of such systems \cite{mostafazadeh2002pseudo,mostafazadeh2002pseudo2,mostafazadeh2002pseudo3,mostafazadeh2003pseudo,znojil1999non,bender1999pt,bender2002complex,bender2003calculation,bender2003must,mostafazadeh2004physical,mostafazadeh2006metric,swanson2004transition,geyer2004quasi,bender2007making,moiseyev2011non,bender2012special,znojil2015special}. Furthermore, during this time, the idea of non-Hermitian Hamiltonians sparked great interest for applications to other areas such as supersymmetry \cite{znojil2000supersymmetry,znojil2002non,znojil2004relativistic,mostafazadeh2002pseudosuper}, quantum field theory \cite{bender2004extension} (this has seen much more development in recent years \cite{mannheim2019goldstone,alexandre2019gauge,alexandre2018spontaneous}) and most drastically, classical optics \cite{alexeeva2012optical,el2007theory,hayward2018complex,makris2010pt,makris2008beam,makris2011mathcal,musslimani2008optical,ramezani2010unidirectional}. This particular area has grown rapidly in the past 20 years and is seen by many as one of the most exciting areas in experimental physics \cite{jorge2015top10}. As this area has grown, there has developed a strong link to optics through the paraxial approximation which draws a comparison between the Schr\"odinger equation and the Helmholtz equation under certain restrictions. In this setting the refractive index is the equivalent to the quantum mechanical potential and is naturally taken to be non-Hermitian. The complex parameters then represent gain and loss in the system. This has been realised experimentally in classical optics \cite{guo2009observation,ruter2010obs,zeuner2015observation,regensburger2013observation} with demonstrations of spontaneous $\mathcal{PT}$-symmetry breaking and the predictions of far more exotic effects \cite{goldzak2018light}. Furthermore, there has been recent work applying the framework of non-Hermitian quantum mechanics to entropy \cite{fring2019eternal,dey2019controlling} and the Berry phase \cite{liang2013topological,lieu2018topological,dangel2018topological,hayward2018complex} with some interesting and potentially far reaching results. 

The early development of non-Hermitian quantum mechanics was primarily concerned with the study of time-independent systems but it did not take long for the question of time-dependent non-Hermitian quantum mechanics to be raised and tackled \cite{fringfaria,faria2007non,mostafazadeh2007time,fring2016non,fring2016unitary,znojil2006construction,znojil2007time,znojil2008time,gong2013time,khantoul2017invariant,maamache2015periodic,maamache2017pseudo}. The regime of time-dependence still posed an interpretational difficulty that was the cause of much disagreement and dispute within the community. The main origin of these disputes was the realisation that in the time-dependent regime, the Hamiltonian ceases to be an observable for the energy. However, with the publication of several papers \cite{fring2016unitary,AndTom2,AndTom3,AndTom4} on the subject this dispute has been resolved and the community is now in general agreement. Having the Hamiltonian as observable is in fact not a necessary condition in quantum mechanics. Instead one finds a new observable energy operator that reverts to the Hamiltonian in the time-independent setting.

The starting point for a proper treatment of time-independent non-Hermitian quantum systems is the time-independent Schr\"odinger equation. We begin with one equation for the non-Hermitian Hamiltonian for which we will use the symbol $H$, and one for the Hermitian Hamiltonian which we will denote with the symbol $h$.

\begin{equation}\label{TISE2}
h\ket{\phi}=E\ket{\phi},\quad H\ket{\psi}=E\ket{\psi}.
\end{equation}
These systems share the same real energy eigenvalues if we can relate the eigenstates via a time-independent Dyson map $\eta$ \cite{dyson1956},

\begin{equation}
\ket{\phi}=\eta\ket{\psi}.
\end{equation}
Under this mapping, the Hamiltonians are related by a similarity transformations (often referred to as the time-independent Dyson equation)

\begin{equation}\label{TIDE}
h=\eta H \eta^{-1}.
\end{equation}
Furthermore, if we now take the Hermitian conjugate of both sides, we obtain the time-independent quasi-Hermiticity equation \cite{dieudonne}

\begin{equation}
H^\dagger=\rho H \rho^{-1},
\end{equation}
where $\rho=\eta^\dagger \eta$ is the time-independent metric. There have been many papers written on the subject of solving the quasi-Hermiticity equation \cite{jones2006equivalent,jones2007scattering,jones2008interface,musumbu2006choice,assis2008metrics,assis2008non,bagchi2009minimal,scholtz2006operator,mostafazadeh2008metric,mostafazadeh2013pseudounbounded,siegl2012metric} to name a few. The term that has arisen to describe operators that obey a relation such as this is quasi-Hermitian (first coined by Dieudonn\'e in 1961 \cite{dieudonne}). This metric is the central object in the framework as it achieves normality of the inner product for the non-Hermitian Hamiltonian 

\begin{equation}
\braket{\phi|\phi}=\bra{\psi}\rho\psi\rangle=\braket{\psi|\psi}_{\rho}=1.
\end{equation}
For this to hold, $\rho$ must be positive definite. This allows us to construct a well defined Hilbert space for the Hamiltonian. Without the metric, the inner product of the eigenstates is in general indefinite. Furthermore, it allows one to calculate observables in the same way as in the Hermitian system

\begin{equation}
O=\left\langle\psi| \mathcal{O}\psi\right\rangle_{\rho}.
\end{equation}
Observables in the non-Hermitian system $\mathcal{O}$ are related to those in the Hermitian system $o$ via a similarity transform

\begin{equation}
o=\eta\mathcal{O}\eta^{-1},
\end{equation}
and therefore must also be quasi-Hermitian

\begin{equation}
\mathcal{O}^\dagger=\rho\mathcal{O}\rho^{-1}.
\end{equation}
In the absence of time, we can describe non-Hermitian Hamiltonians if we are able to calculate the metric $\rho$. However, this framework is incomplete as it does not allow for any time-dependence in the non-Hermitian Hamiltonian. Therefore we need to extend the analysis in complete generality to the time-dependent Schr\"odinger equation. Once again we begin with one for the non-Hermitian Hamiltonian and one for the Hermitian Hamiltonian.

\begin{equation}
h\left(t\right)\ket{\Phi\left(t\right)}=i\partial_t\ket{\Phi\left(t\right)}, \quad H\left(t\right)\ket{\Psi\left(t\right)}=i\partial_t\ket{\Psi\left(t\right)}.
\end{equation}
In analogy to the time-independent case, we now introduce a Dyson map between the wave functions. However, in this case the Dyson map is now a time-dependent operator.

\begin{equation}
\ket{\Phi\left(t\right)}=\eta\left(t\right)\ket{\Psi\left(t\right)}.
\end{equation}
Substituting this expression for the wave function into the time-dependent Schr\"odinger equation results in the time-dependent Dyson equation, which relates the two Hamiltonians

\begin{equation}\label{DysonE}
h\left(t\right)=\eta\left(t\right)H\left(t\right)\eta^{-1}\left(t\right)+i\partial_t\eta\left(t\right) \eta^{-1}\left(t\right).
\end{equation}
Taking the complex conjugate of both sides, we obtain the time-dependent quasi-Hermiticity equation,

\begin{equation}\label{QHE}
H^\dagger\left(t\right)\rho\left(t\right)-\rho\left(t\right) H\left(t\right)=i\partial_t\rho\left(t\right),
\end{equation}
where the metric is now also time-dependent $\rho\left(t\right)=\eta^\dagger\left(t\right)\eta\left(t\right)$. Even if $H$ is time-independent, these equations are already different from the time-independent treatment as we pick up time-derivative terms. We notice immediately that the Hamiltonian $H\left(t\right)$ is no longer quasi-Hermitian with the addition of the $i\partial_t\rho$ term. Observables in the time-dependent non-Hermitian regime are related to their Hermitian counterparts in the same way as in the time-independent regime

\begin{equation}\label{timeobservables}
o\left(t\right)=\eta\left(t\right)\mathcal{O}\left(t\right)\eta^{-1}\left(t\right),
\end{equation}
and are therefore also quasi-Hermitian

\begin{equation}
\mathcal{O}^\dagger\left(t\right)=\rho\left(t\right)\mathcal{O}\left(t\right)\rho^{-1}\left(t\right).
\end{equation}
In both the time-independent and time-dependent cases, the metric $\rho$ and the Dyson map $\eta$ are not uniquely defined, however they can be made unique by choosing two operators as observables \cite{scholtz1992quasi}. This requirement is the same as in the Hermitian case, but is more explicit. In the Hermitian case we choose energy and position to be observable. This may seem obvious, but it is still a choice.

As $H\left(t\right)$ is not quasi-Hermitian, it is not observable and we must define a new energy operator using (\ref{timeobservables}),

\begin{equation}
\tilde{H}(t)=\eta ^{-1}(t)h(t)\eta (t)=H(t)+i \eta ^{-1}(t)\partial_{t}\eta (t).
\end{equation}
This is already a departure from standard quantum mechanics and even time-independent non-Hermitian quantum mechanics. The Hamiltonian $H\left(t\right)$ loses its dual nature as the generator of time-evolution and the energy observable operator. $\tilde{H}\left(t\right)$ replaces $H\left(t\right)$ as the energy observable operator.

The ability to perform consistent analysis on non-Hermitian quantum systems is dependent on the ability to find a metric operator $\rho\left(t\right)$ that forms a well-defined inner product. Furthermore, one needs the Dyson map $\eta\left(t\right)$ in order to relate the system to its corresponding Hermitian system. Therefore, all problems in non-Hermitian quantum mechanics must start with calculating $\rho\left(t\right)$ and $\eta\left(t\right)$. In the time-independent case, the problem reverts to solving a similarity transform for either $\rho$ or $\eta$. This is a non-trivial problem and the process of finding such quantities is highly technical. Therefore, even in the time-independent regime finding any new solutions for $\rho$ and $\eta$ can be considered a worthy task as we have already shown with the number of publications on the subject \cite{jones2006equivalent,jones2007scattering,jones2008interface,musumbu2006choice,assis2008metrics,assis2008non,bagchi2009minimal,scholtz2006operator,mostafazadeh2008metric,mostafazadeh2013pseudounbounded,siegl2012metric}. The problem becomes even harder once the metric and the Dyson map are made time-dependent. The task extends to solving a differential equation in $t$. However, the form of equations (\ref{DysonE}) and (\ref{QHE}) hides the true complexity of the problem as often one obtains a series of coupled non-linear differential equations in $t$ in terms of the various parameters contained within $\rho\left(t\right)$ or $\eta\left(t\right)$. As expected there are far fewer known solutions to the time-dependent Dyson equation and the time-dependent quasi-Hermiticity equations. 

Solving the time-dependent equations has further reaching implications than simply the ability to analyse time-dependent non-Hermitian Hamiltonians. The other main important consequence of the time-dependent framework is the ability to analyse systems in which the $\mathcal{PT}$-symmetry is spontaneously broken. In the time-independent setting there does not exist a metric or a Dyson map when the $\mathcal{PT}$-symmetry is spontaneously broken. In this case the energy eigenvalues become complex (as shown in figure \ref{complexeigen}) and so clearly there is no corresponding Hermitian Hamiltonian as both Hamiltonians share the same energy eigenvalues (from equation (\ref{TISE2})). This is not the case for the time-dependent regime. In the time-dependent regime it is still possible to make sense of the spontaneously broken regime because the Hamiltonians are no longer related by a similarity transform. Furthermore, the eigenvalues of the non-Hermitian Hamiltonian no longer correspond to the energy observables. For this we have a new energy observable operator $\tilde{H}\left(t\right)$. Making sense of the spontaneously broken $\mathcal{PT}$ regime opens a whole new area of quantum mechanics previously discarded as unphysical. This regime shows itself to give new and exotic effects for many applicable areas of quantum mechanics.

In this thesis we will be exploring time-dependent non-Hermitian quantum systems in detail. As we have discussed, this is a vast area that is growing rapidly and this thesis represents a large contribution towards the understanding of the subject. We will begin with a comparison of the multiple approaches that are possible to employ in order to calculate the central objects $\rho\left(t\right)$ and $\eta\left(t\right)$. We will use these approaches frequently in the subsequent chapters and so it is important to understand the procedures involved. The calculation of $\rho\left(t\right)$ and $\eta\left(t\right)$ is vital in order to proceed with more advanced analysis (although the parameters themselves are of great interest), therefore we must be confident in evaluating them. Once we have an established framework for solving for these quantities, we will then analyse matrix models \cite{AndTom1,AndTom2,AndTom3} consisting of 2, 3 and 4 level systems and demonstrate how the spontaneously broken $\mathcal{PT}$ regime can be mended with the introduction of time. We will also show that our analysis extends to an inverted simple harmonic oscillator. We will then move onto coupled oscillator systems \cite{AndTom4} with spontaneously broken $\mathcal{PT}$-symmetry and demonstrate the applicability of the Lewis Riesenfeld invariants \cite{lewis1969exact}. Next we apply the time-dependent framework to three areas: quasi-exactly solvable systems \cite{AndTom5}, von Neumann entropy \cite{fring2019eternal} and Darboux-Crum transformations \cite{cen2019time}.

\chapter{Approaches}

In this chapter we will analyse in detail the various approaches used to solve non-Hermitian Hamiltonians. In order to make sense of these systems we need to calculate the metric operator $\rho\left(t\right)$ and the Dyson operator $\eta\left(t\right)$ related by $\rho\left(t\right)=\eta^\dagger\left(t\right)\eta\left(t\right)$. As outlined in the introduction, these operators are needed to calculate observables in the non-Hermitian setting and there are differing approaches to solve for them. The approaches we will consider in this chapter are the time-dependent quasi-Hermiticity equation

\begin{equation}\label{quasiH}
H^\dagger\left(t\right)\rho\left(t\right)-\rho\left(t\right) H\left(t\right)=i\partial_t\rho\left(t\right),
\end{equation} 
the time-dependent Dyson equation 

\begin{equation}\label{dyson}
h\left(t\right)=\eta\left(t\right)H\left(t\right)\eta^{-1}\left(t\right)+i\partial_t\eta\left(t\right) \eta^{-1}\left(t\right),
\end{equation}
and the Lewis-Riesenfeld invariants (for a detailed definition, see \nameref{LRAppendix})

\begin{equation}\label{LR}
\frac{dI_{\mathcal{H}}(t)}{dt}=\partial _{t}I_{\mathcal{H}}(t)-i \left[
I_{\mathcal{H}}(t),\mathcal{H}(t)\right] =0,\quad ~~~\ \ \text{for~\ }%
\mathcal{H}=h=h^{\dagger },H\neq H^{\dagger }. 
\end{equation}%
where the invariants $I_h$ and $I_H$ are related by a similarity transform with the Dyson operator

\begin{equation}\label{LRi}
I_{h}(t)=\eta (t)I_{H}(t)\eta ^{-1}(t).
\end{equation}
The three approaches have their own advantages and disadvantages that will become clear as we work through an example.\\

In order to compare the solution approaches we will study a non-Hermitian 2 level matrix model. Furthermore, we will consider two separate approaches to the above equations. The first will be a straight forward matrix technique in which we consider each component of the metric/Dyson map in matrix form. The second will be the algebraic technique using the Baker-Campbell-Hausdorff (BCH) relation. For this technique we construct the metric in terms of generators in the algebra of the Hamiltonian. Our metric/Dyson map is a series of exponentials such that the adjoint action on an algebraic element is

\begin{equation}\label{BCHeq}
e^ABe^{-A}=B+\left[A,B\right]+\frac{1}{2!}\left[A,\left[A,B\right]\right]+\frac{1}{3!}\left[A,\left[A,\left[A,B\right]\right]\right]+...
\end{equation}
In this way we can extend our findings in this chapter to generic algebras beyond matrix models. For now, we will study the following matrix model in order to understand the various approaches

\begin{equation}\label{Hamiltonian2Level}
H=-\frac{1}{2}\left[ \Omega \mathbb{I}+\lambda \sigma _{z}+i\kappa \sigma
_{x}\right] ,
\end{equation}
with $\sigma _{x}$, $\sigma _{y}$, $\sigma _{z}$ denoting the Pauli
matrices, $\mathbb{I}$ the identity matrix and $\omega $, $\lambda $, $%
\kappa \in \mathbb{R}$

\begin{equation}
\mathbb{I}=\left(\begin{array}{cc}
1 & 0\\
0 & 1
\end{array}\right), \quad \sigma_{x}=\left(\begin{array}{cc}
0 & 1\\
1 & 0
\end{array}\right), \quad \sigma_{y}=\left(\begin{array}{cc}
0 & -i\\
i & 0
\end{array}\right), \quad \sigma_{z}=\left(\begin{array}{cc}
1 & 0\\
0 & -1
\end{array}\right).
\end{equation}
The two eigenvalues and eigenvectors for this
Hamiltonian are simply

\begin{equation}
E_{\pm }=-\frac{1}{2}\Omega \pm \frac{1}{2}\sqrt{\lambda ^{2}-\kappa ^{2}}%
\quad \text{and}\quad \varphi _{\pm }=\left( 
\begin{array}{c}
i(-\lambda \pm \sqrt{\lambda ^{2}-\kappa ^{2}}) \\ 
\kappa%
\end{array}%
\right).
\end{equation}
The eigenvalues are real provided $|\lambda|>|\kappa|$. The symmetry properties of the Hamiltonian are analysed in detail in chapter 3 where we consider the regime in which the eigenvalues become complex $|\lambda|<|\kappa|$. 

Now we introduce time into this model by setting $\lambda\rightarrow\alpha\kappa\left(t\right)$ and $\kappa\rightarrow\kappa\left(t\right)$. This choice is made to simplify the Hamiltonian so the time-dependence is an overall factor. Solving for $\eta\left(t\right)$ and $\rho\left(t\right)$ is still non-trivial in this setting. The Hamiltonian (\ref{Hamiltonian2Level}) takes the form 

\begin{equation}\label{Hamiltonian2Leveltime}
H\left(t\right)=-\frac{1}{2}\left[ \Omega \mathbb{I}+\alpha\kappa\left(t\right) \sigma _{z}+i\kappa\left(t\right) \sigma
_{x}\right].
\end{equation}
We will solve for the metric and the Dyson operator using the three approaches. There are many other quantities that we could go on to calculate, but this chapter is dedicated to solving for the central quantities $\rho\left(t\right)$, $\eta\left(t\right)$ and $h\left(t\right)$. We will begin with the time-dependent quasi-Hermiticity equation, then move onto the time-dependent Dyson equation and finally the Lewis Riesenfeld invariants.

\section{Time-dependent quasi-Hermiticity equation}\label{qHsection}

On initial inspection, the time-dependent quasi-Hermiticity equation appears to be the most simple starting point of the three approaches as it only contains one unknown (the metric operator $\rho\left(t\right)$). This assumes that we always take a non-Hermitian Hamiltonian $H$ as our initial quantity. Therefore it seems to be the natural beginning for this chapter. However, as will become clear this is not always the case.

\subsection{Matrix technique}\label{qHsectionM}

In order to solve equation (\ref{quasiH}) for the Hamiltonian (\ref{Hamiltonian2Leveltime}) we make the ansatz 

\begin{equation}
\rho\left(t\right)=\rho_0\left(t\right)\mathbb{I}+\sum_{i=x,y,z}\rho_i\left(t\right)\sigma_i, 
\end{equation}
where $\rho_0\left(t\right)$, $\rho_i\left(t\right)\in \mathbb{R}$. As $\rho\left(t\right)$ must be Hermitian, this ansatz is the most general form it can take for any $2\times2$ Hermitian matrix. Substituting this and the Hamiltonian (\ref{Hamiltonian2Leveltime}) into (\ref{quasiH}) results in the following differential equations

\begin{eqnarray}
\dot{\rho}_0&=&\kappa\rho_x,\label{rho0}\\
\dot{\rho}_x&=&\kappa\left(\rho_0+\alpha\rho_y\right),\label{rhox}\\
\dot{\rho}_y&=&-\alpha\kappa\rho_x,\label{rhoy}\\
\dot{\rho}_z&=&0,
\end{eqnarray}
where the overdot signifies differentiation with respect to time. Differentiating equation (\ref{rhox}) and substituting (\ref{rho0}) and (\ref{rhoy}), yields the second order differential equation

\begin{equation}
\ddot{\rho}_x-\dot{\rho}_x\frac{\dot{\kappa}}{\kappa}+\omega^2\kappa^2\rho_x=0,
\end{equation}
where $\omega=\sqrt{\alpha^2-1}$. This is solved with the function

\begin{equation}
\rho_x\left(t\right)=-\frac{c_1}{\omega}\sin\left[\omega\left(\mu\left(t\right)+c_2\right)\right],
\end{equation}
with $\mu\left(t\right)=\int^t\kappa\left(s\right)ds$. This leads to solutions for the other functions
\begin{eqnarray}
\rho_0\left(t\right)&=&\frac{ c_1}{\omega^2}\cos\left[\omega\left(\mu\left(t\right)+c_2\right)\right]+\alpha c_3,\\
\rho_y\left(t\right)&=&-\frac{\alpha c_1}{\omega^2}\cos\left[\omega\left(\mu\left(t\right)+c_2\right)\right]-c_3,\\
\rho_z\left(t\right)&=&c_4,
\end{eqnarray}
where $c_{1,2,3,4}$ are constants of integration. We therefore have a solution for $\rho\left(t\right)$. The determinant is

\begin{equation}
\det\left[\rho\left(t\right)\right]=c_3^2\omega^2-c_4^2-\frac{c_1^2}{\omega^2}.
\end{equation}
Therefore in order to be positive definite, $c_3^2>\frac{c_4^2}{\omega^2}+\frac{c_1^2}{\omega^4}$. The process to calculate the metric above is quite straightforward, however, now we must calculate the Dyson map $\eta\left(t\right)$. To do this, we assume that $\eta\left(t\right)$ is Hermitian and take the square root of the metric, $\eta=\sqrt{\rho}$. To perform the square root we must first diagonalise the metric in the form $\rho=UDU^{-1}$, where $U$ is the matrix formed of the eigenvectors of $\rho$ as columns. In this way the Dyson map is then $\eta=UD^{1/2}U^{-1}$.
\begin{equation}
\eta\left(t\right)=\frac{1}{2}\left[\zeta_++\zeta_-\right]\mathbb{I}+\frac{1}{\zeta_++\zeta_-}\left[\rho_x\sigma_{x}+\rho_y\sigma_{y}\right]+\frac{\rho_z}{2|\vec{\zeta}_0|}\left[\zeta_+-\zeta_-\right]\sigma_{z},
\end{equation}
where the abbreviated functions are

\begin{eqnarray}
\zeta_\pm&=&\sqrt{\rho_0\pm|\vec{\zeta}_0|},\\
\vec{\zeta}_0&=&\rho_x\vec{i}+\rho_y\vec{j}+\rho_z\vec{k},
\end{eqnarray}
with $\det\left[\eta\left(t\right)\right]=\zeta_+\zeta_-$. This calculation is rather lengthy but results in the most general Hermitian form of $\eta\left(t\right)$. The solution for $\eta\left(t\right)$ allows us to calculate the corresponding Hermitian Hamiltonian using the time-dependent Dyson equation (\ref{dyson}),

\begin{equation}
h\left(t\right)=-\frac{1}{2}\left[\Omega\mathbb{I}-\frac{2\rho_z\kappa\left(t\right)}{\zeta_+\left(t\right)+\zeta_-\left(t\right)}\sigma_{y}+\frac{\left(\rho_y\left(t\right)+\alpha\rho_0\left(t\right)+\alpha\zeta_+\left(t\right)\zeta_-\left(t\right)\right)\kappa\left(t\right)}{\rho_0\left(t\right)+\zeta_+\left(t\right)\zeta_-\left(t\right)}\sigma_{z}\right],
\end{equation}
where the time-dependence has been written out explicitly. In order to determine the constants of integration we must restrict the initial conditions. For this we set $\rho\left(0\right)=\mathbb{I}$ and assume that $\int^{t}\kappa\left(s\right)ds+c_2=0$ at $t=0$ so the integral becomes $\mu\left(t\right)=\int^{t}_0\kappa\left(s\right)ds$. Under these initial conditions, we find the constants to be 
\begin{equation}
c_1=-1,\qquad c_2=0,\qquad c_3=\frac{\alpha}{\omega^2},\qquad c_4=0.
\end{equation}
This means that $\det\left[\rho\left(t\right)\right]=1$ and the components of $\rho\left(t\right)$ are

\begin{eqnarray}
\rho_0\left(t\right)&=&\frac{\alpha^2}{\omega^2}-\frac{ 1}{\omega^2}\cos\left[\omega\mu\left(t\right)\right],\\
\rho_x\left(t\right)&=&\frac{1}{\sqrt{\omega^2}}\sin\left[\omega\mu\left(t\right)\right]\\
\rho_y\left(t\right)&=&\frac{\alpha}{\omega^2}\cos\left[\omega\mu\left(t\right)\right]-\frac{\alpha}{\omega^2},\\
\rho_z\left(t\right)&=&0.
\end{eqnarray}
This completes our solution of the matrix technique for the time-dependent quasi-Hermiticity equation. We found that the central equation was fairly straightforward to solve. However, the process of then finding $\eta\left(t\right)$ was quite lengthy even assuming $\eta\left(t\right)=\eta^\dagger\left(t\right)$.

\subsection{Algebraic technique}\label{qHsectionB}

We now wish to use the algebraic technique to solve the quasi-Hermiticity equation. For this we need to consider a metric composed of Pauli matrices,

\begin{equation}\label{algebraicansatz}
\rho\left(t\right)=e^{\left[\beta\left(t\right)+i\gamma\left(t\right)\right]\sigma_+}e^{\log\left[\delta\left(t\right)\right] \sigma_z}e^{\left[\beta\left(t\right)-i\gamma\left(t\right)\right]\sigma_-},
\end{equation}
where $\beta$, $\gamma$ and $\delta \in\mathbb{R}$ and the raising and lowering operators $\sigma_\pm=1/2\left(\sigma_{x}\pm i\sigma_{y}\right)$ obey the commutation relations

\begin{equation}
\left[\sigma_z,\sigma_+\right]=2\sigma_+, \qquad \left[\sigma_z,\sigma_-\right]=-2\sigma_-, \qquad \left[\sigma_+,\sigma_-\right]=\sigma_z.
\end{equation}
We can see that our ansatz for $\rho\left(t\right)$ is Hermitian. We could have formed the ansatz using a single exponential containing a linear combination of the generators, however this makes calculating the time derivative extremely difficult. Therefore we form our ansatz as a product of three separate exponentials. Now we can calculate the result of acting adjointly with $\rho\left(t\right)$ on the elements of this algebra using the BCH relation (\ref{BCHeq}),

\begin{eqnarray}
\rho \sigma_- \rho^{-1}&=&\frac{1}{\delta^2}\sigma_-+\frac{1}{\delta^2}\left(\beta+i\gamma\right)\sigma_z-\frac{1}{\delta^2}\left(\beta+i\gamma\right)^2\sigma_+,\\
\rho \sigma_z \rho^{-1}&=&\frac{2}{\delta^2}\left(\beta-i\gamma\right)\sigma_-+\left[1+\frac{2}{\delta^2}\left(\beta^2+\gamma^2\right)\right]\sigma_z\\
&-&2\left(\beta+i\gamma\right)\left(1+\frac{1}{\delta^2}\left(\beta^2+\gamma^2\right)\right)\sigma_+,\nonumber\\
\rho \sigma_+ \rho^{-1}&=&-\frac{2}{\delta^2}\left(\beta-i\gamma\right)^2\sigma_--\left(\beta-i\gamma\right)\left[1+\frac{2}{\delta^2}\left(\beta^2+\gamma^2\right)\right]\sigma_z\\
&+&\left(\delta^2+2\left(\beta^2+\gamma^2\right)+\frac{1}{\delta^2}\left(\beta^2+\gamma^2\right)^2\right)\sigma_+\nonumber.
\end{eqnarray}
We now arrange the quasi-Hermiticity equation such that $\rho\left(t\right)$ acts on $H\left(t\right)$ adjointly.

\begin{equation}\label{quasiadjoint}
H^\dagger\left(t\right)-\rho\left(t\right) H\left(t\right)\rho^{-1}\left(t\right)=i\partial_t\rho\left(t\right)\rho^{-1}\left(t\right).
\end{equation} 
The time-derivative term $\dot{\rho}\rho^{-1}$ is 
\begin{eqnarray}
\dot{\rho}\rho^{-1}&=&\left[\dot{\beta}+i\dot{\gamma}\right]\sigma_+\\
&+&e^{\left[\beta\left(t\right)+i\gamma\left(t\right)\right]\sigma_+}\frac{\dot{\delta}}{\delta}\sigma_ze^{-\left[\beta\left(t\right)+i\gamma\left(t\right)\right]\sigma_+}\nonumber\\
&+&e^{\left[\beta\left(t\right)+i\gamma\left(t\right)\right]\sigma_+}e^{\log\left[\delta\left(t\right)\right] \sigma_z}\left[\dot{\beta}-i\dot{\gamma}\right]\sigma_-e^{-\log\left[\delta\left(t\right)\right]\sigma_z}e^{-\left[\beta\left(t\right)+i\gamma\left(t\right)\right]\sigma_+}\nonumber.
\end{eqnarray}
Finally, we express the Hamiltonian (\ref{Hamiltonian2Leveltime}) in terms of $\sigma_\pm$

\begin{equation}
H\left(t\right)=-\frac{1}{2}\left[ \Omega \mathbb{I}+\alpha\kappa\left(t\right) \sigma _{z}+i\kappa\left(t\right)\frac{1}{2}\left(\sigma_++\sigma_-\right)\right].
\end{equation}
Now we substitute our ansatz for $\rho\left(t\right)$ and use the BCH relation to act on the Hamiltonian and to calculate $\dot{\rho}\rho^{-1}$. The resulting differential equations that need to be satisfied for equation (\ref{quasiadjoint}) are

\begin{eqnarray}
\dot{\beta}&=&\frac{1}{2}\kappa\left[1-\beta^2+\delta^2+\gamma^2-2\alpha\gamma\right],\label{BCHeq1}\\
\dot{\gamma}&=&\kappa\beta\left(\alpha-\gamma\right),\label{BCHeq2}\\
\dot{\delta}&=&-\kappa\beta\delta.\label{BCHeq3}
\end{eqnarray}
In order to solve these, we notice that we can write equations (\ref{BCHeq2}) and (\ref{BCHeq3}) as 

\begin{equation}
\delta\dot{\gamma}+\left(\alpha-\gamma\right)\dot{\delta}=0,
\end{equation}
and so can eliminate $dt$ and integrate with respect to $d\gamma$ and $d\delta$, this gives

\begin{equation}
\gamma=\alpha+c_1\delta,
\end{equation}
where $c_1$ is a constant of integration. Substituting this into (\ref{BCHeq2}) and solving for $\beta$ gives

\begin{equation}
\beta=-\frac{\dot{\delta}}{\delta\kappa}.
\end{equation}
Finally, substituting the expressions for $\beta$ and $\gamma$ into (\ref{BCHeq1}) gives the following differential equation in terms of $\delta$

\begin{equation}
\ddot{\delta}-\frac{\dot{\kappa}}{\kappa}\dot{\delta}-\frac{3 \dot{\delta}^2}{2\delta}+\frac{\kappa^2}{2}\left(\left(c_1^2+1\right)\delta^3-\omega^2\delta\right)=0,
\end{equation}
where once again $\omega=\sqrt{\alpha^2-1}$. This looks rather daunting at first, however with the substitution $\delta=1/\sigma^2$, the equation reduces to the Ermakov Pinney (EP) \cite{Ermakov,Pinney} equation with a dissipative term,

\begin{equation}\label{EP1st}
\ddot{\sigma}-\frac{\dot{\kappa}}{\kappa}\dot{\sigma}+\frac{1}{4}\omega^2\kappa^2\sigma=\frac{\left(1+c_1^2\right)\kappa^2}{4\sigma^3}.
\end{equation} 
The EP equation emerges in many scenarios of time-dependent quantum mechanics and various areas in mathematics, see for instance \cite{leach2008ermakov} for an overview. The general solution for (\ref{EP1st}), as reported by Pinney \cite{Pinney}, is%
\begin{equation}
\sigma (t)=\left( Au^{2}+Bv^{2}+2Cuv\right) ^{1/2},
\end{equation}
where $u(t)$ and $v(t)$ are the two fundamental solutions to the equation $\ddot{\sigma}-\frac{\dot{\kappa}}{\kappa}\dot{\sigma}+\frac{1}{4}\omega^2\kappa^2\sigma=0$ and the constants $A$, $B$, $C$ are constrained as $C^{2}=AB-\frac{\left(1+\text{C}_1^2\right)\kappa^2}{4}W^{-2}$ with $W=u\dot{v}-v\dot{u}$ denoting the
corresponding Wronskian. We find the functions $u\left(t\right)$ and $v\left(t\right)$ to be
\begin{equation}
u\left(t\right)=\frac{1}{\omega}\sin\left(\frac{1}{2}\omega\mu\left(t\right)\right), \qquad v\left(t\right)=\cos\left(\frac{1}{2}\omega\mu\left(t\right)\right), \qquad \mu\left(t\right)=\int^t\kappa\left(s\right).
\end{equation}
These functions give the Wronskian to be $W=1/2$ and so we find the solution to equation (\ref{EP1st}) to be 

\begin{equation}
\begin{split}
\begin{aligned}
\sigma\left(t\right)=\left[\frac{A}{\omega^2}\sin^2\left(\frac{1}{2}\omega\mu\left(t\right)\right)+B\cos^2\left(\frac{1}{2}\omega\mu\left(t\right)\right)\right.\\
\left.+2\sqrt{\frac{AB-\left(1+c_1^2\right)}{\omega^2}}\sin\left(\frac{1}{2}\omega\mu\left(t\right)\right)\cos\left(\frac{1}{2}\omega\mu\left(t\right)\right)\right]^{1/2}.
\end{aligned}
\end{split}
\end{equation}
This can be written in a more compact and aesthetic form whilst still being general

\begin{equation}
\sigma\left(t\right)=\sqrt{c_2\cos\left[\omega\left(\mu\left(t\right)+c_3\right)\right]+\sqrt{c_2^2+\frac{1+c_1^2}{\omega^2}}}.
\end{equation}
where
\begin{equation}
c_2=\sqrt{\frac{1}{4}\left(\frac{A}{\omega^2}+B\right)^2-\frac{1+c_1^2}{\omega^2}}, \qquad \tan\left(c_3\omega\right)=2\frac{\omega\sqrt{AB-\left(1+c_1^2\right)}}{A-B\omega^2}.
\end{equation}
Therefore the components of the metric are

\begin{eqnarray}
\beta&=&-\frac{c_2\omega\sin\left[\omega\left(\mu\left(t\right)+c_3\right)\right]}{c_2\cos\left[\omega\left(\mu\left(t\right)+c_3\right)\right]+\sqrt{c_2^2+\frac{1+c_1^2}{\omega^2}}},\\
\delta&=&\frac{1}{c_2\cos\left[\omega\left(\mu\left(t\right)+c_3\right)\right]+\sqrt{c_2^2+\frac{1+c_1^2}{\omega^2}}},\\
\gamma&=&\alpha+\frac{c_1}{c_2\cos\left[\omega\left(\mu\left(t\right)+c_3\right)\right]+\sqrt{c_2^2+\frac{1+c_1^2}{\omega^2}}},
\end{eqnarray}
We are now in a place to calculate the Dyson map from $\rho\left(t\right)=\eta^\dagger\left(t\right)\eta\left(t\right)$. Unlike when using the matrix technique, this is incredibly straight forward as we can just read it off from the metric,

\begin{equation}
\eta\left(t\right)=e^{1/2\log\left[\delta\left(t\right)\right] \sigma_z}e^{\left[\beta\left(t\right)-i\gamma\left(t\right)\right]\sigma_-}.
\end{equation}
Unlike the results obtained from matrix technique, this is not Hermitian. We could have again solved $\rho\left(t\right)=\eta\left(t\right)^2$ by assuming $\eta\left(t\right)$ is Hermitian, however this is much harder here. The resulting Hermitian Hamiltonian is

\begin{equation}
h\left(t\right)=-\frac{1}{2}\left[\Omega\mathbb{I}+\delta\left(t\right)\kappa\left(t\right)\sigma_y+\left(\gamma\left(t\right)-\alpha\right)\sigma_z\right].
\end{equation}
Once again we wish to impose the initial condition $\rho\left(0\right)=\mathbb{I}$ assuming that  $\int^{t}\kappa\left(s\right)ds+c_3=0$ such that the integral becomes $\mu\left(t\right)=\int^{t}_0\kappa\left(s\right)ds$. Under these conditions the constants of integration are

\begin{equation}
c_1=-\alpha, \qquad c_2=-\frac{1}{\omega^2}, \qquad c_3=0.
\end{equation}
This means the components of $\rho\left(t\right)$ are

\begin{eqnarray}
\beta&=&\frac{\omega\sin\left[\omega\mu\left(t\right)\right]}{\alpha^2-\cos\left[\omega\mu\left(t\right)\right]},\\
\delta&=&\frac{\omega^2}{\alpha^2-\cos\left[\omega\mu\left(t\right)\right]},\\
\gamma&=&\alpha-\frac{\alpha\omega^2}{\alpha^2-\cos\left[\omega\mu\left(t\right)\right]}.
\end{eqnarray}
This completes our solution of the algebraic technique of the time-dependent quasi-Hermiticity equation. We have found that in contrast to the matrix technique the central differential equation to be solved is rather more technical. However, once we have solved this we are able to easily obtain the Dyson map $\eta\left(t\right)$.

\section{Time-dependent Dyson equation}\label{DysonSection}

We now turn our attention to the time-dependent Dyson equation (\ref{dyson}). We used this equation in section \ref{qHsection} in order to calculate the corresponding Hermitian Hamiltonian, however in this section we will use it as the central equation in order to solve directly for the Dyson map $\eta\left(t\right)$ and subsequently calculate the metric $\rho\left(t\right)$. This is a substantially different order in which to proceed and it will be interesting to compare how practical each approach proves to be. We will again compare between the matrix technique and the algebraic technique.

\subsection{Matrix technique}\label{DysonSectionM}

We intend to solve the non-Hermitian system (\ref{Hamiltonian2Leveltime}) by making a similar Hermitian ansatz for $\eta\left(t\right)$ as we did for $\rho$ in section \ref{qHsectionM}. Unlike in the previous section, we also need to make an ansatz for $h\left(t\right)$,

\begin{equation}\label{etaanstaz}
\eta\left(t\right)=\eta_0\left(t\right)\mathbb{I}+\sum_{i=x,y,z}\eta_i\left(t\right)\sigma_i, \quad h\left(t\right)=-\frac{1}{2}\left[\Omega\mathbb{I}+\chi\left(t\right)\sigma_{z}\right],
\end{equation}
where $\eta_0\left(t\right)$, $\eta_i\left(t\right)\in \mathbb{R}$ and $\chi\left(t\right)$ is a generic time-dependent real function. Substituting these ans\"atze into equation (\ref{dyson}) results in the following differential equations and equivalence relations

\begin{eqnarray}\label{metricdiffs}
\dot{\eta}_{0} &=&\frac{\kappa }{2}\eta _{x},~~~~\dot{\eta}_{x}=\frac{\chi
	+\kappa }{2}\eta _{y}+\frac{\kappa }{2}\eta _{0},~~~~\dot{\eta}_{y}=-%
\frac{\chi +\alpha \kappa }{2}\eta _{x},~~~~\dot{\eta}_{z}=0,\\
\eta _{z} &=&0,~~~~\chi =\kappa \left( \frac{\eta _{y}}{\eta _{0}}%
+\alpha \right) .\nonumber
\end{eqnarray}
The overdot denotes here as usual a differentiation with respect to time.
The equations (\ref{metricdiffs}) are solved by
\begin{equation}
\eta _{0}=c\sqrt{\frac{\kappa }{\chi }},\qquad \eta _{x}=\frac{c}{%
	\sqrt{\kappa \chi }}\left( \frac{\dot{\kappa}}{\kappa }-\frac{\dot{\chi}}{%
	\chi }\right) ,\qquad \eta _{y}=c\left( \sqrt{\frac{\chi }{\kappa }}-\alpha 
\sqrt{\frac{\kappa }{\chi }}\right) ,\qquad \eta_{z}=0,
\end{equation}
with $c$ denoting an integration constant. Using equations (\ref{metricdiffs}), $\chi (t)$ is found to satisfy the
nonlinear second order equation%
\begin{equation}
\ddot{\chi}-\frac{3}{2}\frac{\dot{\chi}^{2}}{\chi }+\left[ \frac{3}{2}\left( 
\frac{\dot{\kappa}}{\kappa }\right) ^{2}-\frac{\ddot{\kappa}}{\kappa }+\frac{%
	1}{2}\kappa ^{2}(1-\alpha ^{2})\right] \chi +\frac{\chi ^{3}}{2}=0.
\label{xi}
\end{equation}%
Using the parameterizations $\chi =2/\sigma ^{2}$ or $\kappa =2/(\sigma ^{2}%
\sqrt{\alpha ^{2}-1})$ this equation is converted into the Ermakov-Pinney equation for $\sigma $ 
\begin{equation}\label{ErmakovP}
\ddot{\sigma}+\lambda (t)\sigma =\frac{1}{\sigma ^{3}},
\end{equation}%
with time-dependent coefficient%
\begin{equation}
\lambda (t)=\frac{1}{2}\frac{\ddot{\kappa}}{\kappa }-\frac{3}{4}\left( \frac{%
	\dot{\kappa}}{\kappa }\right) ^{2}+\frac{1}{4}\kappa ^{2}\omega^2~~~~~~\text{or~~~~~}\lambda (t)=\frac{1}{2}\frac{\ddot{\chi}}{\chi }-%
\frac{3}{4}\left( \frac{\dot{\chi}}{\chi }\right) ^{2}+\frac{1}{4}\chi ^{2},
\end{equation}%
respectively. Once again $\omega=\sqrt{\alpha^2-1}$. Thus either way given the time-dependent field $\kappa (t)$ in 
$H(t)$ or $\chi (t)$ in $h(t)$ the remaining field is constrained by the EP
equation with almost identical coefficients. Thus from the solution of the EP equation for fixed 
$\alpha $ we can obtain now a specific solution for the Dyson map (\ref{etaanstaz}). For definiteness, we assume that $\kappa\left(t\right)$ is given and we must determine $\chi\left(t\right)$ as we have initially defined a non-Hermitian Hamiltonian in terms of $\kappa\left(t\right)$. We find the solution to be

\begin{equation}
\sigma\left(t\right)=\frac{1}{\sqrt{\kappa\left(t\right)}}\left[\tilde{c}_1\cos\left[\omega\left(\mu\left(t\right)+\tilde{c}_2\right)\right]+\sqrt{\tilde{c}_1^2+\frac{4}{\omega^2}}\right]^{1/2},
\end{equation}
with $\tilde{c}_{1,2}$ as constants of integration and $\mu\left(t\right)=\int^t\kappa\left(s\right)ds$. Therefore the components of $\eta\left(t\right)$ are

\begin{eqnarray}
\eta_{0}&=&\frac{c}{\sqrt{2}}\left[\tilde{c}_1\cos\left[\omega\left(\mu\left(t\right)+\tilde{c}_2\right)\right]+\sqrt{\tilde{c}_1^2+\frac{4}{\omega^2}}\right]^{1/2},\\
\eta_{x}&=&-\frac{c\omega\tilde{c}_1\sin\left[\omega\left(\mu\left(t\right)+\tilde{c}_2\right)\right]}{\sqrt{2}\left[\tilde{c}_1\cos\left[\omega\left(\mu\left(t\right)+\tilde{c}_2\right)\right]+\sqrt{\tilde{c}_1^2+\frac{4}{\omega^2}}\right]^{1/2}},\\
\eta_{y}&=&-\frac{c\left[\alpha\tilde{c}_1\cos\left[\omega\left(\mu\left(t\right)+\tilde{c}_2\right)\right]+\alpha\sqrt{\tilde{c}_1^2+\frac{4}{\omega^2}}-2\right]}{\sqrt{2}\left[\tilde{c}_1\cos\left[\omega\left(\mu\left(t\right)+\tilde{c}_2\right)\right]+\sqrt{\tilde{c}_1^2+\frac{4}{\omega^2}}\right]^{1/2}},\\
\eta_{z}&=&0.
\end{eqnarray}
As we have $\eta\left(t\right)$, we can straightforwardly calculate the metric $\rho\left(t\right)=\eta^\dagger\left(t\right)\eta\left(t\right)$.

\begin{equation}
\rho=\left[\eta_{0}^2+\eta_{x}^2+\eta_{y}^2\right]\mathbb{I}+2\eta_{0}\left[\eta_{x}\sigma_{x}+\eta_{y}\sigma_{y}\right].
\end{equation}
By construction we already have the corresponding Hermitian Hamiltonian $h\left(t\right)$ and so we are in a position to compare the results of this approach with that of the quasi-Hermiticity equation in section \ref{qHsection}. To do this we impose the same initial conditions $\rho\left(0\right)=\mathbb{I}$, and find that

\begin{equation}
c=\sqrt{\alpha}, \qquad \tilde{c}_1=-\frac{1}{\alpha\omega^2}, \qquad \tilde{c}_2=0.
\end{equation}
With these constants the components of $\eta\left(t\right)$ are

\begin{eqnarray}
\eta_{0}&=&\frac{1}{\omega}\left[\alpha^2-\cos^2\left(\frac{1}{2}\omega\mu\left(t\right)\right)\right]^{1/2},\\
\eta_{x}&=&\frac{\sin\left(\omega\mu\left(t\right)\right)}{2\left[\alpha^2-\cos^2\left(\frac{1}{2}\omega\mu\left(t\right)\right)\right]^{1/2}},\\
\eta_{y}&=&-\frac{\alpha\left[1-\cos\left(\omega\mu\left(t\right)\right)\right]}{2\omega\left[\alpha^2-\cos^2\left(\frac{1}{2}\omega\mu\left(t\right)\right)\right]^{1/2}},\\
\eta_{z}&=&0,
\end{eqnarray}
and the expression for $\rho\left(t\right)$ matches that in section \ref{qHsectionM}. This completes our solution to the matrix technique of the time-dependent Dyson equation. We have shown that solving directly for the Dyson map $\eta\left(t\right)$ results in a rather technical formulation of the EP equation. However it has a straightforward solution and results in a complete solution for $\eta\left(t\right)$. Furthermore, the process of obtaining $\rho\left(t\right)$ from $\eta\left(t\right)$ is trivial in comparison to the reverse calculation.

\subsection{Algebraic technique}\label{DysonSectionB}

Now we turn our attention to the algebraic technique that was introduced in the previous section to solve the time-dependent quasi-Hermiticity equation. For this we formulate the Hamiltonian and the ansatz in terms of the Pauli raising and lowering operators $\sigma_\pm$ once more. In contrast to the matrix technique, we do not begin with the assumption of $\eta\left(t\right)$ being Hermitian, however this is a choice and we could indeed assume $\eta\left(t\right)=\eta^\dagger\left(t\right)$. In addition we do not need make an ansatz for $h\left(t\right)$

\begin{equation}
\eta\left(t\right)=e^{\left[\epsilon\left(t\right)-i\tau\left(t\right)\right]\sigma_+}e^{\log\left[\vartheta\left(t\right)\right] \sigma_z}e^{\left[\beta\left(t\right)-i\gamma\left(t\right)\right]\sigma_-}.
\end{equation}
where $\epsilon$, $\tau$, $\vartheta$, $\beta$ and $\gamma\in\mathbb{R}$. Similarly to the time-dependent quasi-Hermiticity equation, we substitute this expression for $\eta$ in the time-dependent Dyson equation and use the BCH relation to expand the expression in terms of the algebra. The resulting expression must be $h\left(t\right)$ and therefore is Hermitian. This creates restrictions on the parameters in $\eta\left(t\right)$ in the form of differential equations. What is clear when substituting in this ansatz, is that either the parameters $\epsilon$ and $\tau$ are superfluous or $\beta$ and $\gamma$, therefore we can set either pair to zero. We could also choose them such that $\eta\left(t\right)$ is Hermitian, but in this case the resulting equations are significantly more complicated. In order to differ from the Dyson map in section \ref{qHsectionB} we set the parameters $\beta$ and $\gamma$ to zero such that our Dyson map takes the form

\begin{equation}
\eta\left(t\right)=e^{\left[\epsilon\left(t\right)-i\tau\left(t\right)\right]\sigma_+}e^{\log\left[\vartheta\left(t\right)\right] \sigma_z}.
\end{equation}
In this setting, we obtain the following differential equations when requiring $h\left(t\right)$ to be Hermitian.

\begin{eqnarray}
\dot{\epsilon}&=&\frac{\left(1+\epsilon^2+\vartheta^2+\tau^2-2\alpha\tau\vartheta\right)\kappa}{2\vartheta},\\
\dot{\tau}&=&\epsilon\kappa,\\
\dot{\vartheta}&=&\alpha\epsilon\kappa.
\end{eqnarray}

The final two equations can be combined in order to eliminate $dt$ and find $\tau$ in terms of $\vartheta$,

\begin{equation}
d\tau=\alpha d\vartheta,
\end{equation}
therefore

\begin{equation}
\tau=\tilde{c}_1+\alpha\vartheta,
\end{equation}
where $\tilde{c}_1$ is a constant of integration. Furthermore, we have

\begin{equation}
\epsilon=\frac{\dot{\tau}}{\kappa}.
\end{equation}
Finally, substituting these expressions for $\epsilon$ and $\tau$ into the first equation gives us the underlying differential equation to be solved
\begin{equation}
\ddot{\vartheta}-\frac{\dot{\kappa}}{2\kappa}\dot{\vartheta}-\frac{\dot{\vartheta}^2}{2\vartheta}+\frac{1}{2}\omega^2\kappa^2\vartheta-\frac{\left(1+\tilde{c}_1^2\right)\kappa^2}{2\vartheta}=0,
\end{equation}
where $\omega=\sqrt{\alpha^2-1}$. Now if we make the change of variable $\vartheta=\sigma^2$ we once again obtain the EP equation with a dissipative term.

\begin{equation}
\ddot{\sigma}-\frac{\dot{\kappa}}{\kappa}\dot{\sigma}+\frac{1}{4}\omega^2\kappa^2\sigma-\frac{\left(1+\tilde{c}_1^2\right)\kappa^2}{4\sigma^3}=0.
\end{equation}
This is solved with the function

\begin{equation}
\sigma\left(t\right)=\sqrt{\tilde{c}_2\cos\left[\omega\left(\mu\left(t\right)+\tilde{c}_3\right)\right]+\sqrt{\frac{1+\tilde{c}_1^2}{\omega^2}+\tilde{c}_2^2}},
\end{equation}
with $\tilde{c}_{2,3}$ as constants of integration and $\mu\left(t\right)=\int^t\kappa\left(s\right)ds$. Therefore the expressions in our Dyson map are
\begin{eqnarray}
\epsilon&=&-\tilde{c}_2\omega\sin\left[\omega\left(\mu\left(t\right)+\tilde{c}_3\right)\right],\\
\tau&=&\alpha\tilde{c}_2\cos\left[\omega\left(\mu\left(t\right)+\tilde{c}_3\right)\right]+\tilde{c}_1+\alpha\sqrt{\frac{1+\tilde{c}_1^2}{\omega^2}+\tilde{c}_2^2},\\
\vartheta&=&\tilde{c}_2\cos\left[\omega\left(\mu\left(t\right)+\tilde{c}_3\right)\right]+\sqrt{\frac{1+\tilde{c}_1^2}{\omega^2}+\tilde{c}_2^2}.
\end{eqnarray}
The corresponding Hermitian Hamiltonian is

\begin{equation}
h\left(t\right)=-\frac{1}{2}\left[\Omega\mathbb{I}+\frac{\kappa}{\vartheta}\sigma_{y}+\left(\alpha-\frac{\tau}{\vartheta}\right)\sigma_z\right].
\end{equation}
Finally, the metric operator $\rho\left(t\right)=\eta^\dagger\left(t\right)\eta\left(t\right)$ is

\begin{equation}
\rho\left(t\right)=e^{\log\left[\vartheta\left(t\right)\right] \sigma_z}e^{\left[\epsilon\left(t\right)+i\tau\left(t\right)\right]\sigma_-}e^{\left[\epsilon\left(t\right)-i\tau\left(t\right)\right]\sigma_+}e^{\log\left[\vartheta\left(t\right)\right] \sigma_z},
\end{equation}
which we can see clearly differs from the form of the metric in section \ref{qHsectionB}. However they are equivalent expressions when taking into account the initial condition $\rho\left(0\right)=\mathbb{I}$ that fixes the constants of integration to

\begin{equation}
\tilde{c}_1=-\alpha,\qquad \tilde{c}_2=-\frac{1}{\omega^2}, \qquad \tilde{c}_3=0
\end{equation}
The components of the Dyson map then become

\begin{eqnarray}
\epsilon&=&\frac{1}{\omega}\sin\left[\omega\mu\left(t\right)\right],\\
\tau&=&\frac{\alpha}{\omega^2}-\frac{\alpha}{\omega^2}\cos\left[\omega\mu\left(t\right)\right],\\
\vartheta&=&\frac{\alpha^2}{\omega^2}-\frac{1}{\omega^2}\cos\left[\omega\mu\left(t\right)\right].
\end{eqnarray}
This completes our solution for the algebraic technique to the time-dependent Dyson equation. We have shown that in this setting the EP equation emerges once again. In addition the difference in approaches highlights the fact that the Dyson map $\eta\left(t\right)$ is not unique and can take many forms, both Hermitian and non-Hermitian. Furthermore, the differing Dyson maps result in different corresponding Hermitian Hamiltonians $h\left(t\right)$. However, all approaches are correct, equivalent and lead to a consistent description of the time-dependent non-Hermitian quantum system.

\section{Lewis Riesenfeld invariants}

The final solution approach we will consider in this chapter is the Lewis Riensenfeld (LR) invariant approach. The LR invariants are operators used in time-dependent quantum mechanics in order to break down the process of solving the time-dependent Schr\"odinger equation into more manageable steps \cite{lewis1969exact}. The invariant $I_h\left(t\right)$ is defined for Hermitian systems via Heisenberg's equation of motion as follows 

\begin{equation}\label{Invarianth}
\frac{dI_h(t)}{dt}=\partial _{t}I_h(t)-i \left[I_h(t),h\left(t\right)\right] =0, \qquad I_h^\dagger(t)=I_h(t).
\end{equation}
Therefore, given a Hamiltonian $h\left(t\right)$, we can solve for the invariant using (\ref{Invarianth}). Once we have the invariant, we can then construct the time-dependent wave functions of the Hamiltonian $h\left(t\right)$ from the equations (for more detail see \nameref{LRAppendix})
\begin{eqnarray}
I_h(t)\ket{\phi_n\left(t\right)}
&=&\Lambda_n \ket{\phi_n\left(t\right)} ,~~~~~~~~~~~~~~\ \
\ \ \ \ \ ~\ ~\ket{\Phi_n(t)} =e^{i
	\alpha_n (t)} \ket{\phi_n\left(t\right)} ,~~~~~ 
\\
\dot{\alpha}_n &=& \bra{\phi_n\left(t\right)} i
\partial _{t}-h(t) \ket{\phi_n\left(t\right)}
,\qquad \dot{\Lambda}_n=0.
\end{eqnarray}%
Where the wave function can then be constructed from the dynamical modes $\ket{\Phi\left(t\right)}=\sum_nc_n\ket{\Phi_n\left(t\right)}$. We can see that the utility of the invariant is that it has time-independent eigenvalues by construction. We can therefore solve for the eigenstates of $I_h\left(t\right)$ with less difficulty than solving directly for the wave functions of the Hamiltonian $\ket{\Phi\left(t\right)}$.

In order to see how invariants are related between Hermitian and non-Hermitian systems, we substitute the time-dependent Dyson equation into (\ref{Invarianth}). After some manipulation, we find an equivalent relation for the invariant $I_H\left(t\right)$ of the non-Hermitian system $H\left(t\right)$
\begin{equation}\label{InvariantH}
\frac{dI_H(t)}{dt}=\partial _{t}I_H(t)-i \left[I_H(t),H\left(t\right)\right] =0, \qquad I_H^\dagger(t)\neq I_H(t),
\end{equation}
when the invariants are related by a similarity transform in $\eta\left(t\right)$

\begin{equation}\label{InvarSim}
I_{h}(t)=\eta (t)I_{H}(t)\eta ^{-1}(t).
\end{equation}
This is a remarkable property in the time-dependent setting as we do not encounter any time derivatives in the relation between the invariants \cite{maamache2017pseudo}. The time derivatives are in fact hidden in the process of determining the invariant from the Hamiltonian (\ref{Invarianth}), (\ref{InvariantH}). Furthermore, it is easy to see that $I_H\left(t\right)$ is quasi-Hermitian with respect to the metric $\rho\left(t\right)$
\begin{equation}\label{qHinvariants}
I^\dagger_H\left(t\right)=\rho\left(t\right)I_H\left(t\right)\rho^{-1}\left(t\right).
\end{equation}
Therefore the problem reduces to solving a similarity transform or a quasi-Hermitian relation much like the time-independent case once the invariants are known.\\

We once again consider the time-dependent Hamiltonian (\ref{Hamiltonian2Leveltime}) as our starting point. From here we now wish to calculate the invariant $I_H\left(t\right)$. As this can also be non-Hermitian, we must take our ansatz to be a general non-Hermitian matrix

\begin{eqnarray}
I_H\left(t\right)&=&\frac{1}{2}\left[a^r\left(t\right)+ia^i\left(t\right)\right]\mathbb{I}+\frac{1}{2}\left[b^r\left(t\right)+ib^i\left(t\right)\right]\sigma_x\\
&+&\frac{1}{2}\left[c^r\left(t\right)+ic^i\left(t\right)\right]\sigma_y+\frac{1}{2}\left[d^r\left(t\right)+id^i\left(t\right)\right]\sigma_z,\nonumber
\end{eqnarray}
where $a^{r,i},b^{r,i},c^{r,i},d^{r,i} \in \mathbb{R}$. Substituting this into the invariant equation (\ref{InvariantH}) and collecting real and imaginary terms, we obtain the following differential equations

\begin{eqnarray}
\dot{a}^r&=&\dot{a}^i\;\, =\;\, 0,\\
\dot{b}^r&=&\alpha \kappa c^r, \qquad\qquad\qquad \dot{b}^i \;\,=\;\, \alpha \kappa c^i,\label{invariantrow2}\\
\dot{c}^r&=&-\kappa \left(\alpha b^r +  d^i \right), \qquad \dot{c}^i\;\, =\;\, -\kappa \left(\alpha b^i -  d^r \right),\label{invariantrow3}\\
\dot{d}^i&=&-\kappa c^r, \qquad\qquad\quad\;\;\; \dot{d}^r\;\, =\;\, \kappa c^i,\label{invariantrow4}
\end{eqnarray}
where the variables separate into two independent sets. If we look closely we see that these equations take the same form as the equations (\ref{metricdiffs}) for the metric. We see therefore an equivalence in these approaches already. Differentiating the equations (\ref{invariantrow3}) and substituting in equations (\ref{invariantrow2}) and (\ref{invariantrow4}) we get the two separate governing differential equations.

\begin{equation}\label{invariantdiffs}
\ddot{c}^r-\dot{c}^r\frac{\dot{\kappa}}{\kappa}+\omega^2\kappa^2c^r=0, \qquad \ddot{c}^i-\dot{c}^i\frac{\dot{\kappa}}{\kappa}+\omega^2\kappa^2c^i=0,
\end{equation}
where again $\omega=\sqrt{\alpha^2-1}$. From these equations, we find the solutions for the parameters of the invariant.
\begin{eqnarray}
a^r\left(t\right)&=&\hat{c}_4,\\
b^r\left(t\right)&=&\frac{\alpha \hat{c}_1}{\omega}\sin\left[\omega\left(\mu\left(t\right)+\hat{c}_2\right)\right]-\hat{c}_3,\\
c^r\left(t\right)&=&\hat{c}_1\cos\left[\omega\left(\mu\left(t\right)+\hat{c}_2\right)\right], \\
d^i\left(t\right)&=&-\frac{ \hat{c}_1}{\omega}\sin\left[\omega\left(\mu\left(t\right)+\hat{c}_2\right)\right]+\alpha \hat{c}_3.
\end{eqnarray}
\begin{eqnarray}
a^i\left(t\right)&=&\hat{c}_8,\\
b^i\left(t\right)&=&\frac{\alpha \hat{c}_5}{\omega}\sin\left[\omega\left(\mu\left(t\right)+\hat{c}_6\right)\right]-\hat{c}_7,\\
c^i\left(t\right)&=&\hat{c}_5\cos\left[\omega\left(\mu\left(t\right)+\hat{c}_6\right)\right], \\
d^r\left(t\right)&=&\frac{ \hat{c}_5}{\omega}\sin\left[\omega\left(\mu\left(t\right)+\hat{c}_6\right)\right]-\alpha \hat{c}_7,
\end{eqnarray}
with $\mu\left(t\right)=\int^t\kappa\left(s\right)ds$. This looks like a large amount of information initially, however we can make some vast simplifications if we consider the eigenvalues of this non-Hermitian invariant. Because the invariants of the Hermitian and non-Hermitian system are related via a similarity transform, it follows that the eigenvalues are the same, therefore the eigenvalues of $I_H$ must be real to match the real eigenvalues of $I_h$. We compute

\begin{equation}
\Lambda_\pm=\hat{c}_4+i\hat{c}_8\pm\frac{1}{2}\sqrt{\hat{c}_1^2-\hat{c}_5^2+2i\hat{c}_1\hat{c}_5\cos\left[\left(\hat{c}_2-\hat{c}_6\right)\omega\right]-\omega^2\left(\hat{c}_3+i\hat{c}_7\right)^2}.
\end{equation}
Therefore to ensure the reality of these eigenvalues, we set $\hat{c}_{3,5,6,7,8}=0$. We of course may choose these constants differently, but that is not of great importance at this stage as we are mainly interested in the solution for $\rho\left(t\right)$. Under these choices, the invariant becomes

\begin{equation}
I_H\left(t\right)=-\frac{1}{2}\left[\hat{c}_4\mathbb{I}+b^r\left(t\right)\sigma_{x}+c^r\left(t\right)\sigma_{y}+id^i\left(t\right)\sigma_{z}\right].
\end{equation}
We now will proceed to solve the quasi-Hermitian invariant relation using the matrix technique, and the algebraic technique. 

\subsection{Matrix technique}
Next we make the ansatz for $\rho\left(t\right)$ as we did earlier,

\begin{equation}
\rho\left(t\right)=\rho_0\left(t\right)\mathbb{I}+\sum_{i=x,y,z}\rho_i\left(t\right)\sigma_i, 
\end{equation}
where $\rho_0\left(t\right)$, $\rho_i\left(t\right)\in \mathbb{R}$. From here we solve the quasi-Hermiticity relation (\ref{qHinvariants}) for $\rho\left(t\right)$,

\begin{eqnarray}
\rho_x\left(t\right)&=&\frac{\left(\alpha\rho_y\left(t\right)+\rho_0\left(t\right)\right)\tan\left[\omega\left(\mu\left(t\right)+\hat{c}_2\right)\right]}{\omega},\\
\rho_z\left(t\right)&=&0.
\end{eqnarray}
As is clear we do not have a full solution, only $\rho_x$ in terms of $\rho_0$ and $\rho_y$. Therefore, to complete the solution we substitute our partial solution into the time-dependent quasi-Hermiticity equation and obtain the following relations,

\begin{eqnarray}
\rho_y&=&-\alpha\rho_0+q_2,\\
\dot{\rho}_0&=&\frac{\kappa\tan\left[\omega\left(\mu\left(t\right)+\hat{c}_2\right)\right]\left(\alpha q_2-\omega^2\rho_0\right)}{\omega},
\end{eqnarray}
the latter of which is solved with 

\begin{equation}
\rho_0=\frac{\alpha q_2}{\omega^2}+\frac{q_1}{\omega^2}\cos\left[\omega\left(\mu\left(t\right)+\hat{c}_2\right)\right].
\end{equation}
Therefore the other components are

\begin{eqnarray}
\rho_x&=&-\frac{q_1}{\omega}\sin\left[\omega\left(\mu\left(t\right)+\hat{c}_2\right)\right],\\
\rho_y&=&-\frac{ q_2}{\omega^2}-\frac{\alpha q_1}{\omega^2}\cos\left[\omega\left(\mu\left(t\right)+\hat{c}_2\right)\right].
\end{eqnarray}
If we again fix the initial condition at $\rho\left(0\right)=\mathbb{I}$ then we find the constants of integration are

\begin{equation}
\hat{c}_2=0,\qquad q_1=-1,\qquad q_2=\alpha.
\end{equation}
Under these conditions it is easy to see we match the metric calculated in section \ref{qHsectionM}. This completes the solution for the matrix technique of the LR invariants. This calculation is rather long but not so technical in comparison the other approaches considered in this chapter. The most difficult differential equations we were required to solve were (\ref{invariantdiffs}), which are on the same level as those in section \ref{qHsectionM}. However, we are still required to take the square root of the metric to obtain the the Dyson map, so in this sense we do not avoid the additional technicality.

\subsection{Algebraic technique}

We now wish to solve the quasi-Hermitian relation (\ref{qHinvariants}) using the alegbraic technique. For this we make the same ansatz as in section \ref{qHsectionB}

\begin{equation}
\rho\left(t\right)=e^{\left[\beta\left(t\right)+i\gamma\left(t\right)\right]\sigma_+}e^{\log\left[\delta\left(t\right)\right] \sigma_z}e^{\left[\beta\left(t\right)-i\gamma\left(t\right)\right]\sigma_-},
\end{equation}
where $\beta$, $\gamma$ and $\delta\in \mathbb{R}$. Furthermore, we write the invariant $I_H\left(t\right)$ in terms of the raising and lowering operators

\begin{equation}
I_H\left(t\right)=-\frac{1}{2}\left[\hat{c}_4\mathbb{I}+b^r\left(t\right)\frac{1}{2}\left(\sigma_++\sigma_-\right)+ic^r\left(t\right)\frac{1}{2}\left(\sigma_--\sigma_+\right)+id^i\left(t\right)\sigma_{z}\right].
\end{equation}
Now we substitute our ansatz for $\rho\left(t\right)$ into the relation (\ref{qHinvariants}) and solve for the components $\beta$, $\gamma$ and $\delta$. We obtain the following expressions

\begin{eqnarray}
\beta&=& \frac{\left(1-\alpha\gamma\right)\tan\left[\omega\left(\mu\left(t\right)+\hat{c}_2\right)\right]}{\omega},\\
\delta&=&\sqrt{\frac{\left(\alpha-\gamma\right)^2-\left(1-\alpha\gamma\right)\sec^2\left[\omega\left(\mu\left(t\right)+\hat{c}_2\right)\right]}{\omega^2}}.
\end{eqnarray}
Once again we do not have the full solution. To obtain the expression for $\gamma$ we substitute our partial solution into the time-dependent quasi-Hermiticity equation and obtain the differential equation

\begin{equation}
\frac{\dot{\gamma}-\left(\alpha-\gamma\right)\left(1-\alpha\gamma\right)\kappa\tan\left[\omega\left(\mu\left(t\right)\hat{c}_2\right)\right]}{\omega}=0,
\end{equation}
which, when taking the initial condition $\rho\left(0\right)=\mathbb{I}$ gives the solution

\begin{equation}
\gamma=\alpha-\frac{\alpha\omega^2}{\alpha^2-\cos\left[\omega\mu\left(t\right)\right]},
\end{equation}
and therefore the other components of the metric are
\begin{eqnarray}
\beta&=&\frac{\omega\sin\left[\omega\mu\left(t\right)\right]}{\alpha^2-\cos\left[\omega\mu\left(t\right)\right]},\\
\delta&=&\frac{\omega^2}{\alpha^2-\cos\left[\omega\mu\left(t\right)\right]},
\end{eqnarray}
which matches the results from section \ref{qHsectionB}. This completes our solution of the algebraic technique for the LR invariants. In comparison to the matrix technique here we are able to read off the Dyson map trivially in the same way as in section \ref{qHsectionB}. In this sense we reduce the complexity of the problem as we do not encounter the EP equation at any point here. However, the trade off is a calculation with many more steps. We have used the Lewis Riesenfeld invariants here to solve the quasi-Hermiticity relation $I_H^\dagger=\rho I_H\rho^{-1}$. However, we could of course just as well solve the similarity transform $I_h=\eta I_H\eta^{-1}$ for the invariant of the Hermitian counterpart and the Dyson map. We will demonstrate this approach in chapter 4 when we consider coupled harmonic oscillator systems.

\section{Comparison}

We have analysed in detail six different ways to derive the metric and the Dyson operator with the initial condition $\rho\left(0\right)=\mathbb{I}$. Each approach and technique gives a valid metric and we demonstrate the non-uniqueness of the Dyson map by explicitly taking varying anz\"atze. However, all approaches are indeed equivalent. The origin of this variation is the fact that the corresponding Hermitian Hamiltonian is not fixed. We have only analysed a matrix model in this chapter and so we have not compared the applicability of each approach to a variety of systems (as we will come across in the later chapters). Therefore an approach that seems disadvantageous for this particular matrix model may be the most applicable approach for a system with infinite dimensional Hilbert space.

Solving the time-dependent quasi-Hermiticity equation using the matrix technique results in the simplest differential equation (we do not encounter the EP equation) in terms of $\rho\left(t\right)$. However, when we come to calculate the Dyson map $\eta\left(t\right)$, the calculation is lengthy and tedious. When we solve the same equation using the algebraic, we obtain a version of the EP equation with a dissipative term. This is a much more technical equation to solve but the return is that we are able to easily deduce the Dyson map.

Moving onto the time-dependent Dyson equation, we approach the problem from a new direction by solving for the Dyson map $\eta\left(t\right)$ directly. When using the matrix technique, we make an ansatz for the corresponding Hermitian Hamiltonian $h\left(t\right)$ which results once again in a version of the EP equation. When using the algebraic technique we do not require an ansatz for $h\left(t\right)$ but do obtain an EP equation. In comparison to the time-dependent quasi-Hermiticity equation, we are able to obtain $\eta\left(t\right)$ directly so that we can easily calculate $\rho\left(t\right)$. However, the problem of making a useful ansatz becomes apparent as we are not restricted to a Hermitian operator. This is apparent as we see that $\eta\left(t\right)$ from the matrix technique is Hermitian, whereas from the algebraic technique it is not. Therefore if one wants to make a totally general ansatz, the problem can become vastly complex. 

Approaching the problem using the Lewis Riesenfeld invariants is a slightly more lengthy approach. The advantage is that the differential equations are not as technical as in the other approaches (we do not encounter the EP equation). Furthermore at the end of the calculation, we already have defined important quantities that lead to the solutions for the wave functions for our system. However, the process involves significantly more steps and this can create difficulty keeping track of all the quantities involved.

Overall, there is no one approach that has a clear advantage over the others. Each has its own merit depending on the system under analysis. In the particular 2-level system studied in this chapter the Lewis Riesenfeld invariants do not show any clear advantage, however as we will see in chapter 4, they work well for a specific 2 dimensional coupled harmonic oscillator. As we move on from matrix models, the algebraic technique becomes the necessary approach in each case. Ultimately however, the choice between the 3 approaches is a preference rather than a directive. 

\begin{table}[H]
	\begin{center}
		\begin{tabular}{|>{\columncolor{Dandelion}}m{4.3cm}|m{4.3cm}|m{4.3cm}|}
			\rowcolor{Dandelion}
			\hline\large 
			\multirow{2}{8cm}{Approach \\} & \large \multirow{2}{8cm}{Advantages \\ } & \large \multirow{2}{8cm}{Disadvantages \\ }  \\ 
			\hline		
			\large Time-dependent quasi-Hermiticity equation &  Simpler differential equation &  Taking square root of $\rho$   \\
			& No ansatz for $h$  &\\ 
			& Ansatz for $\rho$ is Hermitian & \\
			\hline
			\large Time-dependent Dyson equation & Immediate solution for $\eta$ ($\eta\rightarrow\rho$ simple) & Ansatz for $h$ \\ 
			& Not restricted to an Hermitian ansatz & Unclear form of $\eta$ as not restricted to be Hermitian \\
			\hline
			\large Lewis Riesenfeld invariants & Simpler differential equations & Greater number of steps  \\ 
			& Can choose to solve for $\eta$ or $\rho$ &  \\
			& Already have tools to solve for $\psi$ & \\
			\hline
		\end{tabular}
	\end{center}
\caption{Comparison of the solution approaches: time-dependent quasi-Hermiticity equation, time-dependent Dyson equation and the Lewis-Riesenfeld invariants.}\label{methodcomparison}
\end{table}

\section{Summary}

We have used a simple but non-trivial matrix model as an example system in this chapter. Using this example, we demonstrated the various solution approaches available to solve for the metric $\rho\left(t\right)$ and the Dyson map $\eta\left(t\right)$. Furthermore, we elaborated on each approach by solving the equations using both a matrix approach and an algebraic approach. Finally, we showed that each approach is indeed equivalent when the same initial conditions are applied to the metric operator. The Dyson map is not unique and we have seen this explicitly with it taking many different forms and resulting in a variety of corresponding Hermitian Hamiltonians. However, as outlined in the introduction, the Dyson map can made unique by forcing two operators to be observable. Table \ref{methodcomparison} highlights the differences between the approaches compactly.

\chapter{Mending the broken $\mathcal{PT}$ regime}

It is well known that non-Hermitian Hamiltonians that commute with an antilinear operator and for which its eigenfunctions are eigenstates \cite{EW} possess real eigenvalue spectra. This concept was introduced and explained in the introduction. $\mathcal{PT}$-symmetry \cite{bender1998real} is a specific example of such an antilinear symmetry for which many examples have been worked out in detail, see e.g. \cite{bender2012special}. As we have seen in chapter 2, it is possible to make sense of such non-Hermitian systems in a quantum mechanical framework with the introduction of a metric operator that defines the inner product on the specific Hilbert space. However, it is possible for this $\mathcal{PT}$-symmetry to be spontaneously broken for some region of the system's parameter set. In this case the wave functions in the broken regime become unbounded, with exponential growth in the time evolution and complex energy eigenvalues. While such a situation is the most interesting one in optical settings \cite{musslimani2008optical,makris2010pt,Guo}, where different channels of gain and loss may be constructed, the development of infinite growth in energy means it is usually discarded as being non-physical in a quantum mechanical framework. 

The spontaneously broken regime arises when the wave functions cease to simultaneously be eigenfunctions of the $\mathcal{PT}$ operator and satisfy the time-dependent Schr\"odinger equation for $H\left(t\right)$. In this chapter we will provide an explanation and interpretation for the spontaneously broken regime by introducing a time-dependence into the central equations and ultimately find a time-dependent metric operator $\rho\left(t\right)$. This metric allows us to construct a well-defined Hilbert space with the inner product $\left\langle\cdot|\cdot\right\rangle_{\rho}=\left\langle\cdot|\rho\left(t\right)\cdot\right\rangle$.
This regularisation of the inner product opens up the broken regime for analysis as all quantities involved become well defined. 

The introduction of time leads to another remarkable property, that is the non-Hermitian Hamiltonian ceases to be observable. This follows from asserting that observable operators $\mathcal{O}$ in the non-Hermitian system need to be related to a self-adjoint operator $o(t)$ in the Hermitian system as $o(t)=\eta (t)\mathcal{O}(t)\eta ^{-1}(t)$. Under this assumption the observable energy operator is in fact

\begin{equation}\label{HenergyO}
\tilde{H}(t)=\eta ^{-1}(t)h(t)\eta (t)=H(t)+i\hbar \eta ^{-1}(t)\partial_{t}\eta (t).
\end{equation}
The original Hamiltonian $H\left(t\right)$ defines the time-dependent Schr\"odinger equation and generates the time-evolution, $\tilde{H}\left(t\right)$ is the energy operator and does not define this equation. We will show that even in the spontaneously broken $\mathcal{PT}$ regime, the energy operator has real expectation values and obeys a new $\widetilde{PT}$-symmetry that remains unbroken.

\section{A two-level system with spontaneously broken $\mathcal{PT}$-symmetry}\label{ModelPT}

To illustrate our point we revisit the simple time-dependent two-level spin model analysed in chapter 2, described by the non-Hermitian Hamiltonian,

\begin{equation}\label{H2}
H(t)=-\frac{1}{2}\left[ \Omega \mathbb{I}+\alpha \kappa (t)\sigma_{z}+i\kappa (t)\sigma _{x}\right],
\end{equation}
with $\sigma _{x}$, $\sigma _{y}$, $\sigma _{z}$ denoting the Pauli
matrices, $\mathbb{I}$ the identity matrix and $\alpha$, $\omega$, $
\kappa\left(t\right)\in \mathbb{R}$. Solving the time-dependent Schr\"odinger equation using the Lewis Riesenfeld invariants (as outlined in \nameref{LRAppendix}) for this Hamiltonian we obtain the time-dependent wave functions
\begin{equation}\label{EH}
\varphi _{\pm }\left(t\right)=\frac{1}{\sqrt{2\alpha\left(\alpha\mp\sqrt{\alpha^2-1}\right)}}e^{{\frac{1}{2}i\left(\Omega t\pm\sqrt{\alpha^2-1}\int^t\kappa\left(s\right)ds\right)}}\left(\begin{array}{c}
i \\ 
-\alpha \pm \sqrt{\alpha^2-1}
\end{array}\right) . 
\end{equation}
from which we can form a general wave function. Using Wigner's argument \cite{EW,bender1998real} the reality of the energy
spectrum for $\left\vert \alpha \right\vert >1 $
is easily explained by identifying an antilinear symmetry operator, denoted
here as $\mathcal{PT}$ , that commutes with the Hamiltonian and for which $%
\varphi _{\pm }\left(t\right)$ are eigenstates of $\mathcal{PT}$ and solutions to the time-dependent Schr\"odinger equation
\begin{equation}
\left[ \mathcal{PT},H\left(t\right)\right] =0,\qquad \text{and\qquad }\mathcal{PT}\varphi
_{\pm }\left(t\right)=e^{i\phi\left(t\right) }\varphi _{\pm }\left(t\right),  \label{PTH}
\end{equation}%
with $\phi \in \mathbb{R}$. When $\left\vert \alpha \right\vert >1$ in our example the symmetry operator is easily
identified as $\mathcal{PT}=\tau \sigma _{z}$ with $\tau $ denoting complex
conjugation. When $\left\vert \alpha \right\vert <1$ the last relation in (\ref{PTH}) no longer holds and the
eigenvalues become complex conjugate to each other, this is precisely what we described above as spontaneously broken $\mathcal{PT}$-symmetry. For the
parameter range of the latter situation this Hamiltonian would be regarded
as non-physical from a quantum mechanical point of view as it possesses
channels of infinite grows in probability, such that the corresponding time
evolution operators would be unbounded.

However, when one introduces an explicit time-dependence into the
Hamiltonian, $H(t)$, it no longer plays the role of the
observable energy operator and so we are not presented with an interpretational obstacle. For a meaningful physical picture one only
needs to guarantee now that the expectation values of $\tilde{H}(t)$, as
defined in (\ref{HenergyO}), are real and instead identify a new $\widetilde{%
\mathcal{PT}}$-symmetry to be responsible for this property%
\begin{equation}
\left[ \widetilde{\mathcal{PT}},\tilde{H}(t)\right] =0,\qquad \text{and\qquad }%
\widetilde{\mathcal{PT}}\tilde{\varphi}_{\pm }\left(t\right)=e^{i\tilde{\phi}\left(t\right)}\tilde{%
\varphi}_{\pm }\left(t\right),  \label{24}
\end{equation}%
with $\tilde{\varphi}_{\pm }\left(t\right)$ denoting the wave functions of $\tilde{H}$ and $%
\tilde{\phi}\in \mathbb{R}$. Notice that $\mathcal{PT}$ and $\widetilde{%
\mathcal{PT}}$ are only symbols here to denote different types of antilinear
operators, which however do not send $t$ to $-t$ as the time is only a real
parameter in this context.\\

In order to proceed we use the solution for the Dyson map we obtained by solving the time-dependent Dyson equation directly in section \ref{DysonSection} using the initial condition $\rho\left(0\right)=\mathbb{I}$

\begin{eqnarray}
\eta_{0}&=&\frac{1}{\omega}\left[\alpha^2-\cos^2\left(\frac{1}{2}\omega\mu\left(t\right)\right)\right]^{1/2},\\
\eta_{x}&=&\frac{\sin\left(\omega\mu\left(t\right)\right)}{2\left[\alpha^2-\cos^2\left(\frac{1}{2}\omega\mu\left(t\right)\right)\right]^{1/2}},\\
\eta_{y}&=&-\frac{\alpha\left[1-\cos\left(\omega\mu\left(t\right)\right)\right]}{2\omega\left[\alpha^2-\cos^2\left(\frac{1}{2}\omega\mu\left(t\right)\right)\right]^{1/2}},\\
\eta_{z}&=&0,
\end{eqnarray}
where $\omega=\sqrt{\alpha^2-1}$ and $\mu\left(t\right)=\int^t_0\kappa\left(s\right)ds$. In addition we have the counterpart Hermitian Hamiltonian

\begin{equation}
h\left(t\right)=-\frac{1}{2}\left[\Omega\mathbb{I}+\frac{\alpha\omega^2\kappa}{\alpha^2-\cos^2\left(\frac{1}{2}\omega\mu\left(t\right)\right)}\sigma_z\right]
\end{equation}
and importantly, the energy operator $\tilde{H}\left(t\right)$

\begin{equation}\label{HT}
\tilde{H}\left(t\right)=-\frac{1}{2}\left[\Omega\mathbb{I}+\chi\left(i\eta_{0}\eta_{y}\sigma_{x}-i\eta_{0}\eta_{x}\sigma_{y}+\left(\frac{\eta_{0}\eta_{y}}{\alpha}-\frac{1}{2}\right)\sigma_{z}\right)\right],
\end{equation}
with
\begin{equation}
\chi\left(t\right)=\frac{\alpha\omega^2\kappa}{\alpha^2-\cos^2\left(\frac{1}{2}\omega\mu\left(t\right)\right)}.
\end{equation}
The characteristic $\omega=\sqrt{\alpha^2-1}$ appears prominently in the parameters and it is clear that the behaviour must change as we pass through the exceptional point at $|\alpha|=1$. Therefore we now analyse in detail the qualitatively different regimes $|\alpha|>1$, $|\alpha|<1$ and $\alpha=1$ corresponding to the original $\mathcal{PT}$-symmetry of the non-Hermitian Hamiltonian.

\subsection{The unbroken $\protect\mathcal{PT}$ regime of $H\left(t\right)$}

\begin{figure}[H]
	\centering
	\includegraphics[scale=0.7]{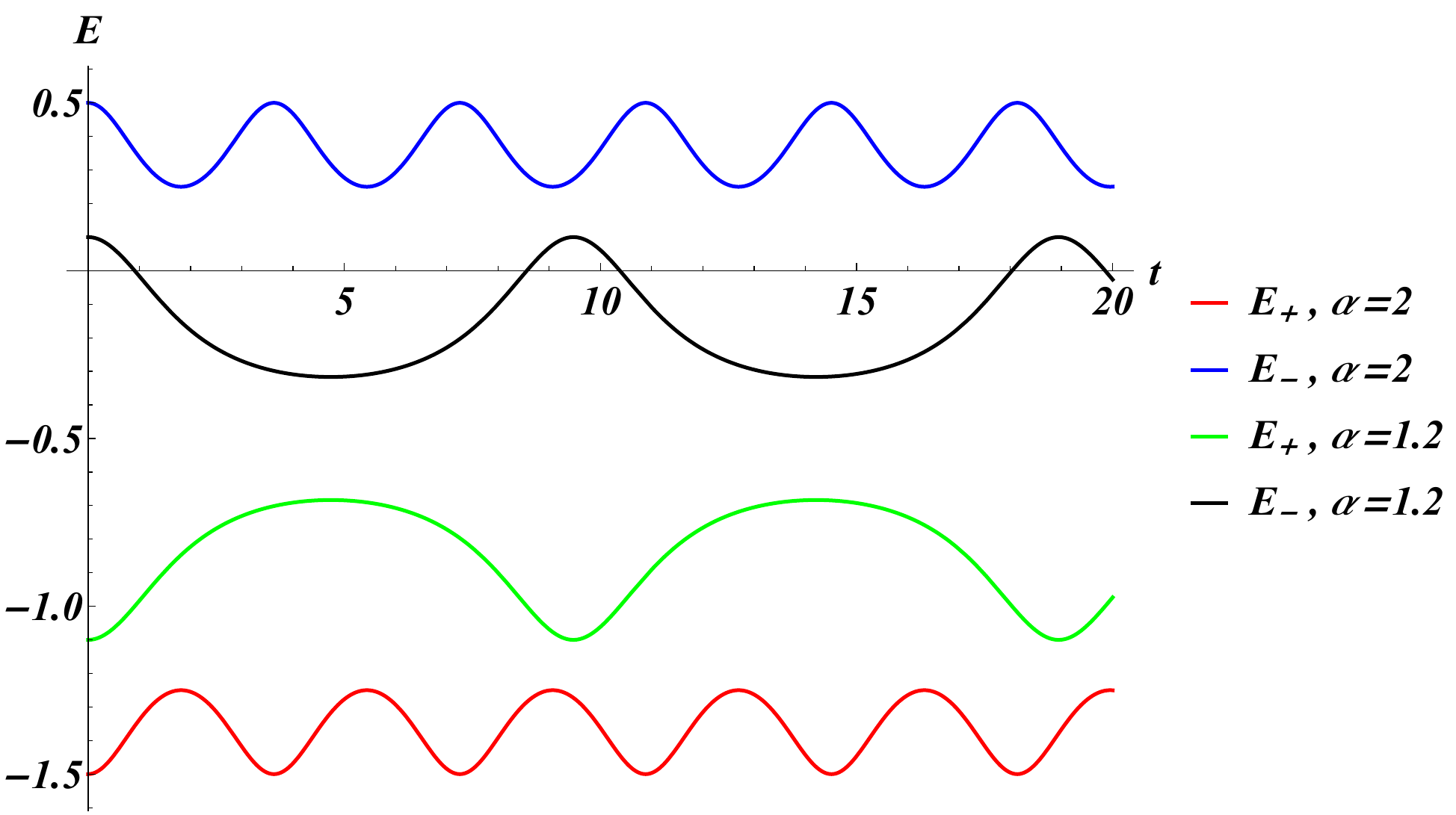}
	\caption{Energy observables $\protect\tilde{E}\protect_{\protect\pm}\left(t\right)$ for $\protect\kappa=1$ and $\protect\omega=1$ with $|\protect\alpha|\protect>1$, corresponding to the $\protect\mathcal{PT}$ unbroken regime.}\label{tunbroken}
\end{figure}

\begin{figure}[H]
	\centering
	\includegraphics[scale=0.7]{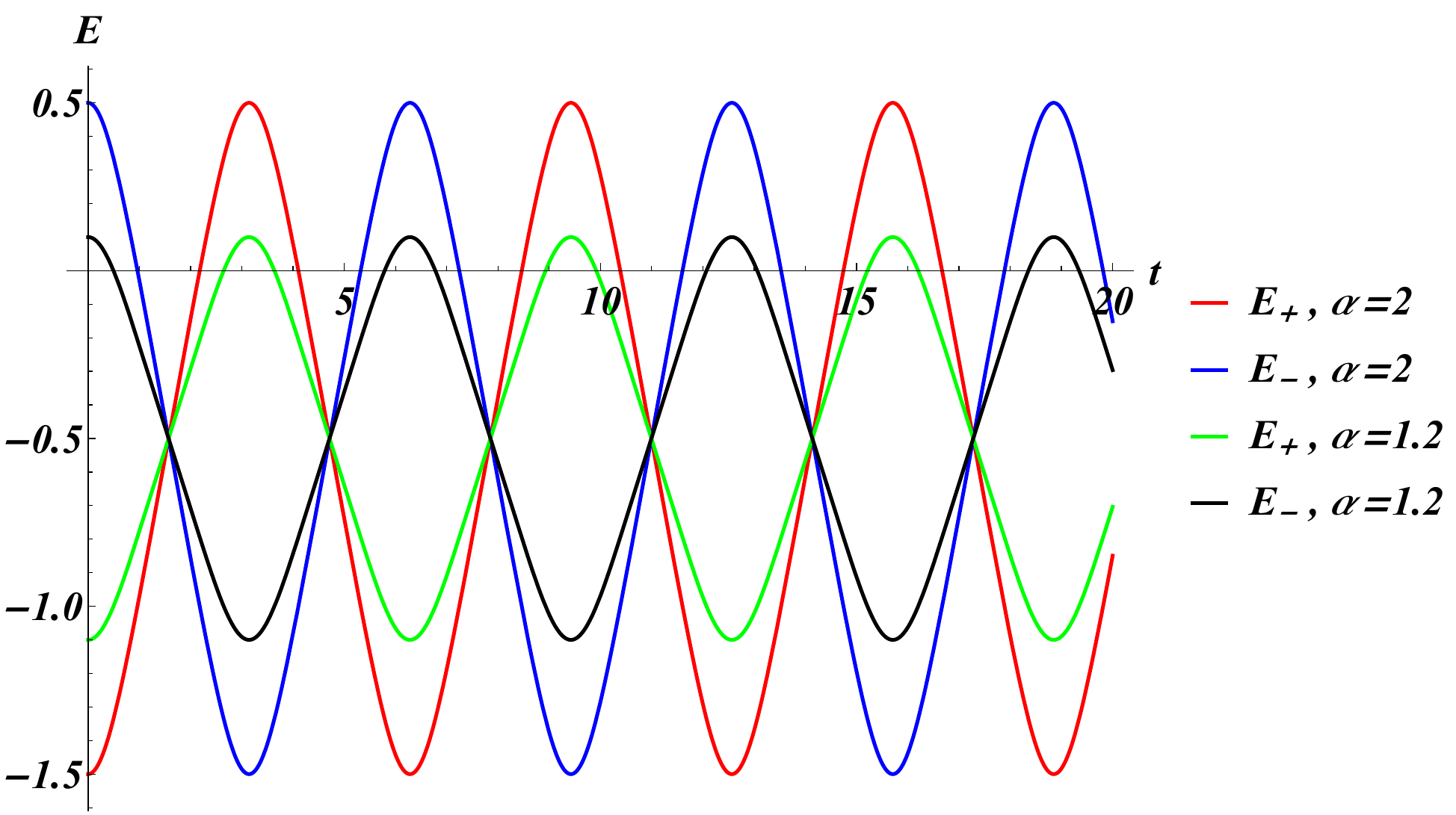}
	\caption{Energy observables $\protect\tilde{E}\protect_{\protect\pm}\left(t\right)$ for $\protect\kappa=\protect\cos t$ and $\protect\omega=1$ with $|\protect\alpha|\protect>1$, corresponding to the $\protect\mathcal{PT}$ unbroken regime.}\label{cosunbroken}
\end{figure}

In this section we will consider the regime with $|\alpha|>1$. As we saw in section \ref{ModelPT}, under these conditions the $\mathcal{PT}$-symmetry of $H\left(t\right)$ is unbroken and we have unitary time-evolution. The Dyson map and the metric are real for any given initial condition. The observable energy eigenvalues coming from the energy operator and wave functions $\bra{\varphi_\pm}\rho\left(t\right)\tilde{H}\left(t\right)\ket{\varphi_\pm}$ are

\begin{equation}
\tilde{E}_\pm\left(t\right)=-\frac{1}{2}\left[\Omega\pm\frac{\alpha\omega^2\kappa}{\alpha^2-\cos^2\left(\frac{1}{2}\omega\mu\left(t\right)\right)}\right].
\end{equation}
This is to be expected as even in the time-independent case the energy expectation values are real when the $\mathcal{PT}$-symmetry is intact. This is however a new time-dependent energy expectation value that differs from the time-independent case. 
Figures (\ref{tunbroken}) and (\ref{cosunbroken}) show plots of these energies for varying values of $\alpha$ and the function $\kappa=1$ and $\kappa=\cos t$. We see that in both cases the behaviour is oscillatory with $\alpha$ determining the period and magnitude of oscillation.

\subsection{The broken $\protect\mathcal{PT}$ regime of $H\left(t\right)$}

\begin{figure}[H]
	\centering
	\includegraphics[scale=0.7]{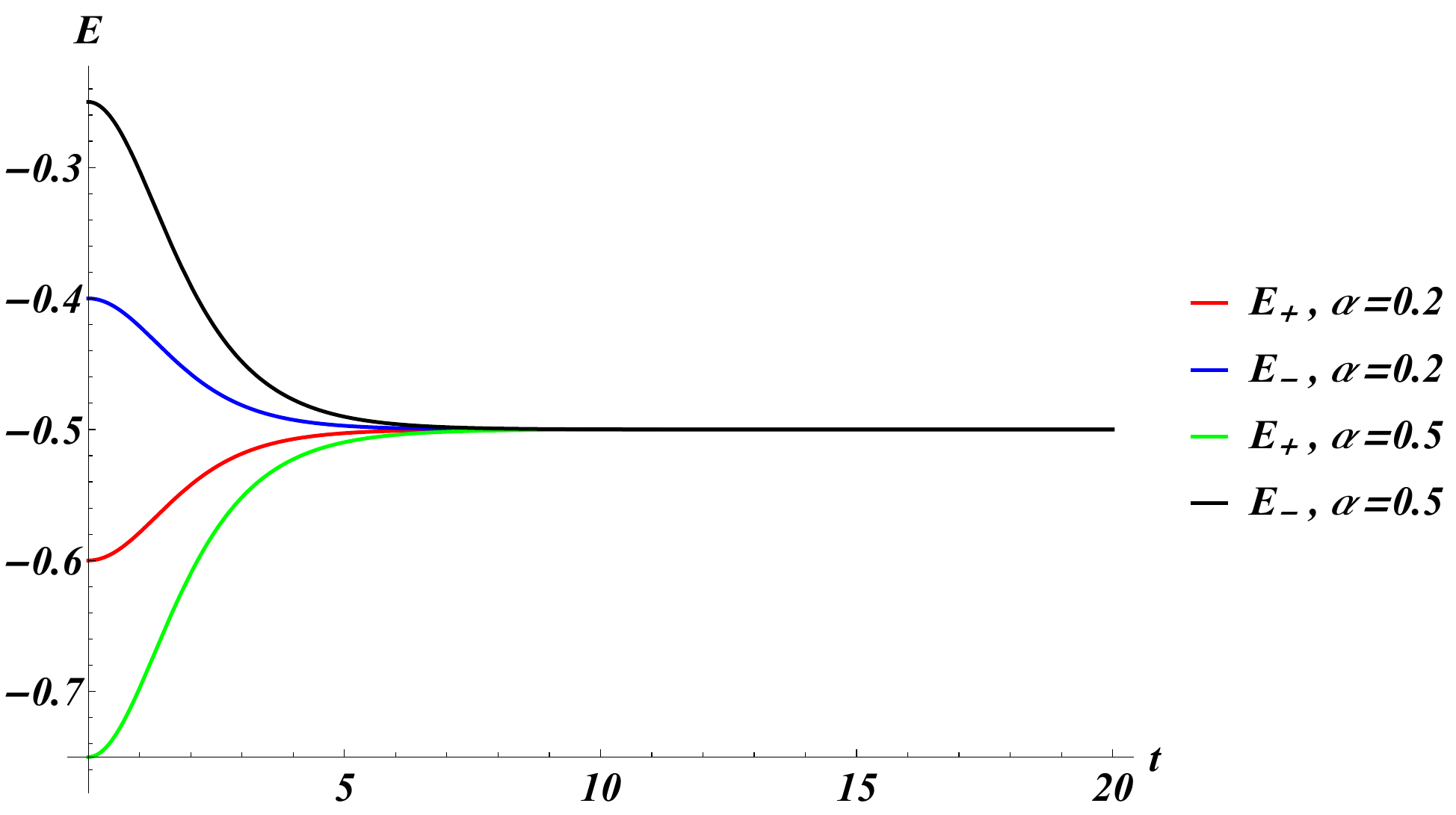}
	\caption{Energy observables $\protect\tilde{E}\protect_{\protect\pm}\left(t\right)$ for $\protect\kappa=1$ and $\protect\omega=1$ with $|\protect\alpha|\protect<1$, corresponding to the $\protect\mathcal{PT}$ broken regime.}\label{tbroken}
\end{figure}

\begin{figure}[H]
	\centering
	\includegraphics[scale=0.7]{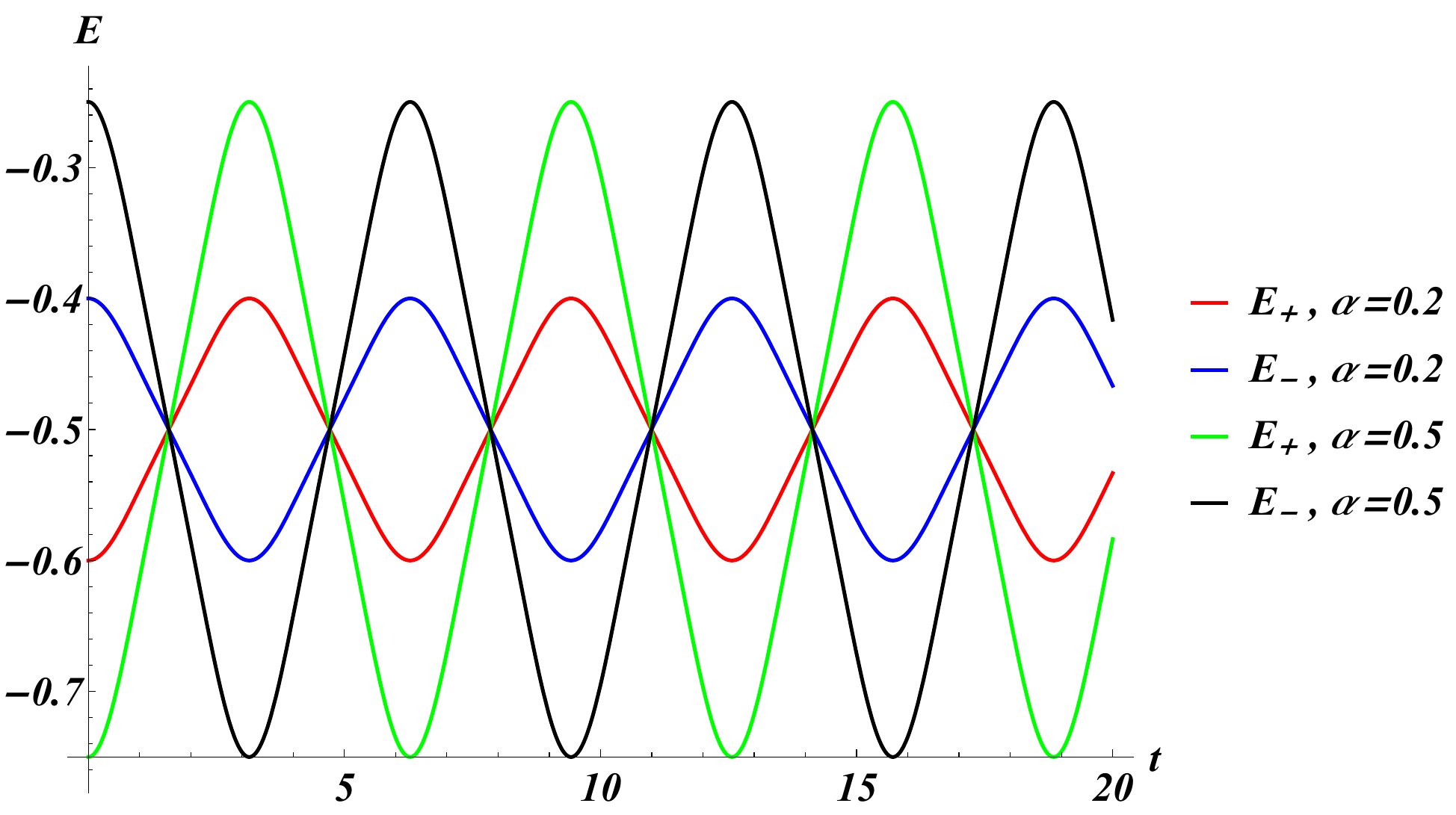}
	\caption{Energy observables $\protect\tilde{E}\protect_{\protect\pm}\left(t\right)$ for $\protect\kappa=\protect\cos t$ and $\protect\omega=1$ with $|\protect\alpha|\protect<1$, corresponding to the $\protect\mathcal{PT}$ broken regime.}\label{cosbroken}
\end{figure}

We now turn our attention to the regime with $|\alpha|<1$. This is the most interesting case because of the fact that time-evolution becomes non-unitary in the non-Hermitian system. As we have discussed this would usually mean the regime has unbounded probability. However we will show in this section that the introduction of time mends the broken $\mathcal{PT}$-symmetry with the introduction of a new $\widetilde{\mathcal{PT}}$-symmetry for the energy operator $\tilde{H}\left(t\right)$. In this case the components of $\eta\left(t\right)$ take the form

\begin{eqnarray}
\eta_{0}&=&\frac{1}{\tilde{\omega}}\left[\cosh^2\left(\frac{1}{2}\tilde{\omega}\mu\left(t\right)\right)-\alpha^2\right]^{1/2},\\
\eta_{x}&=&\frac{\sinh\left(\tilde{\omega}\mu\left(t\right)\right)}{2\left[\cosh^2\left(\frac{1}{2}\tilde{\omega}\mu\left(t\right)\right)-\alpha^2\right]^{1/2}},\\
\eta_{y}&=&-\frac{\alpha\left[1-\cosh\left(\tilde{\omega}\mu\left(t\right)\right)\right]}{2\tilde{\omega}\left[\cosh^2\left(\frac{1}{2}\tilde{\omega}\mu\left(t\right)\right)-\alpha^2\right]^{1/2}}.\\
\eta_{z}&=&0,
\end{eqnarray}
where $\tilde{\omega}=\sqrt{1-\alpha^2}$. These are real for all values of $|\alpha|<1$ and so we can define a metric and therefore an inner product. We see that the behaviour changes from trigonometric to hyperbolic evolution in time. Therefore the reality of the expectation value of $\tilde{H}\left(t\right)$ is preserved and so the energy observables are

\begin{equation}
\tilde{E}_\pm\left(t\right)=\bra{\varphi_\pm}\rho\left(t\right)\tilde{H}\left(t\right)\ket{\varphi_\pm}=-\frac{1}{2}\left[\Omega\pm\frac{\alpha\tilde{\omega}\kappa}{\cosh^2\left(\frac{1}{2}\tilde{\omega}\mu\left(t\right)\right)-\alpha^2}\right].
\end{equation}
Figure (\ref{tbroken}) and (\ref{cosbroken}) show the energy expectation values in the broken regime for various values of $\alpha<1$ and the function $\kappa=1$ and $\kappa=\cos t$. We see substantially different behaviour for $\kappa=1$. In this case the energy expectation values decay to $-\omega/2$. For $\kappa=\cos t$ we once again see oscillations on the energy, but in this case the amplitude is smaller than in the unbroken regime.

\newpage

We now wish to find the $\widetilde{\mathcal{PT}}$-symmetry that explains the reality of this observable. To do this, we make an ansatz for  $\widetilde{\mathcal{PT}}$ and solve the first equation in (\ref{PTH}). Indeed we find as the unique solution the antilinear operator

\begin{equation}\label{PTHT}
\widetilde{\mathcal{PT}}=\frac{1}{\xi}\left[2i\alpha \eta_{0}\eta_{x}\sigma_{y}-\left(2\eta_{0}\eta_{y}-\alpha\right)\sigma_{z}\right]\tau,
\end{equation}
with
\begin{equation}
\xi=\sqrt{\alpha\left(\alpha-4\eta_{0}\eta_{y}\right)+4\eta_{0}^2\left(-\alpha^2\eta_{x}^2+\eta_{y}^2\right)}.
\end{equation} 
We verify that $\widetilde{\mathcal{PT}}$ is involutionary with $\widetilde{\mathcal{PT}}^2=\mathbb{I}$. Furthermore we verify that $\widetilde{\mathcal{PT}}\sigma _{x}\widetilde{\mathcal{PT}}=-\sigma ^{x}$ and $\widetilde{\mathcal{PT}}\sigma _{z}\widetilde{\mathcal{PT}}\neq \sigma _{z}$. Thus when $\alpha \neq 0$ the new $\widetilde{\mathcal{PT}}$-symmetry is not a symmetry of $H(t)$, i.e. we have $\left[\widetilde{\mathcal{PT}},H(t)\right] \neq 0$ but $\left[\widetilde{\mathcal{PT}},\tilde{H}(t)\right] =0$. In order to guarantee that this symmetry is unbroken we also need to
satisfy the second equation in (\ref{24}). We determine the eigenstates of $%
\tilde{H}(t)$ as%
\begin{equation}
\tilde{\varphi}_{\pm }\sim \left( 
\begin{array}{c}
-\left(2\eta_{0}\eta_{y}-\alpha\right)\pm\sqrt{\xi^2-4\alpha^2\eta_{0}^2\eta_{y}^2} \\ 
2\alpha\eta_{0}\left(\eta_{x}+i\eta_{y}\right)%
\end{array}%
\right) ,
\end{equation}%
where the square root is always positive and verify that these vectors are indeed $\widetilde{\mathcal{PT}}$-eigenstates 
\begin{equation}
\widetilde{\mathcal{PT}}\tilde{\varphi}_{\pm }=e^{i\tilde{\omega}_{\pm }}%
\tilde{\varphi}_{\pm },
\end{equation}%
with 
\begin{eqnarray}
\tilde{\omega}_{\pm} &=&\arctan \left[ \frac{4\alpha^2\eta_{0}^2\eta_{x}\eta_{y}}{-\left(2\eta_{0}\eta_{y}-\alpha\right)\sqrt{\xi^2-4\alpha^2\eta_{0}^2\eta_{y}^2}\pm\xi^2}\right].
\end{eqnarray}%
Thus for the regime stated above the $\widetilde{\mathcal{PT}}$-symmetry is
unbroken and the eigenvalues of $\tilde{H}(t)$ are therefore guaranteed to
be real. We notice that for $|\alpha|>1$ the Hamiltonian $H(t)$ is in its $%
\mathcal{PT}$-symmetric phase, but $\widetilde{\mathcal{PT}}$ is still not
a symmetry for $H(t)$.\\

\subsection{The exceptional point of $H\left(t\right)$}

The value $\alpha =1$ is an exceptional point for $H(t)$ as it marks the
transition from real to complex conjugate expectation values and at the same time
the two expectation coalesce. For $\tilde{H}$ it also indicates the
boundary of the expectation values, but they do not become complex conjugate
to each other and the two expectation values remain different. In this case we reconsider our solution to the EP equation (\ref{ErmakovP}) we initially solved in chapter 2

\begin{equation}
\ddot{\sigma}+\lambda (t)\sigma =\frac{1}{\sigma ^{3}},
\end{equation}
with the time-dependent coefficient now 
\begin{equation}
\lambda (t)=\frac{1}{2}\frac{\ddot{\kappa}}{\kappa }-\frac{3}{4}\left( \frac{\dot{\kappa}}{\kappa}\right) ^{2}.
\end{equation}
The solution to this EP equation is radically different to our previous solution, with the initial condition $\rho\left(0\right)=\mathbb{I}$, $\sigma$ is

\begin{equation}
\sigma (t)=\sqrt{\frac{1}{\kappa }}\left[ \frac{1}{2}\mu\left(t\right)^2 +2\right] ^{1/2},
\end{equation}
where $\mu\left(t\right)=\int^t_0\kappa\left(s\right)ds$. This gives the solutions for the components of $\eta\left(t\right)$

\begin{eqnarray}
\eta_{0}&=&\left[ \frac{1}{4}\mu\left(t\right)^2 +1\right] ^{1/2},\\
\eta_{x}&=&-\frac{\frac{1}{2}\mu\left(t\right)}{\left[ \frac{1}{4}\mu\left(t\right)^2 +1\right] ^{1/2}},\\
\eta_{y}&=&-\frac{ \frac{1}{4}\mu\left(t\right)^2}{\left[\frac{1}{4}\mu\left(t\right)^2 +1\right]^{1/2}},\\
\eta_{z}&=&0.
\end{eqnarray}
So rather than being either trigonometric or hyperbolic, at the exceptional point the parameters are linear in the quantity $\int^t_0\kappa\left(s\right)ds$. Using these new Dyson map components we find the $\widetilde{\mathcal{PT}}$ operator to be the same as (\ref{PTHT}) with $\alpha=1$,

\begin{equation}
\widetilde{\mathcal{PT}}=\frac{1}{\xi}\left[2i \eta_{0}\eta_{x}\sigma_{y}-\left(2\eta_{0}\eta_{y}-1\right)\sigma_{z}\right]\tau
\end{equation}
with

\begin{equation}
\xi=\sqrt{\left(1-4\eta_{0}\eta_{y}\right)+4\eta_{0}^2\left(-\eta_{x}^2+\eta_{y}^2\right)}.
\end{equation}  
The energy operator is

\begin{equation}
\tilde{H}\left(t\right)=-\frac{1}{2}\left[\Omega\mathbb{I}+\chi\left(i\eta_{0}\eta_{y}\sigma_{x}-i\eta_{0}\eta_{x}\sigma_{y}+\left(\eta_{0}\eta_{y}-\frac{1}{2}\right)\sigma_{z}\right)\right],
\end{equation}
and the Hermitian Hamiltonian is

\begin{equation}
h\left(t\right)=-\frac{1}{2}\left[\Omega\mathbb{I}+\frac{4\kappa}{4+\mu\left(t\right)^2}\sigma_z\right].
\end{equation}
Therefore we calculate the energy expectation values for the  wave functions $\varphi_\pm\left(t\right)$ to be

\begin{equation}
\tilde{E}_\pm\left(t\right)=-\frac{1}{2}\left[\Omega\pm\frac{4\kappa}{4+\mu\left(t\right)^2}\right].
\end{equation}

\begin{figure}[H]
	\centering
	\includegraphics[scale=0.7]{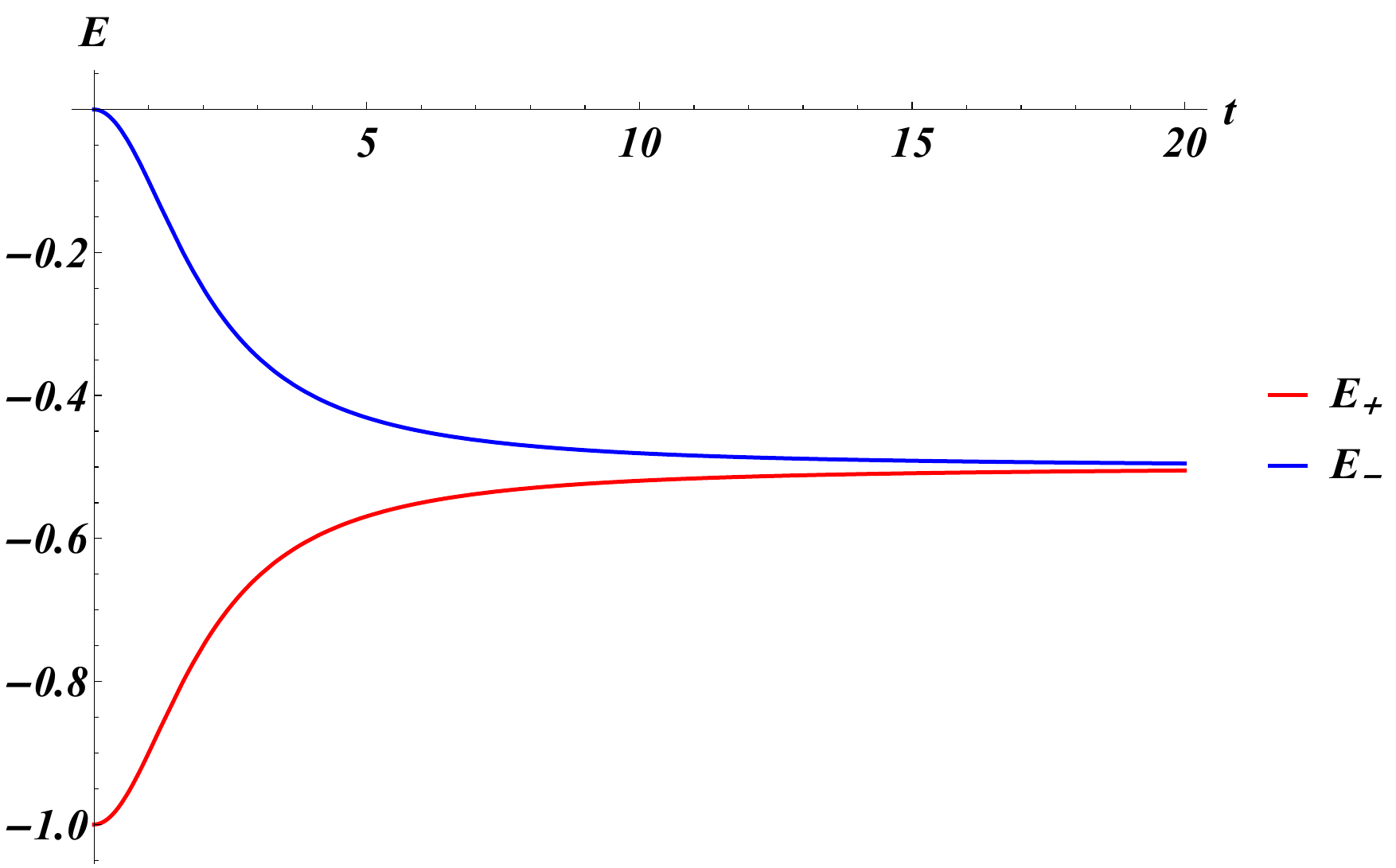}
	\caption{Energy observables $\protect\tilde{E}\protect_{\protect\pm}\left(t\right)$ for $\protect\kappa=1$ and $\protect\omega=1$ with $\protect\alpha=1$, corresponding to the exceptional point.}\label{texceptional}
\end{figure}

\begin{figure}[H]
	\centering
	\includegraphics[scale=0.7]{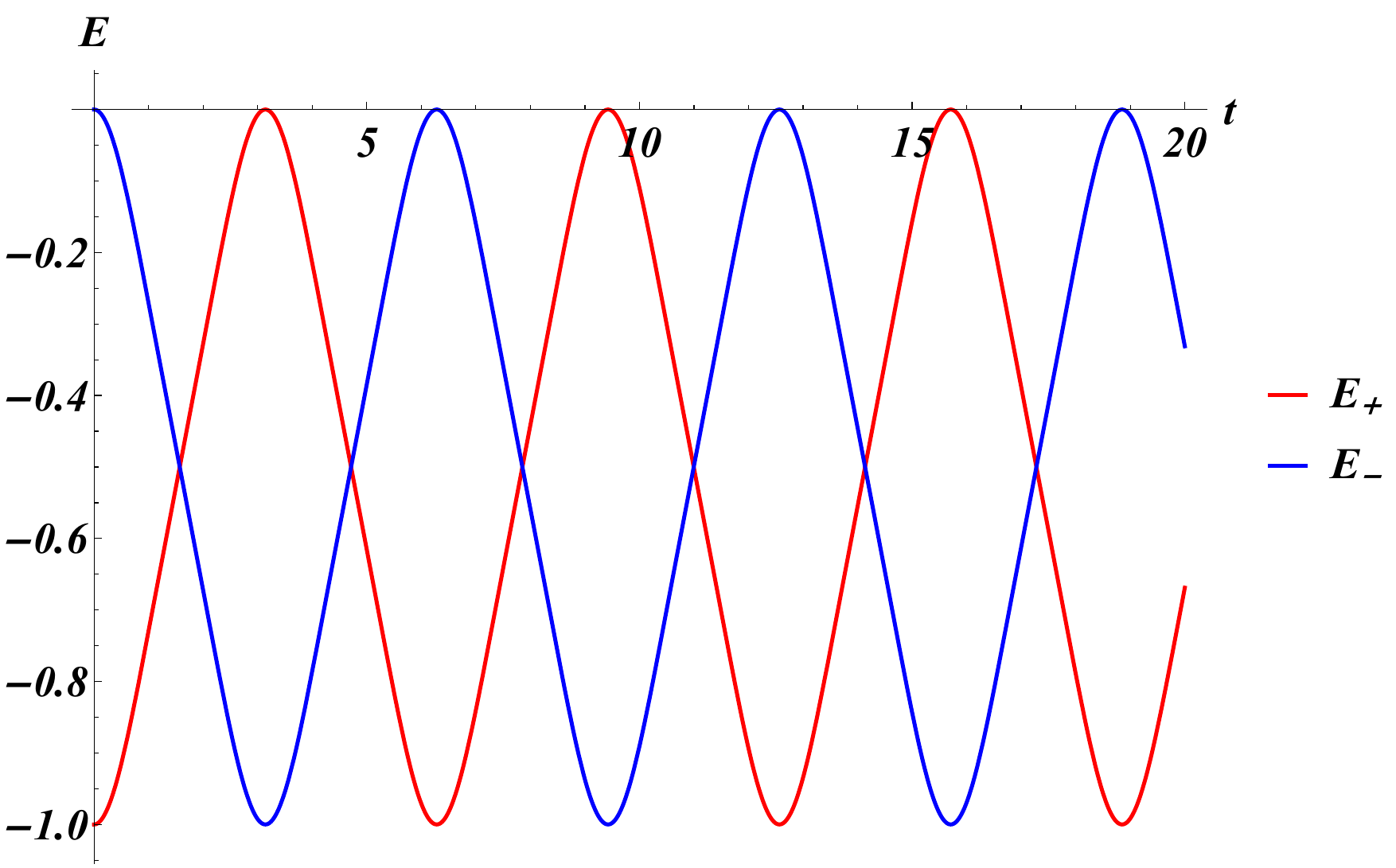}
	\caption{Energy observables $\protect\tilde{E}\protect_{\protect\pm}\left(t\right)$ for $\protect\kappa=\protect\cos t$ and $\protect\omega=1$ with $\protect\alpha=1$, corresponding to the exceptional point.}\label{cosexceptional}
\end{figure}

Figure (\ref{texceptional}) and (\ref{cosexceptional}) show the energy expectation values at the exceptional point $\alpha=1$ for the function $\kappa=1$ and $\kappa=\cos t$. Once again we see some very interesting and unique behaviour here. For $\kappa=1$ the energy decays asymptotically to $-\omega/2$. For $\kappa=\cos t$ the energy oscillates between $0$ and $-1$.

\subsection{The special point at $\protect\alpha=0$}

The value $\alpha =0$ is special as in this case the $\widetilde{\mathcal{PT}%
}$-operator commutes with both $\tilde{H}(t)$ and $H(t)$, but the
eigenvalues of the latter (\ref{H2}) are complex conjugate in this case. In addition, we are unable to satisfy the initial condition of $\rho\left(0\right)=\mathbb{I}$ without the system becoming trivial as this would mean $h\left(t\right)\propto\mathbb{I}$.
This means we expect the wave functions of $H(t)$ not to be eigenstates of the 
$\widetilde{\mathcal{PT}}$-operator. It is instructive to verify this in
detail and since the formulae simplify substantially in this case, it is
also useful to have a simpler example at hand. The energy operator simplifies to%
\begin{equation}
\tilde{H}\left(t\right)=-\frac{1}{2}\left[\omega\mathbb{I}+\frac{\chi}{\delta}\left(i\eta_{0}\eta_{y}\sigma_{x}-i\eta_{0}\eta_{x}\sigma_{y}-\left(\eta_{0}^2+\frac{\delta}{2}\right)\sigma_{z}\right)\right],
\end{equation}
and the $\widetilde{\mathcal{PT}}$-operator reduces to%
\begin{equation}
\widetilde{\mathcal{PT}}=\left[i\sinh\left(\mu\left(t\right)+\tilde{c}_2\right)\sigma _{y}+\cosh\left(\mu\left(t\right)+\tilde{c}_2\right)\sigma _{z}\right] \tau .
\end{equation}%
Now both Hamiltonians are $\widetilde{\mathcal{PT}}$-symmetric, i.e. in
addition to $\left[ \widetilde{\mathcal{PT}},\tilde{H}(t)\right] =0$ we also
have $\left[ \widetilde{\mathcal{PT}},H(t)\right] =0$. However, whereas the
eigenvectors $\tilde{\varphi}_{+}\sim \{-\eta _{0},\eta _{x}+i\eta _{y}\}$, $%
\tilde{\varphi}_{-}\sim \{\eta _{x}-i\eta _{y},\eta_0\}$ of $\tilde{H}(t)$
are $\widetilde{\mathcal{PT}}$-symmetric, the eigenvectors $\varphi _{\pm
}\sim \{\pm 1,1\}$ of $H(t)$ are not eigenstates of the $\widetilde{\mathcal{%
PT}}$-operator. Hence we have 
\begin{equation}
\widetilde{\mathcal{PT}}\varphi _{\pm }\neq e^{i\omega _{\pm }}\varphi _{\pm
}\qquad \text{and\qquad }\widetilde{\mathcal{PT}}\tilde{\varphi}_{\pm }=e^{i%
\tilde{\omega}_{\pm }}\tilde{\varphi}_{\pm }.
\end{equation}%
Concretely we identify 
\begin{equation}
\tilde{\omega}_{\pm }=\arctan\left[\pm\frac{2c^2\eta_{0}^3\eta_{x}}{\delta\eta_{0}^2\eta_{x}^2+c^4\delta -2c^4\eta_{0}^2}\right] .
\end{equation}%
Thus the $H(t)$ system is always in the spontaneously broken $\widetilde{%
\mathcal{PT}}$-symmetry phase, whereas $\tilde{H}(t)$ is $\widetilde{\mathcal{%
PT}}$-symmetric as long as $|\tilde{c}_{1}|>2$.

\section{Higher spin systems with spontaneously broken $\mathcal{PT}$-symmetry}

So far we have only considered a 2 level matrix model characterised by the Pauli spin matrices. These form the generators of the SU(2) Lie Algebra. However, they are only one particular representation and we can in fact write the algebra in terms of more general spin operators

\begin{equation}
\left[S_i,S_j\right]=i\epsilon_{ijk}S_k, \qquad i,j,k=x,y,z.
\end{equation}
The proper representation of $SU\left(2\right)$ in terms of $2\times2$ matrices is $S_i=\frac{1}{2}\sigma_i$. These matrices then describe systems of spin 1/2 particles as we have seen in the previous examples. In addition to the $2\times2$ representation, we can also form this algebra in terms of higher dimensional matrices that correspond to higher spin systems. For example, for spin 1 particles, the spin operators take the form

\begin{equation}
S_x^1=\frac{1}{\sqrt{2}}\left(\begin{array}{ccc}
0 & 1 & 0\\
1 & 0 & 1\\
0 & 1 & 0
\end{array}\right), \quad S_y^1=\frac{1}{\sqrt{2}}\left(\begin{array}{ccc}
0 & -i & 0\\
i & 0 & -i\\
0 & i & 0
\end{array}\right), \quad S_z^1=\left(\begin{array}{ccc}
1 & 0 & 0\\
0 & 0 & 0\\
0 & 0 & -1
\end{array}\right).
\end{equation}
and  for spin 3/2 the spin operators take the form

\begin{eqnarray}
S_x^{3/2}&=&\frac{1}{2}\left(\begin{array}{cccc}
0 & \sqrt{3} & 0 & 0\\
\sqrt{3} & 0 & 2 & 0\\
0 & 2 & 0 & \sqrt{3}\\
0 & 0 & \sqrt{3} & 0
\end{array}\right),\\ S_y^{3/2}&=&\frac{1}{2}\left(\begin{array}{cccc}
0 & -i\sqrt{3} & 0 & 0\\
i\sqrt{3} & 0 & -i2 & 0\\
0 & i2 & 0 & -i\sqrt{3}\\
0 & 0 & i\sqrt{3} & 0
\end{array}\right),\nonumber\\ S_z^{3/2}&=&\frac{1}{2}\left(\begin{array}{cccc}
3 & 0 & 0 & 0\\
0 & 1 & 0 & 0\\
0 & 0 & -1 & 0\\
0 & 0 & 0 & -3
\end{array}\right).\nonumber
\end{eqnarray}
We now show that our analysis works when we extend our representations to higher spin systems \cite{AndTom2}.

\subsection{A spin 1 model}

We look at a spin 1 model built from the SU(2) spin generators

\begin{eqnarray}\label{H1matrix}
H^{1}\left(t\right)&=&-\frac{1}{2}(\Omega\mathbb{I}+\sqrt{2}S^1_{y}\gamma\left(t\right)+i\sqrt{2}\alpha\gamma\left(t\right) S^1_{x})\\
&=&-\frac{1}{2}\left( 
\begin{array}{ccc}
\omega & i(\alpha -1)\gamma\left(t\right) & 0 \\ 
i(\alpha +1)\gamma\left(t\right) & \omega & i(\alpha -1)\gamma\left(t\right) \\ 
0 & i(\alpha +1)\gamma\left(t\right) & \omega%
\end{array}%
\right) . \nonumber 
\end{eqnarray}%
Solving the time-dependent Schr\"odinger equation using the Lewis Riesenfeld invariants for this Hamiltonian, we obtain the time-dependent wave functions

\begin{equation}
\psi _{k}(t)=\frac{1}{2} e^{i\int^t\left(\frac{\Omega }{2}-k\phi\gamma\left(s\right)\right)ds}\left( 
\begin{array}{c}
(-1)^{k}(1-\alpha ) \\ 
2ik\phi \\ 
1+\alpha%
\end{array}%
\right),~~~\qquad 
~~~k=0,\pm 1
\end{equation}%
where $\phi:=\sqrt{(1-\alpha ^{2})/2}$ and from which we can construct a general wave function. Once again in the parameter
region $\left\vert \alpha \right\vert \leq 1$ the non-Hermitian Hamiltonian (%
\ref{H1matrix}) possesses a real eigenvalue spectrum. However, when $|\alpha|>1$ the system is no  longer well defined as we see the functions $\psi_k\left(t\right)$ become unbounded. This is once again the spontaneously broken $\mathcal{PT}$ region. In order to mend this broken symmetry we wish to follow the same procedure as for the 2 level system and solve the time-dependent Dyson equation for $\eta\left(t\right)$. We choose this method so as to avoid taking the square root of a $3\times3$ matrix. We make the most general Hermitian ansatz for $\eta\left(t\right)$

\begin{equation}
\eta (t)=\left( 
\begin{array}{ccc}
\eta _{1}(t) & \eta _{2}(t)-i\eta _{3}(t) & \eta _{4}(t)-i\eta _{5}(t) \\ 
\eta _{2}(t)+i\eta _{3}(t) & \eta _{6}(t) & \eta _{7}(t)-i\eta _{8}(t) \\ 
\eta _{4}(t)+i\eta _{5}(t) & \eta _{7}(t)+i\eta _{8}(t) & \eta _{9}(t)%
\end{array}%
\right) .
\end{equation}
and use the ansatz for $h\left(t\right)$

\begin{equation}
h\left(t\right)=-\frac{1}{2}\left[\Omega\mathbb{I}+X\left(t\right)S_z^1\right]
\end{equation}
Substituting these expressions into the time-dependent Dyson equation yields in principle $18$ equations for the real functions $\eta_{i}(t), $ $i=1,\ldots ,9$. We obtain

\begin{equation}
\begin{array}{l}
\dot{\eta}_{1}=\eta _{2}\frac{\alpha +1}{2}\gamma,\quad\quad  \dot{\eta}_{2}=\eta _{1}%
\frac{\alpha -1}{2}\gamma+\eta _{3}\frac{X}{2}\gamma+\eta _{4}\frac{\alpha +1}{2}\gamma=\eta
_{6}\frac{\alpha +1}{2}\gamma,\\
\begin{array}{c}
\\ 
\\
\end{array}
\hspace{-0.35cm}\dot{\eta}_{3}=-\eta _{2}\frac{X}{2}+\eta _{5}%
\frac{\alpha +1}{2}\gamma=0, \quad \quad \dot{\eta}_{4}=\eta _{2}\frac{\alpha -1}{2}\gamma+\eta _{5}\frac{X}{2}=\eta _{5}%
\frac{X}{2}+\eta _{7}\frac{\alpha +1}{2}\gamma,\\ 
\dot{\eta}_{5}=\eta _{3}%
\frac{\alpha -1}{2}\gamma-\eta _{4}\frac{X}{2}=\eta _{4}\frac{X}{2}-\eta _{8}\frac{%
	\alpha +1}{2}\gamma,%
\begin{array}{c}
\\ 
\\ 
\end{array}
\\ 
\dot{\eta}_{6}=\eta _{2}\frac{\alpha -1}{2}\gamma+\eta _{7}\frac{\alpha +1}{2}\gamma%
,\quad \quad \dot{\eta}_{7}=\eta _{4}\frac{\alpha -1}{2}\gamma+\eta _{8}\frac{X}{2}%
+\eta _{9}\frac{\alpha +1}{2}\gamma=\eta _{6}\frac{\alpha -1}{2}\gamma, \\ 
\dot{\eta}_{8}=\eta _{5}\frac{\alpha -1}{2}\gamma-\eta _{7}\frac{X}{2}=0,\quad
\quad \dot{\eta}_{9}=\eta _{7}\frac{\alpha -1}{2}\gamma,%
\begin{array}{c}
\\ 
\\ 
\end{array}%
\end{array}%
\end{equation}%
and

\begin{equation}
(1+\alpha )\gamma\eta _{3}-X\eta _{1}=(1-\alpha )\gamma\eta _{3}+(1+\alpha )\gamma\eta
_{8}=(1-\alpha )\gamma\eta _{8}+X\eta _{9}=0.  \label{eta10}
\end{equation}%
Unlike the system of equations for the 2 level system this set is highly
overdetermined. Nonetheless, they may be solved by%
\begin{equation}
\begin{array}{l}
\eta _{1}(t)=\frac{c_{1}\gamma}{X},~\quad \eta _{2}(t)=\frac{2c_{1}\left(\dot{\gamma}X-\gamma\dot{X}\right)}{%
	(1+\alpha )\gamma X^{2}},\quad ~\eta _{3}(t)=\frac{c_{1}}{1+\alpha },\\
\eta _{4}(t)=\frac{\left(\left(\alpha^2-1\right)\eta_1+\left(1+\alpha\right)\eta_6\right)\gamma^2-\eta_1X^2}{(1+\alpha )^{2}X^{2}}, \quad \eta _{5}(t)=\frac{2c_{1}\left(\dot{\gamma}X-\gamma\dot{X}\right)}{%
	(1+\alpha )^{2}\gamma^2 X},\\
\eta _{6}(t)=%
\frac{4c_{1}(-X^2\dot{\gamma}^2+\gamma^2\left(2\dot{X}^{2}-X\ddot{X}\right)-\gamma X\left(-\dot{\gamma}\dot{X}+\ddot{\gamma}X\right))}{(1+\alpha )^{2}\gamma^3X^{3}}%
, \quad \eta _{7}(t)=\frac{2c_{1}\left(\alpha-1\right)\left(\dot{\gamma}X-\gamma\dot{X}\right)}{%
	(1+\alpha )^2\gamma X^{2}}%
,\quad \quad 
\begin{array}{c}
\\ 
\\ 
\end{array}
\\ 
\eta _{8}(t)=\frac{c_{1}(\alpha -1)}{(1+\alpha)^{2}},\quad ~\eta _{9}(t)=%
\frac{c_{1}(1-\alpha )^{2}\gamma}{(1+\alpha )^{2}X},%
\end{array}%
\end{equation}%
where $X(t)$ is restricted to obey the second order non-linear differential
equation%

\begin{equation}
\ddot{X}-\frac{3}{2}\frac{\dot{X}^{2}}{X}-X\left(\frac{\ddot{\gamma}}{\gamma}-\frac{3\dot{\gamma}^2}{2\gamma^2}+\frac{1}{4}\left(1-\alpha^2\right)\gamma^2\right)+\frac{%
	X^{3}}{8}=0.  \label{chid}
\end{equation}%
This equations closely resembles (\ref{xi}) and we can once more transform
it to the Ermakov-Pinney equation (\ref{ErmakovP}) by using $X=4/\tilde{\sigma}^{2}$
in this case. Following the same steps as in the previous subsection we obtain
the general solution for (\ref{chid}) as 

\begin{equation}
X(t)=\frac{4\gamma\left(t\right)}{c_{2}\cos\left[\phi\left(\int^t\gamma\left(s\right)ds+c_3\right)\right]+\sqrt{c_{2}^{2}-\frac{8}{\alpha^2-1}}}.
\end{equation}%
This solution then defines the entirety of the Dyson map and so completes our procedure for the spin 1 model. We will now show that the procedure is also completely transferable to a spin 3/2 system.

\subsection{A spin 3/2 model}

We now consider a Hamiltonian built from the spin 3/2 operators,
\begin{eqnarray}
H^{3/2}\left(t\right)&=&-\frac{1}{2}(\omega\mathbb{I}+\frac{2}{3}\gamma\left(t\right) S^{3/2}_y+i\frac{2}{3}\alpha\gamma\left(t\right) S^{3/2}_x)\\
&=&-%
\frac{1}{2}\left( 
\begin{array}{cccc}
\omega & i\frac{\left(\alpha -1\right)\gamma\left(t\right)}{\sqrt{3}} & 0 & 0 \\ 
i\frac{\left(\alpha +1\right)\gamma\left(t\right)}{\sqrt{3}} & \omega & i\frac{2\left(\alpha -1\right)\gamma\left(t\right)}{3} & 0 \\ 
0 & i\frac{2\left(\alpha +1\right)\gamma\left(t\right)}{3} & \omega & \frac{\left(\alpha -1\right)\gamma\left(t\right)}{\sqrt{3}} \\ 
0 & 0 & \frac{\left(\alpha +1\right)\gamma\left(t\right)}{\sqrt{3}} & \omega%
\end{array}%
\right) . \nonumber
\end{eqnarray}
Solving the time-dependent Schr\"odinger equation using the Lewis Riesenfeld invariants for this Hamiltonian, we obtain the time-dependent wave functions

\begin{equation}
\psi _{k}(t)=\frac{\sqrt{3}}{\sqrt{8\left(|k|+\alpha^2\left(3-|k|\right)\right)}}e^{i\int^t\left(\frac{1}{2}\omega-k\hat{\phi}\gamma\left(s\right)\right)ds}\left( 
\begin{array}{c}
i(1-\alpha )^{3/2} \\ 
2\sqrt{3}k\hat{\phi}(1-\alpha )^{1/2} \\ 
2i\sqrt{3}(2\left\vert k\right\vert-k^{2} )\hat{\phi}(1+\alpha )^{1/2} \\ 
\limfunc{sign}(k)(2-\left\vert k\right\vert)(1+\alpha )^{3/2}%
\end{array}\right),
\end{equation}
where $k=\pm1, \pm3$ and $\hat{\phi}:=\sqrt{1-\alpha ^{2}}/6$ and from which we can construct a general wave function. We see that the system becomes ill-defined when $\alpha>1$. This is the region of broken $\mathcal{PT}$-symmetry and we now show that by solving the time-dependent Dyson equation, we mend this regime. We take $\eta (t)$ to be of the most general Hermitian form 

\begin{equation}
\eta (t)=\left( 
\begin{array}{cccc}
\eta _{1}(t) & \eta _{2}(t)-i\eta _{3}(t) & \eta _{4}(t)-i\eta _{5}(t) & 
\eta _{6}(t)-i\eta _{7}(t) \\ 
\eta _{2}(t)+i\eta _{3}(t) & \eta _{8}(t) & \eta _{9}(t)-i\eta _{10}(t) & 
\eta _{11}(t)-i\eta _{12}(t) \\ 
\eta _{4}(t)+i\eta _{5}(t) & \eta _{7}(t)+i\eta _{8}(t) & \eta _{13}(t) & 
\eta _{14}(t)-i\eta _{15}(t) \\ 
\eta _{6}(t)+i\eta _{7}(t) & \eta _{11}(t)+i\eta _{12}(t) & \eta
_{14}(t)+i\eta _{15}(t) & \eta _{16}(t)%
\end{array}%
\right) ,
\end{equation}
and assume $h\left(t\right)$ to be 

\begin{equation}
h\left(t\right)=-\frac{1}{2}\left[\Omega\mathbb{I}+\Xi S^{3/2}_z\right].
\end{equation}
Substituting these expressions into the time-dependent Dyson equation yields in principle $32$ equation for the $\eta _{i}(t),$ $i=1,\ldots,16$. Once again the system is highly overdetermined, but remarkably it can be solved similarly as in the previous sections. Here we only present the
solutions to these equations. We find%
\begin{equation}
\begin{array}{l}
\eta _{1}(t)=\frac{c_{1}\gamma^{3/2}}{\Xi ^{3/2}},~\quad \eta _{2}(t)=\frac{3\sqrt{3}%
		c_{1}\left(\Xi\dot{\gamma}-\dot{\Xi}\gamma\right)}{(1+\alpha )\Xi ^{5/2}\gamma^{1/2}},\quad ~\eta _{3}(t)=\frac{3\sqrt{3}c_{1}\gamma^{1/2}%
	}{2(1+\alpha )\Xi ^{1/2}},\quad ~\eta _{4}(t)=\frac{\left(\alpha+1\right)\eta_{11}}{(\alpha-1 )}, \\ 
	\eta _{5}(t)=\frac{9\sqrt{3}%
		c_{1}\left(\Xi\dot{\gamma}-\dot{\Xi}\gamma\right)}{(1+\alpha )^2\Xi ^{3/2}\gamma^{3/2}}
	,\quad ~\eta _{6}(t)=\eta_9+\frac{2\left(1-\alpha^2\right)\eta_{14}\gamma^2-9\Xi^2\eta_{14}}{\sqrt{3}(\alpha -1)^2\gamma^2},\\
	\eta_{7}(t)=\eta_{10}+\frac{\Xi\left(\sqrt{3}\left(\alpha-1\right)\eta_{11}-3\left(\alpha+1\right)\eta_{16}\right)}{(\alpha -1)^2\gamma},\quad \quad	\eta _{8}(t)=\frac{\left(\alpha+1\right)\eta_{13}}{\left(\alpha-1\right)},
		\begin{array}{c}
	\\ 
	\\ 
	\end{array}
	\\ 
	\eta_{9}(t)=\frac{\left(6\left(\alpha-1\right)\Xi^{5/2}\dot{\eta}_{13}-9c_1\left(\alpha+1\right)\gamma^{1/2}\left(\Xi\dot{\gamma}-\dot{\Xi}\gamma\right)\right)}{2(\alpha -1)^2\Xi^{5/2}\gamma}
	, 
	\eta _{10}(t)=\frac{3\Xi\left(\left(\alpha-1\right)\eta_{13}+3\left(\alpha+1\right)\eta_{16}\right)}{4(\alpha -1)^2\gamma},\\ \eta _{11}(t)=-\frac{\sqrt{3}c_1\left(\alpha-1\right)\left(27\Xi^4\gamma^2-180\dot{\Xi}^2\gamma^2+72\Xi\gamma\left(2\dot{\Xi}\dot{\gamma}+\ddot{\Xi}\gamma\right)+4\Xi^2\left(\left(\alpha^2-1\right)\gamma^4+9\dot{\gamma}^2-18\gamma\ddot{\gamma}\right)\right)}{8(1+\alpha )^{3}\Xi ^{7/2}\gamma^{5/2}},\quad ~%
	\begin{array}{c}
	\\ 
	\\ 
	\end{array}
	\\ 
	\eta _{12}(t)=\frac{9\sqrt{3}%
		c_{1}\left(\alpha-1\right)\left(\Xi\dot{\gamma}-\dot{\Xi}\gamma\right)}{(1+\alpha )^3\Xi ^{3/2}\gamma^{3/2}},\quad \eta _{13}(t)=\frac{9c_1\left(\alpha-1\right)\left(\Xi^4\gamma^2+20\dot{\Xi}^2\gamma^2-8\Xi\gamma\left(2\dot{\Xi}\dot{\gamma}+\ddot{\Xi}\gamma\right)-4\Xi^2\left(\dot{\gamma}^2-2\gamma\ddot{\gamma}\right)\right)}{4(1+\alpha )^{3}\Xi ^{7/2}\gamma^{5/2}},\quad ~ \\ 
	\eta _{14}(t)=\frac{3\sqrt{3}%
		c_{1}\left(\alpha-1\right)^2\left(\Xi\dot{\gamma}-\dot{\Xi}\gamma\right)}{(1+\alpha )^3\Xi ^{5/2}\gamma^{1/2}},\quad ~\eta _{15}(t)=\frac{3\sqrt{3}c_{1}(\alpha -1)^3\gamma^{1/2}}{2		(1+\alpha )^{3}\Xi ^{1/2}},\quad \eta _{16}(t)=\frac{c_{1}(\alpha -1)^3\gamma^{3/2}}{%
		(1+\alpha )^{3}\Xi ^{3/2}},%
	\begin{array}{c}
	\\ 
	\\ 
	\end{array}%
	\end{array}%
	\end{equation}%
where $\Xi (t)$ has to obey the second order non-linear differential equation%
\begin{equation}
\ddot{\Xi}-\frac{3}{2}\frac{\dot{\Xi}^{2}}{\Xi }-\Xi\left(\frac{\ddot{\gamma}}{\gamma}-\frac{3\dot{\gamma}^2}{2\gamma^2}+\frac{1}{18}\left(1-\alpha^2\right)\gamma^2\right) +\frac{\Xi ^{3}}{8}=0.
\end{equation}%
As in the previous subsection we can transform this equation to the Ermakov Pinney
equation (\ref{ErmakovP}) using $\Xi =4/\hat{\sigma}^{2}$ in this case and
therefore we have 
\begin{equation}
\Xi (t)=\frac{4\gamma\left(t\right)}{\hat{c}_{2}\cos\left[\phi\left(\int^t\gamma\left(s\right)ds+\hat{c}_3\right)\right]+\sqrt{\hat{c}_{2}^{2}-\frac{36}{\alpha^2-1}}}.
\end{equation}%
The solution for $\Xi\left(t\right)$ completes the solution for the Dyson map and allows us to compute the components of $\eta\left(t\right)$. We have shown that the framework of time-dependent analysis extends beyond our simple 2 level system and is clearly valid for higher representations. We will now go on to show how it works when applied to a harmonic oscillator system with infinite Hilbert space.

\section{An inverted harmonic oscillator with spontaneously broken $\mathcal{PT}$-symmetry}\label{1dharmonicoscillator}

So far we have only considered matrix models with explicit time-dependence. These models have finite Hilbert space and are extremely useful in approaching the time-dependent problem. In this section we will show that the utility of the time-dependent analysis extends to time-independent non-Hermitian Hamiltonians. In the time-independent regime, when the $\mathcal{PT}$-symmetry is spontaneously broken there is no way to interpret the system as the eigenvalues become complex and the time-evolution becomes non-unitary. However, when we use time-dependent analysis we can map the time-independent non-Hermitian Hamiltonian to a time-dependent Hermitian Hamiltonian using a time-dependent Dyson map.

In this section we will show that our analysis of spontaneously broken $\mathcal{PT}$-symmetry extends to systems with infinite Hilbert space. To do this we consider the time-independent inverted harmonic oscillator system with an additional $\mathcal{PT}$-symmetric term

\begin{equation}\label{1DHamiltonian}
H=\frac{1}{2m}p^2-\frac{1}{2}m\omega^2x^2+i\frac{g}{2}\{x,p\},
\end{equation}
where we work in position space such that $p=-i\partial_x$. Here the curly brackets $\{,\}$ denote the anticommutator between $x$ and $p$. Also $m$, $\omega$ and $g \in \mathbb{R}$. This system represents an inverted harmonic oscillator with an additional non-Hermitian term that regularises the system. Without this non-Hermitian term, the system is unbounded from below and therefore unphysical, even though it is Hermitian. The $\mathcal{PT}$-symmetry that leaves the Hamiltonian invariant is

\begin{equation}
\mathcal{PT}: \qquad p\rightarrow p, \qquad x\rightarrow -x, \qquad i\rightarrow -i.
\end{equation}
Solving the time-independent Schr\"odinger equation, we obtain the energy eigenvalues

\begin{equation}
E_n=\left(n+\frac{1}{2}\right)\Lambda,
\end{equation}
where $\Lambda=\sqrt{g^2-\omega^2}$ and the eigenfunctions

\begin{equation}
\phi_n\left(x\right)=\exp\left[-\frac{1}{2}m\left(\Lambda-g\right)x^2\right]\mathcal{H}_n\left(x\right),
\end{equation}
where $\mathcal{H}_n$ are the Hermite polynomials. It is clear that in the regime $|g|>|\omega|$ the $\mathcal{PT}$-symmetry is preserved and the eigenvalues are real. However, in the regime $|g|<|\omega|$, the symmetry is spontaneously broken as the eigenfunctions are no longer invariant under the $\mathcal{PT}$-symmetry. Therefore in the broken regime we are unable to make sense of the system without the use of time-dependent analysis. We will show this explicitly by first looking at the time-independent mapping and then proceeding to the time-dependent mapping.

\subsection{Time-independent Dyson map}

We wish to find a metric and a Dyson map for the Hamiltonian (\ref{1DHamiltonian}). To do this we solve the time-independent Dyson equation

\begin{equation}
h=\eta H \eta^{-1},
\end{equation}
with the ansatz 

\begin{equation}
\eta=e^{\frac{1}{2}\alpha x^2}.
\end{equation}
Using the BCH relation, we find the resulting Hermitian Hamiltonian to be

\begin{equation}
h=\frac{1}{2m}p^2+\frac{1}{2}m\Lambda^2x^2.
\end{equation}
with $\alpha=-mg$. $h$ is a Harmonic oscillator with frequency $\Lambda=\sqrt{g^2-\omega^2}$. Therefore when $|g|<|\omega|$ the frequency becomes complex and the wave functions become unbounded and therefore unphysical. This matches the condition for the $\mathcal{PT}$-symmetry to be spontaneously broken. This means that whilst we are able to define a Dyson map and therefore a metric $\rho$ in the unbroken regime, the spontaneously broken regime remains elusive without using time-dependent analysis. Therefore we now move onto the time-dependent Dyson equation.

\subsection{Time-dependent Dyson map}

Now we wish to solve the time-dependent Dyson equation (\ref{DysonE}) for the Hamiltonian (\ref{1DHamiltonian}). In this instance we use the creation and annihilation operators to simplify the problem, these are defined as follows

\begin{equation}
a=\frac{ip}{\sqrt{2m\omega}}+\sqrt{\frac{m\omega}{2}} x, \quad a^\dagger=-\frac{ip}{\sqrt{2m\omega}}+\sqrt{\frac{m\omega}{2}} x.
\end{equation}
Using these operators, we can rewrite the Hamiltonian as

\begin{equation}
H=-\frac{1}{2}\left(g+\omega\right){a^\dagger}^2+\frac{1}{2}\left(g-\omega\right)a^2.
\end{equation}
Now we use the non-Hermitian ansatz

\begin{equation}
\eta\left(t\right)=e^{\alpha\left(t\right)a^\dagger a}e^{\left[\beta\left(t\right)+i\gamma\left(t\right)\right]a^2},
\end{equation}

and substitute into the time-dependent Dyson equation. In order to make the resulting Hamiltonian Hermitian, the following differential equations must be satisfied
\begin{eqnarray}
\dot{\alpha}&=&\frac{1}{2}\left(g+\omega\right)\gamma,\label{alphad}\\
\dot{\beta}&=&\frac{1}{2}\left(g+\omega\right)\beta\gamma,\label{betad}\\
\dot{\gamma}&=&\frac{1}{2}\left(g-\omega+\left(g+\omega\right)e^{2\alpha}\right)-\frac{1}{4}\left(g+w\right)\left(\beta^2-\gamma^2\right).\label{gammad}
\end{eqnarray}
The Hermitian Hamiltonian is computed

\begin{equation}
h\left(t\right)=-\frac{1}{2}\left(g+\omega\right)\beta a^\dagger a-\frac{1}{2}\left(g+\omega\right)e^{\alpha}\left({a^\dagger}^2+a^2\right).
\end{equation}
In order to solve the coupled differential equations (\ref{alphad}), (\ref{betad}) and (\ref{gammad}) we note that

\begin{equation}
\dot{\alpha}\beta-\dot{\beta}=0.
\end{equation}
We can integrate this to find a relation between the two variables

\begin{equation}\label{betaex}
\beta=c_1e^{\alpha},
\end{equation}
where $c_1$ is a constant of integration. Substituting this into equation (\ref{betad}) and solving for $\gamma$ we obtain 

\begin{equation}\label{gammaex}
\gamma=\frac{2\dot{\alpha}}{g+\omega}.
\end{equation}
Now we make the change of variable $\alpha=-2\ln\sigma$. Substituting this into equation (\ref{gammad}) along with the expressions (\ref{betaex}) and (\ref{gammaex}) results in $\sigma$ being restricted by the following Ermakov Pinney equation

\begin{equation}
\ddot{\sigma}+\frac{\Lambda^2}{8}\sigma+\frac{\left(g+\omega\right)^2\left(2-c_1^2\right)}{16\sigma^3}=0.
\end{equation}
This is solved with the function

\begin{equation}
\sigma\left(t\right)=\sqrt{c_2\cos\left[\frac{1}{\sqrt{2}}\Lambda\left(t+c_3\right)\right]-\sqrt{\frac{2c_2^2\left(g-\omega\right)-\left(2-c_1^2\right)\left(g+\omega\right)}{2\left(g-\omega\right)}}},
\end{equation}
where $c_2$ and $c_3$ are constants of integration. We now rewrite the Hermitian Hamiltonian in terms of $p$ and $x$

\begin{equation}
h\left(t\right)=-\frac{g+\omega}{\omega\sigma\left(t\right)^2}\left(\frac{p^2}{2m}\left(c_1-2\right)+\frac{m\omega^2x^2}{2}\left(c_1+2\right)-\frac{c_1}{2}\right).
\end{equation}
We see that in order to ensure $h\left(t\right)$ is bounded from below, $|c_1|>2$. We also note that the time dependence is an overall factor. The parameters of the Dyson map are

\begin{eqnarray}
\alpha\left(t\right)&=&-\ln\left[c_2\cos\left[\frac{1}{\sqrt{2}}\Lambda\left(t+c_3\right)\right]-\sqrt{\frac{2c_2^2\left(g-\omega\right)-\left(2-c_1^2\right)\left(g+\omega\right)}{2\left(g-\omega\right)}}\right],\\
\beta\left(t\right)&=&\frac{c_1}{c_2\cos\left[\frac{1}{\sqrt{2}}\Lambda\left(t+c_3\right)\right]-\sqrt{\frac{2c_2^2\left(g-\omega\right)-\left(2-c_1^2\right)\left(g+\omega\right)}{2\left(g-\omega\right)}}},\\
\gamma\left(t\right)&=&\frac{\sqrt{2}}{g+\omega}\left[\frac{\Lambda\sin\left[\frac{1}{\sqrt{2}}\Lambda\left(t+c_3\right)\right]}{c_2\cos\left[\frac{1}{\sqrt{2}}\Lambda\left(t+c_3\right)\right]-\sqrt{\frac{2c_2^2\left(g-\omega\right)-\left(2-c_1^2\right)\left(g+\omega\right)}{2\left(g-\omega\right)}}}\right].
\end{eqnarray}

Now we have a solution for $\eta\left(t\right)$ and therefore also $\rho\left(t\right)$. This means we have can form a consistent theory for the non-Hermitian Hamiltonian (\ref{1DHamiltonian}) even in the broken regime $|g|<|\omega|$. We see a similar transition between trigonometric and hyperbolic functions at the exceptional point $g=\omega$ as we did in the matrix models. We have shown that a time-independent non-Hermitian system with a region of spontaneously broken $\mathcal{PT}$-symmetry can be made physically meaningful by using a time-dependent Dyson map in order to map the system to a time-dependent Hermitian system. The Hermitian system is then well-defined even in the spontaneously broken regime because of the explicit time-dependence in the Dyson map. The ability to make sense of such systems has great implications as we will see in the subsequent chapters, particularly in the analysis of entropy.

\section{Summary}

We have demonstrated that it is entirely possible to make physical sense of time-dependent non-Hermitian Hamiltonians. Furthermore, even when these Hamiltonians are in the spontaneously broken $\mathcal{PT}$-symmetric regime, explicit time-dependence in the Dyson map and the metric allows for a self-consistent quantum mechanical description. This is possible as the Hamiltonian that satisfies the time-dependent Schr\"{o}dinger equation becomes unobservable and instead the energy operator develops real eigenvalues at any instance in time. We identified the new antilinear operator $\widetilde{\mathcal{PT}}$ that explains the reality of the spectrum of the energy operator in parts of the parameter regime. We calculated this new symmetry operator in addition to the energy operator $\tilde{H}\left(t\right)$ for a 2-level matrix model. We showed that as we cross the exceptional point, the behaviour in observable parameters becomes significantly different in character. Specifically, the behaviour changes from trigonometric to hyperbolic evolution.

Following the 2 level matrix model, we calculated the Dyson map for higher spin systems (1, 3/2) corresponding to 3 and 4 level matrix models, both with spontaneously broken $\mathcal{PT}$-symmetry. 

Finally, we moved onto a non-Hermitian harmonic oscillator with infinite dimensional Hilbert space, also with spontaneously broken $\mathcal{PT}$-symmetry. In this example we kept the non-Hermitian Hamiltonian time-independent in order to demonstrate how time-dependent analysis is extremely vital for such systems in order to make sense of the broken regime. The utility is in the ability to investigate the spontaneously broken regime of time-independent systems previously believed to be unphysical and therefore inaccessible in this regime. 

Interestingly, we see that the Ermakov-Pinney equation arises in both situations we have considered.

\chapter{Coupled Oscillators with Spontaneously Broken  $\mathcal{PT}$-Symmetry}

In our analysis so far, we have considered time-dependent matrix models and a time-independent inverted harmonic oscillator. We showed that for the matrix model it was necessary to use time-dependent analysis in order to analyse such a system. The time-independent analysis for the harmonic oscillator was not enough to make sense of the spontaneously broken $\mathcal{PT}$ regime. The introduction of time into the central equations had the effect of mending the broken regime. 

In this chapter we extend the previous analysis of the broken $\mathcal{PT}$ regime from a one dimensional two-level system \cite{AndTom3} and an inverted harmonic oscillator to two-dimensional systems with infinite Hilbert space. These take the form of coupled harmonic oscillators with spontaneously broken $\mathcal{PT}$-symmetry. Studying such systems is of great importance as we wish to ultimately connect our theory to experimental results. Having solutions for coupled systems in more than one dimension is incredibly useful as this easily relates to many experimental scenarios for which a system is coupled to the environment.
 
Furthermore, in this chapter we will demonstrate the utility of the Lewis-Riesenfeld invariants (see \nameref{LRAppendix}) for solving complicated systems. So far we have only encountered LR invariants in chapter 1 where we found the method to involve a large number of steps. In this setting they are particularly useful as we are able to avoid the complicated differential equation that arises when using the time-dependent Dyson equation.

As we begin to investigate more complicated systems, the description of the problem becomes more technical. In chapter 2 we demonstrated the algebraic technique for solving the various central equations while using the algebra of the Pauli matrices SU(2) and in fact the algebra of the raising and lowering operators Sl(2,R). We also used the algebraic technique to solve for the Dyson map $\eta\left(t\right)$ when we considered the inverted harmonic oscillator in chapter 3, however we did not quote an algebra for this system. In the examples presented in this chapter, we will be working primarily in terms of generators of a closed algebra that relate to the Hamiltonians under consideration. This enables us to compute the BCH relation (\ref{BCHeq}) using only the closed algebra. The first coupled oscillator we consider belongs to an algebra consisting of 4 generators. The second system consists of 10 generators. 

\section{$i\left(xy+p_xp_y\right)$ coupled oscillator}

We begin our analysis in this chapter with two harmonic oscillators coupled with an $i\left(xy+p_xp_y\right)$ term. Initially we present the model with time-independent parameters and will introduce time into the system as the chapter progresses,

\begin{equation}
H_{xyp}=\frac{a}{2}\left( p_{x}^{2}+x^{2}\right) +\frac{b}{2}\left(
p_{y}^{2}+y^{2}\right) +i\frac{\lambda }{2}\left( xy+p_{x}p_{y}\right)
,\qquad a,b,\lambda \in \mathbb{R}.  \label{xyp}
\end{equation}
This non-Hermitian Hamiltonian is symmetric with regard to the antilinear
transformations \cite{EW} $\mathcal{PT}_{\pm }:x\rightarrow \pm x$, $%
y\rightarrow \mp y$, $p_{x}\rightarrow \mp p_{x}$, $p_{y}\rightarrow \pm
p_{y}$, $i\rightarrow -i$, i.e. $\left[ \mathcal{PT}_{\pm },H_{xyp}\right] =0$. Thus we expect the eigenvalues to be real or to be grouped in pairs of complex conjugates when the symmetry is broken for the
eigenfunctions. The energy eigenvalues are

\begin{equation}
E_{n,m}=\frac{1}{2}\left(1+n+m\right)\left(a+b\right)+\frac{1}{2}\left(n-m\right)\sqrt{(a-b)^2-\lambda^2}
\end{equation}
and we see that they are real for $|\left(a-b\right)|>|\lambda|$, or when $n=m$. Therefore there is an exceptional point in the parameter space at $|\left(a-b\right)|=|\lambda|$ below which the symmetry is spontaneously broken and the eigenvalues become complex conjugate pairs.
It is convenient to express this Hamiltonian in a more generic algebraic
fashion as 
\begin{equation}
H_{K}=aK_{1}+bK_{2}+i\lambda K_{3},  \label{Hk}
\end{equation}%
where we defined Lie algebraic generators 
\begin{equation}
K_{1}=\frac{1}{2}\left( p_{x}^{2}+x^{2}\right) ,~~K_{2}=\frac{1}{2}\left(
p_{y}^{2}+y^{2}\right) ,~~K_{3}=\frac{1}{2}\left( xy+p_{x}p_{y}\right)
,~~K_{4}=\frac{1}{2}\left( xp_{y}-yp_{x}\right) .  \label{om}
\end{equation}%
Besides the generators already appearing in the Hamiltonian we added one
more generator, $K_{4}=L_{z}/2$, to ensure the closure of the algebra, i.e.
we have%
\begin{equation}
\begin{array}{lll}
\left[ K_{1},K_{2}\right] =0,~ & \left[ K_{1},K_{3}\right] =iK_{4}, & \left[
K_{1},K_{4}\right] =-iK_{3}, \\ 
\left[ K_{2},K_{3}\right] =-iK_{4},~~ & \left[ K_{2},K_{4}\right] =iK_{3},~~
& \left[ K_{3},K_{4}\right] =i(K_{1}-K_{2})/2.%
\end{array}
\label{alg}
\end{equation}%
Notice that $K_{i}^{\dagger }=K_{i}$ for $i=1,\ldots ,4$. In what follows we
mostly use the algebraic formulation so that our results also hold for
representations different from (\ref{om}). 

Now that we have our Hamiltonian set up in a closed algebraic form, we can begin our analysis. We proceed by initially solving the time-independent model, as did with the single inverted harmonic oscillator presented in section \ref{1dharmonicoscillator}. First we will solve the time-independent Dyson equation and show that the procedure breaks down at the exceptional point. We will then mend the broken regime by solving the time-dependent Dyson equation. Finally we will introduce time into the Hamiltonian (\ref{xyp}) and solve for the Dyson map using the time-dependent Dyson equation and then the Lewis-Riesenfeld invariants. 

\subsection{Time-independent Dyson map}

We start our analysis on the non-Hermitian Hamiltonian (\ref{Hk}) by solving the time-independent Dyson equation. For this we use the ansatz
\begin{equation}
\eta =e^{\theta K_{4}},
\end{equation}
and substitute into (\ref{TIDE}). We find that the corresponding Hamiltonian is Hermitian if

\begin{equation}
\tanh\theta =\frac{\lambda}{(b-a)}.
\end{equation}
As the mapping is time-independent, we see the signature exceptional point present in the Dyson map. The map is only valid for  $|\left(a-b\right)|>|\lambda|$ and so we are unable to make sense of the broken regime. The Hermitian Hamiltonian becomes a system of two decoupled harmonic oscillators
\begin{equation}
h_{K}=\eta H_{K}\eta ^{-1}=\frac{1}{2}(a+b)\left( K_{1}+K_{2}\right) +\frac{1%
}{2}\sqrt{(a-b)^{2}-\lambda ^{2}}\left( K_{1}-K_{2}\right) ,
\end{equation}%
for $\left\vert \lambda \right\vert <\left\vert a-b\right\vert $. In order to proceed, it is clear we need to introduce time into the Dyson map and solve the time-dependent Dyson equation.

\subsection{Time-dependent Dyson map}

\begin{figure}[H]
\centering

\includegraphics[scale=0.7]{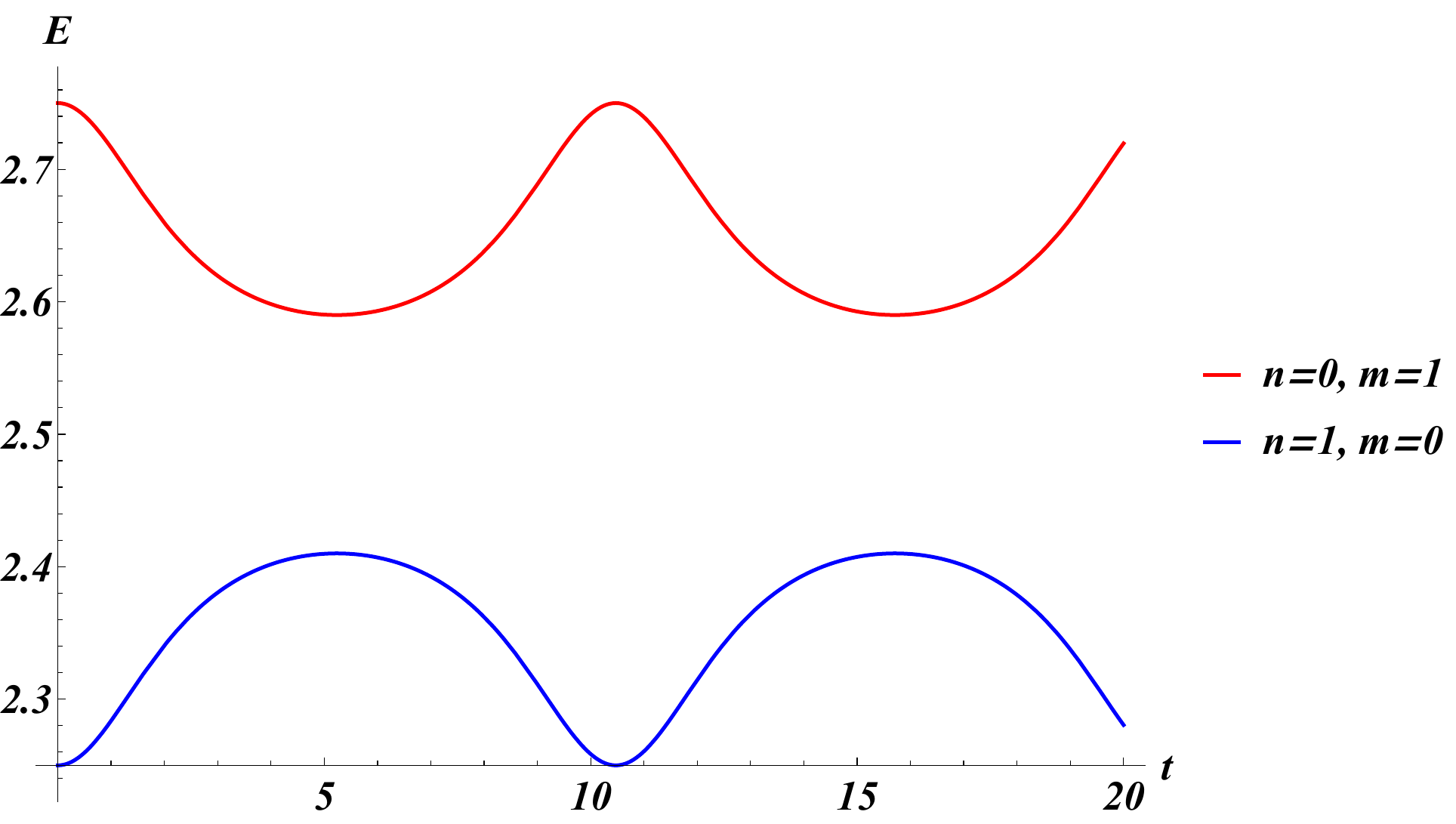}

\caption{Energy observables $\protect\tilde{E}\left(t\right)$ for $a=1.5$, $b=1$ and $\protect\lambda=0.4$, corresponding to the $\protect\mathcal{PT}$ unbroken regime.}\label{coupledun}

\end{figure}

\begin{figure}[H]
	\centering
	
	\includegraphics[scale=0.7]{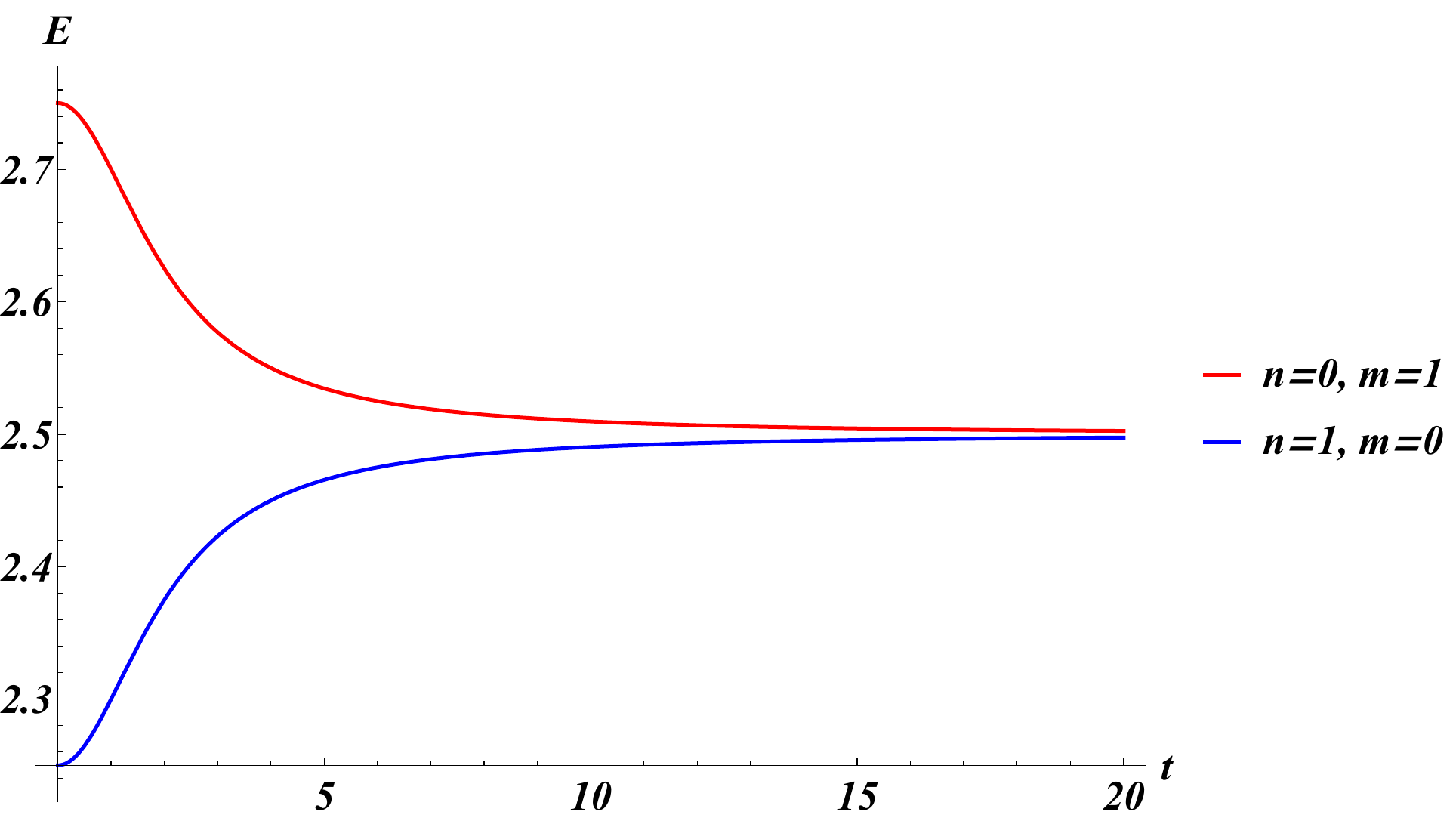}
	
	\caption{Energy observables $\protect\tilde{E}\left(t\right)$ for $a=1.5$, $b=1$ and $\protect\lambda=0.5$, corresponding to the exceptional point.}\label{coupledex}
	
\end{figure}

\begin{figure}[H]
	
	\centering
	
	\includegraphics[scale=0.7]{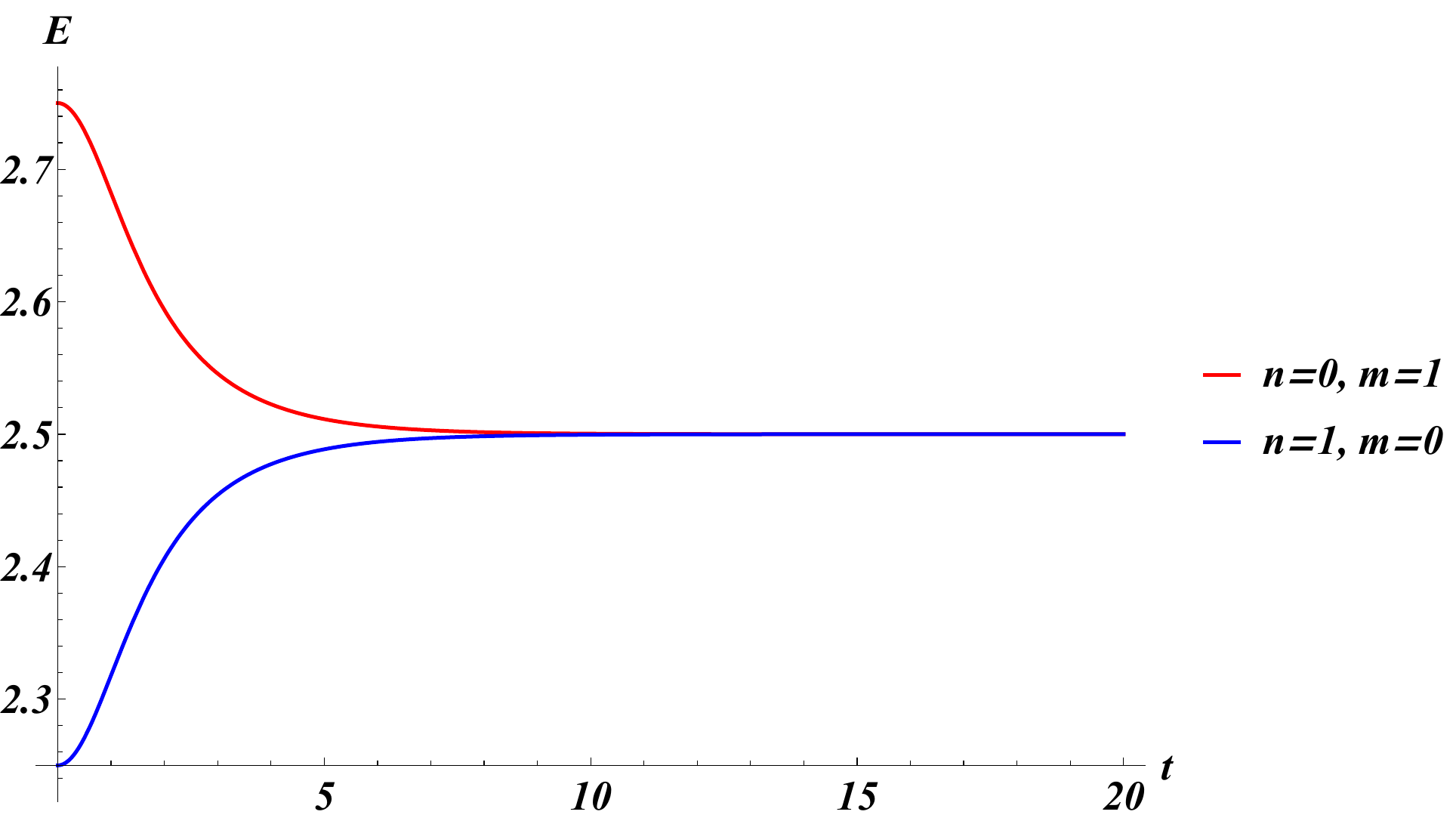}
	
	\caption{Energy observables $\protect\tilde{E}\left(t\right)$ for $a=1.5$, $b=1$ and $\protect\lambda=0.6$, corresponding to the $\protect\mathcal{PT}$ broken regime.}\label{coupledbroke}
	
\end{figure}

We now extend our analysis of Hamiltonian (\ref{Hk}) to the time-dependent Dyson equation. At this stage the Hamiltonian is still time-independent and we only introduce time into the Dyson map. Our ansatz for $\eta\left(t\right)$ now takes the form

\begin{equation}\label{ansatzetaHK}
\eta\left(t\right)=e^{\alpha_3\left(t\right)K_3}e^{\alpha_4\left(t\right)K_4}.
\end{equation}
We substitute this into the Dyson equation and eliminate the non-Hermitian terms. In order to remove these terms we are required to solve the following coupled differential equations
\begin{eqnarray}
\dot{\alpha}_3&=&-\left(a-b\right)\sinh\alpha_4-\lambda\cosh\alpha_4,\label{coupled1}\\
\dot{\alpha}_4&=&\left[\left(a-b\right)\cosh\alpha_4+\lambda\sinh\alpha_4\right]\tanh\alpha_3.\label{coupled2}
\end{eqnarray}
We can decouple these equations by first solving (\ref{coupled1}) for $\alpha_{4}$

\begin{equation}
\alpha_4=\log\left[\frac{-\dot{\alpha_3}+\sqrt{\left(a-b\right)^2-\lambda^2+\dot{\alpha_3}^2}}{a-b+\lambda}\right].
\end{equation}
Substituting this into the second equation (\ref{coupled2}) results in a non-linear differential equation in terms of $\alpha_3$

\begin{equation}
\ddot{\alpha_3}+\left(\left(a-b\right)^2-\lambda^2+\dot{\alpha_3}^2\right)\tanh\alpha_3=0.
\end{equation}
We can rewrite $\alpha_3\left(t\right)=f\left(\omega t\right)$, where $\omega=\sqrt{\left(a-b\right)^2-\lambda^2} $
and so substituting this in we reduce the equation to

\begin{equation}
f''+\left(1+f'^2\right)\tanh f=0, \quad f'=\frac{df}{d \left(\omega t\right)}
\end{equation}
Finally we make the variable change $f=\arcsinh\sigma$ which gives us the differential equation

\begin{equation}
\sigma''+\sigma=0
\end{equation}
which is solved with
 
\begin{equation}
\sigma\left(\omega t\right)=\frac{c_1}{\omega} \sin\left[\omega\left(t+c_2\right)\right].
\end{equation}
Therefore the paramters of the Dyson map are,
\begin{eqnarray}
\sinh\alpha_{3}&=&\frac{c_1}{\omega}\sin\left[\omega\left(t+c_2\right)\right],\\
\exp\alpha_{4}&=&\frac{\sqrt{c_1^2+\omega^2}-c_1\cos\left[\omega\left(t+c_2\right)\right]}{\left(a-b+\lambda\right)\sqrt{1+\frac{c_1^2\sin^2\left[\omega\left(t+c_2\right)\right]}{\omega^2}}},
\end{eqnarray}
which are always real when $|\left(a-b\right)|>|\lambda|$, corresponding to the unbroken regime, and are real for $|c_1|>|\omega|$ in the broken regime. The resulting Hermitian Hamiltonian is
\begin{equation}\label{hxyp}
h\left(t\right)=\frac{1}{2}\left(a+b\right)\left[K_1+K_2\right]+\frac{1}{2}\delta\left(t\right)\left[K_1-K_2\right]
\end{equation}
with $\delta\left(t\right)=\left(\left(a-b\right)\cosh\alpha_4+\lambda\sinh\alpha_4\right)\sech\alpha_3$, that is
\begin{equation}
\delta\left(t\right)=\frac{\omega^2\sqrt{c_1^2+\omega^2}}{\omega^2+c_1^2\sin\left[\omega\left(t+c_2\right)\right]^2}.
\end{equation}
This is now a system of two decoupled harmonic oscillators, much like the case in the time-independent Dyson map. However, now the system is valid in the spontaneously broken $\mathcal{PT}$ regime. We can now find the energy expectation values when $|\left(a-b\right)|<|\lambda|$. Solving the time-dependent Schr\"odinger equation using the Lewis Riesenfeld invariants for the Hamiltonian (\ref{hxyp}), we obtain the wave functions

\begin{equation}
\phi\left(x,y,t\right)=\phi_{-,n}\left(x,t\right)\phi_{+,m}\left(y,t\right),
\end{equation} 
from which we can construct a general wave function and where 

\begin{equation}\label{wavefunctionshxyp}
\phi_{\pm,n}\left(z,t\right)=\frac{e^{i\alpha_\pm\left(t\right)}}{\sqrt{\chi_\pm\left(t\right)}}\exp\left[\left(\frac{i}{f_\pm\left(t\right)}\frac{\dot{\chi}_\pm\left(t\right)}{\chi_\pm\left(t\right)}-\frac{1}{\chi_\pm\left(t\right)^2}\right)\frac{z^2}{2}\right]H_n\left[\frac{z}{\chi_\pm\left(t\right)}\right],
\end{equation}
with
\begin{equation}
f_\pm\left(t\right)=\frac{1}{2}\left(a+b\right)\pm\frac{1}{2}\delta\left(t\right),
\end{equation}
and $\chi_\pm\left(t\right)$ satisfying a dissipative EP equation

\begin{equation}
\ddot{\chi}_\pm-\frac{\dot{f}_\pm}{f_\pm}\dot{\chi}_\pm+f_\pm^2\chi_\pm=\frac{f_\pm^2}{\chi_\pm^3},
\end{equation}
and $\alpha_\pm\left(t\right)$ satisfying the integral

\begin{equation}
\alpha_\pm\left(t\right)=-\left(n+\frac{1}{2}\right)\int^t_0\frac{f_\pm\left(s\right)}{\chi_\pm\left(s\right)^2}ds.
\end{equation}
Using these solutions we calculate the energy expectation values

\begin{equation}
\tilde{E}\left(t\right)=\bra{\psi\left(t\right)}\rho\left(t\right)\tilde{H}\left(t\right)\psi\left(t\right)\rangle=\bra{\phi\left(x,y,t\right)}h\left(t\right)\phi\left(x,y,t\right)\rangle.
\end{equation}
As the corresponding Hermitian system is decoupled it is significantly easier to work in this regime. We obtain

\begin{equation}\label{Eobservable}
\tilde{E}\left(t\right)=\left(n+\frac{1}{2}\right)f_-\left(t\right)+\left(m+\frac{1}{2}\right)f_+\left(t\right).
\end{equation}
From this we can see that $\tilde{E}\left(t\right)$ is static when $n=m$. In order to define the integration constants $c_{1,2}$ we enforce the initial condition $\alpha_{3}\left(0\right)=\alpha_{4}\left(0\right)=0$ such that $\eta\left(0\right)=\mathbb{I}$. Under this condition $c_1=-\lambda$ and $c_2=0$. We can see the reality of the energy operator spectrum across the unbroken and broken $\mathcal{PT}$ regime in figures \ref{coupledun} to \ref{coupledbroke}.

Figure \ref{coupledun} shows the unbroken $\mathcal{PT}$ regime for the first two excited states. In this regime the first excited states oscillate.

Figure \ref{coupledex} shows the exceptional point for $\left(a-b\right)=\lambda$ for the first two excited states. At this point the first two excited states decay asymptotically to $\left(a+b\right)$. All states at the exceptional point where $n\neq m$ decay to $\frac{1}{2}\left(1+n+m\right)\left(a+b\right)$. When $n=m$ the energy is constant at $\left(\frac{1}{2}+m\right)\left(a+b\right)$.

Figure \ref{coupledbroke} shows the broken $\mathcal{PT}$ regime for the first two excited states. At this point the first two excited states decay asymptotically to $\left(a+b\right)$. Like at the exceptional point, all states in the broken regime where $n\neq m$ decay to $\frac{1}{2}\left(1+n+m\right)\left(a+b\right)$. When $n=m$ the energy is constant at $\left(\frac{1}{2}+m\right)\left(a+b\right)$.

It is also interesting to plot the variation of the energy expectation values as $\lambda$ is varied across the exceptional point at different times.

\begin{figure}[H]
\centering

\includegraphics[scale=0.7]{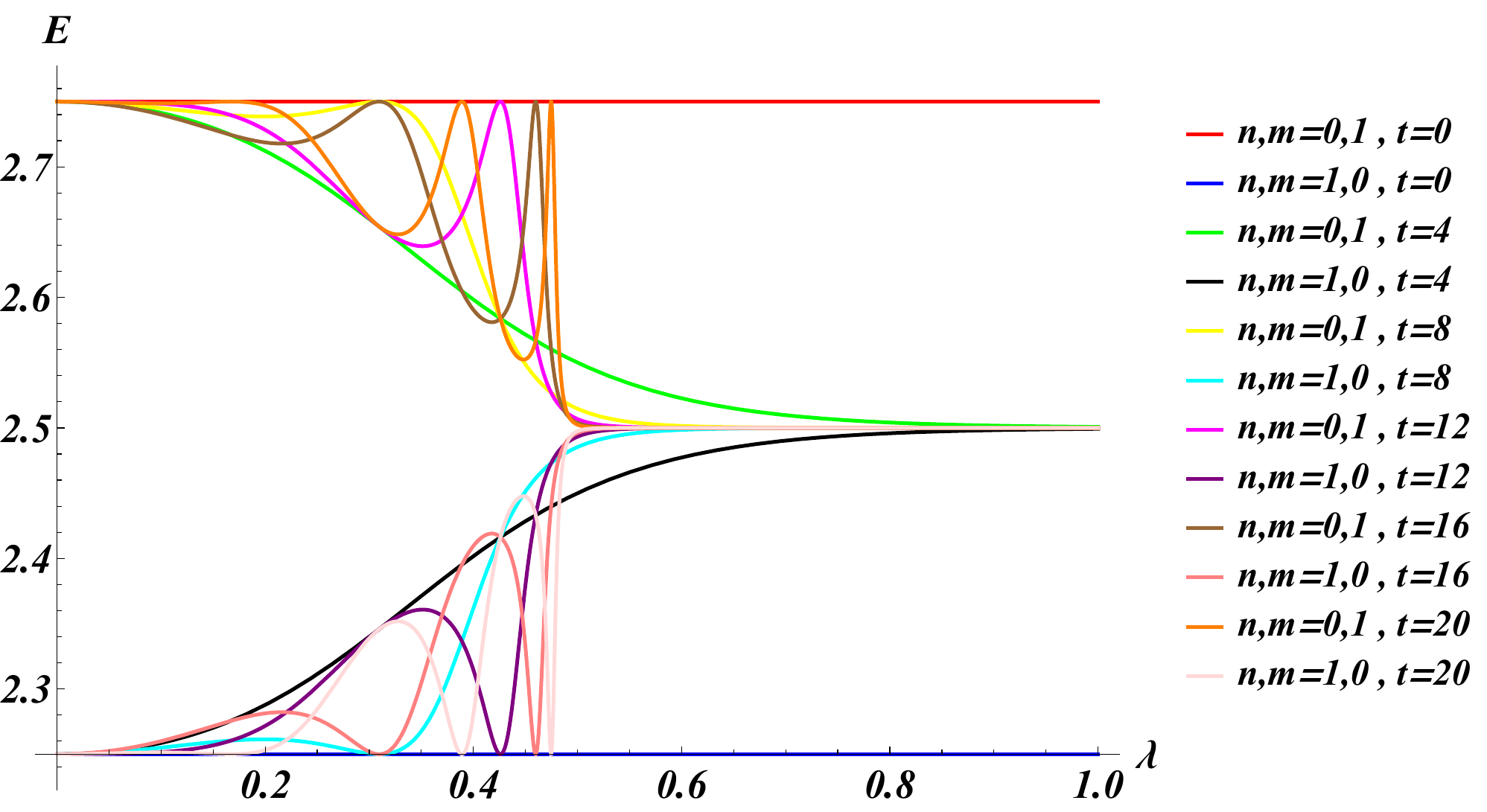}

\caption{Energy observables $\protect\tilde{E}\left(t\right)$ with varying $\protect\lambda$ at different times demonstrating the transition from the $\protect\mathcal{PT}$-symmetric to the $\protect\mathcal{PT}$ broken regime. $a=1.5$ and $b=1$. }\label{ExceptVar}

\end{figure}

Figure \ref{ExceptVar} shows the energy expectation values for $a=1.5$ and $b=1$ as we vary $\lambda$ through the exceptional point at $\lambda=0.5$. At $\lambda<0.5$ the energies are clearly oscillatory, which corresponds to figure \ref{coupledun}. But we see a clear transition at the exceptional point $\lambda=0.5$ from oscillatory to smooth decay, this corresponds to figure \ref{coupledbroke}. At $\lambda>0.5$ the energy expectation values decay at increasing times with no oscillation. This a clear phase transition rather elegantly displayed for a time-dependent system.

\subsection{Fully time-dependent model}

The final step in this model is to introduce time into the non-Hermitian Hamiltonian. To do this we take $\lambda\rightarrow\lambda\left(t\right)$. Furthermore, in order to simplify the resulting equations, we take $b=a$ such that the Hamiltonian is permanently in the broken regime. This is particularly interesting as we would ordinarily discard such a system as non physical. However, using a time-dependent Dyson map, we will be able to make the system meaningful. Firstly, we will solve this system using the time-dependent Dyson equation and then we will go on to show that in this particular case, employing the LR invariants is actually a far simpler path to the solution for $\eta\left(t\right)$.

\subsubsection{Time-dependent Dyson equation}

We use the same ansatz as in (\ref{ansatzetaHK}), which when substituted into the time-dependent Dyson equation results in a Hermitian Hamiltonian if the following coupled differential equations are satisfied
\begin{eqnarray}
\dot{\alpha}_3&=&-\lambda\left(t\right)\cosh\alpha_4,\label{coupledtime1}\\
\dot{\alpha}_4&=&\lambda\left(t\right)\sinh\alpha_4\tanh\alpha_3.\label{coupledtime2}
\end{eqnarray}
In order to solve these, we eliminate $\lambda $ and $dt$ from the equations, so that 
\begin{equation}
d\alpha_{4}=-\tanh \alpha_{3}\tanh \alpha_{4}d\alpha_{3},
\end{equation}
hence obtaining $\alpha_{4}$ as a function of $\alpha_{3}$ 
\begin{equation}
\sinh\alpha_{4}=\tilde{c}_1\sech\alpha_{3}
\label{43}
\end{equation}%
with integration constant $\tilde{c}_1$. Next we define $\chi=\cosh \alpha_{3}$ and use (\ref{coupledtime1}) and (\ref{coupledtime2}) to derive that the central equation that
needs to be satisfied is once more the Ermakov-Pinney equation \cite{Ermakov,Pinney} with a dissipative term
\begin{equation}
\ddot{\chi}-\frac{\dot{\lambda}}{\lambda }\dot{\chi}-\lambda ^{2}\chi =\frac{\tilde{c}_1^{2}\lambda ^{2}}{\chi ^{3}}.  \label{DEP}
\end{equation}
This equation is ubiquitous in the context of solving time-dependent Hermitian systems, even in the Hermitian setting, see e.g. \cite{leach2008ermakov}. To solve this, we rewrite $\chi\left(t\right)=f\left[\mu\left(t\right)\right]$, where $\mu\left(t\right)=\int^t\lambda\left(s\right)ds$ and substitute this into (\ref{DEP}), which gives

\begin{equation}
f''-f=\frac{\tilde{c}_1^2}{f^3}, \quad f'=\frac{df}{d\mu}.
\end{equation}
This is now a much simpler Ermakov Pinney equation to solve, and we find the solution to be

\begin{equation}
f\left[\mu\left(t\right)\right]=\frac{1}{\sqrt{2}}\sqrt{\left(1+\tilde{c}_2^2\right)\cosh\left[2\tilde{c}_3-2\mu\left(t\right)\right]+\left(1-\tilde{c}_2^2\right)}.
\end{equation}
Therefore we have the solutions for the components of the Dyson map

\begin{eqnarray}
\cosh\alpha_{3}&=&\frac{1}{\sqrt{2}}\sqrt{\left(1+\tilde{c}_1^2\right)\cosh\left[2\tilde{c}_2-2\mu\left(t\right)\right]+\left(1-\tilde{c}_1^2\right)},\label{DysonSolsHxyp1}\\
\sinh\alpha_{4}&=&\frac{\tilde{c}_1\sqrt{2}}{\sqrt{\left(1+\tilde{c}_1^2\right)\cosh\left[2\tilde{c}_2-2\mu\left(t\right)\right]+\left(1-\tilde{c}_1^2\right)}}.\label{DysonSolsHxyp2}
\end{eqnarray}
This is real for all values of $\tilde{c}_1$. The corresponding Hermitian Hamiltonian is

\begin{equation}
h(t)=a(t)\left( K_{1}+K_{2}\right) +\frac{\lambda (t)}{2}\frac{\sinh \alpha_{4}}{\cosh \alpha_{3}}\left( K_{1}-K_{2}\right) .  \label{hher}
\end{equation}
Once again, this Hamiltonian is Hermitian and the Dyson map is valid even though the non-Hermitian Hamiltonian we started with is permanently sitting in the broken regime. The solutions to the Schr\"odinger equation for (\ref{hher}) take he same form as (\ref{wavefunctionshxyp}) with $f_\pm\left(t\right)$ replaced with

\begin{equation}
\tilde{f}_\pm\left(t\right)=a\left(t\right)\pm\frac{\lambda\left(t\right)}{2}\frac{\sinh\alpha_4}{\cosh\alpha_3}.
\end{equation}

\subsubsection{Lewis Riesenfeld Invariants}

We will now show that the same solution for the Dyson map (\ref{DysonSolsHxyp1}),  (\ref{DysonSolsHxyp2}) can be obtained with far fewer technical equations using the Lewis Riesenfeld invariants. The first step is to find the non-Hermitian invariant from the equation

\begin{equation}\label{InvarDefining}
\partial_tI_H\left(t\right)=i\left[I_H\left(t\right),H\left(t\right)\right].
\end{equation}
For this we take the most general form of $I_H\left(t\right)$ for our ansatz

\begin{equation}
I_H\left(t\right)=u_1\left(t\right)K_1+u_2\left(t\right)K_2+u_3\left(t\right)K_1+u_4\left(t\right)K_4.
\end{equation}
Substituting this into equation (\ref{InvarDefining}) along with the Hamiltonian (\ref{Hk}) results in the coupled differential equations

\begin{equation}
\dot{u}_1=\frac{i}{2}\lambda u_4, \qquad \dot{u}_2=-\frac{i}{2}\lambda u_4, \qquad \dot{u}_3=0, \qquad \dot{u}_4=i\lambda\left(u_2-u_1\right).
\end{equation}
These equations are easily solved with the functions,
\begin{eqnarray}
u_1&=&\frac{q_1}{2}+q_3\cosh\left[q_4-\mu\left(t\right)\right], \\
u_2&=&q_1-u_1,\\
u_3&=&q_2,\\
u_4&=&2iq_3\sinh\left[q_4-\mu\left(t\right)\right],
\end{eqnarray}
with complex integration constants $q_j=q^r_j+iq^i_j$, $q^r_j,q^i_j\in\mathbb{R}$. Now we have defined the non-Hermitian invariant, we move on to solve the similarity transform between the non-Hermitian invariant $I_H$ and the Hermitian invariant $I_h$

\begin{equation}
I_h\left(t\right)=\eta\left(t\right)I_H\left(t\right)\eta^{-1}\left(t\right).
\end{equation}
In order to solve this we use the same ansatz for the Dyson map (\ref{ansatzetaHK}) and use the BCH relation to expand in terms of the algebraic elements. When we restrict the resulting invariant to be Hermitian, we obtain the following constraints

\begin{equation}
q^i_1=0, \qquad 4q^r_3q^i_3=-q^r_2q^i_2.
\end{equation}
Furthermore, the parameters of the Dyson map are found to be

\begin{eqnarray}
\cosh\alpha_{3}&=&\frac{\sqrt{2\left(q^r_3\right)^2\cosh\left[2q^r_4-2\mu\left(t\right)\right]+2\left(q^r_3\right)^2-\left(q^i_2\right)^2}}{\sqrt{4\left(q^r_3\right)^2-\left(q^i_2\right)^2}},\\
\sinh\alpha_{4}&=&\frac{-q^i_2}{\sqrt{2\left(q^r_3\right)^2\cosh\left[2q^r_4-2\mu\left(t\right)\right]+2\left(q^r_3\right)^2-\left(q^i_2\right)^2}}.
\end{eqnarray}
Therefore identifying $q^r_3=\sqrt{\frac{\tilde{c}_1^2+1}{2}}$, $q^i_2=-\tilde{c}_1\sqrt{2}$ and $q^r_4=\tilde{c}_4$ our solutions match those obtained in the previous section. Notice that we never had to solve the EP equation using this method. The Hermitian invariant is 

\begin{equation}\label{IHermitian}
I_h\left(t\right)=v_1\left(t\right)K_1+v_2\left(t\right)K_2+v_3\left(t\right)K_3+v_4\left(t\right)K_4,
\end{equation}
where the time-dependent coefficients are

\begin{eqnarray}
v_1&=&\frac{1}{2}q^r_1+\frac{1}{2}\sqrt{4\left(q^r_3\right)^2-\left(q^i_2\right)^2},\\
v_2&=&\frac{1}{2}q^r_1-\frac{1}{2}\sqrt{4\left(q^r_3\right)^2-\left(q^i_2\right)^2},\\
v_3&=&\frac{q^r_2\left(4\left(q^r_3\right)^2-\left(q^i_2\right)^2\right)\cosh\left[q^r_4-\mu\left(t\right)\right]}{2q^r_3\sqrt{2\left(q^r_3\right)^2\cosh\left[2q^r_4-2\mu\left(t\right)\right]+2\left(q^r_3\right)^2-\left(q^i_2\right)^2}},\\
v_4&=&\frac{q^i_2q^r_2\sqrt{4\left(q^r_3\right)^2-\left(q^i_2\right)^2}\sinh\left[q^r_4-\mu\left(t\right)\right]}{2q^r_3\sqrt{2\left(q^r_3\right)^2\cosh\left[2q^r_4-2\mu\left(t\right)\right]+2\left(q^r_3\right)^2-\left(q^i_2\right)^2}}.
\end{eqnarray}
We can then compute the Hermitian Hamiltonian from the invariant (\ref{IHermitian}). Typically the invariant is computed from the Hamiltonian, but in this case we use equation (\ref{Invarianth}) the opposite way round. The Hermitian Hamiltonian is precisely of the form (\ref{hher}). Therefore we have shown the equivalence of the Lewis Riesenfeld invariant method and the time-dependent Dyson equation. However, it is clear that in this context the Lewis Riesenfeld invariant method exhibits simpler central equations to solve. The trade off is that there are more steps to the solution process. Now we have analysed a coupled harmonic oscillator, we move on to a significantly more complicated system.

\section{$ixy$ coupled oscillator}

We follow our analysis of the $i\left(xy+p_xp_y\right)$ coupling by now investigating the more popular non-Hermitian $ixy$ coupling term. In this case the Hamiltonian takes the form

\begin{equation}
H_{xy}=\frac{1}{2m}\left( p_{x}^{2}+p_{y}^{2}\right) +\frac{1}{2}m\left(
\Omega _{x}^{2}x^{2}+\Omega _{y}^{2}y^{2}\right) +i\lambda xy,~~~~~~m,\kappa
,\Omega _{x},\Omega _{y}\in \mathbb{R}.  \label{Napkin}
\end{equation}
This non-Hermitian Hamiltonian is symmetric with regard to the antilinear
transformations \cite{EW} $\mathcal{PT}_{\pm }:x\rightarrow \pm x$, $%
y\rightarrow \mp y$, $p_{x}\rightarrow \mp p_{x}$, $p_{y}\rightarrow \pm
p_{y}$, $i\rightarrow -i$, i.e. $\left[ \mathcal{PT}_{\pm },H_{xy}\right] =0$. Clearly this Hamiltonian is also symmetric with regard to the same antilinear symmetry as $H_{xyp}$. Thus we expect the eigenvalues to be real or to be
grouped in pairs of complex conjugates when the symmetry is broken for the
eigenfunctions. It has the eigenvalues

\begin{equation}
E_{n_1,n_2}=\left(n_1+\frac{1}{2}\right)\omega_x+\left(n_2+\frac{1}{2}\right)\omega_y,
\end{equation} 
with
\begin{equation}
\omega_{x,y}^2=\frac{1}{2m}\left(m\Omega_+^2\pm\sqrt{m^2\Omega_-^4-4\lambda^2}\right),
\end{equation}
and $\Omega_\pm^2=\Omega_y^2\pm\Omega_x^2$. This system has been studied in detail in the time-independent regime \cite{AndTom4,PhysRevA.91.062101,MANDAL20131043} and it is clear from the energy eigenvalues that the $\mathcal{PT}$-symmetry is spontaneously broken when $|m\Omega_-^2|<2|\lambda|$ as they become complex. We show in this section that by using a time-dependent metric, we can mend the broken regime and return a physical meaning to it.

We wish to express the Hamiltonian (\ref{Napkin}) in terms of a closed algebra. This allows us to formulate our Ansatz for $\eta$ in terms of the generators of the algebra and guarantees that the resulting Hermitian Hamiltonian will also be expressible in terms of these generators. The algebra for our Hamiltonian is comprised of the ten Hermitian generators
\begin{equation}
\begin{aligned}
\begin{split}
K_{\pm}^{z}=&\frac{1}{2}\left(p_z^2\pm z^2\right),\quad K_0^z=\frac{1}{2}\{z,p_z\},\quad J_\pm=\frac{1}{2}\left(xp_y\pm yp_x\right), \quad I_\pm=\frac{1}{2}\left(xy\pm p_xp_y\right),\\
\end{split}
\end{aligned}
\end{equation}
where $z=x,y$. The commutation relations for these generators are
\begin{eqnarray}
\left[K_0^z,K_\pm^z\right]\!\!\!\!&=&\!\!\!\!2iK_\mp^z,\quad \left[K_+^z,K_-^z\right]=2iK_0^z,\quad \left[K_{\mu}^x,K_{\nu}^y\right]=0,\\
\left[K_0^x,J_\pm\right]\!\!\!\!&=&\!\!\!\!-iJ_\mp, \,\,\,\,\quad\left[K_0^y,J_\pm\right]=iJ_\mp, \qquad\,\, \left[K_0^z,I_\pm\right] =- iI_\mp,  \\
\left[K_\pm^x,J_{+}\right]\!\!\!\!&=&\!\!\!\!\pm iI_{\mp}, \!\!\!\qquad\qquad\qquad\qquad\qquad\quad \left[K_\pm^y,J_{+}\right]=\pm iI_{\mp}\\
\left[K_\pm^x,J_{-}\right]\!\!\!\!&=&\!\!\!\!\mp iI_{\pm}, \!\!\!\qquad\qquad\qquad\qquad\qquad\quad \left[K_\pm^y,J_{-}\right]=\pm iI_{\pm}\\
\left[K_\pm^x,I_{+}\right]\!\!\!\!&=&\!\!\!\!\pm iJ_{\mp}, \!\!\!\qquad\qquad\qquad\qquad\qquad\quad \left[K_\pm^y,I_{+}\right]=-iJ_{\mp},\\ 
\left[K_\pm^x,I_{-}\right]\!\!\!\!&=&\!\!\!\!\mp iJ_{\pm}, \!\!\!\qquad\qquad\qquad\qquad\qquad\quad \left[K_\pm^y,I_{-}\right]=-iJ_{\pm},\\ 
\left[J_+,J_-\right]\!\!\!\!&=&\!\!\!\!\frac{i}{2}\left(K_0^x-K_0^y\right), \!\!\qquad\qquad\qquad\qquad \left[I_+,I_-\right]=-\frac{i}{2}\left(K_0^x+K_0^y\right),\\
\left[J_{+},I_\pm\right]\!\!\!\!&=&\!\!\!\!\pm\frac{i}{2}\left(K_{\mp}^x+K_{\mp}^y\right), \!\!\!\qquad\qquad\qquad\quad
\left[J_{-},I_\pm\right]=\mp\frac{i}{2}\left(K_{\pm}^x-K_{\pm}^y\right),
\end{eqnarray}
with $\mu,\nu=+,-,0$. As is clear, this is a rich algebra containing many closed sub-algebras contained within. We can rewrite the Hamiltonian (\ref{Napkin}) as
\begin{equation}
H_{xy}=\sum_{z,\sigma=\pm}\Lambda^z_{\sigma} K^z_{\sigma}+i\lambda\left(I_++I_-\right),
\end{equation}
where $\Lambda^z_\pm=\frac{1}{2m}\left(1\pm m^2\Omega_z^2\right)$. As the generators are all Hermitian, the Hamiltonian is non-Hermitian due to the contribution from the last term. With our Hamiltonian expressed in this form we are now able to proceed with solving the time-dependent Dyson equation using the BCH relation to evaluate the adjoint action of $\eta$ on the Hamiltonian and solving the resulting differential equations. These equations arise when we enforce the condition of Hermiticity on the resulting Hamiltonian. However, we first recall the solution to the time-independent Dyson equation \cite{AndTom4} in order to emphasise that the mapping and the metric breaks down as the $\mathcal{PT}$-symmetry is spontaneously broken in the absence of time.

\subsection{Time-independent Dyson map}

In the time-independent case, the time-dependent Dyson equation (\ref{DysonE}) reduces to a similarity transformation and is solved with \cite{AndTom4}

\begin{equation}
\eta=e^{\theta J_-}, \quad \tanh2\theta=\frac{2\lambda}{m\Omega_-^2}.
\end{equation}
This mapping is only valid for $|m\Omega_-^2|>2|\lambda|$ which matches the results from \cite{PhysRevA.91.062101,MANDAL20131043}. The resulting Hermitian Hamiltonian is

\begin{equation}
h=\frac{1}{2m}\left(p_x^2+p_y^2\right)+\frac{1}{2}m\left(\omega_x^2x^2+\omega_y^2y^2\right),
\end{equation}

\begin{equation}
\omega_x^2=\frac{\Omega_x^2\cosh^2\theta+\Omega_y^2\sinh^2\theta}{\cosh2\theta}, \quad \omega_y^2=\frac{\Omega_x^2\sinh^2\theta+\Omega_y^2\cosh^2\theta}{\cosh2\theta}.
\end{equation}
In this time-independent setting, when the $\mathcal{PT}$-symmetry is broken we cannot construct a metric $\varrho=\eta^\dagger\eta$ and therefore cannot make sense of the broken regime. To progress, we must acknowledge that our choice for $\eta$, and therefore $\varrho$, is not restricted to be time-independent. Introducing an explicit time-dependence into these parameters means we are led to solve the TDDE resulting in a time-dependent Hermitian Hamiltonian.

\subsection{Time-dependent Dyson map}

We now use a time-dependent Dyson map of the form
\begin{equation}\label{dysonmap}
\eta\left(t\right)=e^{\alpha_-\left(t\right)L_-}e^{\theta_+\left(t\right) J_+}e^{\alpha_+\left(t\right)L_+}e^{\theta_-\left(t\right) J_-}, \quad L_+=\frac{1}{2}\left(I_++I_-\right),\quad L_-=\frac{1}{2}\left(I_+-I_-\right).
\end{equation}
This Ansatz is of course not the most general choice. We could use all ten generators in our Ansatz in order to find the most general form of $\eta$. However, it is well known that $\eta$ is not unique \cite{scholtz1992quasi} and so we are content here to find \textbf{a} solution. We choose (\ref{dysonmap}) to be comprised of the interaction generators between the two oscillators as this produces a comprehensible solution. The quantities $\alpha_+$, $\alpha_-$, $\theta_+$ and $\theta_-$ could all be chosen to be fully complex, however this substantially increases the complexity of problem. We wish to avoid all quantities being purely imaginary, as in this case $\eta$ is just a gauge transformation. Therefore we restrict $\alpha_+$, $\alpha_-$, $\theta_+\in\mathbb{R}$ and $e^{\theta_-}\in\mathbb{R}$. The comparatively relaxed restriction $e^{\theta_-}\in\mathbb{R}$ allows for $\theta_-$ to contain an imaginary term of the form $in\pi, \; n\in\mathcal{Z}$. This is included as we see that $\theta_-$ only appears in hyperbolic form in the resulting expressions and therefore the combinations are always real. Substituting $\eta$ into the time-dependent Dyson equation (\ref{DysonE}), the imaginary terms are eliminated when the following differential equations hold
\begin{equation}\label{diffeq}
\begin{aligned}
\begin{split}
\dot{\alpha}_-=&-\frac{2}{m}\sin\theta_+,\\
\dot{\theta}_+=&\frac{1}{4m}\alpha_-\left(2m^2\Omega_+^2-\alpha_+^2\right)\sec\theta_+-\frac{\alpha_+}{m},\\
\dot{\alpha}_+=&\frac{1}{2m}\left(2m^2\Omega_+^2-\alpha_+^2\right)\tan\theta_++m\Omega_-^2\sinh\theta_--2\lambda\cosh\theta_-,\\
\dot{\theta}_-=&\frac{\alpha_-\left(2\lambda\sinh\theta_--m\Omega_-^2\cosh\theta_-\right)}{2\cos\theta_++\alpha_+\alpha_-},
\end{split}
\end{aligned}
\end{equation}
We solve these coupled differential equations by differentiating the first equation three times and at each stage substituting in the expressions for $\dot{\theta}_+$, $\dot{\theta}_-$, $\dot{\alpha}_+$ and $\dot{\alpha}_-$
\begin{equation}\label{alphameq}
\begin{aligned}
\begin{split}
\ddot{\alpha}_-&=\frac{1}{2m^2}\alpha_+\left(4\cos\theta_++\alpha_+\alpha_-\right)-\Omega_+^2\alpha_-,\\
\dddot{\alpha}_-&=\frac{1}{m^2}\left(m\Omega_-^2\sinh\theta_--2\lambda\cosh\theta_-\right)\left(2\cos\alpha_++\alpha_+\alpha_-\right)+\frac{4\Omega_+^2}{m}\sin\alpha_+,\\
\ddddot{\alpha}_-&=-2\Omega_+^2\left[\frac{1}{2m^2}\alpha_+\left(4\cos\theta_++\alpha_+\alpha_-\right)-\Omega_+^2\alpha_-\right]-\left(\Omega_-^4-4\frac{\lambda^2}{m^2}\right)\alpha_-.
\end{split}
\end{aligned}
\end{equation}
In the final, fourth order equation we can clearly replace the bracket by $\ddot{\alpha}_-$ to obtain a fourth order equation solely in terms of $\alpha_-$
\begin{equation}\label{FourthOrder}
\ddddot{\alpha}_-+2\Omega_+^2\ddot{\alpha}_-+\delta\alpha_-=0,
\end{equation}
where $\delta=\Omega_-^4-4\frac{\lambda^2}{m^2}$. The solution to equation (\ref{FourthOrder}) allows us to go back and calculate $\theta_+$, $\theta_-$ and $\alpha_+$ using the equations (\ref{diffeq}) and (\ref{alphameq}) in terms of $\alpha_-$ together with its derivatives. 
\begin{equation}\label{dysonparameters}
\theta_+=-\arcsin \frac{m\dot{\alpha}_-}{2}, \quad \alpha_+=\frac{-\sqrt{4-m^2\dot{\alpha}^2_-}\pm\beta}{\alpha_-}, \quad e^{\theta_-}=\frac{\pm m^2\gamma\left(\pm\right) m\sqrt{m^2\gamma^2+\delta\beta^2}}{\beta\left(m\Omega_-^2-2\lambda\right)},
\end{equation}
where $\beta=\sqrt{4+2m^2\Omega^2_+\alpha^2_--m^2\left(\dot{\alpha}^2_--2\alpha_-\ddot{\alpha}_-\right)}$ and $\gamma=2\Omega_+^2\dot{\alpha}_-+\dddot{\alpha}_-$. The $\left(\pm\right)$ indicates that we can take either the positive or negative square root independently of the choice of sign for $\pm m^2\gamma$. These expressions require some further analysis in order to ensure they remain real and smooth for all values of $\alpha_-$ and its derivatives. It is clear that there is the possibility of singularities in $\alpha_+$ when $\alpha_-=0$ and in $\theta_-$ when $m\Omega_-^2=2\lambda$. We therefore analyse these solutions in the corresponding limits. In order to ensure there are no singularities in $\alpha_+$, we must take the sign of $\beta$ to be positive. This can be seen by expanding $\alpha_+$ when $\alpha_-\rightarrow0$

\begin{eqnarray}
\begin{aligned}
\begin{split}
\alpha_{+\left(\alpha_-\rightarrow0\right)}=\frac{-\sqrt{4-m^2\dot{\alpha}^2_-}}{\alpha_-}\pm\frac{\sqrt{4-m^2\dot{\alpha}^2_-}}{\alpha_-}\left[1+\frac{\alpha_-m^2\left(\Omega_+^2\alpha_-+\ddot{\alpha}_-\right)}{4-m^2\dot{\alpha}_-^2}\right.\\
\left.-\frac{1}{8}\left(\frac{2\alpha_-m^2\left(\Omega_+^2\alpha_-+\ddot{\alpha}_-\right)}{4-m^2\dot{\alpha}_-^2}\right)^2+...\right].
\end{split}
\end{aligned}
\end{eqnarray}
Therefore it is clear we must take the sign of $\beta$ to be positive in order to eliminate the singularity when $\alpha_-=0$. In this case $\alpha_{+\left(\alpha_-\rightarrow0\right)}$ becomes

\begin{equation}
\alpha_{+\left(\alpha_-\rightarrow0\right)}=\frac{m^2\ddot{\alpha}_-}{\sqrt{4-m^2\dot{\alpha}^2_-}},
\end{equation}
and so we see there is no singularity at $\alpha_-=0$. This also means the sign $\pm m^2\gamma$ must be taken to be positive. When $m\Omega_-^2=2\lambda$ there appears to be a singularity in $\theta_-$. In order to investigate this we expand the square root in $\theta_-$ in terms of $\delta$,
\begin{equation}
e^{\theta_-}=\frac{ m^2\gamma}{\beta\left(m\Omega_-^2-2\lambda\right)}\left(\pm\right)\frac{m^2\gamma}{\beta\left(m\Omega_-^2-2\lambda\right)}\left[1+\frac{1}{2}\frac{\delta\beta^2}{m^2\gamma^2}-\frac{1}{8}\left(\frac{\delta\beta^2}{m^2\gamma^2}\right)^2+...\right].
\end{equation}
Recalling that $\delta=\Omega_-^4-4\frac{\lambda^2}{m^2}=\left(m\Omega_-^2-2\lambda\right)\left(m\Omega_-^2+2\lambda\right)/m^2$, we must therefore choose the sign of the square root $\left(\pm\right)$ to be negative. This ensures the singularity is eliminated when $m\Omega_-^2=2\lambda$. The resulting expression is
\begin{equation}
e^{\theta_-}=-\frac{\Omega_-^2\beta}{m\gamma}.
\end{equation}
When $m\Omega_-^2=-2\lambda$ we must choose the sign of the square root $\left(\pm\right)$ to be positive in order to avoid a logarithm of zero. In this case the expression is
\begin{equation}
e^{\theta_-}=\frac{m\gamma}{\Omega_-^2\beta}.
\end{equation}
Now we rewrite our solutions for $\theta_+$, $\theta_-$ and $\alpha_+$ using this new information
\begin{equation}\label{dysonparameterssingularity}
\theta_+=-\arcsin \frac{m\dot{\alpha}_-}{2}, \quad \alpha_+=\frac{-\sqrt{4-m^2\dot{\alpha}^2_-}+\beta}{\alpha_-},
\end{equation}
\begin{equation}\label{dysonparameterssingularity2}
e^{\theta_-}=\frac{ m^2\gamma\pm m\sqrt{m^2\gamma^2+\delta\beta^2}}{\beta\left(m\Omega_-^2-2\lambda\right)} \quad \text{for}\;\; m\Omega_-^2\neq\pm2\lambda,
\end{equation}
\begin{equation}
e^{\theta_-}=-\frac{\Omega_-^2\beta}{m\gamma} \quad \text{for} \;\; m\Omega_-^2=2\lambda, \quad e^{\theta_-}=\frac{m\gamma}{\Omega_-^2\beta} \quad \text{for}\;\; m\Omega_-^2=-2\lambda.
\end{equation}
As we can see in the equations (\ref{dysonparameterssingularity}) and (\ref{dysonparameterssingularity2}), there are restrictions placed on the magnitude of $\alpha_-$ and its derivatives in order for the parameters $\alpha_+$, $\theta_+$ and $e^{\theta_-}$ to be real. From the first equation in (\ref{dysonparameterssingularity}) we must have $|m\dot{\alpha_-}/2|\leq1$ and from equation (\ref{dysonparameterssingularity2}) we must have $m^2\gamma^2+\delta\beta^2\geq0$ when $\delta<0$. We address these restrictions as we calculate $\alpha_-$. 

We can solve (\ref{FourthOrder}) without consideration to the sign of $\delta$ and obtain a valid solution. However, we wish to preserve the reality of $\alpha_-$ in order to prevent $\eta$ from becoming unitary and leading to a simple gauge transformation. Therefore we must consider three separate regimes arising from the time-independent analysis, these are: the unbroken regime where $|m\Omega_-^2|>2|\lambda|$ ($\delta>0$), the spontaneously broken regime with $|m\Omega_-^2|<2|\lambda|$ ($\delta<0$) and the exceptional point where $|m\Omega_-^2|=2|\lambda|$ ($\delta=0$). These regimes must be treated separately as they lead to qualitatively different solutions. In all three cases $\alpha_-\in\mathbb{R}$ as required.\\

\noindent\underline{For $\delta>0$}, the solution is
\begin{equation}
\alpha_-=c_1\cos\left(\Delta_+ t\right)+c_2\sin\left(\Delta_+ t\right)+c_3\cos\left(\Delta_- t\right)+c_4\sin\left(\Delta_- t\right),
\end{equation}
where  $c_{1,2,3,4}$ are constants of integration. The number of constants reflects the number of first order differential equations we started with in (\ref{diffeq}), and so we get four as expected. However, they are not free to take any value as we must have $|m\dot{\alpha_-}/2|\leq1$. The choice

\begin{equation}
c_1=\frac{1}{m\Delta_+}, \quad c_3=\frac{1}{m\Delta_-}, \quad c_2=c_4=0
\end{equation}
ensures the condition is met. There are many other choices that satisfy this condition as the Dyson map and the metric is not unique as already stated. However we choose the constants above for simplicity and as a demonstration. The frequencies are
\begin{equation}
\Delta_\pm=\sqrt{\Omega_+^2\pm2\sqrt{\Omega_x^2\Omega_y^2+\frac{\lambda^2}{m^2}}}.
\end{equation}
The condition for $\Delta_-$ to be real is $\delta>0$, so in the unbroken regime both $\Delta_\pm$ are real. When $\delta<0$, $\Delta_-$ becomes imaginary and so solving (\ref{FourthOrder}) in broken regime must be considered separately.\\

\noindent\underline{For $\delta<0$}, the solution is
\begin{equation}
\alpha_-=\tilde{c}_1\cos\left(\tilde{\Delta}_+ t\right)+\tilde{c}_2\sin\left(\tilde{\Delta}_+ t\right)+\tilde{c}_3\cosh\left(\tilde{\Delta}_- t\right)+\tilde{c}_4\sinh\left(\tilde{\Delta}_- t\right)
\end{equation} 
with $\tilde{c}_{1,2,3,4}$ being the constants of integration and 
\begin{equation}
\tilde{\Delta}_\pm=\sqrt{2\sqrt{\Omega_x^2\Omega_y^2+\frac{\lambda^2}{m^2}}\pm\Omega_+^2}.
\end{equation} 
We have both $\tilde{\Delta}_\pm$ being real when $\delta<0$. In order to satisfy the condition $|m\dot{\alpha_-}/2|\leq1$ we must choose our constants such that we eliminate exponential growth from the hyperbolic functions. The choice 

\begin{equation}
\tilde{c}_4=-\tilde{c}_3
\end{equation}
restricts the solution to exponential decay. The condition $m^2\gamma^2+\delta\beta^2\geq0$ is satisfied with the choice $\tilde{c}_1^2+\tilde{c}_2^2\geq\frac{1}{\tilde{\Delta}_+^2m}\left(2/m+\Omega_+/\sqrt{m\Omega_x^2\Omega_y^2+\lambda^2}\right)$. Therefore we set

\begin{eqnarray}
\tilde{c}_1&=&\sqrt{\frac{1}{m\tilde{\Delta}_+^2}\left(\frac{2}{m}+\frac{\Omega_+}{\sqrt{m\Omega_x^2\Omega_y^2+\lambda^2}}\right)}, \\ \tilde{c}_3&=&-\tilde{c}_4\;\;=\;\;-\frac{1}{2m\tilde{\Delta}_-}\left(\frac{2}{m}-\frac{\Omega_+}{\sqrt{m\Omega_x^2\Omega_y^2+\lambda^2}}\right), \\
\tilde{c}_2&=&0.
\end{eqnarray}
With these choices, even in the broken regime, we obtain a real solution for $\alpha_-$ and consequently for $\eta$ as we satisfy all the conditions imposed on the integration constants. Once again there are other choices that satisfy these conditions but we make these choices for simplicity and as a demonstration.\\

\noindent\underline{For $\delta=0$}, the solution is 
\begin{equation}
\alpha_-=\hat{c}_1\cos\left(\sqrt{2}\Omega_+ t\right)+\hat{c}_2\sin\left(\sqrt{2}\Omega_+ t\right)+\hat{c}_3t+\hat{c}_4,
\end{equation}
where $\hat{c}_{1,2,3,4}$ are the constants of integration. Once again we must enforce the condition $|m\dot{\alpha_-}/2|\leq1$. Setting 

\begin{equation}
\hat{c}_1=\frac{1}{m\sqrt{2}\Omega_+}, \quad \hat{c}_2=0, \quad \hat{c}_3=\hat{c}_4=\frac{1}{m},
\end{equation}
satisfies this constraint. Of course, there are many choices that satisfy the conditions but we pick simple constants here for demonstration purposes.\\

We have obtained a real solution for $\alpha_-$ for all values of $\delta$ and have fixed the integration constants such that the parameters $\alpha_+$ and $\theta_+$ and $e^{\theta_-}$ are real for all $t$. This means we have well-defined metric, $\varrho\left(t\right)=\eta\left(t\right)^\dagger\eta\left(t\right)$, for all values of $\Omega_x$, $\Omega_y$, $\lambda$ and $m$.
\begin{equation}
\varrho\left(t\right)=e^{\theta_-\left(t\right)J_-}e^{\alpha_+\left(t\right) L_+}e^{\theta_+\left(t\right)J_+}e^{2\alpha_-\left(t\right)L_-}e^{\theta_+\left(t\right) J_+}e^{\alpha_+\left(t\right)L_+}e^{\theta_-\left(t\right) J_-}.
\end{equation}
Thus importantly we have a time-independent, non-Hermitian system exhibiting spontaneously broken $\mathcal{PT}$-symmetry that ordinarily (in the time-independent regime) only has a well-defined metric in the unbroken regime. However, we have shown that by introducing time-dependence into this metric, the system becomes well-defined over the \textit{entire} parameter set, including the broken regime. 

The resulting Hermitian Hamiltonian is
\begin{equation}
h\left(t\right)=h_{x,-}\left(t\right)+h_{y,+}\left(t\right),
\end{equation}
where
\begin{equation}
h_{z,\pm}\left(t\right)=\frac{1}{2M_\pm\left(t\right)}p_z^2+\frac{1}{2}M_\pm\left(t\right)\omega_\pm\left(t\right)^2z^2\pm g\left(t\right)\{z,p_z\}, \quad z=x,y,
\end{equation}
%\begin{equation}
%h_y\left(t\right)=\frac{1}{2M_y\left(t\right)}p_y^2+\frac{1}{2}M_y\left(t\right)\tilde{\omega}_y\left(t\right)^2y^2-g\left(t\right)\{y,p\}
%\end{equation}
are Swanson type \cite{swanson2004transition} Hamiltonians with time-dependent mass and frequency. The time-dependent terms can be expressed in terms of the Dyson map parameters.

\begin{equation}
M_\pm\left(t\right)=m\left[\cos\theta_++m\alpha_-^2\Gamma_\pm\right]^{-1},
\end{equation}

\begin{equation}
\omega_\pm\left(t\right)^2=\frac{4\Gamma_\mp}{M_\pm},
\end{equation}

\begin{equation}
g\left(t\right)=\frac{\alpha_-\Theta\sin\theta_+}{4},
\end{equation}
where 
\begin{equation}
\Theta\left(t\right)=\frac{2\lambda\sinh\theta_--m\Omega_-^2\cosh\theta_-}{2\cos\theta_++\alpha_+\alpha_-}, \quad \Gamma_\pm\left(t\right)=\frac{1}{16m}\sec\theta_+\left(2m^2\Omega_+^2-\alpha_+^2\right)\pm\frac{\Theta\cos\theta_+}{4}.
\end{equation}
As a consistency check, we see that $\theta_-$ does indeed only occur in hyperbolic form and therefore all quantities above are real for all $t$. We can recover the time-independent solution by setting $\theta_+\left(t\right)=\alpha_-\left(t\right)=\alpha_+\left(t\right)=0$. In this case $M_\pm\left(t\right)=m$, $\omega_-\left(t\right)=\omega_x$, $\omega_+\left(t\right)=\omega_y$ and $g\left(t\right)=0$. Finally $\theta_-\left(t\right)=\theta$.

As the resulting Hermitian Hamiltonian $h\left(t\right)$ is decoupled, we can solve each system separately following \cite{pedrosa2004complete}. Therefore for $h_{z,\pm}\left(t\right)$ we can solve the time-dependent Schr\"odinger equation using the Lewis Riesenfeld invariants for the time-dependent wave functions

\begin{equation}
\phi_{z,n,\pm}\left(t\right)=\frac{e^{i\alpha_{n,\pm}\left(t\right)}}{\sqrt{\rho_\pm\left(t\right)}}\exp\left[iM_\pm\left(t\right)\left(\frac{i}{M_\pm\left(t\right)\rho_\pm\left(t\right)^2}+\frac{\dot{\rho}_\pm\left(t\right)}{\rho_\pm\left(t\right)}\mp2g\left(t\right)\right)\frac{z^2}{2}\right]\mathcal{H}_n\left[\frac{z}{\rho_\pm\left(t\right)}\right],
\end{equation}
where $\mathcal{H}_n$ are the Hermite polynomials of order $n$ and
\begin{equation}
\alpha_{n,\pm}\left(t\right)=-\left(n+\frac{1}{2}\right)\int^{t}\frac{1}{M_\pm\left(s\right)\rho_\pm\left(s\right)^2}ds.
\end{equation}
$\rho_\pm$ obeys the dissipative Ermakov-Pinney equation
\begin{equation}
\ddot{\rho}_\pm+\frac{\dot{M}_\pm}{M_\pm}\dot{\rho}_\pm+\left(\omega^2_\pm\mp2\dot{g}-4g^2\mp2g\frac{\dot{M}_\pm}{M_\pm}\right)\rho_\pm=\frac{1}{M^2_\pm\rho^3_\pm}.
\end{equation}
This is a rather technical equation as it contains complicated expressions involving the Dyson map parameters. The wave functions that satisfy the time-dependent Schr\"odinger equation for $h\left(t\right)$ coming from the Lewis Riesenfeld invariants are therefore
\begin{equation}
\phi_{n,m}\left(t\right)=\phi_{x,n,-}\left(t\right)\phi_{y,m,+}\left(t\right),
\end{equation}
from which we can form a general wave function. We see that the final expressions become quite technical but are nonetheless explicitly calculated and manageable. We have shown that even a highly technical system comprised of 10 generators elicits a Dyson map and a metric when careful attention is paid to the solution procedure (here the choice of solving the time-dependent Dyson equation rather than the quasi-Hermiticity equation or the Lewis Riesenfeld invariants).

\section{Summary}

We have presented the first higher dimensional solution of the
time-dependent Dyson relation (\ref{xyp}) relating a non-Hermitian and a
Hermitian Hamiltonian system with infinite dimensional Hilbert space. As for
the one dimensional case studied in Chapter 3, we have demonstrated
that the time-independent non-Hermitian system in the spontaneously broken $%
\mathcal{PT}$-regime becomes physically meaningful when including an
explicit time-dependence into the parameters of the model and allowing the
metric operator also to be time-dependent. The energy operator has perfectly well-defined real expectation values (\ref{Eobservable}). Furthermore, we have solved two higher dimensional systems, one comprised of an algebra of 4 generators, and the other with 10 generators. This shows that we can solve even highly technical systems for the Dyson operator and the metric. It will be of great importance when we come to consider systems coupled to the environment (in chapter 6) in order to analyse the entanglement entropy.

Technically we have compared two equivalent solution procedures, solving the
time-dependent Dyson relation directly for the Dyson map or alternatively
computing Lewis-Riesenfeld invariants first and subsequently constructing
the Dyson map from the similarity relation that related the Hermitian and
non-Hermitian invariants. The latter approach was found to be simpler as the
similarity relation is far easier than the differential version (\ref{DysonE}).
The price one pays in this approach is that one needs to compute the two
invariants first. However, the differential equations for these quantities
turned out to be easier than equation (\ref{DysonE}). In particular, it was possible
to entirely bypass the dissipative Ermakov-Pinney equation in the
computation of $\eta (t)$. Nonetheless, this ubiquitous equation re-emerged
in the evaluation of the eigenfunctions involving different time-dependent
fields and with a changed sign.

\chapter{Quasi-Exactly Solvable Systems}

Quasi-exactly solvable (QES) quantum systems are characterized by the
feature that only part of their infinite energy spectrum and corresponding
eigenfunctions can be calculated analytically. Systematic studies of such
type of systems have been carried out by casting them into the form of Lie
algebraic quantities \cite{OP2,OP4}. QES systems that can be cast into such a form are usually
referred to as QES models of Lie algebraic type \cite{Turbiner00,Tur0}. The
relevant underlying algebras are either of $sl_{2}(\mathbb{C})$-type, with
their compact and non-compact real forms $su(2)$ and $su(1,1)$, respectively 
\cite{Hum}, or of Euclidean Lie algebras type \cite%
{E2Fring,E2Fring2,fring2016unifying}. The latter class was found to be
particularly useful when dealing with certain types of non-Hermitian systems.

While many QES models have been studied in stationary settings, little was
known for time-dependent systems before our work \cite{AndTom5}. So far a time-dependence has only been introduced into the eigenfunctions in form of a dynamical phase \cite%
{hou1999quasi,mayer2000time}. However, no QES systems with explicitly
time-dependent Hamiltonians have been considered up to now. The main purpose
of this chapter is to demonstrate how they can be dealt with and to initiate
further studies of such type of systems. We provide the analytical solutions
to a QES Hamiltonian quantum system with explicit time-dependence. As a
concrete example we consider QES systems of $E_{2}$-Lie algebraic type.
Technically we make use of the time-dependent Dyson equation. It will allow us to solve a Hermitian time-dependent Hamiltonian
system by solving first a static non-Hermitian system as an auxiliary
problem with a time-dependence in the metric operator.

Systems built up from Euclidean Lie algebras, in particular of $E_{2}$, have
a wide range of physical applications. They have been employed for instance
in the formal quantisation of strings on tori \cite{Isham}. Depending on the
chosen representation of the algebra one can describe a large number of
concrete physical systems. Common representations for $E_{2}$ may lead to
two dimensional systems or most commonly in optical settings, the
trigonometric representation, see below, correspond to Mathieu potentials
and variations thereof. The latter have proven to be useful and accurate in the
decription of energy band structures in crystals \cite{MatLongo} and
especially in the experimental and theoretical study of optical solitons 
\cite{musslimani2008optical,makris2010pt,MatHugh,MatHughEva,Guo,OPMidya}. Here we consider
explicitly time-dependent versions of these type of systems and keep our
discussion generic, that is independent of the choice of a concrete
representation for the underlying algebra.

\section{$E_2$-Hamiltonian Systems}\label{E2Intro}
The Hermitian Hamiltonian systems we study here are of the general form%
\begin{equation}
h(t)=\mu _{JJ}(t)J^{2}+\mu _{J}(t)J+\mu _{u}(t)u+\mu _{v}(t)v+\mu
_{uu}(t)u^{2}+\mu _{vv}(t)v^{2}+\mu _{uv}(t)uv,  \label{1}
\end{equation}%
where the time-dependent coefficient functions $\mu _{i}$, $i\in
\{J,JJ,u,v,uu,vv,uv\}$, are real and $u$, $v$ and $J$ denote the three
generators that span the Euclidean-algebra $E_{2}$. They obey the
commutation relations%
\begin{equation}
\left[ u,J\right] =iv,\qquad \left[ v,J\right] =-iu,\qquad \text{and\qquad }%
\left[ u,v\right] =0.  \label{E2}
\end{equation}%
Considering here only Hermitian representations with $J^{\dagger }=J$, $%
v^{\dagger }=v$ and $u^{\dagger }=u$, the Hamiltonian in equation (\ref{1})
is clearly Hermitian. Standard representation are for instance the
trigonometric representation $J:=-i\partial _{\theta }$, $u:=\sin \theta $
and $v:=\cos \theta $ or a two-dimensional representation $J:=yp_{x}-xp_{y}$%
, $u:=x$ or $v:=y$ with $x$, $y$, $p_{x}$, $p_{y}$ denoting Heisenberg
canonical variables with non-vanishing commutators $\left[ x,p_{x}\right] =%
\left[ y,p_{y}\right] =i$.

Before we solve a concrete system in a quasi-exactly solvable fashion we
consider first the fully time-dependent Dyson relation with time-dependent
non-Hermitian Hamiltonian $H(t)$ and investigate which type of Hamiltonians
can be related to the Hermitian Hamiltonian $h(t)$ in (\ref{1}). We will see
that in some cases we are even forced to take $H(t)$ or part of it to be
time-independent. As not many explicit solutions to the time-dependent Dyson
relation are known, this will be a valuable result in itself.

This chapter is organized as follows: In section \ref{E2DE} we explore various
types of $\mathcal{PT}$-symmetries that leave the Euclidean $E_{2}$-algebra
invariant and investigate time-dependent non-Hermitian Hamiltonians in terms of
$E_{2}$-algebraic generators that respect these symmetries. We find new
solutions to the time-dependent Dyson relation for those type of
Hamiltonians by computing the corresponding Hermitian Hamiltonians and the
Dyson map. In section \ref{TDQE} we provide analytical solutions for a
concrete model respecting a particular $\mathcal{PT}$-symmetry. We compute
the eigenstates of the Lewis-Riesenfeld invariants and the time-dependent
Hermitian Hamiltonian in a quasi-exactly solvable fashion. A three-level
system is then presented in more detail. 

\section{Solutions to the time-dependent Dyson equation for $E_{2}$%
-Hamiltonians}\label{E2DE}

A key property in the study and classification of Hamiltonian systems
related to the $E_{2}$-algebra are the antilinear symmetries \cite{EW} that
leave the algebra (\ref{E2}) invariant. Given the general context of $%
\mathcal{PT}$-symmetric/quasi-Hermitian systems we call these symmetries $%
\mathcal{PT}_{i},i=1,2,\ldots $ As discussed in more detail in \cite%
{DFM,DFM2}, there are many options which all give rise to models with
qualitatively quite distinct features. It is easy to see that each of the
following antilinear maps leave all the commutation relations (\ref{E2})
invariant 
\begin{equation}
\begin{array}{lllll}
\mathcal{PT}_{1}:~~ & J\rightarrow -J,~~ & u\rightarrow -u,~~ & v\rightarrow
-v,~~~ & i\rightarrow -i, \\ 
\mathcal{PT}_{2}: & J\rightarrow -J, & u\rightarrow u, & v\rightarrow v, & 
i\rightarrow -i, \\ 
\mathcal{PT}_{3}: & J\rightarrow J, & u\rightarrow v, & v\rightarrow u, & 
i\rightarrow -i, \\ 
\mathcal{PT}_{4}: & J\rightarrow J, & u\rightarrow -u, & v\rightarrow v, & 
i\rightarrow -i, \\ 
\mathcal{PT}_{5}: & J\rightarrow J, & u\rightarrow u, & v\rightarrow -v, & 
i\rightarrow -i.%
\end{array}
\label{PT}
\end{equation}%
Next we seek non-Hermitian Hamiltonians that respect either of these
symmetries. Focussing here on time-dependent Hamiltonians consisting
entirely of linear and quadratic combinations of $E_{2}$-generators they can
all be cast into the general form 
\begin{eqnarray}
H_{\mathcal{PT}_{i}}(t) &=&\mu _{JJ}(t)J^{2}+\mu _{J}(t)J+\mu _{u}(t)u+\mu
_{v}(t)v+\mu _{uJ}(t)uJ+\mu _{vJ}(t)vJ  \label{H} \\
&&+\mu _{uu}(t)u^{2}+\mu _{vv}(t)v^{2}+\mu _{uv}(t)uv.  \notag
\end{eqnarray}%
Demanding that $\left[ H_{\mathcal{PT}_{i}}(t),\mathcal{PT}_{i}\right] =0$,
the symmetries are implemented by taking the coefficient functions to be
either real, purely imaginary or relate different functions to each other by
conjugation. For the different symmetries in (\ref{PT}) we are forced to
take 
\begin{equation}
\begin{array}{lll}
\mathcal{PT}_{1}:~~ & (\mu _{J},\mu _{u},\mu _{v})\in i\mathbb{R},~~ & (\mu
_{JJ},\mu _{uJ},\mu _{vJ},\mu _{uu},\mu _{vv},\mu _{uv})\in \mathbb{R},~~ \\ 
\mathcal{PT}_{2}: & (\mu _{J},\mu _{uJ},\mu _{vJ})\in i\mathbb{R}, & (\mu
_{u},\mu _{v},\mu _{JJ},\mu _{uu},\mu _{vv},\mu _{uv})\in \mathbb{R}, \\ 
\mathcal{PT}_{3}: & (\mu _{JJ},\mu _{J},\mu _{uv})\in \mathbb{R}, & \mu
_{u}=\mu _{v}^{\ast },\mu _{uJ}=\mu _{vJ}^{\ast },\mu _{uu}=\mu _{vv}^{\ast }
\\ 
\mathcal{PT}_{4}: & (\mu _{u},\mu _{uJ},\mu _{uv})\in i\mathbb{R}, & (\mu
_{J},\mu _{v},\mu _{JJ},\mu _{vJ},\mu _{uu},\mu _{vv})\in \mathbb{R}, \\ 
\mathcal{PT}_{5}: & (\mu _{v},\mu _{vJ},\mu _{uv})\in i\mathbb{R},~ & (\mu
_{J},\mu _{u},\mu _{JJ},\mu _{uJ},\mu _{uu},\mu _{vv})\in \mathbb{R}.%
\end{array}
\label{coe}
\end{equation}%
Except for very specific combinations of the coefficient functions, the
Hamiltonians $H_{\mathcal{PT}_{i}}(t)$ are non-Hermitian in general.

We now solve the time-dependent Dyson relation (\ref{DysonE}) for $\eta (t)$
by mapping different $\mathcal{PT}_{i}$-symmetric versions of $H(t)$ to a
Hermitian Hamiltonian $h(t)$ of the form (\ref{1}). For the time-dependent
Dyson map we make an ansatz in terms of all the $E_{2}$-generators 
\begin{equation}
\eta (t)=e^{\tau (t)v}e^{\lambda (t)J}e^{\rho (t)u}.  \label{etaQH}
\end{equation}%
At this point we allow $\lambda ,\tau ,\rho \in \mathbb{C}$, keeping in mind
that $\eta (t)$ does not have to be Hermitian. However, we exclude here unitary
operators as in that case $\eta
(t)$ just becomes a gauge transformation. The adjoint action of this
operator on the $E_{2}$-generators is computed by using the BCH relation
\begin{eqnarray}
\eta J\eta ^{-1} &=&J+i\rho \cosh (\lambda )v-[i\tau +\rho \sinh (\lambda
)]u,  \label{ad1} \\
\eta u\eta ^{-1} &=&\cosh (\lambda )u-i\sinh (\lambda )v,  \label{ad2} \\
\eta v\eta ^{-1} &=&\cosh (\lambda )v+i\sinh (\lambda )u.  \label{ad3}
\end{eqnarray}%
The gauge-like term in (\ref{DysonE}) acquires the form%
\begin{equation}
i\dot{\eta}\eta ^{-1}=i\dot{\lambda}J+\left[ i\dot{\rho}\cosh \left( \lambda
\right) +\tau \dot{\lambda}\right] u+\left[ \dot{\rho}\sinh \left( \lambda
\right) +i\dot{\tau}\right] v.  \label{gt}
\end{equation}%
For the
computation of the time-dependent energy operator $\tilde{H}(t)$, see below,
we also require the term%
\begin{equation}
i\eta ^{-1}\dot{\eta}=i\dot{\lambda}J+\left[ i\dot{\rho}+\dot{\tau}\sinh
\left( \lambda \right) \right] u+\left[ \rho \dot{\lambda}+i\dot{\tau}\cosh
\left( \lambda \right) \right] v.  \label{auxen}
\end{equation}%
Using (\ref{ad1})-(\ref{ad3}) we calculate next the adjoint action of $\eta $
on $H(t)$ and add the expression in (\ref{gt}). Demanding that the result is
Hermitian will constrain the time-dependent functions $\mu _{i}(t)$, $%
\lambda (t)$, $\tau (t)$ and $\rho (t)$. We need to treat each $\mathcal{PT}$%
-symmetry separately.

\subsection{Time-dependent $\mathcal{PT}_{1}$-invariant Hamiltonians}

For the $\mathcal{PT}_{1}$-invariant Hamiltonian with coefficient functions
as specified in (\ref{coe}) we have to be aware that for $\mu _{J}=\mu _{uJ}=$ $%
\mu _{vJ}=0$ the Hamiltonian $H_{\mathcal{PT}_{1}}(t)$ becomes Hermitian.
Substituting the general form for $H_{\mathcal{PT}_{1}}(t)$ into (\ref{DysonE}),
using (\ref{ad1})-(\ref{ad3}), (\ref{gt}), reading off the coefficients in front of the
generators and demanding that the right hand side becomes Hermitian enforces
to take the functions $\lambda ,\tau ,\rho \in \mathbb{R}$ in (\ref{etaQH}). The
resulting Hermitian Hamiltonian is%
\begin{eqnarray}
h_{\mathcal{PT}_{1}} &=&J^{2}\mu _{JJ}+\frac{\left[ \mu _{vJ}\tanh \lambda
	-\mu _{J}\mu _{vJ}\right] \sinh \lambda }{2\mu _{JJ}}u-\frac{\mu _{J}\mu
	_{uJ}\tanh \lambda \sech\lambda }{2\mu _{JJ}}v \\
&&+\left( \mu _{uu}-\frac{\mu _{uJ}^{2}\tanh ^{2}\lambda }{4\mu _{JJ}}%
\right) u^{2}+\left( \mu _{uu}+\frac{\cosh ^{2}(\lambda )\mu _{vJ}^{2}-\mu
	_{uJ}^{2}}{4\mu _{JJ}}\right) v^{2}+\mu _{uv}uv,  \notag \\
&&+\frac{\mu _{uJ}}{2}\sech\lambda \{u,J\}+\frac{\mu _{vJ}}{2}\cosh
\lambda \{v,J\}  \notag
\end{eqnarray}%
with 7 constraining relations%
\begin{eqnarray}
\lambda  &=&-\int\nolimits^{t}\mu _{J}(s)ds,~~\tau =\frac{\mu _{vJ}\sinh
	\lambda }{2\mu _{JJ}},~~\rho =\frac{\mu _{uJ}\tanh \lambda }{2\mu _{JJ}}%
,~~\mu _{vv}=\mu _{uu}+\frac{\mu _{vJ}^{2}-\mu _{uJ}^{2}}{4\mu _{JJ}}, \\
\mu _{uv} &=&\frac{\mu _{uJ}\mu _{vJ}}{2\mu _{JJ}},~~\mu _{u}=\frac{\mu
	_{J}\mu _{uJ}-\dot{\mu}_{uJ}\tanh \lambda }{2\mu _{JJ}}+\frac{\mu _{vJ}}{2}%
,~~\mu _{v}=\frac{\mu _{J}\mu _{vJ}-\dot{\mu}_{vJ}\tanh \lambda }{2\mu _{JJ}}%
-\frac{\mu _{uJ}}{2}.  \notag
\end{eqnarray}%
Thus from the original 12 free parameters, i.e. the 9 coefficient functions $%
\mu _{i}$ and the 3 functions $\lambda ,\tau ,\rho $ in the Dyson map, we
can still freely choose 5. In comparison with the other $\mathcal{PT}_{i}$%
-symmetries, this is the most constrained case. We also note that this
system is the only one in which all three functions in the Dyson map are
constrained when we take the coefficient functions $\mu _{i}$ as primary
quantities.

\subsection{Time-dependent $\mathcal{PT}_{2}$-invariant Hamiltonians}\label{PT2}

For convenience we take the coefficient function $\mu _{JJ}$ to be
time-independent. Of course the general scenario with $\mu _{JJ}(t)$ is also
possible to consider, but leads to more cumbersome expressions. The
Hamiltonian $H_{\mathcal{PT}_{2}}(t)$ becomes Hermitian for $\mu _{J}=0$, $%
\mu _{uJ}=2\mu _{u}$, $\mu _{vJ}=-2\mu _{u}$, but is non-Hermitian
otherwise. Proceeding as in the previous section the implementation of (\ref%
{DysonE}) enforces to take $\tau ,\rho \in \mathbb{R}$ and $\lambda \in i%
\mathbb{R}$ in (\ref{etaQH}), which makes the Dyson map $\mathcal{PT}_{2}$%
-symmetric. The Hermitian Hamiltonian is computed to be
\begin{eqnarray}
h_{\mathcal{PT}_{2}} &=&\mu _{JJ}J^{2}+\dot{\lambda}J+\left[ \left( \mu _{u}+%
\frac{\mu _{vJ}}{2}\right) \cos \lambda +\left( \frac{\mu _{uJ}}{2}-\mu
_{v}\right) \sin \lambda \right] u \\
&&+\left[ \left( \mu _{v}-\frac{\mu _{uJ}}{2}\right) \cos \lambda +\left(
\mu _{u}+\frac{\mu _{vJ}}{2}\right) \sin \lambda \right] v+\left[ \left( 
\frac{\mu _{uJ}^{2}-\mu _{vJ}^{2}}{8\mu _{JJ}}+\frac{\mu _{uu}-\mu _{vv}}{2}%
\right) \cos (2\lambda )\right.  \notag \\
&&-\left. \left( \frac{\mu _{uJ}\mu _{vJ}}{4\mu _{JJ}}+\frac{\mu _{uv}}{2}%
\right) \sin (2\lambda )+\frac{\mu _{uJ}^{2}+\mu _{vJ}^{2}}{8\mu _{JJ}}+%
\frac{\mu _{uu}+\mu _{vv}}{2}\right] u^{2}  \notag \\
&&+\left[ \left( \frac{\mu _{uJ}^{2}}{4\mu _{JJ}}+\mu _{uu}\right) \sin
^{2}\lambda +\left( \frac{\mu _{uJ}\mu _{vJ}}{4\mu _{JJ}}+\frac{\mu _{uv}}{2}%
\right) \sin 2\lambda +\left( \frac{\mu _{vJ}^{2}}{4\mu _{JJ}}+\mu
_{vv}\right) \cos ^{2}\lambda \right] v^{2}  \notag \\
&&+\left[ \left( \frac{\mu _{uJ}^{2}-\mu _{vJ}^{2}}{4\mu _{JJ}}+\mu
_{uu}-\mu _{vv}\right) \sin (2\lambda )+\left( \frac{\mu _{uJ}\mu _{vJ}}{%
2\mu _{JJ}}+\mu _{uv}\right) \cos (2\lambda )\right] uv,  \notag
\end{eqnarray}%
with 5 constraining relations 
\begin{equation}
\tau =\frac{\mu _{uJ}}{2\mu _{JJ}}\sec \lambda ,\qquad \rho =-\frac{\mu
_{vJ}+\mu _{uJ}\tan \lambda }{2\mu _{JJ}},\qquad \mu _{J}=\dot{\mu}_{uJ}=%
\dot{\mu}_{vJ}=0.  \label{2con}
\end{equation}%
One of the three functions in the Dyson map, e.g. $\lambda $, can be freely
chosen. Compared to the other cases this is the only one for which $\eta $
has the same $\mathcal{PT}_{i}$-symmetry as the corresponding non-Hermitian
Hamiltonian $H_{\mathcal{PT}_{i}}(t)$ when taking the constraints on $\tau
,\rho ,\lambda $ into account. In comparison we note that
we have less constraints as for instance in the case for the $\mathcal{PT}%
_{1}$-symmetry, however, they are quite different as some of the coefficient
functions can not be taken to be time-dependent and one even has to vanish.

\subsection{Time-dependent $\mathcal{PT}_{3}$-invariant Hamiltonians}

The Hamiltonian $H_{\mathcal{PT}_{3}}(t)$ becomes Hermitian for $\mu
_{vJ}=\mu _{uu}=0$ and $\mu _{uJ}=2\mu _{v}$. Using the same arguments as
above, we are forced to take $\tau ,\rho \in \mathbb{R}$ and $\lambda \in i%
\mathbb{R}$ in (\ref{etaQH}). The Hermitian Hamiltonian is computed to 
\begin{eqnarray}
h_{\mathcal{PT}_{3}} &=&J^{2}\mu _{JJ}+\left( \mu _{J}-\text{$\dot{\lambda}$}%
\right) J+\cos \lambda \left( \mu _{u}-\frac{\mu _{vJ}}{2}\right) (u+v)+\sin
\lambda \left( \mu _{u}-\frac{\mu _{vJ}}{2}\right) (v-u)~~~~~~~~ \\
&&+\left( \mu _{vv}+\frac{\mu _{vJ}^{2}}{4\mu _{JJ}}\right) \left(
u^{2}+v^{2}\right) +\left( \frac{\mu _{vJ}^{2}}{4\mu _{JJ}}-\frac{\mu _{uv}}{%
	2}\right) \sin (2\lambda )\left( u^{2}-v^{2}\right)   \notag \\
&&+\frac{\mu _{uJ}}{2}\cos \lambda \left[ \{v,J\}+\{u,J\}\right] +\frac{\mu
	_{uJ}}{2}\sin \lambda \left[ \{v,J\}-\{u,J\}\right]   \notag \\
&&+\cos (2\lambda )\left( \mu _{uv}-\frac{\mu _{vJ}^{2}}{2\mu _{JJ}}\right)
uv,  \notag
\end{eqnarray}%
with 5 constraining relations 
\begin{equation}
\tau =\frac{\mu _{vJ}}{2\mu _{JJ}}\sec \lambda ,~~\rho =\frac{\mu _{vJ}-\mu
	_{vJ}\tan \lambda }{2\mu _{JJ}},~~\mu _{v}=\frac{\mu _{vJ}}{2}+\frac{\mu
	_{J}\mu _{vJ}}{2\mu _{JJ}},~~\mu _{uv}=-\frac{\mu _{vJ}\mu _{uJ}}{2\mu _{JJ}}%
,~~\dot{\mu}_{vJ}=0.
\end{equation}%
Once again one of the coefficient functions has to be time-independent and
one of the three functions in the Dyson map can be chosen freely.

\subsection{Time-dependent $\mathcal{PT}_{4}$-invariant Hamiltonians}

The Hamiltonian $H_{\mathcal{PT}_{4}}(t)$ becomes Hermitian for $\mu
_{uJ}=\mu _{uv}=0$ and $\mu _{vJ}=2\mu _{u}$. By the same reasoning as above
we have to take $\tau ,\rho \in \mathbb{R}$ and $\lambda \in i\mathbb{R}$ in
(\ref{etaQH}). The Hermitian Hamiltonian results to to 
\begin{eqnarray}
h_{\mathcal{PT}_{4}} &=&J^{2}\mu _{JJ}+\left( \mu _{J}-\text{$\dot{\lambda}$}%
\right) J+\sin \lambda \left( \frac{\mu _{uJ}}{2}-\mu _{v}\right) u+\cos
\lambda \left( \mu _{v}-\frac{\mu _{uJ}}{2}\right) v \\
&&+\left( \mu _{uu}-\mu _{vv}+\frac{\mu _{uJ}^{2}}{4\mu _{JJ}}\right) \sin
(2\lambda )uv-\frac{\mu _{vJ}}{2}\sin \lambda \{u,J\}+\frac{\mu _{vJ}}{2}%
\cos \lambda \{v,J\}  \notag \\
&&+\left[ \left( \frac{\mu _{uu}-\mu _{vv}}{2}+\frac{\mu _{uJ}^{2}}{8\mu
	_{JJ}}\right) \cos (2\lambda )+\left( \frac{\mu _{uu}+\mu _{vv}}{2}\right) +%
\frac{\mu _{uJ}^{2}}{8\mu _{JJ}}\right] u^{2}  \notag \\
&&+\left[ \left( \mu _{uu}+\frac{\mu _{uJ}^{2}}{4\mu _{JJ}}\right) \sin
^{2}\lambda +\cos ^{2}\lambda \mu _{vv}\right] v^{2},  \notag
\end{eqnarray}%
with 5 constraining relations 
\begin{equation}
\tau =\frac{\mu _{uJ}}{2\mu _{JJ}}\sec \lambda ,\quad \rho =-\frac{\mu
	_{uJ}\tan \lambda }{2\mu _{JJ}},\quad \mu _{u}=\frac{\mu _{vJ}}{2}+\frac{\mu
	_{J}\mu _{uJ}}{2\mu _{JJ}},\quad \mu _{uv}=\frac{\mu _{vJ}\mu _{uJ}}{2\mu
	_{JJ}},\quad \dot{\mu}_{uJ}=0.
\end{equation}%
This case is similar to the previous one with one of the coefficient
functions forced to be time-independent and one of the three functions in
the Dyson map being freely choosable.

\subsection{Time-dependent $\mathcal{PT}_{5}$-invariant Hamiltonians}

The Hamiltonian $H$ becomes Hermitian for $\mu _{vJ}=\mu _{uv}=0$ and $\mu
_{uJ}=-2\mu _{v}$. Here we have to take $\rho \in \mathbb{R}$ and $\lambda
,\tau \in i\mathbb{R}$ in (\ref{etaQH}). The Hermitian Hamiltonian is computed to 
\begin{eqnarray}
h_{\mathcal{PT}_{5}} &\!\!\!=\!\!\!&J^{2}\mu _{JJ}+\left( \mu _{J}-\text{$\dot{\lambda}$}%
\right) J+\left( \tau \mu _{J}+\frac{\mu _{uJ}}{2}\cos \lambda \right)
\{u,J\}+\frac{\mu _{uJ}}{2}\sin \lambda \{v,J\} \\
&&+\left[ \tau \left( \mu _{J}-\text{$\dot{\lambda}$}\right) +\cos \lambda
\left( \mu _{u}+\frac{\mu _{vJ}}{2}\right) \right] u+\left[ \sin \lambda
\left( \mu _{u}+\frac{\mu _{vJ}}{2}\right) -\dot{\tau}\right] v  \notag \\
&&+\left[ \tau ^{2}\mu _{JJ}+\sin ^{2}\lambda \left( \frac{\mu _{vJ}^{2}}{%
	4\mu _{JJ}}+\mu _{vv}\right) +\tau \cos \lambda \mu _{uJ}+\cos ^{2}\lambda
\mu _{uu}\right] u^{2}  \notag \\
&&+\sin \lambda \left[ 2\cos \lambda \left( \mu _{uu}-\mu _{vv}-\frac{\mu
	_{vJ}^{2}}{4\mu _{JJ}}\right) +\tau \tau \mu _{uJ}\right] uv  \notag \\
&&+\left[ \left( \frac{\mu _{vJ}^{2}}{4\mu _{JJ}}+\mu _{vv}\right) \cos
^{2}\lambda +\mu _{uu}\sin ^{2}\lambda \right] v^{2},  \notag
\end{eqnarray}%
with only 4 constraining relations 
\begin{equation}
\rho =-\frac{\mu _{vJ}}{2\mu _{JJ}},\quad \mu _{v}=-\frac{\mu _{uJ}}{2}+%
\frac{\mu _{J}\mu _{vJ}}{2\mu _{JJ}},~~\dot{\mu}_{vJ}=0,~~\mu _{uv}=\frac{%
	\mu _{vJ}\mu _{uJ}}{2\mu _{JJ}}.
\end{equation}%
In comparison with the other symmetries, this is the least constraint case.
From the three functions in the Dyson map only one is constraint and the
others can be chosen freely. However, one of the coefficient functions needs
to be time-independent.

\section{Time-dependent quasi-exactly solvable systems}\label{TDQE}

We will now specify one particular model and show how it can be
quasi-exactly solved in the metric picture. Since the $\mathcal{PT}_{2}$-symmetry appears to be somewhat special, in the sense that it is the only
case for which the Dyson map respects the same symmetry as the Hamiltonian,
we focus here on that case to present some features in more detail. A
particular non-Hermitian $\mathcal{PT}_{2}$-symmetric time-independent
Hamiltonian of the form

\begin{equation}
H=m_{JJ}J^{2}+m_{v}v+m_{vv}v^{2}+im_{uJ}uJ.  \label{HH}
\end{equation}%
Given the constraining equations (\ref{2con}), we could in principle take $%
m_{v}$, $m_{vv}$ to be time dependent, but to enforce the metric picture we
take here all four coefficients $m_{JJ}$, $m_{v}$, $m_{vv}$ and $m_{uJ}$ to
be time-independent real constants. According to the analysis in section
\ref{PT2}, the time-dependent Dyson map%
\begin{equation}
\eta (t)=e^{\tau (t)v}e^{i\lambda (t)J}e^{\varrho (t)u},\quad \tau (t)=\frac{%
\mu _{uJ}}{2\mu _{JJ}}\sec \lambda (t),\quad \varrho (t)=-\frac{\mu _{uJ}}{%
2\mu _{JJ}}\tan \lambda (t),  \label{etas}
\end{equation}%
with $\lambda ,\tau ,\rho \in \mathbb{R}$, maps the time-independent
non-Hermitian Hamiltonian $\hat{H}$ to the time-dependent Hermitian
Hamiltonian%
\begin{eqnarray}
h(t) &=&m_{JJ}J^{2}-\dot{\lambda}J+\sin \lambda \left( \frac{m_{uJ}}{2}%
-m_{v}\right) u+\cos \lambda \left( m_{v}-\frac{m_{uJ}}{2}\right) v
\label{hh} \\
&&+\left[ \cos (2\lambda )\left( \frac{m_{uJ}^{2}}{8\mu _{JJ}}-\frac{m_{vv}}{%
2}\right) +\frac{m_{uJ}^{2}}{8\mu _{JJ}}+\frac{m_{vv}}{2}\right] u^{2} 
\notag \\
&&+\left[ \frac{m_{uJ}^{2}}{4\mu _{JJ}}\sin ^{2}\lambda +m_{vv}\cos
^{2}\lambda \right] v^{2}+\sin (2\lambda )\left( \frac{m_{uJ}^{2}}{4\mu _{JJ}%
}-m_{vv}\right) uv.  \notag
\end{eqnarray}%
Here we are free to chose the time-dependent function $\lambda (t)$. As
previously pointed out for non-Hermitian systems with time-dependent metric,
one needs to distinguish between the Hamiltonian, that is a non-observable
operator, and the observable energy operator. This feature remains also true
when the non-Hermitian Hamiltonian is time-independent, but the metric is
dependent on time. In reverse, it simply means that when one identifies the
non-Hermitian Hamiltonian with the energy operator one has made the choice
for the metric to be time-independent. With $\eta (t)$ as specified in (\ref%
{etas}), the energy operator is computed with the help of (\ref{auxen}) to 
\begin{eqnarray}
H(t) &=&\eta ^{-1}(t)h(t)\eta (t)=\hat{H}+i\hbar \eta
^{-1}(t)\partial _{t}\eta (t) \\
&=&m_{JJ}J^{2}+m_{v}v+m_{vv}v^{2}+im_{uJ}uJ-\text{$\dot{\lambda}J~$}-i\frac{%
m_{uJ}}{m_{JJ}}\text{$\dot{\lambda}$}u.
\end{eqnarray}%
We note that $\tilde{H}(t)$ is also $\mathcal{PT}_{2}$-symmetric when we
include $\partial _{t}\rightarrow -\partial _{t}$ into the symmetry
transformation. In order to demonstrate that this system is quasi-exactly
solvable we specify the constants in the Hamiltonian (\ref{HH}) further to $%
m_{JJ}=4$, $m_{uJ}=2(1-\beta )\zeta $, $m_{vv}=-\beta \zeta ^{2}$, $%
m_{v}=2\zeta N$ so that it becomes 
\begin{equation}
H(N,\zeta ,\beta )=4J^{2}+i2(1-\beta )\zeta uJ-\beta \zeta ^{2}v^{2}+2\zeta
Nv,\qquad \beta ,\zeta ,N\in \mathbb{R}.  \label{newH}
\end{equation}%
This Hamiltonian can be obtained from one discussed in \cite%
{fring2016unifying} by transforming $\theta \rightarrow \theta /2$, $%
J\rightarrow 2J$ in the trigonometric representation. The constants in $%
H(N,\zeta ,\beta )$ are chosen so that it exhibits an interesting double
scaling limit $\lim_{\zeta \rightarrow 0,N\rightarrow \infty }H(N,\zeta
,\beta )=4J^{2}+2gv$ when assuming that $g:=\zeta N$. In the trigonometric
representation this limiting Hamiltonian is the Mathieu Hamiltonian.

The Hermitian Hamiltonian (\ref{hh}) simplifies in this case to%
\begin{equation}
h(t,N,\zeta ,\beta )=4J^{2}-\dot{\lambda}J+\zeta \left( 2N+\beta -1\right)
\left( \cos \lambda v-\sin \lambda u\right) +\frac{\gamma ^{2}}{4}\left(
\cos \lambda u+\sin \lambda v\right) ^{2}+\beta \zeta ^{2}C  \label{hhat}
\end{equation}%
where we denoted the Casimir operator by $C:=v^{2}+u^{2}$ and abbreviated $%
\gamma :=(1+\beta )\zeta $. In the aforementioned double scaling limit we
obtain a time-dependent Hamiltonian of the form $\lim_{\zeta \rightarrow
0,N\rightarrow \infty }h(t,N,\zeta ,\beta )=4J^{2}-\dot{\lambda}J+2g\left(
\cos \lambda v-\sin \lambda u\right) $.

\subsection{Quasi-exactly solvable Lewis-Riesenfeld invariants}

The most efficient way to solve the time-dependent Dyson equation (\ref{DysonE}%
) is to use the Lewis-Riesenfeld approach \cite{lewis1969exact} and compute at
first the respective time-dependent invariants $I_{h}(t)$ and $I_{H}(t)$ for
the Hamiltonian $h(t)$ and $H(t)$ as outlined in chapter 2.

Taking now $H$ to be time-independent, we may assume $I_{H}=H+c\mathbb{I}$
with $c$ being some constant. The Lewis-Riesenfeld phase then just becomes a
dynamical phase factor 
\begin{equation}
\dot{\alpha}=\left\langle \tilde{\psi}\right\vert \rho (t)\left[ i\hbar
\partial _{t}-H\right] \left\vert \tilde{\psi}\right\rangle =\left\langle 
\tilde{\psi}\right\vert \rho (t)\left[ c\mathbb{I-}I_{H}\right] \left\vert 
\tilde{\psi}\right\rangle =c-\Lambda =-E,
\end{equation}%
such that $\alpha (t)=-Et$.

Next we quasi-exactly construct the Lewis-Riesenfeld invariants together
with its eigenstates for the time-dependent Hermitian and time-independent
non-Hermitian systems (\ref{hh}) and (\ref{HH}), respectively.

\subsubsection{The quasi-exactly solvable symmetry operator $I_{H}$}

We make a general Ansatz for the invariant of $H$ of the form%
\begin{equation}
I_{H}=\nu _{JJ}J^{2}+\nu _{J}J+\nu _{u}u+\nu _{v}v+\nu _{uJ}uJ+\nu
_{vJ}vJ+\nu _{uu}u^{2}+\nu _{vv}v^{2}+\nu _{uv}uv,  \label{IHQH}
\end{equation}%
with unknown constants $\nu _{i}$. The invariant for the time-independent
system is of course just a symmetry and we only need to compute the
commutator of $I_{H}$ with $H$ to determine the coefficients in (%
\ref{IHQH}). We find the most general symmetry or invariant to be 
\begin{eqnarray}
I_{H} &=&\nu _{JJ}J^{2}+m_{v}\frac{\nu _{JJ}}{m_{JJ}}v+im_{uJ}\frac{%
\nu _{JJ}}{m_{JJ}}uJ+\left( \nu _{vv}-m_{vv}\frac{\nu _{JJ}}{m_{JJ}}\right)
u^{2}+\nu _{vv}v^{2}  \label{IH2} \\
&=&H+(\beta \zeta ^{2}+\nu _{vv})C,
\end{eqnarray}%
where in the last equation we have taken $\nu _{JJ}=m_{JJ}$. Since the last
term only produces an overall shift in the spectrum we set $\nu _{vv}=0$ for
convenience.

Next we choose a trigonometric representation for the $E_2$ as described in section \ref{E2Intro} and  compute the eigensystem for $I_{H}$ by solving (\ref{InvariantH}).
Assuming the two linear independent eigenfunctions to be of the general
forms 
\begin{equation}
\tilde{\psi}_{H}^{c}(\theta )=\psi _{0}\sum_{n=0}^{\infty
}c_{n}P_{n}(\Lambda )\cos (n\theta ),\quad \text{and\quad }\tilde{\psi}_{%
H}^{s}(\theta )=\psi _{0}\sum_{n=1}^{\infty }c_{n}Q_{n}(\Lambda )\sin
(n\theta ),  \label{FS}
\end{equation}%
with constants $c_{n}=1/\zeta ^{n}(N+\beta )(1+\beta )^{n-1}\left[
(1+N+2\beta )/(1+\beta )\right] _{n-1}$ where $\left[ a\right] _{n}:=\Gamma
\left( a+n\right) /\Gamma \left( a\right) $ denotes the Pochhammer symbol.
The ground state $\psi _{0}=e^{-\frac{1}{2}\zeta \cos (\theta )}$ is taken
to be $\mathcal{PT}_{2}$-symmetric. The constants $c_{n}$ are chosen
conveniently to ensure the simplicity of the polynomials $P_{n}(\Lambda )$, $%
Q_{n}(\Lambda )$ in the eigenvalues $\Lambda $. We then find that the
functions $\tilde{\psi}_{H}^{c}$ and $\tilde{\psi}_{H}^{s}$
satisfy the eigenvalue equation provided the coefficient functions $%
P_{n}(\Lambda )$ and $Q_{n}(\Lambda )$ obey the three-term recurrence
relations%
\begin{eqnarray}
P_{2} &=&(\Lambda -4)P_{1}+2\zeta ^{2}\left( N-1\right) \left(
N+\beta \right) P_{0},  \label{r1} \\
P_{n+1} &=&(\Lambda -4n^{2})P_{n}-\zeta ^{2}\left[ N+n\beta +(n-1)\right] %
\left[ N-(n-1)\beta -n\right] P_{n-1},  \label{r2} \\
Q_{2} &=&(\Lambda -4)Q_{1},  \label{r3} \\
Q_{m+1} &=&(\Lambda -4m^{2})Q_{m}-\zeta ^{2}\left[ N+m\beta +(m-1)\right] %
\left[ N-(m-1)\beta -m\right] Q_{m-1},~~~~~  \label{r4}
\end{eqnarray}%
for $n=0,2,\ldots $ and for $m=2,3,4,\ldots $ Setting $P_{0}=1$ and $Q_{1}=1$%
, the first solutions for (\ref{r1}) - (\ref{r4}) are found to be 
\begin{eqnarray}
P_{1} &=&\Lambda , \\
P_{2} &=&\Lambda ^{2}-4\Lambda -2\zeta ^{2}(N-1)(\beta +N),  \notag \\
P_{3} &=&\Lambda ^{3}-20\Lambda ^{2}+\left[ \zeta ^{2}\left( 2\beta
^{2}+7\beta -3N^{2}-3(\beta -1)N+2\right) +64\right] \Lambda +32\zeta
^{2}(N-1)(\beta +N),  \notag
\end{eqnarray}%
and%
\begin{eqnarray}
Q_{2} &=&\left( \Lambda -4\right) , \\
Q_{3} &=&(\Lambda -20)\Lambda +\zeta ^{2}(\beta -N+2)(2\beta +N+1)+64, 
\notag \\
Q_{4} &=&\Lambda ^{3}-56\Lambda ^{2}+\left[ 2\zeta ^{2}\left( 4\beta
^{2}+9\beta -N^{2}-\beta N+N+4\right) +784\right] \Lambda  \notag \\
&&+8\zeta ^{2}\left[ 5N^{2}+5(\beta -1)N-12-\beta (12\beta +29)\right] -2304.
\notag
\end{eqnarray}%
The well-known and crucial feature responsible for a system to be
quasi-exactly solvable is the occurrence of the three-term recurrence
relations and that they can be forced to terminate at certain values of $n$.
This is indeed the case for our relations (\ref{r2}), (\ref{r4}) and can
be achieved for some specific values $n=\hat{n}$ or $m=\hat{n}$,
respectively. To see this we take $N=\hat{n}+(\hat{n}-1)\beta $ and note
that the polynomials $P_{n}$ and $Q_{m}$ factorize for $n\geq \hat{n}$, $%
m\geq \hat{n}$ as%
\begin{equation}
P_{\hat{n}+\ell }=P_{\hat{n}}R_{\ell }\qquad \text{and\qquad }Q_{\hat{n}%
+\ell }=Q_{\hat{n}}R_{\ell }\text{,}  \label{fact}
\end{equation}%
where the first $R_{\ell }$-polynomials are%
\begin{eqnarray}
R_{1} &=&\Lambda -4\hat{n}^{2}, \\
R_{2} &=&16\hat{n}^{2}(\hat{n}+1)^{2}+\Lambda \left[ \Lambda -4-8\hat{n}(%
\hat{n}+1)\right] +2\hat{n}\gamma ^{2}.
\end{eqnarray}%
Since according to (\ref{fact}) the polynomials $P_{\hat{n}}$ and $Q_{\hat{n}%
}$ are factor in all $P_{n}$ and $Q_{m}$ for $n\geq \hat{n}$ and $m\geq \hat{%
n}$, respectively, all higher order polynomial vanish when setting $P_{\hat{n%
}}(\Lambda )=Q_{\hat{n}}(\Lambda )=0$. These latter constraints are the
quantization conditions for $\Lambda $. Thus setting $P_{\hat{n}}(\Lambda
)=0 $ at the different levels $\hat{n}$, we find the real eigenvalues%
\begin{eqnarray}
\hat{n} &=&1:\quad \Lambda _{1}^{c}=0, \\
\hat{n} &=&2:\quad \Lambda _{2}^{c,\pm }=2\pm 2\sqrt{1+\gamma ^{2}}, \\
\hat{n} &=&3:\quad \Lambda _{3}^{c,\ell =0,\pm 1}=\frac{4}{3}\left\{
5+2\kappa \cos \left[ \frac{\ell \pi }{3}-\frac{1}{3}\arccos \left( \frac{%
35-18\gamma ^{2}}{\kappa ^{3}}\right) \right] \right\} ,
\end{eqnarray}%
with $\kappa =\sqrt{13+3\gamma ^{2}}$, and from $Q_{\hat{n}}(\Lambda )=0$ we
find the real eigenvalues%
\begin{eqnarray}
\hat{n} &=&2:\quad \Lambda _{2}^{s}=4, \\
\hat{n} &=&3:\quad \Lambda _{3}^{s,\pm }=10\pm 2\sqrt{9+\gamma ^{2}}, \\
\hat{n} &=&4:\quad \Lambda _{4}^{s,\ell =0,\pm 1}=\frac{8}{3}\left\{ 7+%
\tilde{\kappa}\cos \left[ \frac{\ell \pi }{3}-\frac{1}{3}\arccos \left( 
\frac{143-18\gamma ^{2}}{\tilde{\kappa}^{3}}\right) \right] \right\} ,
\end{eqnarray}%
with $\tilde{\kappa}=\sqrt{49+3\gamma ^{2}}$.

Thus $H$ is a QES system with eigenfunctions identical to those in (%
\ref{FS}) and energies $E=\Lambda -\beta \zeta ^{2}$.

\subsubsection{The quasi-exactly solvable invariant $I_{h}$}

Next we construct the invariant $I_{h}$ together with their
eigenfunctions. In principle we have to solve the equation in (\ref%
{Invarianth}) for this purpose, however, since we already know the Dyson map we
can simply use (\ref{InvarSim}) and act adjointly with $\eta (t)$, as given in (%
\ref{etas}), on $I_{\hat{H}}$ as specified in (\ref{IH2}). This yields the
time-dependent invariant for $h(t)$ as%
\begin{equation}
I_{h}=\eta (t)I_{H}(t)\eta ^{-1}(t)=h+\dot{\lambda}J+\beta
\zeta ^{2}C  \label{Ih}
\end{equation}%
We convince ourselves that the relation (\ref{Invarianth}) is indeed satisfied by $%
I_{h}$ as given in (\ref{Ih}) and $h(t)$ as in (\ref{hhat}). The
eigenfunctions for $I_{h}$ are then simply obtained as $\tilde{\phi}%
=\eta \tilde{\psi}$. From (\ref{FS}) we compute 
\begin{equation}
\tilde{\phi}_{h}^{c}(\theta,t )=\phi _{0}\sum_{n=0}^{\infty
}c_{n}P_{n}(\Lambda )\cos \left[ n(\theta +\lambda\left(t\right) )\right] ,\quad \tilde{%
\phi}_{h}^{s}(\theta,t )=\phi _{0}\sum_{n=1}^{\infty }c_{n}Q_{n}(\Lambda
)\sin \left[ n(\theta +\lambda\left(t\right) )\right] .  \label{sumh}
\end{equation}%
with ground state wave function $\phi _{0,t}=e^{-\frac{1}{4}\zeta (1+\beta
)\cos (\theta +\lambda\left(t\right) )}$ and coefficients $c_{n}$, $P_{n}(\Lambda )$, $%
Q_{n}(\Lambda )$ as defined above. According to the above arguments, the
solutions to the time-dependent Schr\"{o}dinger equation are $\phi _{\hat{h}%
}^{c,s}(\theta,t )=e^{-iEt/\hbar }\tilde{\phi}_{\hat{h}}^{c,s}(\theta,t )$.

\subsection{A time-dependent three level system}

For each integer value of $\hat{n}$ we have now obtained a time-dependent
QES system with a finite dimensional Hilbert space. We present here the case
for $\hat{n}=2$ in more detail, since it is the easiest non-trivial example
and three level systems are of course of essential importance in the
description and understanding of basic physical effects such as population
inversion that is vital for lasing to occur. For relatively recent survey
that include more sophisticated effects that can be understood from
time-dependent three-level systems see for instance \cite{drei}. From (\ref%
{sumh}) we obtain three orthonormal wave functions%
\begin{eqnarray}
\phi _{\pm }(\theta ,t) &=&\frac{\sqrt{\gamma }}{2\sqrt{\pi N_{\pm }}}e^{-%
\frac{1}{4}\gamma \cos \left[ \theta +\lambda (t)\right] -iE_{\pm }t}\left[
\gamma +(1\pm \sqrt{1+\gamma ^{2}})\right] \cos \left[ \theta +\lambda (t)%
\right] ,  \label{f1} \\
\phi _{0}(\theta ,t) &=&\frac{\sqrt{\gamma }}{2\sqrt{\pi N_{0}}}e^{-\frac{1}{%
4}\gamma \cos \left[ \theta +\lambda (t)\right] -iE_{0}t}\sin \left[ \theta
+\lambda (t)\right] ,  \label{f2}
\end{eqnarray}%
with normalization constants%
\begin{eqnarray}
N_{\pm } &=&\gamma \left( 1+\gamma ^{2}\pm \sqrt{1+\gamma ^{2}}\right)
I_{0}\left( \gamma /2\right)\\
&& -\left[ 2+2\gamma ^{2}\pm (2+\gamma ^{2})\sqrt{%
1+\gamma ^{2}}\right] I_{1}\left( \gamma /2\right) ,\nonumber \\
N_{0} &=&I_{1}\left( \gamma /2\right) ,
\end{eqnarray}%
and eigenenergies $E_{0}=4-\beta \zeta ^{2}$, $E_{\pm }=2-\beta \zeta
^{2}\pm 2\sqrt{1+\gamma ^{2}}$. The $I_{n}\left( z\right) $ denotes here the
modified Bessel functions of the first kind. The functions in (\ref{f1}) and (%
\ref{f2}) solves the time-dependent Schr\"{o}dinger equation for $h(t)$
and are orthonormal on any interval $[\theta _{0},\theta _{0}+2\pi ]$%
\begin{equation}
\left\langle \phi _{n}(\theta ,t)\right. \left\vert \phi _{m}(\theta
,t)\right\rangle =:\int\nolimits_{\theta _{0}}^{\theta _{0}+2\pi }\phi
_{n}^{\ast }(\theta ,t)\phi _{m}(\theta ,t)d\theta =\delta _{n,m}\qquad
n,m\in \{0,\pm \}.
\end{equation}%
We may now compute analytically all time-dependent quantities of physical
interest. For instance, the expectation values for the generators in the
trigonometric representation result to%
\begin{eqnarray}
\left\langle \phi _{\pm }(\theta ,t)\right\vert u\left\vert \phi _{\pm
}(\theta ,t)\right\rangle  &=&-\frac{M_{\pm }}{N_{\pm }}\sin \left[ \lambda
(t)\right] ,~~\left\langle \phi _{0}(\theta ,t)\right\vert u\left\vert \phi
_{0}(\theta ,t)\right\rangle =\frac{I_{2}\left( \gamma /2\right) }{%
I_{1}\left( \gamma /2\right) }\sin \left[ \lambda (t)\right] ,~~~~~~~~ \\
\left\langle \phi _{\pm }(\theta ,t)\right\vert v\left\vert \phi _{\pm
}(\theta ,t)\right\rangle  &=&\frac{M_{\pm }}{N_{\pm }}\cos \left[ \lambda
(t)\right] ,~~\left\langle \phi _{0}(\theta ,t)\right\vert v\left\vert \phi
_{0}(\theta ,t)\right\rangle =-\frac{I_{2}\left( \gamma /2\right) }{%
I_{1}\left( \gamma /2\right) }\cos \left[ \lambda (t)\right] , \\
\left\langle \phi _{\ell }(\theta ,t)\right\vert J\left\vert \phi _{\ell
}(\theta ,t)\right\rangle  &=&0,~~~~\ell \in \{0,\pm \},
\end{eqnarray}%
where we abbreviated%
\begin{equation}
M_{\pm }=\gamma \left( 1-\gamma ^{2}\pm \sqrt{1+\gamma ^{2}}\right)
I_{1}\left( \gamma /2\right) +\left[ 2+2\gamma ^{2}\pm (2+\gamma ^{2})\sqrt{%
1+\gamma ^{2}}\right] I_{2}\left( \gamma /2\right) .
\end{equation}%
Similarly we may obtain any kind of $n$-level system from (\ref{sumh}).

\section{Summary}

We have provided new analytical solutions for the time-dependent Dyson
equation. The time-dependent non-Hermitian Hamiltonians (\ref{H}) considered
are expressed in terms of linear and quadratic combinations of the generators
for an Euclidean $E_{2}$-algebra respecting the $\mathcal{PT}_{i}$%
-symmetries defined in (\ref{coe}). Restricting the coefficient functions
appropriately, the corresponding time-dependent Hermitian Hamiltonians were
constructed. We expect a different qualitative behaviour for Hamiltonians
belonging to different symmetry classes.

A specific $\mathcal{PT}_{2}$-symmetric system was analyzed in more detail.
For that model we assumed the non-Hermitian Hamiltonian to be
time-independent so that we could employ the metric picture. This enabled us
to compute the corresponding eigensystems in a quasi-exactly solvable
fashion using Lewis-Riesenfeld invariants. Thus we found for the first time
quasi-exactly solvable systems for Hamiltonians with explicit
time-dependence. A time-dependent Hermitian three-level system is presented
in more detail.

Evidently there are many open issues and problems for further investigations
left. Having solved the time-dependent Dyson equation for a large class of
models in section \ref{E2DE}, it would be interesting to solve their corresponding
time-dependent Schr\"{o}dinger equation as carried out for the model in
section \ref{TDQE}. Furthermore, it is desirable in this type of analysis to allow an
explicit time-dependence also in the non-Hermitian Hamiltonians. Clearly one
may also generalize these studies to Euclidean algebras of higher rank and
other types of Lie algebras.

\chapter{Eternal Life of Entropy}
The information contained within a quantum system is of great importance for various practical implementations of quantum mechanics, most importantly for the development of quantum computers, e.g.  \cite{nielsen2002quantum,bennett2000quantum,raussendorf2001one,steane1998quantum}. In order to understand the quantum information, one must find a way of measuring the entanglement of a state. Entanglement is a defining feature of quantum mechanics that distinguishes it from classical mechanics and there has been much work in recent years into the evolution of entanglement with time, particularly the observation of the abrupt decay of entangled states, coined as "sudden death" \cite{yu2009sudden,yonacc2006sudden}. The decoherence of entanglement \cite{unruh1995maintaining,palma1996quantum} is a problem for the operation of quantum computers and so understanding the mechanism behind this is an important contribution to the development of future machines. One particular measure of entanglement and quantum information is the Von Neumann entropy. This is well-understood in the standard quantum mechanical setting, however to date there has only been a small amount of work done concerning the proper treatment of entropy in non-Hermitian, $\mathcal{PT}$-symmetric systems \cite{scolarici2006time,croke2015pt,kawabata2017information,jones2010quantum}. These differ from open quantum systems as the energy eigenvalues are real or appear as complex conjugate pairs and do not describe decay. 
 
In this chapter, we extend the current understanding of Von Neumann entropy to Non-Hermitian, $\mathcal{PT}$-symmetric quantum mechanics. In the unbroken $\mathcal{PT}$ regime it has been shown that such systems exhibit real energy eigenvalues and unitary time evolutions. This is possible due to the existence of a non-trivial metric operator. Of particular interest are non-Hermitian systems with spontaneously broken $\mathcal{PT}$-symmetry. These systems possess an exceptional point above which the $\mathcal{PT}$-symmetry is broken. In this regime the system exhibits complex energy eigenvalues, becoming ill-defined and is therefore ordinarily discarded as non-physical. However, it has been shown \cite{AndTom1,AndTom3,AndTom2,fring2018tdm} that when a time-dependence is introduced into the central equations it is possible to make sense of the broken regime via a time-dependent metric. This allows for the definition of a Hilbert space and therefore a well-defined inner product. This will be central to our analysis in non-Hermitian systems as we will be showing how the evolution of entropy changes significantly as we vary the system parameters through the exceptional point. 

We will first set up the framework for analysing the Von Neumann entropy for non-Hermitian systems in section \ref{Theory}. In section \ref{Model} we will apply the framework to a simple model consisting of a bosonic system coupled to a bath and solve the time-dependent Dyson equation. Finally, in section \ref{EntropyEvolution} we demonstrate how the evolution of entropy differs depending on the state of the $\mathcal{PT}$-symmetry of the non-Hermitian system.

\section{Entanglement Von Neumann Entropy}\label{Theory}

In order to make calculations of the quantum entropy for non-Hermitian systems, we must first introduce some new quantities when compared to the Hermitian case. In what follows we use natural units, setting $\hbar=1$. The density matrix for Hermitian systems is defined as an Hermitian operator describing the statistical ensemble of states

\begin{equation}
\varrho_h=\sum_{i}p_i\ket{\phi_i}\bra{\phi_i},
\end{equation} 
where the subscript $h$ indicates it relates to an Hermitian system. $\ket{\phi_i}$ are general pure states, and $p_i$ is the probability that the system is in the pure state $\ket{\phi_i}$, with  $0\leq p_i\leq 1$ and $\sum_ip_i=1$. Therefore $\varrho_h$ represents a mix of pure states (a mixed state). If the system is comprised of subsystems $A$ and $B$ one can define the reduced density operator of these subsystems as the partial trace over the opposing subsystem's Hilbert space

\begin{equation}
\varrho_{h,A}=Tr_B\left[\varrho_h\right]=\sum_{i}\bra{n_{i,B}}\varrho_h\ket{n_{i,B}},
\end{equation}
\begin{equation}
\varrho_{h,B}=Tr_A\left[\varrho_h\right]=\sum_{i}\bra{n_{i,A}}\varrho_h\ket{n_{i,A}},
\end{equation}
where $\ket{n_{i,A}}$ and $\ket{n_{i,B}}$ are the eigenstates of the subsystems $A$ and $B$, respectively. In this way one can isolate the density matrix for each subsystem and perform entropic analysis on them individually. We now want to find the relationship between the $\varrho_h$ and $\varrho_H$, where the subscript $H$ indicates a non-Hermitian system. The clearest starting point is the Von Neumann equation which governs the time evolution of the density matrix. For the Hermitian system it is

\begin{equation}\label{VNHermitian}
i\partial_t\varrho_h=\left[h,\varrho_h\right],
\end{equation}
where $h$ is the Hermitian Hamiltonian. We now wish to find the equivalent relation in the non-Hermitian setting. In order to do this we substitute the time-dependent Dyson equation (\ref{DysonE}) into the Von Neumann equation.  After some manipulation, this results in the following equation

\begin{equation}
i\partial_t\varrho_H=\left[H,\varrho_H\right],
\end{equation}
when assuming that the density matrix in the Hermitian system is related to that of the non-Hermitian system via a similarity transformation

\begin{equation}\label{SimTrans}
\varrho_h=\eta\varrho_H\eta^{-1}.
\end{equation}
Recalling that $\ket{\phi}=\eta\ket{\psi}$, this leads us to the definition of the density matrix $\varrho_H$ for non-Hermitian systems,

\begin{equation}
\varrho_H=\sum_{i}p_i\ket{\psi_i}\bra{\psi_i}\rho,
\end{equation}
%This must be included as we are performing a spectral decomposition with respect to the inner product and the non-Hermitian inner product contains the metric. 
where $\ket{\psi_i}$ are general pure states for the non-Hermitian system and $\rho=\eta^\dagger\eta$ is the metric. Notice that $\varrho_H$ is a Hermitian operator in the Hilbert space related to the metric $\bra{\cdot}\rho\ket{\cdot}$. It is therefore clear that the existence of a well defined metric is essential for the calculation of entropy in non-Hermitian systems. These results match those from \cite{scolarici2006time}. Having defined the density matrix for non-Hermitian systems and found the relation to Hermitian systems we can now consider the entropy. For the total system, the Von Neumann entropy is defined as

\begin{equation}
S_h=-tr\left[\varrho_h\ln\varrho_h\right].
\end{equation}
This can also be expressed as a sum of the eigenvalues $\lambda_i$ of the density matrix $\varrho_h$ as it is an Hermitian operator. This allows us to write $\varrho$ and functions of $\varrho$ as a spectral decomposition, and so the Von Neumann entropy is

\begin{equation}
S_h=-\sum_{i}\lambda_i\ln\lambda_i.
\end{equation}
As the density matrix for the Hermitian and non-Hermitian systems are related by a similarity transform, they share the same eigenvalues, therefore

\begin{equation}
S_H=S_h.
\end{equation}
Is is important to recall, however, that this relation only holds true for the existence of a well-defined Dyson operator $\eta$ and metric $\rho$. Without them, we are unable to form the relation (\ref{SimTrans}). For closed systems, the Von Neumann entropy is constant with time. However, we wish to consider the entropy for particular subsystems and for this we must consider the partial trace of the density matrix. In this setting the entropy for subsystem $A$ becomes

\begin{equation}
S_{h,A}=-tr\left[\varrho_{h,A}\ln\varrho_{h,A}\right]=-\sum_{i}\lambda_{i,A}\ln\lambda_{i,A},
\end{equation}
where once again the entropy of the Hermitian subsystem is equal to that of the non-Hermitian subsystem $S_{h,A}=S_{H,A}$ with the existence of $\eta$ and $\rho$. The entropy of a particular subsystem is not confined to be constant and we show that it exhibits some very interesting and novel properties when evolved in time.

%This is in contrast to \cite{sergi2016quantum}.

\section{System bath coupled model}\label{Model}

We now consider a time-independent non-Hermitian Hamiltonian consisting of coupled harmonic oscillators. We have a system composed of $a$, $a^\dagger$ bosonic operators coupled to a bath of $N$ $q_i$, $q_i^\dagger$ bosonic operators, this is equivalent to a system of coupled harmonic oscillators. The Hamiltonian takes the form

\begin{equation}\label{NHHamiltonian}
\begin{split}
\begin{aligned}
H=\nu a^\dagger a+\nu\sum_{n=1}^{N}q^\dagger_nq_n+\left(g+\kappa\right)a^\dagger\sum_{n=1}^{N}q_n
+\left(g-\kappa\right)a\sum_{n=1}^{N}q^\dagger_n,
\end{aligned}
\end{split}
\end{equation}
with $\nu$, $g$ and $\kappa$ being real time-independent parameters and the bosonic operators obeying the commutation relations

\begin{equation}
\left[a,a^\dagger\right]=1, \quad \left[q_i,q_j^\dagger\right]=\delta_{ij}, \quad \left[a,q_i^\dagger\right]=0.
\end{equation}
This is similar to the coupled oscillators studied in chapter 4.

\subsection{$\mathcal{PT}$-symmetry}
The Hamiltonian (\ref{NHHamiltonian}) is $\mathcal{PT}$-symmetric under the anti-linear transformation 
\begin{equation}
\begin{split}
\begin{aligned}
\mathcal{PT}: \quad i\rightarrow -i, \quad a\rightarrow-a, \quad a^\dagger\rightarrow-a^\dagger, \quad q_n\rightarrow-q_n, \quad q_n^\dagger\rightarrow-q_n^\dagger,
\end{aligned}
\end{split}
\end{equation}
as it commutes with the $\mathcal{PT}$-operator for all values of $\nu$, $g$ and $\kappa$

\begin{equation}
\left[\mathcal{PT},H\right]=0.
\end{equation}
The energy eigenvalues are 

\begin{equation}
E_{m,N}^\pm=m\left(\nu\pm\omega\sqrt{N}\right), \quad m\in\mathbb{N},
\end{equation}
where $\omega=\sqrt{g^2-\kappa^2}$. In order to ensure boundedness from below the system must have $\nu>\omega\sqrt{N}$. Note that there is an exceptional point at $g=\kappa$ and when $\kappa>g$ this system is in the broken $\mathcal{PT}$ regime. This is clear when studying the first excited state ($m=1$) expanded in terms of creation operators acting on a tensor product of Fock states. The general state consists of one Fock state for the system of $a$ and $a^\dagger$ bosonic operators and $N$ Fock states for the bath of $q_i$ and $q_i^\dagger$ bosonic operators

\begin{equation}
\ket{\psi}=\ket{n_a}\otimes\ket{n_{q_1}}\otimes\ket{n_{q_2}}....=\ket{n_a}\bigotimes_{i=1}^{N}\ket{n_{q_i}}.
\end{equation}
When considering the first excited state, we will be dealing with very few non-zero states, and as such we can make some simplifications to the notation. If all the states in the $q$ bath are in the ground state we will represent this with $\ket{\boldmath{0}_q}$. Similarly, if the $i$th state in the $q$ bath is in the first excited state with the rest in the ground state, we will represent this with a $\ket{\boldmath{1}_{i}}$

\begin{equation}
\begin{split}
\begin{aligned}
\ket{\boldmath{0}_q}=\bigotimes_{i=1}^{N}\ket{0_{q_i}}, \quad  \ket{\boldmath{1}_{i}}=\left[\bigotimes_{j=1}^{i-1}\ket{0_{q_j}}\right]\otimes\ket{1_{q_i}}\otimes\left[\bigotimes_{k=i+1}^{N}\ket{0_{q_k}}\right].
\end{aligned}
\end{split}
\end{equation}
We can now write down the first excited state, 
\begin{equation}
\begin{aligned}
\begin{split}
\ket{\psi_{1,N}^\pm}=&\sqrt{\frac{g+\kappa}{2g}}\ket{1_a}\otimes\ket{\boldmath{0}_q}\pm\sqrt{\frac{g-k}{2gN}}\ket{0_a}\otimes\sum_{i=1}^{N}\ket{ 1_{i}}\\
=&\sqrt{\frac{g+\kappa}{2g}}\ket{1_a\boldmath{0}_q}\pm\sqrt{\frac{g-\kappa}{2gN}}\sum_{i=1}^{N}\ket{0_a 1_{i}}\\
=&\sqrt{\frac{g+\kappa}{2g}}a^\dagger\ket{0_a\boldmath{0}_q}\pm\sqrt{\frac{g-\kappa}{2gN}}\sum_{i=1}^{N}q_i^\dagger\ket{0_a \boldmath{0}_{q}}.
\end{split}
\end{aligned}
\end{equation}
%We can expand this in terms of creation operators
%
%\begin{equation}
%\ket{\psi_{1,N}}=
%\end{equation}
In order for the $\mathcal{PT}$-symmetry to remain unbroken, the wave function must also remain unchanged up to a phase factor when acted on by the $\mathcal{PT}$-operator

\begin{equation}
\mathcal{PT}\ket{\psi_{1,N}^\pm}= e^{i\phi}\ket{\psi_{1,N}^\pm}.
\end{equation}
However, the wave functions are only eigenfunctions of the $\mathcal{PT}$-operator when $\kappa<g$
%As an example, when $N=2$ the first excited state is

%\begin{equation}
%\ket{\psi_{1,2}}=\sqrt{\frac{g+\kappa}{2g}}\ket{100}+\frac{\sqrt{g-k}}{\sqrt{4g}}\left(\ket{0 10}+\ket{0 01}\right).
%\end{equation}

\begin{equation}
\mathcal{PT}\ket{\psi_{1,N}^\pm}=-\ket{\psi_{1,N}^\pm}.
\end{equation}
When $\kappa>g$, the wave functions are no longer eigenfunctions of the $\mathcal{PT}$-operator,

\begin{equation}
\mathcal{PT}\ket{\psi_{1,N}^\pm}\neq e^{i\phi}\ket{\psi_{1,N}^\pm}.
\end{equation}
Therefore we need to employ time-dependent analysis in order to make sense of the broken regime. To do this we first must solve the time-dependent Dyson equation. 

\subsection{Solving the time-dependent Dyson equation}

We wish to find the time-dependent metric $\rho\left(t\right)$ that allows us to perform entropic analysis on our model (\ref{NHHamiltonian}). In order to do this we must find the Dyson operator $\eta\left(t\right)$ and the equivalent time-dependent Hermitian system $h\left(t\right)$. The model (\ref{NHHamiltonian}) is in fact part of a larger family of Hamiltonians belonging to the closed algebra with Hermitian generators:

\begin{equation}
\begin{aligned}
\begin{split}
N_A&=a^\dagger a, \qquad N_Q=\sum_{n=1}^{N}q^\dagger_nq_n, \qquad N_{AQ}=N_A-\frac{1}{N}N_Q-\frac{1}{N}\sum_{n\neq m}q^\dagger_nq_m \\ A_x&=\frac{1}{\sqrt{N}}\left(a^\dagger\sum_{n=1}^{N}q_n+a\sum_{n=1}^{N}q^\dagger_n\right), \quad A_y=\frac{i}{\sqrt{N}}\left(a^\dagger\sum_{n=1}^{N}q_n-a\sum_{n=1}^{N}q^\dagger_n\right).
\end{split}
\end{aligned}
\end{equation}
The commutation relations are

\begin{eqnarray}
\begin{aligned}
\begin{split}
\left[N_A,N_Q\right]&=0, \quad\qquad \left[N_A,N_{AQ}\right]=0, \\
\left[N_A,A_x\right]&=-iA_y, \quad\quad \left[N_A,A_y\right]=iA_y,\\
\left[N_Q,A_x\right]&=iA_y, \quad\quad\;\;\; \left[N_Q,A_y\right]=-iA_x, \\ 
\left[N_{AQ},A_x\right]&=-2iA_y, \quad \left[N_{AQ},A_y\right]=2iA_x.
\end{split}
\end{aligned}
\end{eqnarray}
In terms of this algebra, our original Hamiltonian (\ref{NHHamiltonian}) can be written as

\begin{equation}
H=\nu N_A+\nu N_Q+ \sqrt{N}gA_x-i\sqrt{N}\kappa A_y.
\end{equation}
We are now in a position to begin solving the time-dependent Dyson equation (\ref{DysonE}). For this we make the ansatz

\begin{equation}\label{eta}
\eta\left(t\right)=e^{\beta\left(t\right)A_y}e^{\alpha\left(t\right)N_{AQ}},
\end{equation}
and use the BCH relation to expand the Dyson equation (\ref{DysonE}) in terms of generators. In order to make the resulting Hamiltonian Hermitian, we must solve two coupled differential equations to eliminate the non-Hermitian terms.

\begin{equation}
\dot{\alpha}=-\tanh\left(2\beta\right)\left[\sqrt{N}g\cosh\left(2\alpha\right)+\sqrt{N}\kappa\sinh\left(2\alpha\right)\right],\label{alphadot}
\end{equation}

\begin{equation}
\hspace{-2.3cm}\dot{\beta}=\sqrt{N}\kappa\cosh\left(2\alpha\right)+\sqrt{N}g\sinh\left(2\alpha\right).\label{betadot}
\end{equation}
Equation (\ref{betadot}) can be solved for $\alpha$,

\begin{equation}\label{alpha}
\tanh \left(2\alpha\right)=\frac{-N g\kappa+\dot{\beta}\sqrt{\dot{\beta}^2+\omega^2N}}{Ng^2+\dot{\beta}^2}.
\end{equation}
In principle this could lead to a restriction to the term on the RHS of equation (\ref{alpha}) as $-1<\tanh\left(2\alpha\right)<1$. However as we will see, this restriction is obeyed with the final solutions for $\alpha$ and $\beta$. Substituting (\ref{alpha}) into equation (\ref{alphadot}) gives

\begin{equation}
\ddot{\beta}+2\tanh\left(2\beta\right)\left[\omega^2N+\dot{\beta}^2\right]=0.
\end{equation}
Now making the substitution $\sinh\left(2\beta\right)=\sigma$, this reverts to an harmonic oscillator equation

\begin{equation}
\ddot{\sigma}+4\omega^2N\sigma=0,
\end{equation}
which is solved with the function 

\begin{equation}
\sigma=\frac{c_1}{\omega}\sin\left(2\omega\sqrt{N}\left(t+c_2\right)\right),
\end{equation}
for all values of $\kappa$, where $c_1$ and $c_2$ are constants of integration. We can now write down expressions for $\alpha$ and $\beta$

\begin{equation}
\hspace{-5.6cm}\tanh\left(2\alpha\right)=\frac{\zeta^2-1}{\zeta^2+1},
\end{equation}

\begin{equation}
\sinh\left(2\beta\right)=\frac{c_1}{\omega}\sin\left(2\omega\sqrt{N}\left(t+c_2\right)\right),
\end{equation}
where $\zeta$ is of the form

\begin{equation}
\begin{aligned}
\begin{split}
\zeta=\sqrt{2}\sqrt{\frac{g-\kappa}{g+\kappa}}\left[
\frac{\sqrt{c_1^2+\omega^2}+c_1\cos\left(2\omega\sqrt{N}\left(t+c_2\right)\right)}{\sqrt{c_1^2+2\omega^2-c_1^2\cos\left(4\omega\sqrt{N}\left(t+c_2\right)\right)}}\right].
\end{split}
\end{aligned}
\end{equation}
Therefore we have a well-defined solution for $\eta\left(t\right)$ from our original ansatz (\ref{eta}) which results in the following time-dependent Hermitian Hamiltonian

\begin{equation}\label{HHamiltonian}
h\left(t\right)=\nu N_A+\nu N_Q+\mu\left(t\right)A_x,
\end{equation}
where 

\begin{equation}\label{mu}
\mu\left(t\right)=\frac{\omega^2\sqrt{N}\sqrt{c_1^2+\omega^2}}{c_1^2+2\omega^2-c_1^2\cos\left(4\omega\sqrt{N}\left(t+c_2\right)\right)}.
\end{equation}
This is real provided $|\frac{c_1}{\omega}|>1$. The general time-dependent first excited state is 

\begin{eqnarray}\label{1State}
\ket{\phi\left(t\right)}&=&e^{-i\nu t}\left(A\sin\mu_I\left(t\right)+B\cos\mu_I\left(t\right)\right)\ket{1_a\boldmath{0}_q}\\
&+&\frac{e^{-i\nu t}}{\sqrt{N}}\left(A\cos\mu_I\left(t\right)-B\sin\mu_I\left(t\right)\right)\sum_{i=1}^{N}\ket{0_a \boldmath{1}_{i}},
\end{eqnarray}
with $A^2+B^2=1$ and

\begin{equation}
\mu_I\left(t\right)=\int^t\mu\left(s\right)ds=
\frac{1}{2}\arctan\left(\frac{\sqrt{c_1^2+\omega^2}\tan\left(2\omega\sqrt{N}\left(t+c_2\right)\right)}{\omega}\right).
\end{equation}
Now we have a full solution for $\eta\left(t\right)$ and therefore $\rho\left(t\right)=\eta\left(t\right)^\dagger\eta\left(t\right)$. This allows us to calculate the entropy for our non-Hermitian system (\ref{NHHamiltonian}). The easiest route to take is to work with the resulting Hermitian system (\ref{HHamiltonian}) as it was shown in section \ref{Theory} that the entropy in both systems is equivalent when $\eta\left(t\right)$ is well-defined. It is important to note that if the $\eta\left(t\right)$ ever becomes ill-defined, then our analysis of the Hermitian system does not correspond to the original non-Hermitian Hamiltonian as we cannot form a metric $\rho\left(t\right)$.

\section{Three types of entropy evolution}\label{EntropyEvolution}

We now calculate the entropy of the system and show how varying the parameters $N$, $g$ and $\kappa$ affects its evolution with time. We prepare our system in an entangled first excited state (\ref{1State}) at time $t=0$,

\begin{equation}
\ket{\phi\left(0\right)}=\sin\gamma\ket{{1_a \boldmath{0}_q}}+\frac{\cos\gamma}{\sqrt{N}}\sum_{i=1}^N\ket{{0_a\boldmath{1}_{i}}},
\end{equation}
for which we choose $A=\sin\gamma$, $B=\cos\gamma$ and $c_2=0$. Therefore the general state at time $t$ is

\begin{eqnarray}
\ket{\phi\left(t\right)}&=&e^{-i\nu t}\left(\sin\gamma\sin\mu_I\left(t\right)+\cos\gamma\cos\mu_I\left(t\right)\right)\ket{1_a\boldmath{0}_q}\\
&+&\frac{e^{-i\nu t}}{\sqrt{N}}\left(\sin\gamma\cos\mu_I\left(t\right)-\cos\gamma\sin\mu_I\left(t\right)\right)\sum_{i=1}^{N}\ket{0_a \boldmath{1}_{i}}.
\end{eqnarray}
We form the density matrix for the system (a) with a partial trace over the external bosonic bath (q),

\begin{equation}
\varrho_a\left(t\right)=Tr_q\left[\varrho_h\left(t\right)\right]=
\left(\begin{array}{cc}
\left(\sin\gamma\sin\mu_I\left(t\right)+\cos\gamma\cos\mu_I\left(t\right)\right)^2\hspace{-3cm} & \hspace{-1cm}0\hspace{-1cm}\\
\hspace{-1cm}0\hspace{-1cm} & \hspace{-3cm} \left(\sin\gamma\cos\mu_I\left(t\right)-\cos\gamma\sin\mu_I\left(t\right)\right)^2
\end{array}
\right).
\end{equation}
Now we calculate the Von Neumann entropy of the system using this reduced density matrix. First we read off the eigenvalues of $\varrho_a\left(t\right)$ as it is diagonal,
\begin{equation}
\begin{split}
\begin{aligned}
\lambda_1\left(t\right)&=\left(\sin\gamma\sin\mu_I\left(t\right)+\cos\gamma\cos\mu_I\left(t\right)\right)^2,\\
\lambda_2\left(t\right)&=\left(\sin\gamma\cos\mu_I\left(t\right)-\cos\gamma\sin\mu_I\left(t\right)\right)^2,\\
\end{aligned}
\end{split}
\end{equation}
and substitute these into the expression for the entropy

\begin{equation}
S_{h,a}\left(t\right)=S_{H,a}\left(t\right)=-\lambda_1\left(t\right)\ln\left[\lambda_1\left(t\right)\right]-\lambda_2\left(t\right)\ln\left[\lambda_2\left(t\right)\right].
\end{equation}
With this expression we are free to choose the initial state of our system with a given value of $\gamma$. If the initial state of our system is a maximally entangled state with $\gamma=\pi/4$, then we observe how the entanglement entropy evolves with time. This is most applicable to quantum computing as in that context one would like to preserve the entangled state. We will now vary the parameters $N$, $g$ and $\kappa$ to see how they affect the evolution of entropy with time. Of particular interest is the exceptional point $g=\kappa$ where the non-Hermitian system enters the broken $\mathcal{PT}$ regime in the time-independent setting. It is in this area that the evolution we see differs from the standard evolution of entropy in Hermitan quantum mechanics. 

\begin{figure}[H]
	\centering
	\includegraphics[scale=0.7]{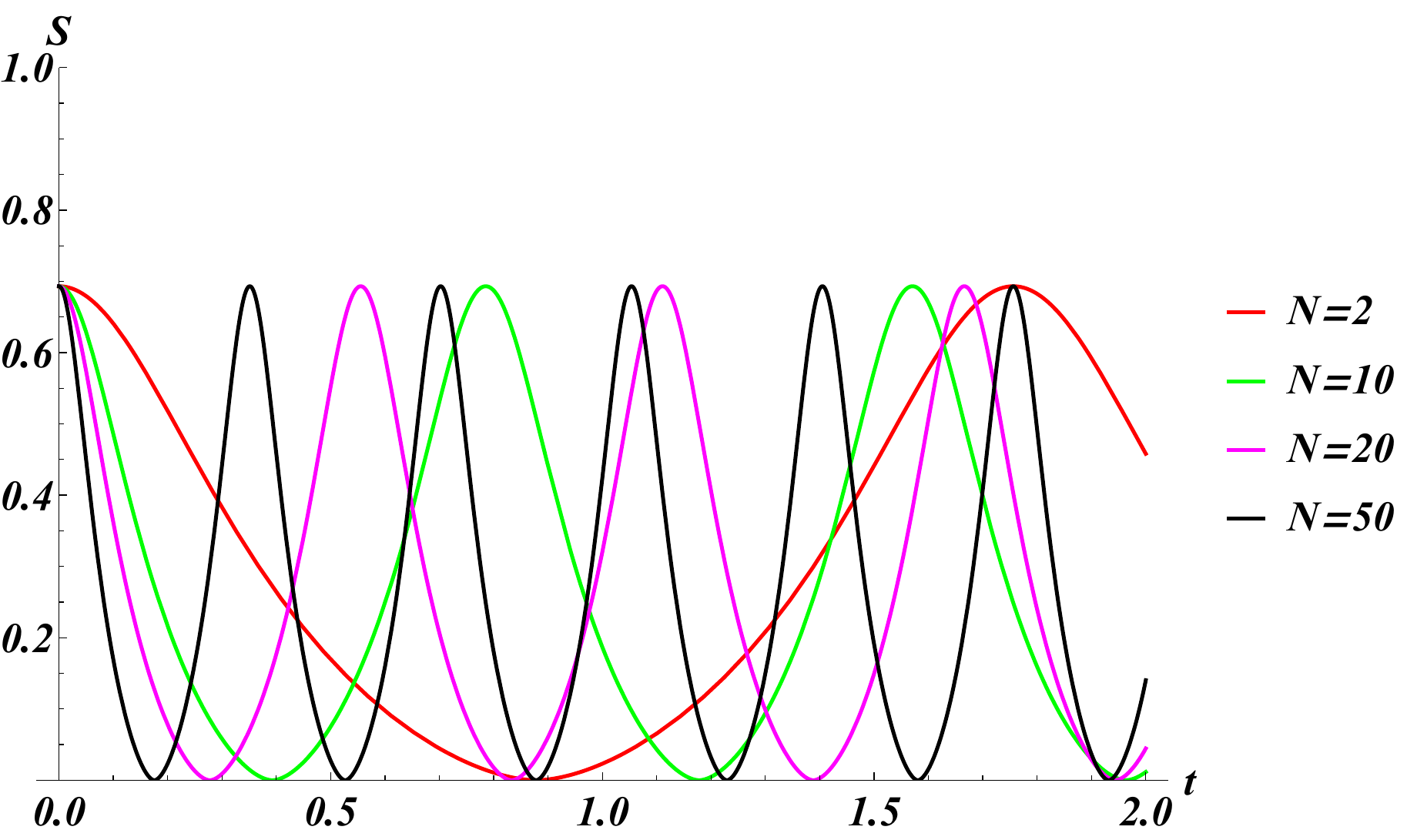}
	\caption{Von Neumann entropy as a function of time and varied bath size, with $c_1=1$, $g=0.7$, $\protect\kappa=0.3$, corresponding to the $\protect\mathcal{PT}$ unbroken regime.}
	\label{fig:unbroken}
\end{figure}

Figure \ref{fig:unbroken} shows how the entropy evolves when $\kappa<g$. This is equivalent to the unbroken $\mathcal{PT}$ regime of the non-Hermitian model. In this setting the entropy experiences so called "sudden death" similar to \cite{yonacc2006sudden}. The entropy rapidly decays from a maximum value to zero with a subsequent revival after the initial death. When the number of oscillators in the bath increases, the moment of vanishing entropy occurs at an earlier time.

\begin{figure}[H]
	\centering
	\includegraphics[scale=0.7]{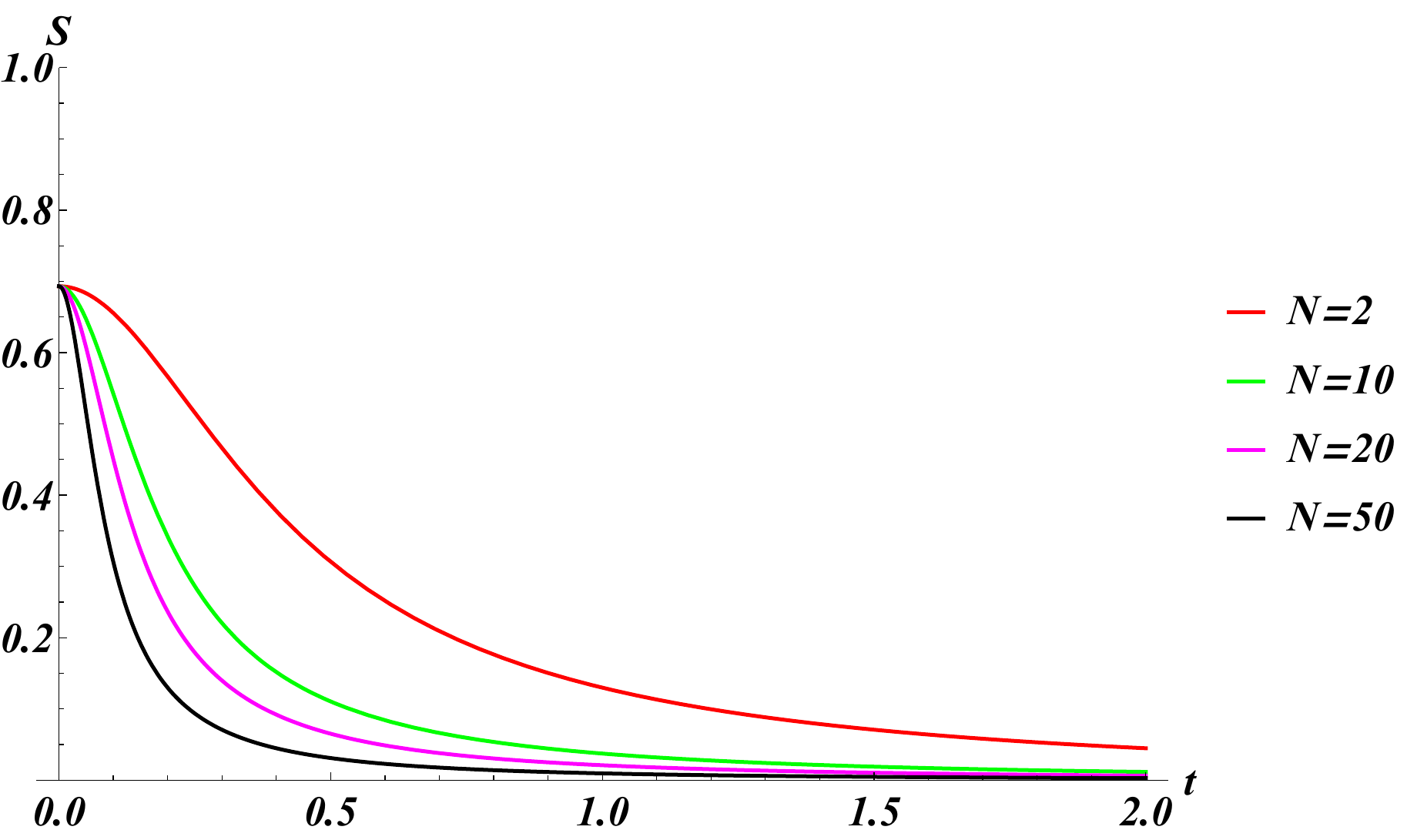}
	\caption{Von Neumann entropy as a function of time and varied bath size, with $c_1=1$, $g=\protect\kappa$, corresponding to the exceptional point.}
	\label{fig:exceptional}
\end{figure}

Figure \ref{fig:exceptional} depicts the entropy evolution when $\kappa=g$. This is equivalent to the exceptional point of the non-Hermitian model. As $\kappa=g$, any dependence on either $\kappa$ or $g$ disappears as they only appear in the combination $g^2-\kappa^2$ in the entropy. In this specific setting, the system decays asymptotically from maximal entropy to zero. The half life of this decay decreases with the number of oscillators in the bath. 

\begin{figure}[H]
	\centering
	\includegraphics[scale=0.7]{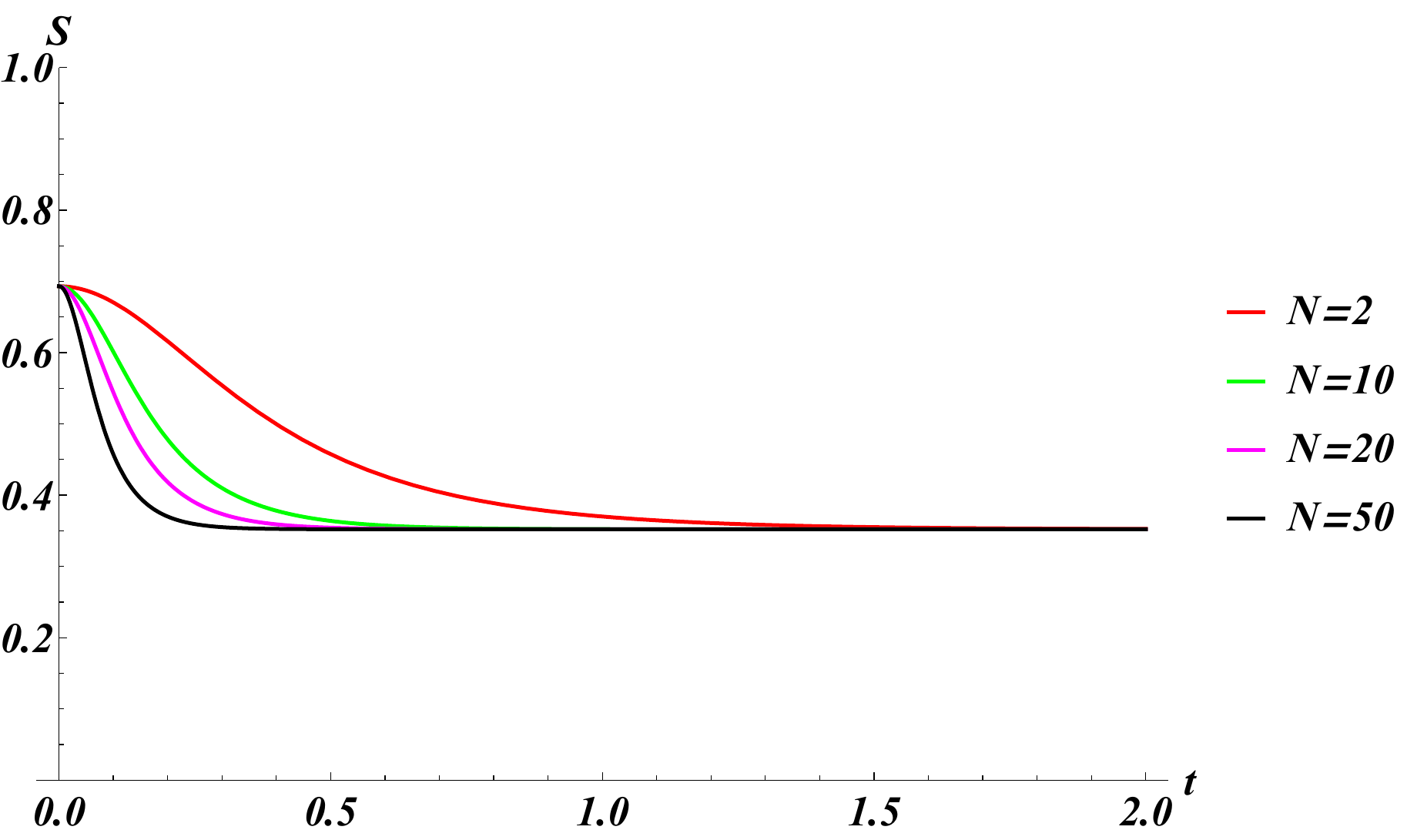}
	\caption{Von Neumann entropy as a function of time and varied bath size, with $c_1=1$, $g=0.3$, $\protect\kappa=0.7$, corresponding to the $\protect\mathcal{PT}$ broken regime. The asymptote is at $S\protect\protect_{t\protect\rightarrow\protect\infty}\protect\approx0.3521$.}
	\label{fig:broken}
\end{figure}

Figure \ref{fig:broken} now shows the results of entropy evolution when $g<\kappa$. This is the spontaneously broken $\mathcal{PT}$ regime of the original time-independent non-Hermitian model. In this case the system once again decays asymptotically but in this instance the decay is to a non-zero value of entropy. In this way, the entropy is preserved eternally. Once again the half life decreases with increasing $N$. The finite value that is asymptotically approached independently of $N$ is

\begin{equation}\label{min_entropy}
\begin{split}
\begin{aligned}
S_{t\rightarrow\infty}=-\frac{1}{2}(1+\xi)\ln\left[\frac{1}{2}(1+\xi)\right]
-\frac{1}{2}(1-\xi)\ln\left[\frac{1}{2}(1-\xi)\right],
\end{aligned}
\end{split}
\end{equation}
where 

\begin{equation}
\xi=\frac{\sqrt{c_1^2+\omega^2}}{c_1}.
\end{equation}
We see the condition for the asymptote to exist is $|\frac{c_1}{\omega}|>1$, which matches the reality condition of $\mu$ in equation (\ref{mu}).

We have found three significantly different phenomena at $\kappa>g$, $\kappa=g$ and $\kappa<g$. Specifically we see a change from rapid decay of entropy to zero, to asymptotic decay to zero through to asymptotic decay to a non-zero entropy. This can be interpreted as crossing the $\mathcal{PT}$ exceptional point into the spontaneously broken regime of the original time-independent non-Hermitian system. However, with the existence of a time-dependent metric, the broken regime is no longer truly broken as we are able to provide a well-defined interpretation.

\section{Summary}

We derived a framework for the Von Neumann entropy in non-Hermitian quantum systems and applied it to a simple system bath coupled bosonic model. In order to analyse the model we were required to find a time-dependent metric and we chose to solve the time-dependent Dyson equation for this. This method also gave us the equivalent Hermitian system which we worked with to perform the analysis as the framework showed the entropy was equivalent in both systems. The $\mathcal{PT}$-symmetry of the non-Hermitian system played an important role for the characterisation of the regimes of different qualitative behaviour in the evolution of the Von Neumann entropy. We found three different types of behaviour depending on whether we are in the $\mathcal{PT}$ unbroken regime, at the exceptional point or in the spontaneously broken $\mathcal{PT}$ regime. In the unbroken regime, the entropy underwent rapid decay to zero. At subsequent times it was revived and continued this oscillatory behaviour indefinitely. At the exceptional point, the entropy decayed asymptotically to zero and in the spontaneously broken regime, the entropy decayed asymptotically from a maximum to a finite minimum (\ref{min_entropy}) that remained constant with time. 

Our findings may have implications for maintaining entanglement in quantum computers when the computer is operated in the spontaneously broken $\mathcal{PT}$ regime. The challenge here is to construct a system in a laboratory that mimics that of the non-Hermitian system presented here. However, non-Hermitian systems have been realised in quantum optical experiments \cite{Guo,ruter2010obs} and so it is certainly possible that the same could be carried in quantum computing. \\\\

\chapter{Darboux Transformations}

Darboux transformations \cite{darboux} are very efficient tools in the study
of exactly or quasi-exactly solvable systems. Formally they map solutions
and coefficient functions of a partial differential equation to new
solutions and a differential equation of similar form with different
coefficient functions. The classic example is a second order differential
equation of Sturm-Liouville type or time-independent Schr\"{o}dinger
equation (TDSE). Since in this context the Darboux transformation relates
two operators that can be identified as isospectral Hamiltonians, this
scenario has been interpreted as the quantum mechanical analogue of
supersymmetry \cite{Witten:1981nf,Cooper,bagrov1}. Many potentials with
direct physical applications may be generated with this technique, such as
for instance complex crystals with invisible defects \cite%
{MatLongo,correa2015p}. By relating quantum mechanical systems to soliton
solutions of nonlinear differential equations, such as for instance the
Korteweg-de Vries equation, the sine-Gordon equation or the nonlinear Schr%
\"{o}dinger equation, Darboux transformations have also been very
efficiently utilized in the construction of multi-soliton solutions \cite%
{matveevdarboux,correahidden,CCFsineG,guilarte2017perfectly,cen2019asymptotic}%
.

Initially Darboux transformations were developed for stationary equations,
so that the treatment of the full TDSE was not possible. Evidently the
latter is a much more intricate problem to solve, especially for
non-autonomous Hamiltonians. Explicitly time-dependent Darboux
transformations for TDSE, rather than the time-independent Schr\"{o}dinger
equation, were first introduced by Bagrov and Samsonov \cite{bagrov2} and
subsequently generalized to other types of time-dependent systems \cite%
{song2003,suzko2009darboux}. The limitations of the generalization from the
time-independent to the time-dependent Schr\"{o}dinger equation were that
the solutions considered in \cite{bagrov2} force the Hamiltonians involved
to be Hermitian. One of the central purposes of this chapter is to
overcome this shortcoming and propose fully time-dependent Darboux
transformations that deal directly with the TDSE involving non-Hermitian
Hamiltonians. We extend our analysis to the entire hierarchy of solvable
time-dependent Hamiltonians constructed from generalized versions of
Darboux-Crum transformations. As an alternative scheme we also discuss the
intertwining relations for Lewis-Riesenfeld invariants for Hermitian as well
as non-Hermitian Hamiltonians. These quantities are constructed as auxiliary
objects to convert the fully TDSE into an eigenvalue equation that is easier
to solve and subsequently allows to tackle the TDSE. The class of
non-Hermitian Hamiltonians we consider here is the one of $\mathcal{PT}$%
-symmetric/quasi-Hermitian ones %\cite{Urubu,Benderrev,Alirev}% 
that are
related to a Hermitian counterpart by means of the time-dependent Dyson
equation (TDDE).
%\cite%
%{CA,time1,time6,time7,fringmoussa,AndTom1,AndTom2,AndTom3,AndTom4,mostafazadeh2018energy,AndTom5}%

Given the interrelations of the various quantities in the proposed scheme
one may freely choose different initial starting points. A quadruple of
Hamiltonians, two Hermitian and two non-Hermitian ones, is related by two
TDDE and two intertwining relations in form of a commutative diagram. This
allows to compute all four Hamiltonians by solving either two intertwining
relations and one TDDE or one intertwining relations and two TDDE, with the
remaining relation being satisfied by the closure of the commutative
diagram. We discuss the working of our proposal by taking two concrete
non-Hermitian systems as our starting points, the Gordon-Volkov Hamiltonian
with a complex electric field and a reduced version of the Swanson model.
From the various solutions to the TDSE we construct explicitly
time-dependent rational, hyperbolic, Airy function and nonlocal potentials.

This chapter is organized as follows: In section \ref{TDDC} we review the
time-dependent Darboux transformations for Hermitian Hamiltonians and stress
the limitations of previous results. We propose a new scheme that allows for
the treatment of non-Hermitian Hamiltonians. Subsequently we extend the
Darboux transformations to Darboux-Crum transformations, that is we
construct two hierarchies from intertwining operators built from solutions
previously ignored. In section \ref{ILR} we discuss the intertwining relations for
Lewis-Riesenfeld invariants. Taking a complex Gordon-Volkov Hamiltonian as
starting point we discuss in section \ref{GVH} various options of how to close the
commutative diagrams constructing the intertwining operators from different
types of solutions for rational, hyperbolic, Airy function potentials. In
section \ref{RS} we start from a reduced version of the Swanson model and carry out
the analysis for two different Dyson maps. In addition we discuss
intertwining relations for Lewis-Riesenfeld invariants for this concrete
system. The solutions to the TDSE discussed in this section depend on the
solutions of an auxiliary equation known as the dissipative Ermakov-Pinney
equation. We discuss in \nameref{EPAppendix} how to obtain explicit solutions to this
nonlinear second order differential equation.

\section{Time-dependent Darboux-Crum transformations}\label{TDDC}

\subsection{Time-dependent Darboux transformations for Hermitian systems}

Before introducing the time-dependent Darboux transformations for
non-Hermitian systems we briefly recall the construction for the Hermitian
setting. This revision will not only establish our notation, but it also
serves to highlight why previous suggestions are limited to the treatment of
Hermitian systems. Here we wish to overcome this shortcoming and extend the
theory of Darboux transformations to include the treatment of time-dependent
non-Hermitian Hamiltonians. Our main emphasis is on non-Hermitian systems
that belong to the class of $\mathcal{PT}$-symmetric Hamiltonians, as specified in the introduction (e.g. \ref{PTSymmetry}). Such type of systems
are of physical interest as potentially they possess energy operators with
real instantaneous eigenvalues, that are different from the Hamiltonians in
the non-Hermitian case.

The time-dependent Hermitian intertwining relation introduced in \cite%
{bagrov2} reads 
\begin{equation}
\ell \left( i\partial _{t}-h_{0}\right) =\left( i\partial _{t}-h_{1}\right)
\ell ,  \label{HI}
\end{equation}%
where the Hermitian Hamiltonians $h_{0}$ and $h_{1}$ involve explicitly
time-dependent potentials $v_{j}\left( x,t\right) $ 
\begin{equation}
h_{j}\left( x,t\right) =p^{2}+v_{j}\left( x,t\right) ,\qquad j=0,1.
\label{HamiltonianForm}
\end{equation}%
The intertwining operator $\ell $ is taken to be a first order differential
operator 
\begin{equation}
\ell \left( x,t\right) =\ell _{0}\left( x,t\right) +\ell _{1}\left(
x,t\right) \partial _{x}.  \label{ll}
\end{equation}%
In general we denote by $\phi _{j}$, $j=0,1$, the solutions to the two
partner TDSEs $i\partial _{t}\phi _{j}=h_{j}\phi _{j}$. Throughout this chapter we use the convention $\hbar =1$. Taking a specific solution $%
u(x,t):=\phi _{0}(x,t)$ to one of these equations, the constraints imposed
by the intertwining relation (\ref{HI}) can be solved by%
\begin{equation}
\ell _{1}\left( x,t\right) =\ell _{1}\left( t\right) ,\quad \ell _{0}\left(
x,t\right) =-\ell _{1}\frac{u_{x}}{u},\quad v_{1}=v_{0}+i\frac{\left( \ell
	_{1}\right) _{t}}{\ell _{1}}+2\left( \frac{u_{x}}{u}\right) ^{2}-2\frac{%
	u_{xx}}{u},  \label{v1}
\end{equation}%
where, as indicated, $\ell _{1}$ must be an arbitrary function of $t$ only.
At this point the new potential $v_{1}$ might still be complex. However,
besides mapping the coefficient functions, the main practical purpose of the
Darboux transformations is that one also obtains exact solutions $\phi _{1}$
for the partner TDSE $i\partial _{t}\phi _{1}=h_{1}\phi _{1}$ by employing
the intertwining operator. In this case the direct application, that is
acting with (\ref{HI}) on $u$, yields just the trivial solution $\phi
_{1}=lu=0$. For this reason different types of nontrivial solutions were
proposed in \cite{bagrov2} 
\begin{equation}
\hat{\phi}_{1}=\frac{1}{\ell _{1}u^{\ast }},\qquad \text{and\qquad }\tilde{%
	\phi}_{1}=\hat{\phi}_{1}\int^{x}\left\vert u\right\vert ^{2}dx^{\prime },
\label{ntsol}
\end{equation}%
which require, however, that one imposes%
\begin{equation}
\ell _{1}(t)=\exp \left[ -\int^{t}\func{Im}\left( v_{0}+2\left( \frac{u_{x}}{%
	u}\right) ^{2}-2\frac{u_{xx}}{u}\right) dt^{\prime }\right] .  \label{h1}
\end{equation}%
It is this assumption on the particular form of the solution that forces the
new potentials in the proposal of \cite{bagrov2} to be real $v_{1}=\func{Re}%
\left( v_{0}+2\left( u_{x}/u\right) ^{2}-2u_{xx}/u\right) $. Notice that one
might not be able to satisfy (\ref{h1}), as the right hand side must be
independent of $x$. If the latter is not the case, the solutions in (\ref%
{ntsol}) and the partner Hamiltonian $h_{1}$ do not exist.

Here we also identify another type of nontrivial solutions. Acting with
equation (\ref{HI}) to the right on a solution of the TDSE $i\partial
_{t}\phi _{0}=h_{0}\phi _{0}$, say $\phi _{0}=\tilde{u}$, that is linearly
independent from $\phi _{0}=u$ used in the construction of the intertwining
operator will in general lead to nontrivial solutions 
\begin{equation}
\phi _{1}=\mathcal{L}\left[ u\right] (\tilde{u}),\quad \text{with \ }%
\mathcal{L}\left[ u\right] (f):=\ell _{1}\left( t\right) \left( \partial
_{x}f-\frac{u_{x}}{u}f\right)   \label{second}
\end{equation}%
to the second TDSE $i\partial _{t}\phi _{1}=h_{1}\phi _{1}$. This type of
solution was overlooked in \cite{bagrov2} and in principle might lead to
complex potentials $v_{1}$ as it is not restricted by any additional
constraints.

\subsection{Time-dependent Darboux transformations for non-Hermitian systems}

In order to extend the previous analysis in the way that allows for other
types of complex potentials, and especially general non-Hermitian
Hamiltonians that are $\mathcal{PT}$-symmetric/quasi-Hermitian%\cite%
%{Urubu,Benderrev,Alirev}%
, we make use of the time-dependent Dyson equation
(TDDE) %\cite%
%{CA,time1,time6,time7,fringmoussa,AndTom1,AndTom2,AndTom3,AndTom4,mostafazadeh2018energy,AndTom5}
for both time-dependent Hermitian Hamiltonians $h_{0}(t)$, $h_{1}(t)$ and
the time-dependent non-Hermitian Hamiltonians $H_{0}(t)$, $H_{1}(t)$ 
\begin{equation}
h_{j}=\eta _{j}H_{j}\eta _{j}^{-1}+i\partial _{t}\eta _{j}\eta
_{j}^{-1},\qquad j=0,1.  \label{Dysoneq}
\end{equation}%
The time-dependent Dyson maps $\eta _{j}(t)$ relate the solutions of the
TDSE $i\partial _{t}\psi _{j}=H_{j}\psi _{j}$ to the previous ones for $\phi
_{j}$ as%
\begin{equation}
\phi _{j}=\eta _{j}\psi _{j},\qquad j=0,1.
\end{equation}%
Using (\ref{Dysoneq}) in the intertwining relation (\ref{HI}) yields 
\begin{equation}
\ell \left( i\partial _{t}-\eta _{0}H_{0}\eta _{0}^{-1}-i\partial _{t}\eta
_{0}\eta _{0}^{-1}\right) =\left( i\partial _{t}-\eta _{1}H_{1}\eta
_{1}^{-1}-i\partial _{t}\eta _{1}\eta _{1}^{-1}\right) \ell .  \label{aux}
\end{equation}%
Multiplying (\ref{aux}) from the left by $\eta _{1}^{-1}$ and acting to the
right on $\eta _{0}f$, with $f(x,t)$ being some arbitrary test function, we
obtain 
\begin{eqnarray}
&&\eta _{1}^{-1}\ell \left[ i\left( \partial _{t}\eta _{0}\right) f+i\eta
_{0}\partial _{t}f-\eta _{0}H_{0}f-i\left( \partial _{t}\eta _{0}\right) f%
\right] =  \notag \\
&&\left( i\eta _{1}^{-1}\ell \eta _{0}\partial _{t}f+i\eta _{1}^{-1}\left(
\partial _{t}\ell \eta _{0}\right) f-H_{1}\eta _{1}^{-1}\ell \eta
_{0}f-i\eta _{1}^{-1}\left( \partial _{t}\eta _{1}\right) \eta _{1}^{-1}\ell
\eta _{0}f\right) .
\end{eqnarray}%
Rearranging the time derivative terms and removing the test function, we
derive the new intertwining relation for non-Hermitian Hamiltonians%
\begin{equation}
L\left( i\partial _{t}-H_{0}\right) =\left( i\partial _{t}-H_{1}\right) L,
\label{IH}
\end{equation}%
where we introduced the new intertwining operator 
\begin{equation}
L:=\eta _{1}^{-1}\ell \eta _{0}.  \label{DarbouxGeneral}
\end{equation}%
We note that $H_{j}-p^{2}$ is in general not only no longer real and might
also include a dependence on the momenta, i.e. $H_{j}$ does not have to be a
potential Hamiltonian and could be nonlocal. Denoting by $\psi _{0}=U=\eta
_{0}^{-1}u~$a particular solution to the TDSE for $H_{0}$, the standard new
solution $\psi _{1}=LU=\eta _{1}^{-1}\ell \eta _{0}\eta _{0}^{-1}u$ remains
trivial. The nontrivial solutions (\ref{ntsol}) generalize to 
\begin{equation}
\hat{\psi}_{1}=\eta _{1}^{-1}\frac{1}{\ell _{1}\left( \eta _{0}U\right)
	^{\ast }},\qquad \text{and\qquad }\tilde{\psi}_{1}=\hat{\psi}%
_{1}\int^{x}\left\vert \eta _{0}U\right\vert ^{2}dx^{\prime }.  \label{ntP}
\end{equation}%
The nontrivial solution (\ref{second}) becomes%
\begin{equation}
\psi _{1}=L\left[ U\right] \left( \tilde{U}\right)
\end{equation}%
in the non-Hermitian case. In summary, our quadruple of Hamiltonians is
related as depicted in the commutative diagram%
\begin{equation}
\begin{array}{ccccc}
& H_{0} & ~~\underrightarrow{\eta _{0}}~~ & h_{0} &  \\ 
\eta _{1}^{-1}\mathcal{L}\left[ u\right] \eta _{0} & \downarrow &  & 
\downarrow & \mathcal{L}\left[ u\right] \\ 
& H_{1} & \underrightarrow{\eta _{1}} & h_{1} & 
\end{array}
\label{DH}
\end{equation}

One may of course also try to solve the intertwining relation (\ref{IH})
directly and build the intertwining operator $L$ from a solution $U=\eta
_{0}^{-1}u$ for the TDSE for $H_{0}$ and ignore initially the fact that the
Hamiltonians $H_{0}$ and $H_{1}$ involved are non-Hermitian. To make sense
of these Hamiltonians one still needs to construct the Dyson maps $\eta _{0}$
and $\eta _{1}$. Considering the diagram 
\begin{equation}
\begin{array}{ccccc}
& H_{0} & ~~\underrightarrow{\eta _{0}}~~ & h_{0} &  \\ 
\mathcal{L}\left[ U\right] & \downarrow &  & \downarrow & \mathcal{L}\left[ u%
\right] \\ 
& H_{1} & \underrightarrow{?} & h_{1} & 
\end{array}
\label{dia}
\end{equation}%
in which the TDDE has been solved for $\eta _{0}$, $H_{0}$, $h_{0}$ and $%
H_{1}$, $h_{1}$ have been constructed with intertwining operators build from
the solutions of the respective TDSE, we address the question of whether it
is possible to close the diagram, that is making it commutative. For this to
be possible we require%
\begin{equation}
\mathcal{L}\left[ U\right] =\eta _{1}^{-1}\mathcal{L}\left[ u\right] \eta
_{0}  \label{LLL}
\end{equation}%
to be satisfied. It is easy to verify that (\ref{LLL}) holds if and only if%
\begin{equation}
\eta _{1}\mathcal{=}\eta _{0},\qquad \text{and\qquad }\frac{\eta
	_{0}^{-1}u_{x}}{\eta _{0}^{-1}u}=\eta _{0}^{-1}\frac{u_{x}}{u}\eta _{0}.
\label{sc}
\end{equation}%
A solution for the second equation in (\ref{sc}) is for instance $\eta
_{0}=f(x)T_{\alpha }(x)$, with $T_{\alpha }=e^{i\alpha p}$ being a standard
shift operator, i.e. $T_{\alpha }g(x)=g(x+\alpha )$, and $f(x)$ an arbitrary 
$x$-dependent function.

\subsection{Time-dependent Darboux-Crum transformations for Hermitian systems%
}

Next we demonstrate that the iteration procedure of the Darboux
transformation, usually referred to as Darboux-Crum (DC) transformations 
\cite{darboux,crum,matveevdarboux}, will lead also in the time-dependent
case to an entire hierarchy of exactly solvable time-dependent Hamiltonians $%
h_{0}$, $h_{1}$, $h_{2}$, \ldots\ for the TDSEs $i\partial _{t}\phi
^{(n)}=h_{n}\phi ^{(n)}$ related to each other by intertwining operators $%
\ell ^{(n)}$\ 
\begin{equation}
\ell ^{(n)}\left( i\partial _{t}-h_{n-1}\right) =\left( i\partial
_{t}-h_{n}\right) \ell ^{(n)},\qquad n=1,2,\ldots  \label{iter}
\end{equation}%
For $n=1$ this is equation (\ref{HI}) with $\ell =\ell ^{(1)}$ and solutions 
$\phi _{0}=\phi ^{(0)}$, $\phi _{1}=\phi ^{(1)}$. Taking a particular
solution $\phi _{0}=u$ to depend on some parameter $\gamma $, continuously
or discretely, we denote the solutions at different values as $%
u_{i}:=u(\gamma _{i})$. Given now $\ell ^{(1)}=\mathcal{L}\left[ u_{0}\right]
$ from (\ref{ll}) we act with (\ref{iter}) for $n=1$ on $u_{1}$, so that we
can cast the intertwining operator and the solution (\ref{second}) in the
form 
\begin{equation}
\ell ^{(1)}(f)=\mathcal{L}\left[ u_{0}\right] (f)=\ell _{1}\left( t\right) 
\frac{W_{2}[u_{0},f]}{W_{1}[u_{0}]},\quad \qquad \phi ^{(1)}=\ell
^{(1)}(u_{1})=\mathcal{L}\left[ u_{0}\right] (u_{1}),
\end{equation}%
with corresponding time-dependent Hamiltonian%
\begin{equation}
h_{1}=h_{0}-2\left[ \ln W_{1}(u_{0})\right] _{xx}+i\left[ \ln \ell _{1}%
\right] _{t}.
\end{equation}%
We employed here the Wronskian $W_{n}[u_{1},u_{2},\ldots ,u_{n}]:=$ $\det
\omega $ with $\omega _{jk}=\partial ^{j-1}u_{k}/\partial x^{j-1}$ for $%
j,k=1,\ldots ,n$, e.g. $W_{1}[u_{0}]=$ $u_{0}$, $W_{2}[u_{0},u_{1}]=$ $%
u_{0}\left( u_{1}\right) _{x}-u_{1}\left( u_{0}\right) _{x}$, etc., which
allows to write the expressions for the intertwining operator and
Hamiltonians in the hierarchy in a very compact form. Iterating these
equations we obtain the compact closed form for the intertwining operator 
\begin{eqnarray}
\ell ^{(n)}(f) &=&\mathcal{L}\left[ \ell ^{(n-1)}(u_{n-1})\right] (\ell
^{(n-1)}(f))  \label{DCa} \\
&=&\ell _{1}^{n}\left( t\right) \frac{W_{n+1}[u_{0},u_{1},\ldots ,u_{n-1},f]%
}{W_{n}[u_{0},u_{1},\ldots ,u_{n-1}]} \\
&=&\ell _{1}^{n}\left( t\right) \left\vert \Omega \right\vert _{(n+1)(n+1)},
\end{eqnarray}%
where $\left\vert \Omega \right\vert _{(n+1)(n+1)}$ denotes a
quasideterminant \cite{gelfand2005} for the (n+1)$\times $(n+1)-matrix $%
\Omega $ with $\Omega _{jk}=\partial ^{j-1}u_{k}/\partial x^{j-1}$, $\Omega
_{j(n+1)}=\partial ^{j-1}f/\partial x^{j-1}$ for $j=1,\ldots ,n+1$, $%
k=1,\ldots ,n$. For the time-dependent Hamiltonians we derive%
\begin{equation}
h_{n}=h_{0}-2\left[ \ln W_{n}\left( u_{0},u_{1},\ldots ,u_{n-1}\right) %
\right] _{xx}+in\left( \ln \ell _{1}\right) _{t}.  \label{hnn}
\end{equation}%
Nontrivial solutions of the type (\ref{second}) to the related TDSE $%
i\partial _{t}\phi ^{(n)}=h_{n}\phi ^{(n)}$ are then obtained as%
\begin{equation}
\phi ^{(n)}=\ell ^{(n)}(u_{n})\text{.}  \label{nsol}
\end{equation}%
Instead of using the same solution $u_{i}$ of the TDSE for $h_{0}$ at
different parameter values in the closed expression, it is also possible to
replace some of the solutions $u_{i}$ by the second linear independent
solutions $\tilde{u}_{i}$ at the same parameter values, see e.g. \cite%
{CorreaFring,CCFsineG,CenFringHir} and references therein for details. This
choice allows for the treatment of degenerate solutions. Closed expressions
for DC-transformation built from the solutions (\ref{ntP}) can be found in 
\cite{bagrov2}. Below we will illustrate the working of the formulae in this
section with concrete examples.

\subsection{Time-dependent DC transformations for non-Hermitian systems}

The iteration procedure for the non-Hermitian system goes along the same
lines as for the Hermitian case, albeit with different intertwining
operators $L$. The iterated systems are\ 
\begin{equation}
L^{(n)}\left( i\partial _{t}-H_{n-1}\right) =\left( i\partial
_{t}-H_{n}\right) L^{(n)},\qquad n=1,2,\ldots  \label{H1}
\end{equation}%
The intertwining operators read in this case 
\begin{equation}
L^{(n)}(f)=\mathcal{L}\left[ L^{(n-1)}(U_{n-1})\right] (L^{(n-1)}(f))=\ell
_{1}^{n}\left( t\right) \frac{W_{n+1}[U_{0},U_{1},\ldots ,U_{n-1},f]}{%
	W_{n}[U_{0},U_{1},\ldots ,U_{n-1}]},
\end{equation}%
and the time-dependent Hamiltonians are%
\begin{equation}
H_{n}=H_{0}-2\left[ \ln W_{n}[U_{0},U_{1},\ldots ,U_{n-1}]\right] _{xx}+in%
\left[ \ln \ell _{1}\right] _{t}.
\end{equation}%
The nontrivial solutions to the related TDSE are then obtained as%
\begin{equation}
\psi ^{(n)}=L^{(n)}(U_{n}).  \label{H4}
\end{equation}%
Notice that in (\ref{H1})-(\ref{H4}) the only Dyson maps involved are $\eta
_{0}$ and $\eta _{1}$. Alternatively we can also express $L^{(n)}=\eta
_{n}^{-1}l^{(n)}\eta _{n-1}$ and $\psi ^{(n)}=\eta _{n}^{-1}\phi ^{(n)}$,
but the computation of the $\eta _{n}$ for $n>1$ is not needed. Since the
solutions (\ref{ntP}) require the Hamiltonians involved to be Hermitian,
hierarchies build on them do not exist in the non-Hermitian case.

\section{Intertwining relations for Lewis-Riesenfeld invariants}\label{ILR}

As previously argued \cite{maamache2017pseudo,AndTom4,AndTom5}, the most
efficient way to solve the TDDE (\ref{Dysoneq}), as well as the TDSE, is to
employ the Lewis-Riesenfeld invariants \cite{lewis1969exact}. The steps in this
approach consists of first solving the evolution equation for the invariants
of the Hermitian and non-Hermitian system separately and subsequently
constructing a similarity transformation between the two invariants. By
construction the map facilitating this transformation is the Dyson map
satisfying the TDDE.

Here we need to find four time-dependent invariants $I_{j}^{h}(t)$ and $%
I_{j}^{H}(t)$, $j=0,1$, that solve the equations%
\begin{equation}
\partial _{t}I_{j}^{H}(t)=i\left[ I_{j}^{H}(t),H_{j}(t)\right] ,\quad \text{%
	and\quad }\partial _{t}I_{j}^{h}(t)=i\left[ I_{j}^{h}(t),h_{j}(t)\right] 
\text{.}  \label{LRin}
\end{equation}%
The solutions $\phi _{j}(t)$, $\psi _{j}(t)$ to the respective TDSEs are
related by a phase factor $\left\vert \phi _{j}(t)\right\rangle =e^{i\alpha
	_{j}(t)}\left\vert \check{\phi}_{j}(t)\right\rangle $, $\left\vert \psi
_{j}(t)\right\rangle =e^{i\alpha _{j}(t)}\left\vert \check{\psi}%
_{j}(t)\right\rangle $ to the eigenstates of the invariants 
\begin{equation}
I_{j}^{h}(t)\left\vert \check{\phi}_{j}(t)\right\rangle =\Lambda
_{j}\left\vert \check{\phi}_{j}(t)\right\rangle
,~~~~~~~I_{j}^{H}(t)\left\vert \check{\psi}_{j}(t)\right\rangle =\Lambda
_{j}\left\vert \check{\psi}_{j}(t)\right\rangle ,~~~~~~~\text{with }\dot{%
	\Lambda}_{j}=0.  \label{LR1}
\end{equation}%
Subsequently the phase factors can be computed from 
\begin{equation}
\dot{\alpha}_{j}=\left\langle \check{\phi}_{j}(t)\right\vert i\partial
_{t}-h_{j}(t)\left\vert \check{\phi}_{j}(t)\right\rangle =\left\langle 
\check{\psi}_{j}(t)\right\vert \eta _{j}^{\dagger }(t)\eta _{j}(t)\left[
i\partial _{t}-H_{j}(t)\right] \left\vert \check{\psi}_{j}(t)\right\rangle .
\end{equation}%
As has been shown \cite{maamache2017pseudo,AndTom4,AndTom5}, the two
invariants for the Hermitian and non-Hermitian system obeying the TDDE are
related to each other by a similarity transformation 
\begin{equation}
I_{j}^{h}=\eta _{j}I_{j}^{H}\eta _{j}^{-1}\text{.}  \label{simhH}
\end{equation}%
Here we show that the invariants $I_{0}^{H}$, $I_{1}^{H}$ and $I_{0}^{h}$, $%
I_{1}^{h}$ are related by the intertwining operators $L$ in (\ref%
{DarbouxGeneral}) and $\ell $ in (\ref{ll}), respectively. We have 
\begin{equation}
LI_{0}^{H}=I_{1}^{H}L,\qquad \text{and}\qquad \ell I_{0}^{h}=I_{1}^{h}\ell .
\label{II}
\end{equation}%
This is seen from computing%
\begin{equation}
i\partial _{t}\left( LI_{0}^{H}-I_{1}^{H}L\right) =H_{1}\left(
LI_{0}^{H}-I_{1}^{H}L\right) -\left( LI_{0}^{H}-I_{1}^{H}L\right) H_{0},
\label{LHH}
\end{equation}%
where we used (\ref{IH}) and (\ref{LRin}) to replace time-derivatives of $L$
and $I_{0}^{H}$, respectively. Comparing (\ref{LHH}) with (\ref{IH}) in the
form $i\partial _{t}L=H_{1}L-LH_{0}$, we conclude that $%
L=LI_{0}^{H}-I_{1}^{H}L$ or $LI_{0}^{H}=I_{1}^{H}L$. The second relation in (%
\ref{II}) follows from the first when using (\ref{DarbouxGeneral}) and (\ref%
{simhH}). Thus schematically the invariants are related in the same manner
as depicted for the Hamiltonians in (\ref{DH}) with the difference that the
TDDE is replaced by the simpler adjoint action of the Dyson map. Given the
above relations we have no obvious consecutive orderings of how to compute
the quantities involved. For convenience we provide a summary of the above
in the following diagram to illustrate schematically how different
quantities are related to each other:

\begin{figure}
	\centering
	\thispagestyle{empty} \setlength{\unitlength}{1.0cm} 
	\begin{picture}(14.48,9.0)(-2.2,6.5)
	\thicklines
	\put(-0.6,12.0){\LARGE{$H_0  \quad \longleftrightarrow \quad h_0 \quad \longleftrightarrow \quad h_1 \quad \longleftrightarrow \quad H_1$}}
	\put(-0.6,10.0){\LARGE{$I_0^H  \quad \longleftrightarrow \quad I_0^h \quad \longleftrightarrow \quad I_1^h \quad \longleftrightarrow \quad I_1^H$}}
	
	\put(-0.6,8.0){\LARGE{$\check{\psi}_0 \quad \longleftrightarrow \quad \check{\phi}_0 \quad \longleftrightarrow \quad \check{\phi}_1 \quad \longleftrightarrow \quad \check{\psi}_1$}}
	\put(-0.6,14.0){\LARGE{${\psi}_0 \quad \longleftrightarrow \quad {\phi}_0 \quad \longleftrightarrow \quad {\phi}_1 \quad \longleftrightarrow \quad {\psi}_1$}}
	
	\put(-0.3,11.8){\vector(0,-1){1.2}}	
	\put(3.,11.8){\vector(0,-1){1.2}}
	\put(6.1,11.8){\vector(0,-1){1.2}}
	\put(9.3,11.8){\vector(0,-1){1.2}}
	
	\put(-0.3,9.7){\vector(0,-1){1.1}}	
	\put(3.,9.7){\vector(0,-1){1.1}}
	\put(6.1,9.7){\vector(0,-1){1.1}}
	\put(9.3,9.7){\vector(0,-1){1.1}}
	
	\put(-0.3,12.5){\vector(0,1){1.1}}	
	\put(3.,12.5){\vector(0,1){1.1}}
	\put(6.1,12.5){\vector(0,1){1.1}}
	\put(9.3,12.5){\vector(0,1){1.1}}
	
	\put(1.2,14.6){\LARGE{$\eta_0 $}}
	\put(1.2,12.6){\LARGE{$\eta_0 $}}
	\put(1.2,9.6){\LARGE{$\eta_0 $}}
	\put(1.2,7.6){\LARGE{$\eta_0 $}}
	
	\put(4.5,14.4){\LARGE{$\ell $}}
	\put(4.5,12.4){\LARGE{$\ell $}}
	\put(4.5,9.5){\LARGE{$\ell $}}
	\put(4.5,7.5){\LARGE{$\ell $}}

	\put(7.6,14.6){\LARGE{$\eta_1 $}}
	\put(7.6,12.6){\LARGE{$\eta_1 $}}
	\put(7.6,9.6){\LARGE{$\eta_1 $}}
	\put(7.6,7.6){\LARGE{$\eta_1 $}}
	
	\thicklines
	\put(0.14,11.67){\vector(-3,2){0.2}}
	\put(0.14,10.72){\vector(-3,-2){0.2}}
	\put(9.0,11.68){\vector(3,2){0.2}}
	\put(9.0,10.71){\vector(3,-2){0.2}}
	
	\put(-0.9,13.8){\vector(2,3){0.2}}
	
	\put(9.91,13.8){\vector(-2,3){0.2}}

	\qbezier(0.1, 11.7)(4.8, 9.9)(9.0,11.7)
	\qbezier(0.1, 10.7)(4.8, 12.5)(9.0,10.7)
	\put(4.5,11.0){\LARGE{$L$}}
	
	\qbezier(9.8, 14.0)(11.1, 11.0)(9.8,8.0)
	\put(10.5,11.0){\LARGE{$\alpha_1 $}}
	
	\qbezier(-0.8, 14.0)(-2.1, 11.0)(-0.8,8.0)
	\put(-2.2,11.0){\LARGE{$\alpha_0 $}}
	
	\end{picture}
	
	\caption{ Schematic representation
		of Dyson maps $\protect\eta\protect _{0}$,$\protect\eta \protect_{1}$ and intertwining operators $\protect\ell $,$L$
		relating quadruples of Hamiltonians $h\protect_{0}$,$h\protect_{1}$,$H\protect_{0}$,$H\protect_{1}$ and
		invariants $I\protect_{0}\protect^{h}$,$I\protect_{1}\protect^{h}$,$I\protect_{0}\protect^{H}$,$I\protect_{1}\protect^{H}$ together with
		their respective eigenstates $\protect\phi \protect_{0}$,$\protect\phi \protect_{1}$,$\protect\psi \protect_{0}$,$\protect\psi \protect_{1}$
		and $\protect\check{\protect\phi}\protect_{0}$,$\protect\check{\protect\phi}\protect_{1}$,$\protect\check{\protect\psi}\protect_{0}$,$\protect\check{\protect\psi}\protect_{1}$ that are related by phases $\protect\alpha \protect_{0}$,$\protect\alpha \protect_{1}$.}
	\label{fig0}
\end{figure}

\section{Solvable potentials from the complex Gordon-Volkov Hamiltonian}\label{GVH}

We will now discuss how the various elements in figure \ref{fig0} can be computed.
Evidently the scheme allows to start from different quantities and compute
the remaining ones by following different indicated pathes, that is we may
solve intertwining relations and TDDE in different orders for different
quantities. As we are addressing here mainly the question of how to make
sense of non-Hermitian systems, we always take a non-Hermitian Hamiltonian $%
H_{0}$ as our initial starting point and given quantity. Subsequently we
solve the TDDE (\ref{Dysoneq}) for $h_{0}$,$\eta _{0}$ and thereafter close
the commutative diagrams in different ways.

We consider a complex version of the Gordon-Volkov Hamiltonian \cite{GV1,GV2}
\begin{equation}
H_{0}=H_{GV}=p^{2}+iE\left( t\right) x,
\end{equation}%
in which $iE\left( t\right) \in i\mathbb{R}$ may be viewed as a complex
electric field. In the real setting $H_{GV}$ is a Stark Hamiltonian with
vanishing potential term around which a perturbation theory can be build in
the strong field regime, see e.g. \cite{Rev1}. Such type of potentials are
also of physical interest in the study of plasmonic Airy beams in linear
optical potentials \cite{plasmonic}. Even though the Hamiltonian $H_{GV}$ is
non-Hermitian, it belongs to the interesting class of $\mathcal{PT}$%
-symmetric Hamiltonians, i.e. it remains invariant under the antilinear
transformation $\mathcal{PT}:$ $x\rightarrow -x$, $p\rightarrow p$, $%
i\rightarrow -i$.

In order to solve the TDDE (\ref{Dysoneq}) involving $H_{0}$ we make the
Ansatz 
\begin{equation}
\eta _{0}=e^{\alpha \left( t\right) x}e^{\beta \left( t\right) p},
\label{e0}
\end{equation}%
with $\alpha \left( t\right) $, $\beta \left( t\right) $ being some
time-dependent real functions. The adjoint action of $\eta _{0}$ on $x$, $p$
and the time-dependent term are easily computed to 
\begin{equation}
\eta _{0}x\eta _{0}^{-1}=x-i\beta ,\quad \eta _{0}p\eta _{0}^{-1}=p+i\alpha
,\quad i\dot{\eta}_{0}\eta _{0}^{-1}=i\dot{\alpha}x+i\dot{\beta}\left(
p+i\alpha \right) .
\end{equation}%
We use now frequently overdots as an abbreviation for partial derivatives
with respect to time. Therefore the right hand side of the TDDE (\ref%
{Dysoneq}) yields 
\begin{equation}
h_{0}=h_{GV}=p^{2}+ip\left( 2\alpha +\dot{\beta}\right) -\alpha
^{2}+ix\left( E+\dot{\alpha}\right) +E\beta -\dot{\beta}\alpha .
\end{equation}%
Thus, for $h_{0}$ to be Hermitian we have to impose the reality constraints 
\begin{equation}
\dot{\alpha}=-E,\quad \dot{\beta}=-2\alpha ,
\end{equation}%
so that $h_{0}$ becomes a free particle Hamiltonian with an added real
time-dependent field 
\begin{equation}
h_{0}=h_{GV}=p^{2}+\alpha ^{2}+E\beta =p^{2}+\left[ \int\nolimits^{t}E%
\left( s\right) ds\right] ^{2}+2E\left( t\right)
\int\nolimits^{t}\int\nolimits^{s}E\left( w\right) dwds.
\label{freeparticle}
\end{equation}%
There are numerous solutions to the TDSE $i\partial _{t}\phi _{0}=h_{GV}\phi
_{0}$, with each of them producing different types of partner potentials $%
v_{1}$ and hierarchies. We will now discuss various ways to construct the
next level in the hierarchy by using different types of solutions.

\subsection{Solvable time-dependent hyperbolic potentials, two separate
	intertwinings}\label{hyperbolic}

We start by considering the scenario as depicted in the commutative diagram (%
\ref{dia}). Thus we start with a solution to the TDDE in form of $h_{0}$, $%
H_{0}$, $\eta _{0}$ as given above and carry out the intertwining relations
separately using the intertwining operators $\mathcal{L}\left[ u\right] $
and $\mathcal{L}\left[ U\right] $ in the construction of $h_{1}$ and $H_{1}$%
, respectively. According to (\ref{sc}), in this case the expression for the
second Dyson map is dictated by the closure of the diagram to be $\eta
_{1}=\eta _{0}$. We construct our intertwining operator from the simplest
solutions to the TDSE for $h_{0}=h_{GV}$%
\begin{equation}
\phi _{0,m}\left( x,t\right) =\cosh (mx)e^{-\alpha x+im^{2}t-i\int^{t}\left(\alpha
	^{2}+E\beta\right) ds}  \label{fm}
\end{equation}%
with continuous parameter $m$. A second linearly independent solution $%
\tilde{\phi}_{0,m}$ is obtained by replacing the $\cosh $ in (\ref{fm}) by $%
\sinh $. Taking $\phi _{0,m}$ as our seed function we compute%
\begin{eqnarray}
\ell &=&\mathcal{L}\left[ \phi _{0,m}\right] =\ell _{1}\left( t\right) \left[
\partial _{x}-m\tanh (mx)\right] \\
h_{1} &=&p^{2}-2m^{2}\func{sech}^{2}(mx)+\alpha ^{2}+E\beta +i\frac{\left(
	\ell _{1}\right) _{t}}{\ell _{1}} \\
\phi _{1,m,m^{\prime }} &=&\ell \lbrack \phi _{0,m}](\phi _{0,m^{\prime }})
\\
&=&\!\!\!\!\ell _{1}\left( t\right) \left[ m^{\prime }\sinh (m^{\prime }x)-m\cosh
(m^{\prime }x)\tanh (mx)\right] e^{im^{2}t-i\int^{t}\left(\alpha
	^{2}+E\beta\right) ds} 
\end{eqnarray}%
Evidently $\ell _{1}(t)$ must be constant for $h_{1}$ to be Hermitian, so
for convenience we set $\ell _{1}(t)=1$. Since $\eta _{0}$ is of the form
that solves the second equation in (\ref{sc}), we can also directly solve
the intertwining relation (\ref{IH}) for $H_{0}$ and $H_{1}$ using an
intertwining operator build from a solution for the TDSE of $H_{0}$, i.e. $%
\mathcal{L}\left[ U\right] =\mathcal{L}\left[ \eta _{0}^{-1}\phi _{0,m}%
\right] $. We obtain%
\begin{eqnarray}
H_{1} &=&p^{2}-2m^{2}\func{sech}^{2}\left[ m(x+i\beta )\right] +iE\left(
t\right) x, \\
\psi _{1,m,\tilde{m}} &=&e^{-\alpha (x+i\beta )}\phi _{1,m,\tilde{m}%
}(x+i\beta ).
\end{eqnarray}%
We verify that the TDDE for $h_{1}$ and $H_{1}$ is solved by $\eta _{1}=\eta
_{0}\,$, which is enforced by the closure of the diagram (\ref{dia}) and the
first relation in (\ref{sc}).

We can extend our analysis to the Darboux-Crum transformation and compute
the two hierarchies of solvable time-dependent hyperbolic Hamiltonians $%
H_{0} $,$H_{1}$,$H_{2}$,$\ldots $ and $h_{0}$,$h_{1}$,$h_{2}$,$\ldots $
directly from the expressions (\ref{DCa})-(\ref{H4}). For instance, we
calculate%
\begin{equation}
H_{2}=p^{2}+\frac{2(m^{2}-\tilde{m}^{2})\left[ \tilde{m}^{2}\cosh (m\hat{x}%
	)-m^{2}\cosh (\tilde{m}\hat{x})\right] }{\left[ m\cosh (\tilde{m}\hat{x}%
	)\sinh (m\hat{x})-\tilde{m}\cosh (m\hat{x})\sinh (\tilde{m}\hat{x})\right]
	^{2}}+iE\left( t\right) x
\end{equation}%
with $\hat{x}=x+i\beta $. The solutions to the corresponding TDSE are
directly computable from the generic formula (\ref{H4}).

\subsection{Solvable time-dependent rational potentials, intertwining and
	TDDE}

Next we start again with a solution to the TDDE in form of $h_{0}$, $H_{0}$, 
$\eta _{0}$, carry out the intertwining to construct $h_{1}$ and
subsequently solve the TDDE for $H_{1}$, $\eta _{1}$ with given $h_{1}$ as
depicted in the commutative diagram 
\begin{equation}
\begin{array}{ccccc}
& H_{0} & ~~\underrightarrow{\eta _{0}}~~ & h_{0} &  \\ 
? & \downarrow &  & \downarrow & \mathcal{L}\left[ u\right] \\ 
& H_{1} & \underleftarrow{\eta _{1}} & h_{1} & 
\end{array}
\label{CD1}
\end{equation}%
In this case the expression for the intertwining operator between $H_{0}$
and $H_{1}$ is dictated by the closure of the diagram to be $\eta _{1}^{-1}%
\mathcal{L}\left[ u\right] \eta _{0}\neq \mathcal{L}\left[ U\right] $. We
discuss this for a more physical solution as in the previous section that
can be found for instance in \cite{miller1977symmetry} for the free
particle, which we modify by an additional phase 
\begin{equation}
\phi _{n}^{(0)}\left( x,t\right) =\frac{1}{\left( t^{2}+1\right) ^{1/4}}%
H_{n}(iz)\exp \left[ \left( 1+it\right) z^{2}+i\kappa _{n}(t)\right]
\end{equation}%
where $z:=x/\sqrt{2+2t^{2}}$ and $\kappa _{n}(t)=\left( n+\frac{1}{2}\right)
\arctan t-\int^{t}\left(\alpha (s)^{2}+E(s)\beta (s)\right)ds$. There exists a more
general solution in terms of parabolic cylinder functions with a continuous
parameter, but we consider here the specialized version that only involves
Hermite polynomials $H_{n}(x)$ as this leads to more interesting potentials
of rational type. Using $\phi _{n}^{(0)}$ allows us to compute the
corresponding intertwining operators $\ell _{n}^{(1)}$ and partner
potentials $v_{n}^{(1)}$. Evaluating the formulae in (\ref{v1}) we obtain 
\begin{eqnarray}
\ell _{n}^{(1)} &=&\ell _{1}\left( t\right) \left[ -\frac{i}{2}\left( \frac{x%
}{i+t}+\frac{2n\sqrt{2}H_{n-1}(iz)}{\sqrt{1+t^{2}}H_{n}(iz)}\right)
+\partial _{x}\right] , \\
v_{n}^{(1)} &=&\frac{4n}{1+t^{2}}\left[ \frac{%
	(n-1)H_{n-2}(iz)H_{n}(iz)-nH_{n-1}^{2}(iz)}{H_{n}^{2}(iz)}\right] +\alpha
^{2}+E\beta -\frac{1+it}{1+t^{2}}+i\frac{\left( \ell _{1}\right) _{t}}{\ell
	_{1}}.  \notag
\end{eqnarray}%
Since the combination of Hermite polynomials in $v_{n}^{(1)}$ is always
real, we notice that $\func{Im}[v_{1}^{(n)}]$ is only a function of $t$ and
can be eliminated by a suitable choice of $\ell _{1}$. The choice (\ref{h1})
yields $\ell _{1}=\sqrt{1+t^{2}}$ for all $n$ and the rational potentials in 
$x$ and $t$ 
\begin{eqnarray}
v_{0}^{(1)} &=&\alpha ^{2}+E\beta -\frac{1}{1+t^{2}}%
,~~v_{1}^{(1)}=v_{0}^{(1)}+\frac{2}{x^{2}},~~v_{2}^{(1)}=v_{0}^{(1)}-\frac{%
	4\left( 1+t^{2}-x^{2}\right) }{\left( 1+t^{2}+x^{2}\right) ^{2}}  \notag \\
v_{3}^{(1)} &=&v_{0}^{(1)}+\frac{6\left[ 3\left( 1+t^{2}\right) ^{2}+x^{4}%
	\right] }{x^{2}\left( 3+3t^{2}+x^{2}\right) ^{2}},~~ \\
v_{4}^{(1)} &=&v_{0}^{(1)}+\frac{8\left[ 3\left( 1+t^{2}\right)
	x^{4}+9\left( 1+t^{2}\right) ^{2}x^{2}-9\left( 1+t^{2}\right) ^{3}+x^{6}%
	\right] }{\left[ 6\left( 1+t^{2}\right) x^{2}+3\left( 1+t^{2}\right)
	^{2}+x^{4}\right] ^{2}},\ldots  \notag
\end{eqnarray}%

\begin{figure}
\includegraphics[width=5cm]{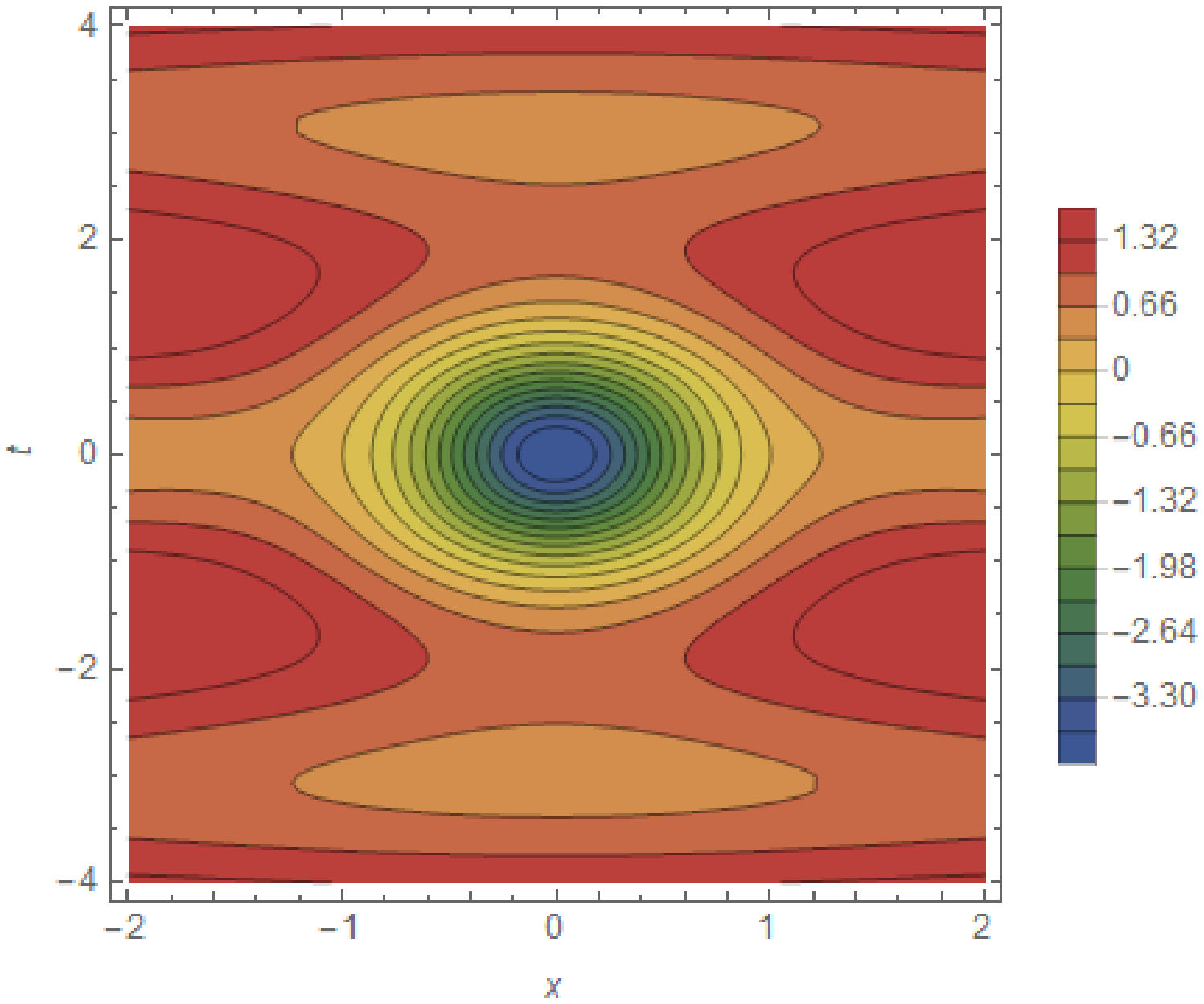}\includegraphics[width=5cm]{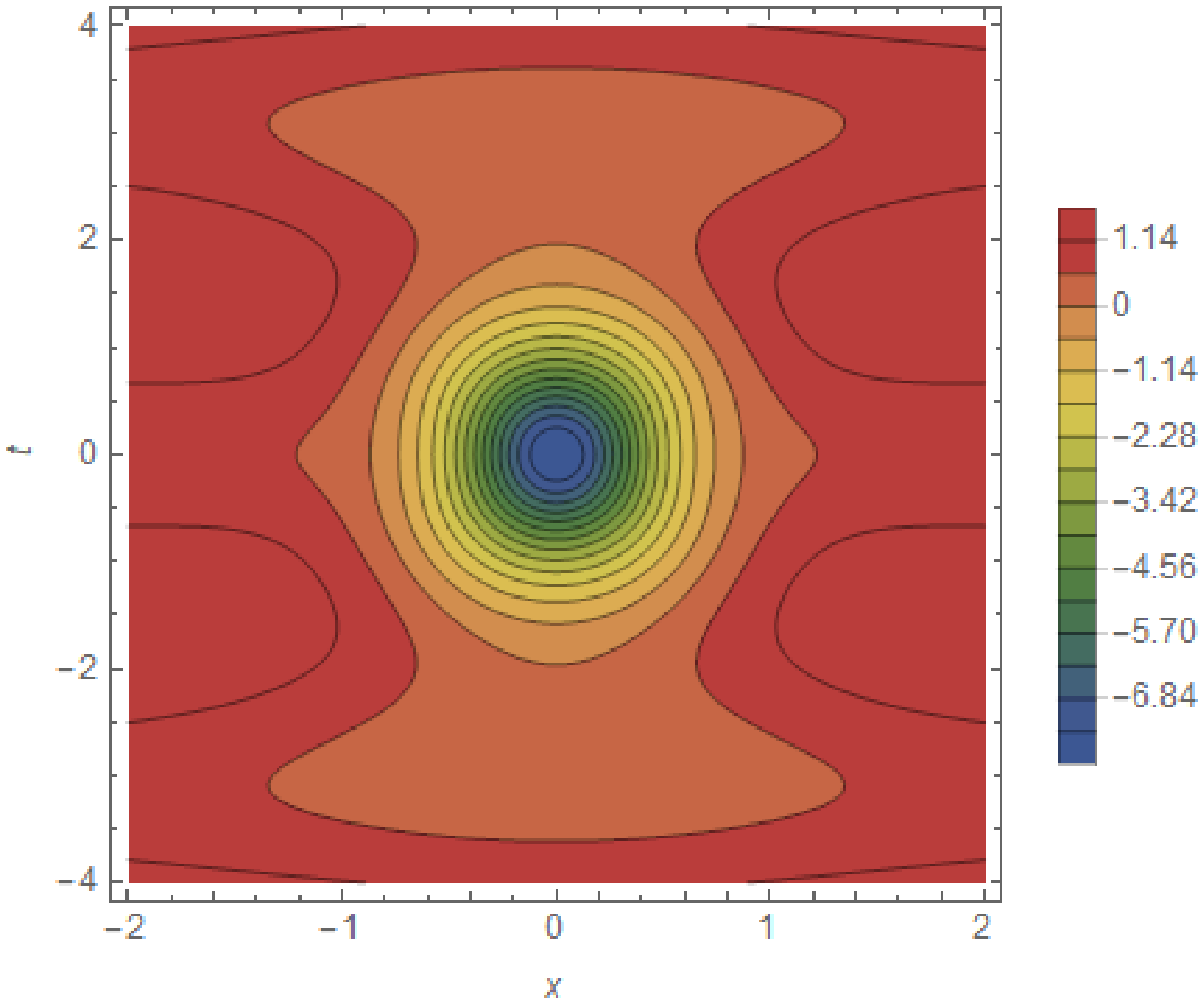}\includegraphics[width=5cm]{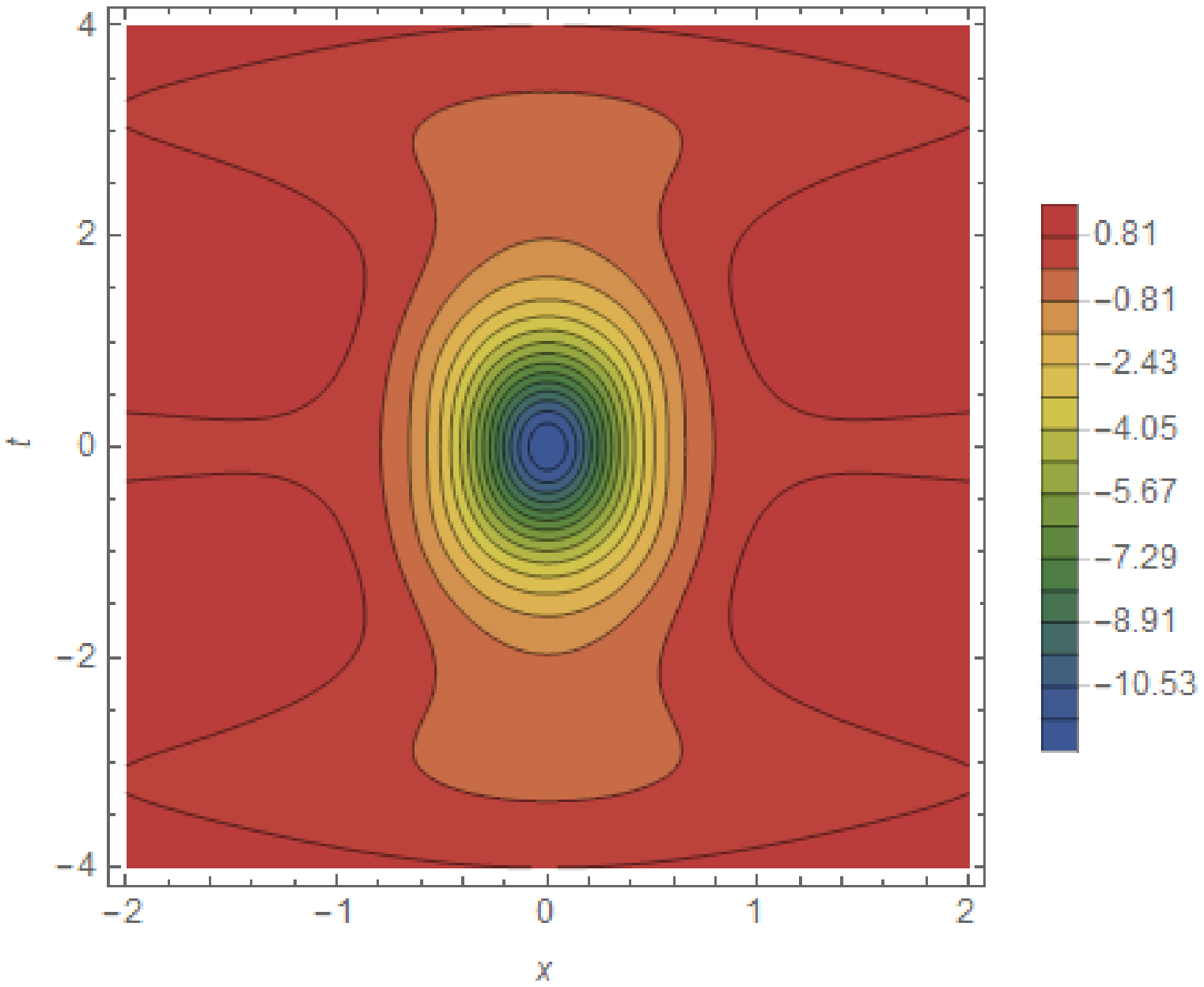} 
	\caption{ Time-dependent rational potentials $v\protect_{2}\protect^{(1)}(x,t)$, $v\protect_{4}\protect^{(1)}(x,t)$ and $v\protect_{6}\protect^{(1)}(x,t)$ with $E(t)=\protect\sin t$.}
	\label{fig1}
\end{figure}

We observe that all potentials $v_{n}^{(1)}$ with $n$ odd are singular at $%
x=0$, whereas those with $n$ even are regular for all values of $x$ and $t$.
We depict some of these finite potentials in figure \ref{fig1}, noting that
they possess well defined minima and finite asymptotic behaviour. The
nontrivial solutions (\ref{second}) to the TDSE for the Hamiltonians
involving $v_{n}^{(1)}$ are%
\begin{eqnarray}
\phi _{n,m}^{(1)} &=&\mathcal{L}\left[ \phi _{n}^{(0)}\right] \left( \phi
_{m}^{(0)}\right) ,\qquad n\neq m \\
&=&i\sqrt{2}\frac{mH_{m-1}(iz)H_{n}(iz)-nH_{m}(iz)H_{n-1}(iz)}{\left(
	1+t^{2}\right) ^{1/4}H_{n}(iz)}e^{i\kappa _{m}(t)}
\end{eqnarray}%
and the nontrivial solutions obtained from (\ref{ntsol}) are 
\begin{eqnarray}
\hat{\phi}_{0}^{(1)} &=&e^{-z^{2}}\phi _{0}^{(0)},\quad \tilde{\phi}%
_{0}^{(1)}=\sqrt{2}F(z)\phi _{0}^{(0)},  \notag \\
\hat{\phi}_{1}^{(1)} &=&\frac{e^{-z^{2}}}{4z^{2}}\phi _{1}^{(0)},\quad 
\tilde{\phi}_{0}^{(1)}=\left[ x\sqrt{1+t^{2}}-\sqrt{2}(1+t^{2})F(z)\right]
\phi _{1}^{(0)}, \\
\hat{\phi}_{2}^{(1)} &=&\frac{(1+t^{2})^{2}e^{-z^{2}}}{4(1+t^{2}+x^{2})^{2}}%
\phi _{2}^{(0)},\quad \tilde{\phi}_{2}^{(1)}=\frac{\left[ x(x^{2}-t^{2}-1)%
	\sqrt{1+t^{2}}-2\sqrt{2}(1+t^{2})^{2}F(z)\right] }{(1+t^{2}+x^{2})^{2}}\phi
_{2}^{(0)},  \notag
\end{eqnarray}%
where $F(z)$ denotes the Dawson integral $F(z):=\exp
(-z^{2})\int\nolimits_{0}^{z}\exp (s^{2})ds$.

Finally we compute the non-Hermitian counterpart $H_{1}$ from the TDDE (\ref%
{Dysoneq}). Taking now $\eta _{1}$ to be of the same form as $\eta _{0}$ but
different time-dependent parameters we make the Ansatz 
\begin{equation}
\eta _{1}=e^{\gamma \left( t\right) x}e^{\delta \left( t\right) p}
\label{eta1}
\end{equation}%
and compute 
\begin{equation}
H_{1}(x,p,t)=h_{1}(x+i\delta ,p-i\gamma ,t)-i\dot{\gamma}x-i\dot{\delta}p+%
\dot{\gamma}\delta .  \label{Ham1}
\end{equation}%
Thus we obtain%
\begin{eqnarray}
H_{1,0} &=&p^{2}-2i\gamma p-\gamma ^{2}+\alpha ^{2}+E\beta -\frac{1}{1+t^{2}}%
-i\dot{\gamma}x+\dot{\gamma}\delta -i\dot{\delta}p \\
H_{1,1} &=&H_{1,0}+\frac{2}{(x+i\delta )^{2}},\quad H_{1,2}=H_{1,0}-\frac{4%
	\left[ 1+t^{2}-(x+i\delta )^{2}\right] }{\left[ 1+t^{2}+(x+i\delta )^{2}%
	\right] ^{2}}, \\
H_{1,3} &=&H_{1,0}+\frac{6\left[ 3\left( 1+t^{2}\right) ^{2}+(x+i\delta )^{4}%
	\right] }{(x+i\delta )^{2}\left[ 3+3t^{2}+(x+i\delta )^{2}\right] ^{2}}%
,\ldots
\end{eqnarray}%
By setting $\dot{\delta}=-2\gamma $ we may remove the linear term in $p$ and
convert the Hamiltonian into a potential one. We notice that the
singularities for $v_{1,n}$ with $n$ odd have been regularized in the
non-Hermitian setting for $\delta \neq 0$. The remaining factors lead to
further restrictions for $\delta $ when demanding regularity for the $%
H_{1,n} $. In this case we require in addition $|\delta |<1$ for $n=2$, $%
|\delta |<\sqrt{3}$ for $n=3$, $|\delta |>\sqrt{3+\sqrt{6}}$ for $n=4$%
,\ldots\ 

We verify that according to the commutative diagram (\ref{CD1}) the
intertwining operator relating $H_{0}$ and $H_{1}$ in (\ref{IH}) is indeed $%
L=\eta _{1}^{-1}\mathcal{L}\left[ u\right] \eta _{0}$. From this we can now
also compute the nontrivial solutions (\ref{ntP}) to the TDSE%
\begin{equation}
\psi _{1}(x,t)=e^{-\gamma x-i\gamma \delta }\hat{\phi}_{1}(x+i\delta
),\qquad \text{and\qquad }\tilde{\psi}_{1}(x,t)=e^{-\gamma x-i\gamma \delta }%
\tilde{\phi}_{1}(x+i\delta ).
\end{equation}%
Hence all of these systems are exactly solvable and the diagram (\ref{CD1})
does indeed close. The two hierarchies of solvable time-dependent rational
Hamiltonians are then directly computed from the expressions (\ref{DCa})-(%
\ref{H4}).

\subsection{Solvable time-dependent Airy function potentials, two
	intertwinings}

Finally we start again with a solution to the TDDE for $h_{0}$, $H_{0}$, $%
\eta _{0}$ and carry out the intertwining relations separately constructing $%
h_{1}$, $H_{1}$, but unlike as in section \ref{hyperbolic} we use the intertwining
operator $L=\eta _{1}^{-1}\mathcal{L}\left[ u\right] \eta _{0}$ involving an
arbitrary operator $\eta _{1}$, 
\begin{equation}
\begin{array}{ccccc}
& H_{0} & ~~\underrightarrow{\eta _{0}}~~ & h_{0} &  \\ 
L=\eta _{1}^{-1}\mathcal{L}\left[ u\right] \eta _{0} & \downarrow &  & 
\downarrow & \mathcal{L}\left[ u\right] \\ 
& H_{1} & \underrightarrow{?} & h_{1} & 
\end{array}
\label{CD2}
\end{equation}%
which, by the closure of the diagram, must be the Dyson map for the system $%
1 $.

We discuss this scenario for a somewhat less well known solution to the free
particle TDSE in terms of Airy wave packet solutions as found forty years ago by
Berry and Balazs \cite{berry1979}, see also \cite{gori1999general} for a
different approach. The interesting feature of these wave packets is that
they continually accelerate in a shape-preserving fashion despite the fact
that no force is acting on them. Only more recently such type of waves have
been realized experimentally in various forms, e.g. \cite%
{siviloglou2007o,siviloglou2007a,baumgartl2008,vettenburg2014light,patsyk2018}%
. As in the previous section we modify the standard solution by a phase so
that it solves the TDSE for $h_{GV}$ 
\begin{equation}
\phi _{0}^{X}\left( x,t\right) =\limfunc{Xi}\left( \gamma x-\gamma
^{4}t^{2}\right) \exp \left[ i\gamma ^{3}t\left( x-\frac{2\gamma ^{3}t^{2}}{3%
}\right) -i\int \alpha ^{2}+E\beta dt\right] .
\end{equation}%
Here $\limfunc{Xi}\left( z\right) $ denotes any of the two Airy functions $%
\limfunc{Ai}\left( z\right) $ or $\limfunc{Bi}\left( z\right) $ and $\gamma
\in \mathbb{C}$ is a free parameter. Using once more the relation in (\ref%
{v1}), we obtain the intertwining operators and new Hamiltonians 
\begin{eqnarray}
\ell ^{X} &=&\ell _{1}\left( t\right) \left[ -i\gamma t^{3}-\gamma \frac{%
	\limfunc{Xi}^{\prime }\left( \gamma x-\gamma ^{4}t^{2}\right) }{\limfunc{Xi}%
	\left( \gamma x-\gamma ^{4}t^{2}\right) }+\partial _{x}\right] , \\
h_{1}^{X} &=&2\gamma ^{3}(x-\gamma ^{3}t^{2})-2\gamma ^{2}\left[ \frac{%
	\limfunc{Xi}^{\prime }\left( \gamma x-\gamma ^{4}t^{2}\right) }{\limfunc{Xi}%
	\left( \gamma x-\gamma ^{4}t^{2}\right) }\right] ^{2}+\alpha ^{2}+E\beta +i%
\frac{\left( \ell _{1}\right) _{t}}{\ell _{1}},  \notag
\end{eqnarray}%
with $\limfunc{Xi}^{\prime }\left( z\right) $ denoting the derivative of the
Airy functions. Taking $\ell _{1}$ to be a constant and $\gamma \in \mathbb{R%
}$ these are indeed Hermitian Hamiltonians. We also note that $h_{1}^{X}$
becomes singular when $\gamma x-\gamma ^{4}t^{2}$ equals a zero of the Airy
functions on the negative real axis. In addition, $h_{1}^{B}$ becomes
singular when $\pi /3<\arg (\gamma x-\gamma ^{4}t^{2})<\pi /2$. The
nontrivial solutions according to (\ref{second}) are computed to%
\begin{equation}
\phi _{1}^{A/B}=\ell ^{A/B}(\phi _{0}^{B/A})=\pm \frac{\ell _{1}(t)\gamma
	\exp \left[ -\frac{1}{3}i\left( 2\gamma ^{6}t^{3}-3\gamma ^{3}tx+3\int^{t}%
	\left[ \alpha (s)^{2}+\beta (s)\kappa (s)\right] \,ds\right) \right] }{\pi 
	\limfunc{Ai}/\limfunc{Bi}\left( \gamma x-\gamma ^{4}t^{2}\right) }.
\end{equation}%
We have constructed these solutions from the two linearly independent
solutions to the original TDSE rather than from one particular solution with
different parameters $\gamma $, i.e.%
\begin{equation}
\phi _{1}^{A/B}(\gamma _{1},\gamma _{2})=\mathcal{L}[\phi _{0}^{A/B}(\gamma
_{1})](\phi _{0}^{A/B}(\gamma _{2}))
\end{equation}%
are also solutions. Additional solutions can also be obtained in a
straightforward manner from (\ref{ntsol}).

\begin{figure}
\includegraphics[width=7cm]{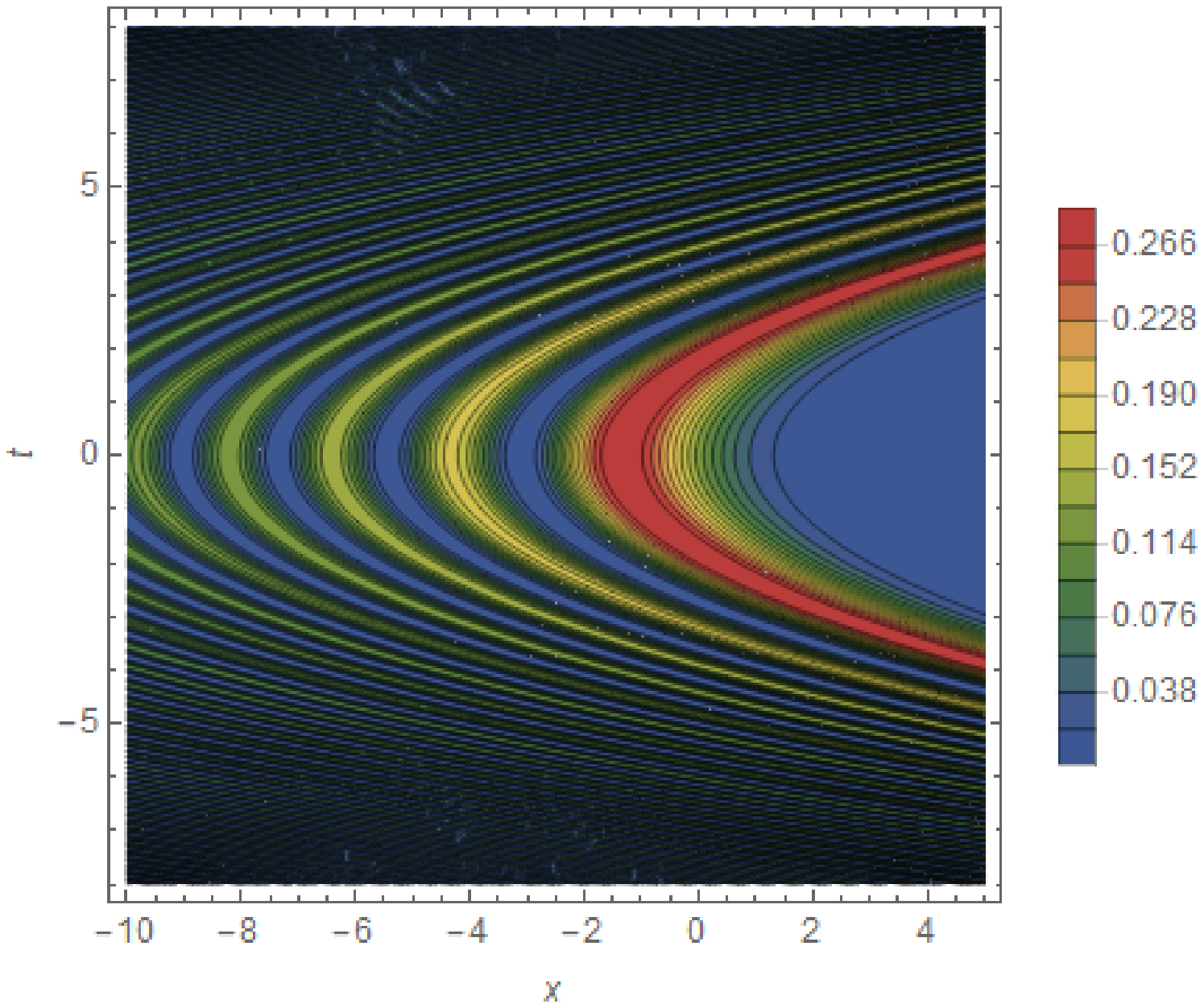} \includegraphics[width=7cm]{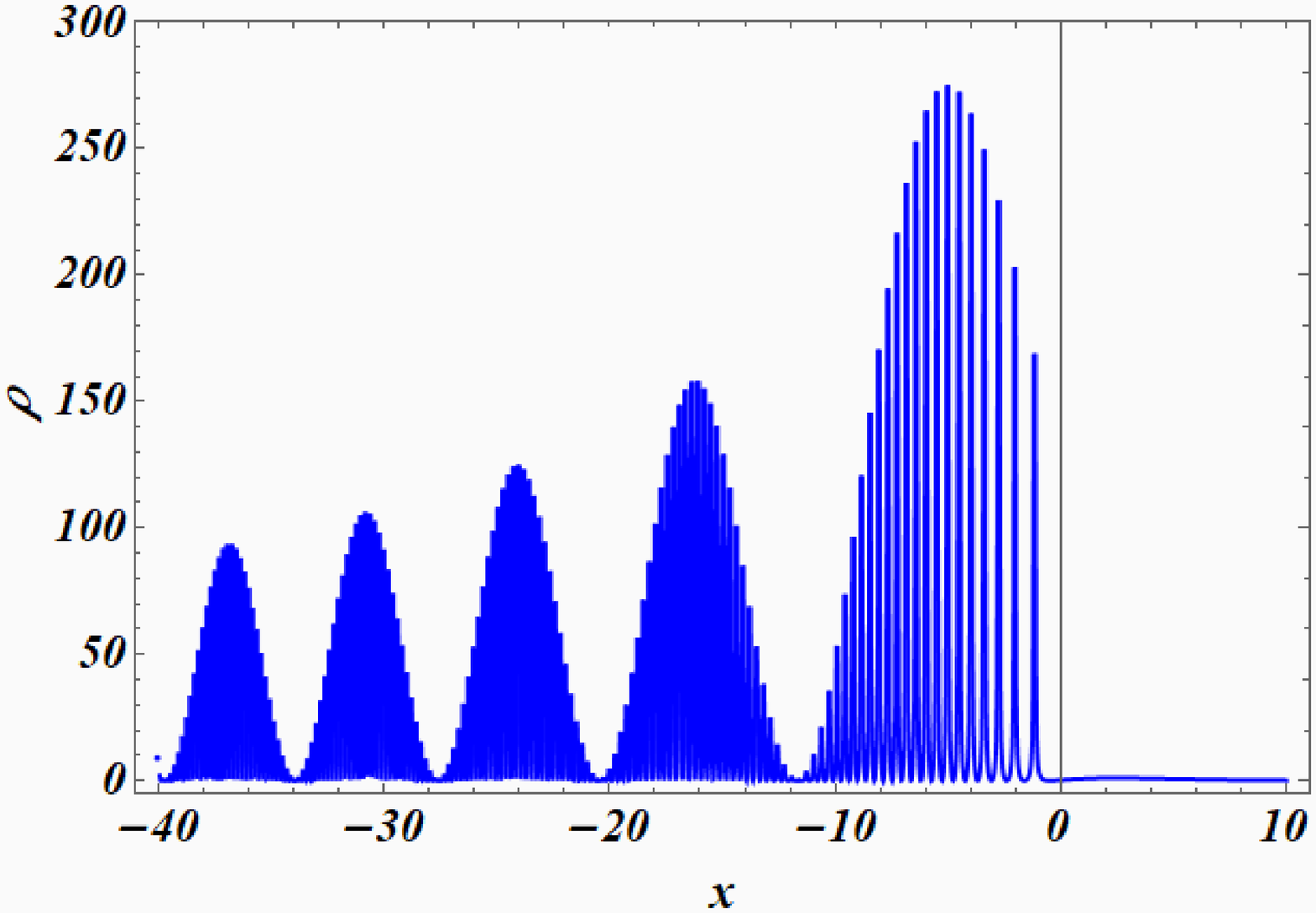}
	\caption{Probability densities for Airy wavepackets for solutions of the level 1 and 2 TDSE $\protect\rho\protect_0 =\protect\left\protect\vert \protect\phi
		\protect_{0}\protect^{A}(\protect\gamma =0.75)\protect\right\protect\vert \protect^{2}$ and $\protect\rho\protect_1 =\protect\left\protect\vert \protect\phi
		\protect_{1}\protect^{A}(t=1, \protect\gamma \protect_{1}=0.2,\protect\gamma \protect_{2}=2.0)\protect\right\protect\vert \protect^{2}$, left and right
		panel, respectively.}
	\label{fig3}
\end{figure}

For fixed values of time we observe in figure \ref{fig3} panel (a) the two
characteristic qualitatively different types of behaviour of the Airy wave
function, that is being oscillatory up to a certain point $x=x_{0}$ and
beyond which the density distribution becomes decaying. We observe further
that for increasing positive time, or decreasing negative time, the wave
packets accelerate. For the density wave function of the partner Hamiltonian
in panel (b) we observe this behaviour for one dominating value of $\gamma $
modulated by the other.

According to our commutative diagram (\ref{CD2}) we calculate next the
non-Hermitian counterpart $H_{1}^{X}$ using the intertwining operator $%
L=\eta _{1}^{-1}\mathcal{L}\left[ \phi _{0}^{X}\right] \eta _{0}$ with $\eta
_{1}$ as specified in (\ref{eta1}). We obtain 
\begin{equation}
H_{1}^{X}(x,p,t)=h_{1}^{X}(x+i\delta ,p-i\gamma ,t)-i\dot{\gamma}x-i\dot{%
	\delta}p+\dot{\gamma}\delta .
\end{equation}%
We verify the closure of the diagram by noting that $H_{1}^{X}$ satisfies
indeed the TDDE with $h_{1}^{X}$, $\eta _{1}$.

The above mentioned singularities on the real axis are now regularized.

\section{Reduced Swanson model hierarchy}\label{RS}

\begin{figure}
	\centering
\includegraphics[width=5cm]{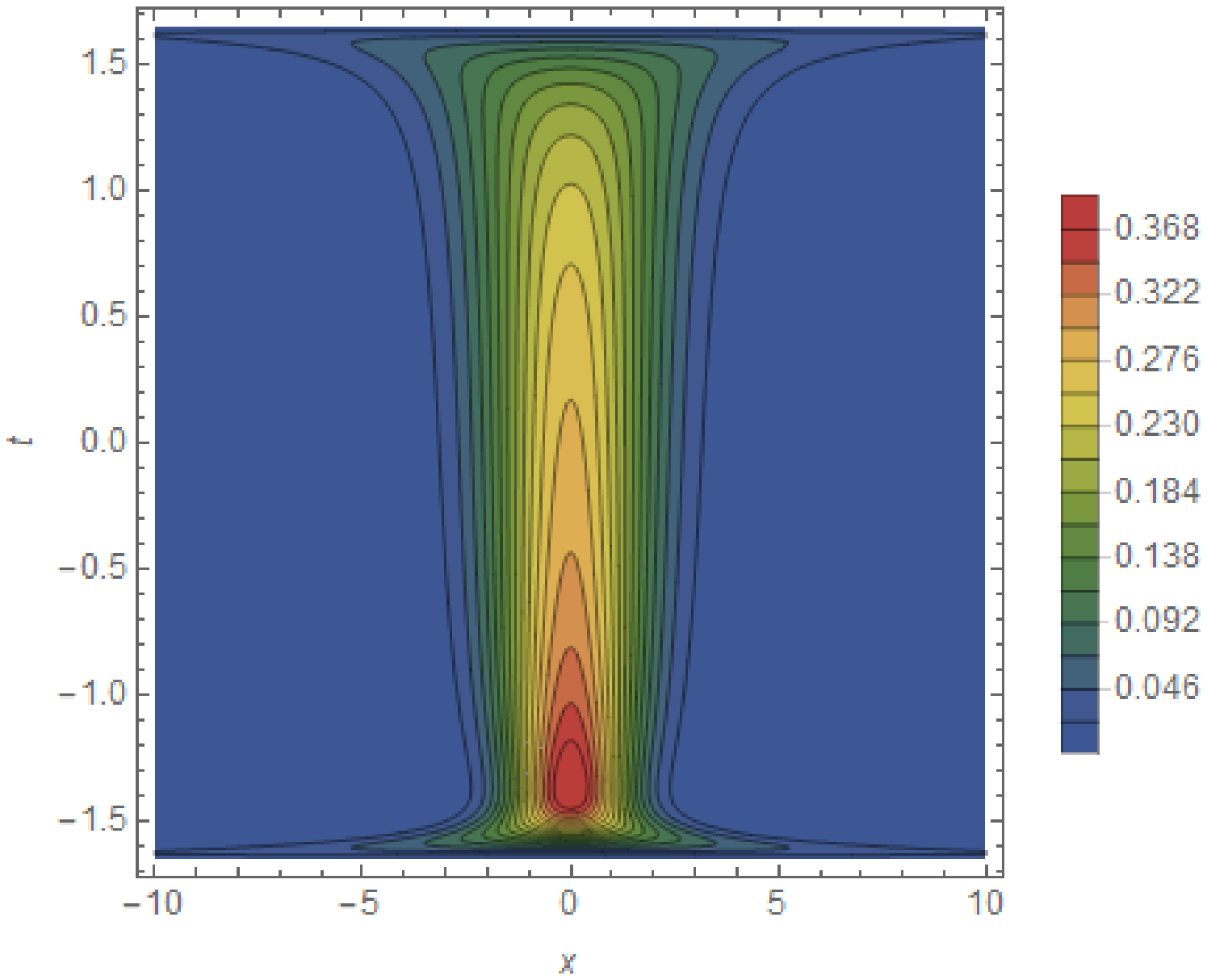}\includegraphics[width=5cm]{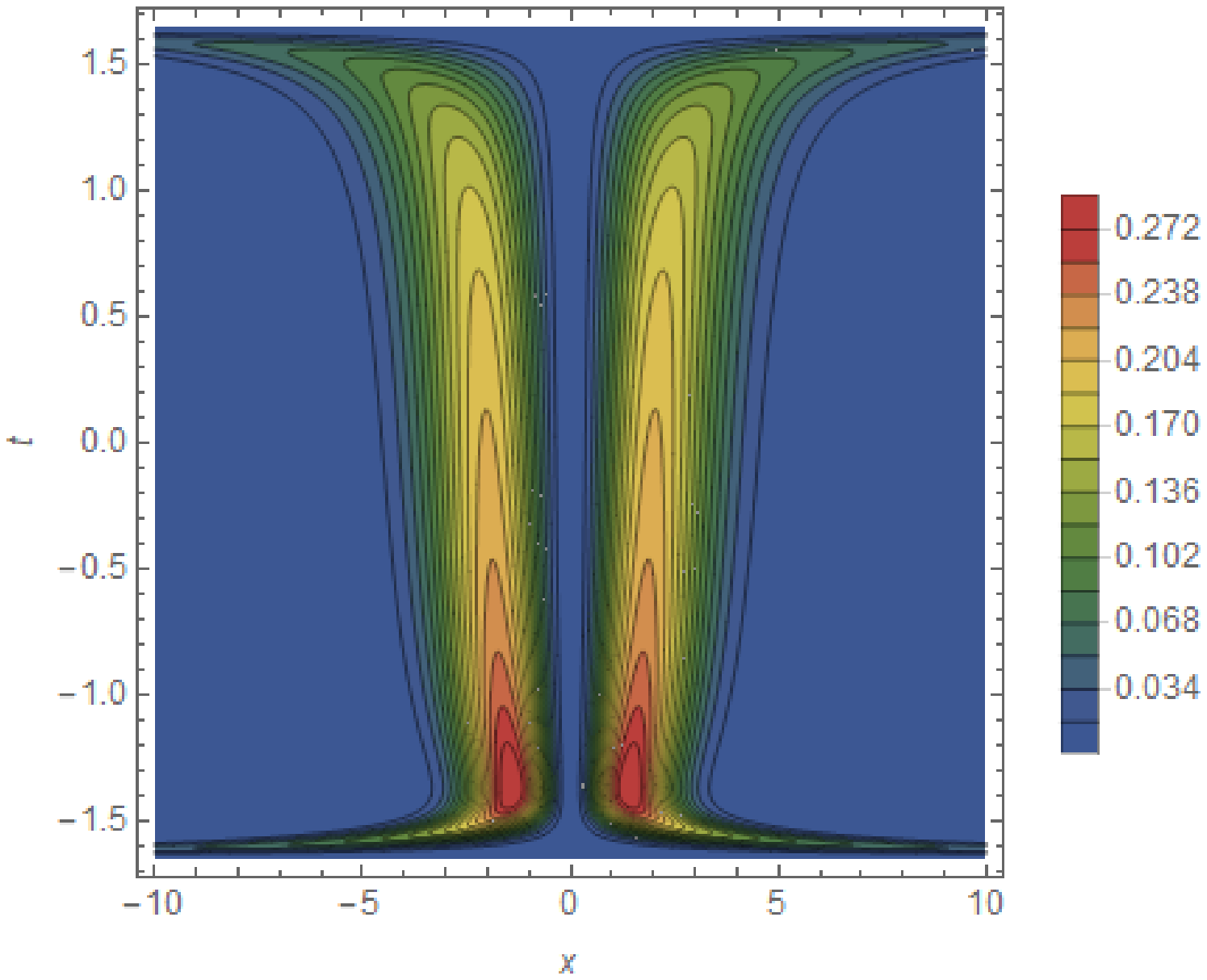}\includegraphics[width=5cm]{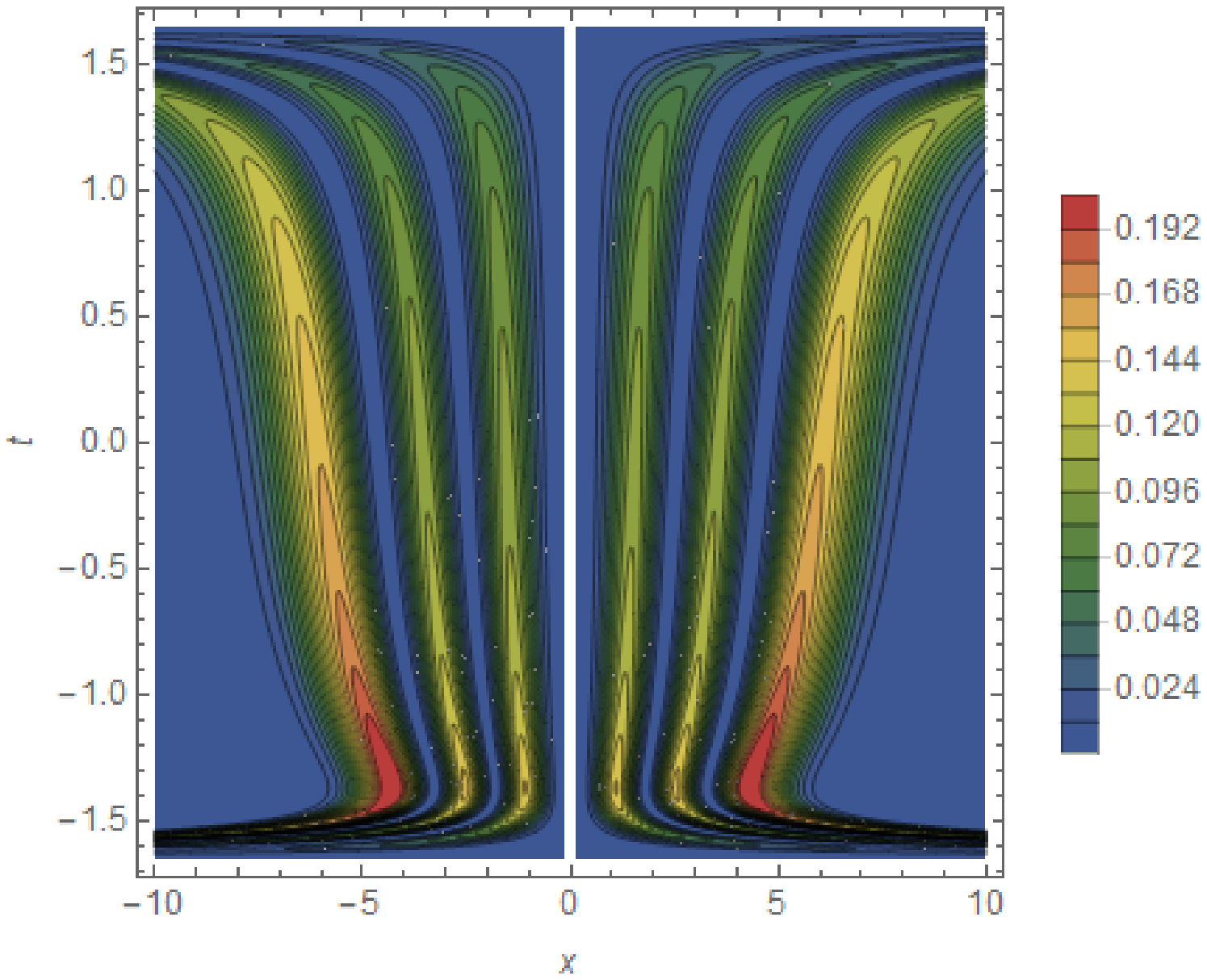} 
	\caption{Probability densities $\protect\left\protect\vert \protect\phi \protect_{0}\protect^{(0)}\protect\right\protect\vert \protect^{2}$, $\protect\left\protect\vert \protect\phi \protect_{1}\protect^{(0)}\protect\right\protect\vert \protect^{2}$, $\protect\left\protect\vert \protect\phi
		\protect_{1,7}\protect^{(1)}\protect\right\protect\vert \protect^{2}$ from left to right for $g(t)=(1+t\protect^{2})/4$, $m(t)=\left[ 1+\protect\cos (t+t\protect^{3}/3)\right] /(1+t\protect^{2})$, $\protect\varrho (t)=\protect\sqrt{1+[C+B\protect\tan (t/2+t\protect^{3}/6)]\protect^{2}}$ with $B=1/2$ and $C=1$. }
	\label{fig4}
\end{figure}

Next we consider a model that is build from a slightly more involved
time-dependent Dyson map. We proceed as outlined in the commutative diagram (%
\ref{CD1}). Our simple starting point is a non-Hermitian, but $\mathcal{PT}$%
-symmetric, Hamiltonian that may be viewed as reduced version of the
well-studied Swanson model \cite{swanson2004transition} 
\begin{equation}
H_{0}=H_{RS}=ig\left( t\right) xp.
\end{equation}%
We follow the same procedure as before and solve at first the TDDE for $\eta
_{0}$ and $h_{0}$ with given $H_{0}$. In this case the arguments in the
exponentials of the time-dependent Dyson map can no longer be linear and we
therefore make the Ansatz 
\begin{equation}
\eta _{0}=e^{\lambda \left( t\right) xp}e^{\zeta \left( t\right) p^{2}/2}.
\label{eta0}
\end{equation}%
The right hand side of the TDDE (\ref{Dysoneq}) is then computed to 
\begin{equation}
h_{0}=h_{RS}=\left[ \left( g\zeta +i\frac{\dot{\zeta}}{2}\right) \cos
(2\lambda )+\left( ig\zeta -\frac{\dot{\zeta}}{2}\right) \sin (2\lambda )%
\right] p^{2}+i(g+\dot{\lambda})xp.
\end{equation}%
Thus for $h_{0}$ to be Hermitian we have to impose 
\begin{equation}
\dot{\lambda}=-g,\quad \dot{\zeta}=-2g\zeta \tan 2\lambda ,  \label{const}
\end{equation}%
so that we obtain a free particle Hamiltonian with a time-dependent mass $%
m(t)$%
\begin{equation}
h_{0}=h_{RS}=\frac{1}{2m(t)}p^{2},\qquad \text{with }m(t)=\frac{1}{2g\zeta
	\sec (2\lambda )}.  \label{freep}
\end{equation}%
Time-dependent masses have been proposed as a possible mechanism to explain
anomalous nuclear reactions which cannot be explained by existing
conventional theories in nuclear physics, see e.g. \cite{masst}. The reality
constraints (\ref{const}) can be solved by 
\begin{equation}
\lambda (t)=-\int^{t}g\left( s\right) ds,\quad \text{and \quad }\zeta
(t)=c\sec \left( 2\int^{t}g\left( s\right) ds\right) ,\quad
\end{equation}%
with constant $c$. Thus the time-dependent mass $m(t)$ can be expressed
entirely in terms of the time-dependent coupling $g(t)$. An exact solution
to the TDSE for $h_{RS}$ can be found for instance in \cite{pedrosa1997exact}
when setting in there the time-dependent frequency to zero%
\begin{eqnarray}
\phi _{n}^{(0)}\left( x,t\right) &=&\!\!\!\!\frac{e^{i\alpha _{0,n}(t)}}{\pi ^{1/4}%
	\sqrt{n!2^{n}\varrho (t)}}\exp \left[ m(t)\left( i\frac{\dot{\varrho}(t)}{%
	\varrho (t)}-\frac{1}{m(t)\varrho ^{2}(t)}\right) \frac{x^{2}}{2}\right]
H_{n}\left[ \frac{x}{\varrho (t)}\right],  \label{Ped} \\
\alpha _{0,n}(t) &=&~\ -\int\nolimits_{0}^{t}\frac{\left( n+1/2\right) }{%
	m(s)\varrho ^{2}(s)}ds.  \label{al}
\end{eqnarray}%
For (\ref{Ped}) to be a solution, the auxiliary function $\varrho (t)$ needs
to obey the dissipative Ermakov-Pinney equation with vanishing linear term 
\begin{equation}
\ddot{\varrho}+\frac{\dot{m}}{m}\dot{\varrho}=\frac{1}{m^{2}\varrho ^{3}}.
\label{EP2}
\end{equation}%
We derive an explicit solution for this equation in \nameref{EPAppendix}. Evaluating
the formulae in (\ref{v1}), with $h_{0}$ and $h_{1}$ divided by $2m(t)$, we
obtain the intertwining operators and the partner Hamiltonians 
\begin{eqnarray}
\ell _{n}^{(1)} &=&\ell _{1}\left( t\right) \left[ \frac{x}{\varrho ^{2}}-%
\frac{2nH_{n-1}\left[ x/\varrho \right] }{\varrho H_{n}\left[ x/\varrho %
	\right] }-ixm\frac{\dot{\varrho}}{\varrho }+\partial _{x}\right] ,
\label{ln} \\
h_{1,n} &=&h_{0}+\frac{4n}{m\varrho ^{2}}\left[ \frac{nH_{n-1}^{2}\left[
	x/\varrho \right] -(n-1)H_{n-2}\left[ x/\varrho \right] H_{n}\left[
	x/\varrho \right] }{H_{n}^{2}\left[ x/\varrho \right] }\right] +\frac{1}{%
	m\varrho ^{2}}+i\left[ \frac{\dot{\ell}_{1}}{\ell _{1}}-\frac{\dot{\varrho}}{%
	\varrho }\right] ,~~~  \notag  \label{hn}
\end{eqnarray}%
respectively. As in the previous section, the imaginary part of the
Hamiltonian only depends on time and can be made to vanish with the suitable
choice of $\ell _{1}=\varrho $. For concrete values of $n$ we obtain for
instance the time-dependent Hermitian Hamiltonians 
\begin{eqnarray}
h_{1,0} &=&\frac{p^{2}}{2m}+\frac{1}{m\varrho ^{2}},\qquad h_{1,1}=h_{1,0}+%
\frac{1}{mx^{2}},\qquad h_{1,2}=h_{1,0}+\frac{4(\varrho ^{2}+2x^{2})}{%
	m(\varrho ^{2}-2x^{2})^{2}}, \\
h_{1,3} &=&h_{1,0}+\frac{3(3\varrho ^{4}+4x^{4})}{m(2x^{3}-3x\varrho
	^{2})^{2}},\; h_{1,4}^{(4)}=h_{1,0}+\frac{8\left( 9\varrho
	^{6}-12x^{4}\varrho ^{2}+18x^{2}\varrho ^{4}+8x^{6}\right) }{m\left(
	3\varrho ^{4}-12x^{2}\varrho ^{2}+4x^{4}\right) ^{2}}.~~~~~~
\end{eqnarray}%
Notice that all these Hamiltonians are singular at certain values of $x$ and 
$t$ as $\varrho $ is real. Solutions to the TDSE for the Hamiltonian $%
h_{1,n} $ can be computed according to (\ref{second})%
\begin{equation}
\phi _{n,k}^{(1)}\left( x,t\right) =\ell _{n}^{(1)}(\phi _{k}^{(0)})=\frac{%
	2^{3/2}}{\sqrt{k-n}}\left[ \frac{kH_{k-1}\left[ x/\varrho \right] }{H_{k}%
	\left[ x/\varrho \right] }-\frac{nH_{n-1}\left[ x/\varrho \right] }{H_{n}%
	\left[ x/\varrho \right] }\right] \phi _{k}^{(0)},~~\ n\neq k.
\end{equation}%
Both $\phi _{n}^{(0)}$ and $\phi _{n,k}^{(1)}$ are square integrable
functions with $L^{2}(\mathbb{R})$-norm equal to $1$. In figure \ref{fig4}
we present the computation for some typical probability densities obtained
from these functions. Notice that demanding $m(t)>0$ we need to impose some
restrictions for certain choices of $g(t)$.

Next we compute the non-Hermitian counterpart $H_{1\text{ }}$with a concrete
choice for the second Dyson map. Taking $\eta _{1}$ for instance to be of
the same form as in (\ref{eta1}) the non-Hermitian Hamiltonian is formally
the same as in equation (\ref{Ham1}). In our concrete case we obtain for
instance 
\begin{equation}
H_{1,1}=\frac{p^{2}}{2m}+\frac{1}{m(x+i\delta )^{2}}-i\dot{\gamma}x+\frac{1}{%
	m\varrho ^{2}}-\frac{\gamma ^{2}}{2m}+\dot{\gamma}\delta ,
\end{equation}%
where we have also imposed the constraint $\dot{\delta}=-\gamma /m$ to
eliminate a linear term in $p$, hence making the Hamiltonian a potential
one. The solutions for the TDSEs for $H_{0}$ and $H_{1,n}$ are%
\begin{equation}
\psi _{n}^{(0)}=\eta _{0}^{-1}\phi _{n}^{(0)},\quad \text{and\quad }\psi
_{n,k}^{(1)}=\eta _{1}^{-1}\phi _{n,k}^{(1)},
\end{equation}%
respectively.

\subsection{Lewis-Riesenfeld invariants}

Having solved the TDDE for $\eta _{0}$ and $\eta _{1}$ we can now also
verify the various intertwining relations for the Lewis-Riesenfeld
invariants as derived in section \ref{ILR}. We proceed here as depicted in the
following commutative diagram 
\begin{equation}
\begin{array}{ccccc}
& I_{0}^{h} & ~~\underrightarrow{\eta _{0}}~~ & I_{0}^{H} &  \\ 
\mathcal{L}\left[ \check{\phi}\right] & \downarrow &  & \downarrow & ? \\ 
& I_{1}^{h} & \underrightarrow{\eta _{1}} & I_{1}^{H} & 
\end{array}
\label{di2}
\end{equation}%
See also the more general schematic representation in figure \ref{fig0}. We
start with the Hermitian invariant $I_{0}^{h}$ from which we compute the
non-Hermitian invariant $I_{0}^{H}$ using the Dyson map $\eta _{0}$ as
specified in (\ref{eta0}). Subsequently we use the intertwining operator $%
\ell _{n}^{(1)}$ in (\ref{ln}) to compute the Hermitian invariants $%
I_{1,n}^{h}$ for the Hamiltonians $h_{1,n}$. The invariant $I_{1}^{H}$ is
then computed from the adjoint action of $\eta _{1}^{-1}$ as specified in (%
\ref{eta1}). Finally, the intertwining relation between the non-Hermitian
invariants $I_{0}^{H}$ and $I_{1}^{H}$ is just given by the closure of the
diagram in (\ref{di2}).

The invariant for the Hermitian Hamiltonian $h_{0}$ has been computed
previously in \cite{pedrosa1997exact}\footnote{%
	We corrected a small typo in there and changed the power $1/2$ on the $%
	x/\rho $-term into $2$.}%
\begin{equation}
I_{0}^{h}=A_{h}(t)p^{2}+B_{h}(t)x^{2}+C_{h}(t)\{x,p\},
\end{equation}%
where the time-dependent coefficients are%
\begin{equation}
A_{h}=\frac{\varrho ^{2}}{2},\quad B_{h}=\frac{1}{2}\left( \frac{1}{\varrho
	^{2}}+m^{2}\dot{\varrho}^{2}\right) ,\quad C_{h}=-\frac{1}{2}m\varrho \dot{%
	\varrho}.
\end{equation}%
It then follows from%
\begin{equation}
\left[ I_{0}^{h},h_{0}\right] =\frac{2i}{m}\left( C_{h}p^{2}+\frac{1}{2}%
B_{h}\{x,p\}\right) ,\quad \dot{A}_{h}=-\frac{2}{m}C_{h},\quad \dot{B}%
_{h}=0,\quad \dot{C}_{h}=-\frac{1}{m}B_{h},
\end{equation}%
that the defining relation (\ref{LRin}) for the invariant is satisfied by $%
I_{0}^{h}$. According to the relation (\ref{simhH}), the non-Hermitian
invariant $I_{0}^{H}$ for the non-Hermitian Hamiltonian $H_{0}$ is simply
computed by the adjoint action of $\eta _{0}^{-1}$ on $I_{0}^{h}$. Using the
expression (\ref{eta0}) we obtain%
\begin{equation}
I_{0}^{H}=\eta _{0}^{-1}I_{0}^{h}\eta
_{0}=A_{H}(t)p^{2}+B_{H}(t)x^{2}+C_{H}(t)\{x,p\},
\end{equation}%
with%
\begin{equation}
A_{H}=\frac{1}{2}e^{-2i\lambda }\rho ^{2}-\zeta ^{2}B_{H}-i\zeta m\rho \dot{%
	\rho},\quad B_{H}=\frac{e^{2i\lambda }\left( 1+m^{2}\rho ^{2}\dot{\rho}%
	^{2}\right) }{2\rho ^{2}},\quad C_{H}=i\zeta B_{H}-\frac{1}{2}m\rho \dot{\rho%
}.
\end{equation}%
We verify that $I_{0}^{H}$ is indeed an invariant for $H_{0}$ according to
the defining relation (\ref{LRin}), by computing%
\begin{equation}
\left[ I_{0}^{H},H_{0}\right] =2g\left( A_{H}p^{2}-B_{H}x^{2}\right) ,\quad 
\dot{A}_{H}=2igA_{H},\quad \dot{B}_{h}=-2igB_{H},\quad \dot{C}_{H}=0,
\end{equation}%
using the constraints (\ref{const}) and (\ref{EP2}).

Given the intertwining operators $\ell _{n}^{(1)}$ in (\ref{ln}) and the
invariant $I_{0}^{h}$, we can use the intertwining relation (\ref{II}) to
compute the invariants $I_{1,n}^{h}$ for the Hamiltonians $h_{1,n}$ in (\ref%
{hn}). Solving (\ref{II}) we find%
\begin{equation}
I_{1,n}^{h}=I_{0}^{h}+1+4n^{2}\frac{H_{n-1}^{2}\left[ x/\varrho \right] ^{2}%
}{H_{n}^{2}\left[ x/\varrho \right] ^{2}}-4n(n-1)\frac{H_{n-2}\left[
	x/\varrho \right] }{H_{n}^{2}\left[ x/\varrho \right] }.
\end{equation}%
We verify that this expression solves (\ref{LRin}). The last invariant in
our quadruple is%
\begin{equation}
I_{1,n}^{H}(x,p)=\eta _{1}^{-1}I_{1,n}^{h}(x,p)\eta
_{1}=I_{1,n}^{h}(x+i\delta ,p-i\gamma )
\end{equation}%
Finally we may also verify the eigenvalue equations for the four invariants.
Usually this is of course the first consideration as the whole purpose of
employing Lewis-Riesenfeld invariants is to reduce the TDSE to the much
easier to solve eigenvalue equations. Here this computation is simply a
consistency check. With%
\begin{eqnarray}
\check{\phi}_{n}^{(0)} &=&e^{-i\alpha _{0,n}}\phi _{n}^{(0)},\quad \check{%
	\phi}_{n,m}^{(1)}=e^{-i\alpha _{0,m}}\phi _{n,m}^{(1)},\quad \\
\check{\psi}_{n}^{(0)} &=&e^{-i\alpha _{0,n}}\psi _{n}^{(0)},\quad \check{%
	\psi}_{n,m}^{(1)}=e^{-i\alpha _{0,m}}\psi _{n,m}^{(1)},
\end{eqnarray}%
and $\alpha _{0,n}$ as specified in equation (\ref{al}) we compute 
\begin{eqnarray}
I_{0}^{h}\check{\phi}_{n}^{(0)} &=&\left( n+1/2\right) \check{\phi}%
_{n}^{(0)},~~~\ \ \ I_{1,n}^{h}\check{\phi}_{n,m}^{(1)}=\left( m+1/2\right) 
\check{\phi}_{n,m}^{(1)},~~ \\
I_{0}^{H}~\check{\psi}_{n}^{(0)} &=&\left( n+1/2\right) \check{\psi}%
_{n}^{(0)},~~~\ \ \ I_{1,n}^{H}\check{\psi}_{n,m}^{(1)}=\left( m+1/2\right) 
\check{\psi}_{n,m}^{(1)}.
\end{eqnarray}%
As expected all eigenvalues are time-independent.

\section{Summary}

We have generalized the scheme of time-dependent Darboux transformations to
allow for the treatment of non-Hermitian Hamiltonians that are $\mathcal{PT}$%
-symmetric/quasi-Hermitian. It was essential to employ intertwining
operators different from those used in the Hermitian scheme previously
proposed. We have demonstrated that the quadruple of Hamiltonians, two
Hermitian and two non-Hermitian ones, can be constructed in alternative
ways, either by solving two TDDEs and one intertwining relation or by
solving one TDDE and two intertwining relations. For a special class of
Dyson maps it is possible to independently carry out the intertwining
relations for the Hermitian and non-Hermitian sector, which, however, forced
the seed function used in the construction of the intertwining operator to
obey certain constraints. We extended the scheme to the construction of the
entire time-dependent Darboux-Crum hierarchies. We also showed that the
scheme is consistently adaptable to construct Lewis-Riesenfeld invariants by
means of intertwining relations. Here we verified this for a concrete system
by having already solved the TDSE, however, evidently it should also be
possible to solve the eigenvalue equations for the invariants first and
subsequently construct the solutions to the TDSE. As in the Hermitian case,
our scheme allows to treat time-dependent systems directly instead of having
to solve the time-independent system first and then introducing time by
other means. The latter is not possible in the context of the Schr\"{o}%
dinger equation, unlike as in the context of nonlinear differential
equations that admit soliton solutions, where a time-dependence can be
introduced by separate arguments, such as for instance using Galilean
invariance. Naturally it will be very interesting to apply our scheme to the
construction of multi-soliton solutions.

\chapter{Conclusion}

\section{Overview}

Over the course of this thesis we have demonstrated the validity and the utility of time-dependent non-Hermitian quantum mechanics. We have found the metric operator $\rho\left(t\right)$ and the Dyson map $\eta\left(t\right)$ for a large number of quantum systems and showed that they allow for a consistent description of time-dependent non-Hermitian quantum systems. Furthermore, we have used this time-dependent analysis to investigate systems with spontaneously broken $\mathcal{PT}$-symmetry. Ordinarily, the broken regime would be discarded as unphysical as the energy eigenvalues become complex and the time-evolution becomes non-unitary. However, when a time-dependence is introduced into the metric and the Dyson map, we are able to provide a consistent description of this broken regime. This is made possible by the introduction of a new observable energy operator $\tilde{H}\left(t\right)$. The Hamiltonian $H\left(t\right)$ becomes unobservable but still governs the time-evolution of the system.

We began by assessing the three approaches available for computing $\rho\left(t\right)$ and $\eta\left(t\right)$. While each has its advantages and drawbacks, there is no overall best approach for every circumstance. Each approach may be more applicable for certain problems and it requires some insight to make the choice. For example, when solving matrix models, the time-dependent Dyson equation seems the most appropriate and we avoid taking the cumbersome square root. However, we found that for a certain 2 dimensional coupled oscillator, the Lewis Riesenfeld invariants were the simplest as we did not have to solve a highly technical non-linear differential equation.

We then showed explicitly how the spontaneously broken $\mathcal{PT}$-symmetric regime could be mended using a time-dependent metric and Dyson map. In addition we showed that the energy operator behind this mending obeyed a new unbroken $\mathcal{PT}$-symmetry. The models we used to illustrate this point ranged from a two level matrix model to higher dimensional matrix models and finally an inverted harmonic oscillator system with infinite dimensional Hilbert space. Following on from this, we demonstrated the utility of the Lewis Riesenfeld invariants for solving coupled harmonic oscillator systems. 

With the establishment of a consistent framework for time-dependent systems, we applied our method to three important topics in mathematical physics: quasi-exactly solvable systems, entropy and Darboux transformations. In doing so, we obtained some new and exciting results. We solved an explicitly time-dependent quasi-exactly solvable system (the first type of solution of this kind). We then demonstrated how entropy in the spontaneously broken $\mathcal{PT}$ regime decays from a maximum to a non-zero value in a finite time. This is in contrast to the unbroken regime in which the entropy decays to zero rapidly with a later revival. Finally, we developed a general framework from performing Darboux and Darboux-Crum transformations on non-Hermitian Hamiltonian type systems. This has far reaching implications in the field of multi-solitons. 

\section{Further study}

As with all scientific endeavours, solving problems and answering questions often opens the door to an ever increasing array of questions and problems. The avenues of investigation multiply and expand each time we make progress. This is certainly the case for time-dependent non-Hermitian quantum systems. There are a number of theoretical points that have arisen from this body of work and a further list of potential research directions to apply the framework to.

The first point of intrigue is the ubiquity of the non linear Ermakov-Pinney equation in all of our analysis. It has arisen a startling number of times and suggests an underlying deeper structure to the central time-dependent Dyson equation. Interestingly we do not observe it if we choose to solve the time-dependent quasi-Hermiticity equation or the Lewis-Riesenfeld invariant relation. However the solutions to all of these methods are indeed solutions of the Ermakov-Pinney equation. Therefore there seems to be a strong link between this framework and the Ermakov-Pinney equation. What this link may be is a very interesting question and deserves being addressed.

The next question that has arisen is the condition for the existence of a metric for a non-Hermitian quantum system. There are certainly examples where the metric cannot be found exactly, such as for the $ix^3$ potential, although perturbative methods can be used \cite{mostafazadeh2006metric}. We can make some guesses as to the probability of its existence if we can find a set of generators that form a closed algebra and represent the Hamiltonian. However, this is not always the case and even with an algebraic representation we may not be able to solve the resulting equation. Therefore we can ask what the condition is on the non-Hermitian Hamiltonian that predicts the existence of the metric.

Following on from the current state of the art, there are many areas to which the framework can be applied. The first is entropy and quantum information. We made an initial stride into this areas, but there is significantly more research to be done. Specifically, the framework can be applied to an array of more technical models such as the Su-Schrieffer-Heeger and the Jaynes-Cumming model. In addition, we only considered the Von Neumann entropy measure and there are many other useful measure to be investigated in the non-Hermitian setting such as the joint entropy \cite{dunkel2005time}.

Optics is such a large part of the non-Hermitian community that applying the time-dependent framework to $\mathcal{PT}$-symmetric optical systems must be considered as one of the top priorities. For example, how does one deal with a time varying refractive index in the optical setting? Furthermore the use of $E_2$ algebraic systems are important in these systems and so our work on quasi-exactly solvable models will be useful in this context.

Next there is a large area of research to be filled following the establishment of time-dependent non-Hermitian Darboux transformations. The main application will be to non-Hermitian potentials corresponding to soliton solutions of non-linear partial differential equations such as the Korteweg-de Vries equation, the sine-Gordon equation and the non-linear Schr\"odinger equation.

Finally, work has already begun applying the knowledge obtained from non-Hermitian quantum mechanics to non-Hermitian quantum field theory \cite{mannheim2019goldstone,alexandre2018spontaneous,alexandre2019gauge,fring2019goldstone}. Is there an equivalent to the Dyson equation in this setting and if so is there an equivalent to time-dependent Dyson equation? Furthermore can broken $\mathcal{PT}$ or $\mathcal{CPT}$-symmetry be mended in a similar fashion?

This body of work has established concretely the framework for time-dependent non-Hermitian quantum systems and now allows for the scientific community to take it further. There is much to build upon and now is the time as the foundations are strong and will hold firm when understood and applied correctly.

\chapter{Appendix}
\section{Appendix A}\label{LRAppendix}

We give an introduction to the Lewis Riesenfeld Invariants used throughout this thesis. These invariants are used in time-dependent quantum mechanics in order to aid the solution of the time-dependent Schr\"odinger equation. Employing them reduces the difficulty of the problem by increasing the number of steps.\\

\noindent The dynamical Lewis Riesenfeld invariant $I\left(t\right)$ satisfies the equation

\begin{equation}\label{InvariantAppendix}
\frac{dI\left(t\right)}{dt}=\partial_t I\left(t\right)-i\hbar\left[I\left(t\right),H\left(t\right)\right]=0,
\end{equation}
where $H\left(t\right)$ is the Hamiltonian of a quantum system, satisfying the time-dependent Schr\"odinger equation

\begin{equation}\label{TDSEAppendix}
i\hbar\partial_t\ket{\psi\left(t\right)}=H\left(t\right)\ket{\psi\left(t\right)}.
\end{equation}
Through this construction, the eigenvalues of $I\left(t\right)$ are time-independent.

\begin{equation}\label{InvariantEquation}
I\left(t\right)\ket{\phi_n\left(t\right)}=\Lambda_n\ket{\phi_n\left(t\right)}.
\end{equation}
Therefore, solving (\ref{InvariantEquation}) for the eigenstates $\ket{\phi_n\left(t\right)}$ is a much simpler problem then solving equation (\ref{TDSEAppendix}) for the wave function given an intial condition. Furthermore, the eigenstates $\ket{\phi_n\left(t\right)}$ allow us to construct a general wave function as a superposition of dynamical modes

\begin{equation}
\ket{\psi\left(t\right)}=\sum_nc_n\ket{\psi_n\left(t\right)},
\end{equation}
where the dynamical modes $\ket{\psi_n\left(t\right)}$ are related to the eigenstates $\ket{\phi_n\left(t\right)}$ via a time-dependent phase

\begin{equation}
\ket{\psi_n\left(t\right)}=e^{i\hbar\alpha_n\left(t\right)}\ket{\phi_n\left(t\right)}.
\end{equation}
Substituting this expression into (\ref{TDSEAppendix}) gives the expression for $\alpha_n\left(t\right)$

\begin{equation}
\dot{\alpha}_n=\frac{1}{\hbar}\bra{\phi_n\left(t\right)}i\hbar\partial_t-H\left(t\right)\ket{\phi_n\left(t\right)}.
\end{equation}
Therefore, with the calculation of $\alpha_n\left(t\right)$ we can construct general solutions to the time-dependent Schr\"odinger equation (\ref{TDSEAppendix}).\\

In this thesis, these invariants have an additional utility. In the time-dependent setting, the invariants for a Hermitian systems $I_h\left(t\right)$ and a non-Hermitian system $I_H\left(t\right)$ are related by a similarity transform

\begin{equation}
I_h\left(t\right)=\eta\left(t\right)I_H\left(t\right)\eta^{-1}\left(t\right),
\end{equation}
where $\eta\left(t\right)$ is the time-dependent Dyson map. This can be derived by substituting the time-dependent Dyson equation into equation (\ref{InvariantAppendix}) The relation differs from the time-dependent Dyson equation and the time-dependent quasi-Hermiticity equation as there is no time derivative term. This simplifies the approach as we only deal with simultaneous equations rather than coupled differential equations.

\newpage

\section{Appendix B}\label{EPAppendix}

We briefly explain how to solve the Ermakov-Pinney equation with dissipative
term (\ref{EP2})%
\begin{equation}
\ddot{\varrho}+\frac{\dot{m}}{m}\dot{\varrho}=\frac{1}{m^{2}\varrho ^{3}}.
\label{A1}
\end{equation}%
The solutions to the standard version of the equation \cite{Ermakov,Pinney} 
\begin{equation}
\ddot{\sigma}+\lambda (t)\sigma =\frac{1}{\sigma ^{3}}  \label{EPtime}
\end{equation}%
are well known to be of the form \cite{Pinney}%
\begin{equation}
\sigma (t)=\left( Au^{2}+Bv^{2}+2Cuv\right) ^{1/2},  \label{EPsol}
\end{equation}%
with $u(t)$ and $v(t)$ denoting the two fundamental solutions to the
equation $\ddot{\sigma}+\lambda (t)\sigma =0$ and $A$, $B$, $C$ are
constants constrained as $C^{2}=AB-W^{-2}$ with Wronskian $W=u\dot{v}-v\dot{u%
}$. The solutions to the equation with an added dissipative term
proportional to $\dot{\sigma}$ are not known in general. However, the
equation of interest here, (\ref{A1}), which has the linear term removed may
be solved exactly. For this purpose we assume $\varrho (t)$ to be of the form%
\begin{equation}
\varrho (t)=f[q(t)],\qquad \ \ \ \text{with }q(t)=\int\nolimits^{t}\frac{1}{%
	m(s)}ds.  \label{fq}
\end{equation}%
Using this, equation (\ref{A1}) transforms into%
\begin{equation}
\frac{d^{2}f}{dq^{2}}=\frac{1}{f^{3}},
\end{equation}%
which corresponds to (\ref{EPtime}) with $\lambda (t)=0$. Taking the linear
independent solutions to that equation to be $u(t)=1$ and $v(t)=q$, we obtain%
\begin{equation}
f(q)=\frac{\pm 1}{\sqrt{B}}\sqrt{1+(Bq+C)^{2}}
\end{equation}%
and hence with (\ref{fq}) a solution to (\ref{A1}).

%\addcontentsline{toc}{section}{References}

\bibliographystyle{frithstyle}
\bibliography{bibliographyThesis}

\end{document}